\documentclass[manuscript,revised]{geophysics}
\usepackage{graphicx}
\usepackage{subfigure}
\usepackage{amsmath}

\renewcommand{\figdir}{Fig} 

\begin{document}

\title{Target-oriented elastic full-waveform inversion through extended image-space redatuming}

\renewcommand{\thefootnote}{\fnsymbol{footnote}} 

\address{
\footnotemark[1] Formerly Stanford University, \\
Geophysics Department, \\
397 Panama Mall Mitchell Building, 3rd Floor \\
Stanford, CA 94305 \\
Currently California Institute of Technology\\
Seismological Laboratory\\
1200 E California Blvd \\
Pasadena, CA 91125\\
\footnotemark[2] Stanford University, \\
Geophysics Department, \\
397 Panama Mall Mitchell Building, 3rd Floor \\
Stanford, CA 94305 \\}
\author{Ettore Biondi\footnotemark[1], Guillaume Barnier\footnotemark[2], Biondo Biondi\footnotemark[2], and Rober G. Clapp\footnotemark[2]}

\footer{Target-oriented elastic FWI}
\lefthead{Biondi et al.}
\righthead{Target-oriented elastic FWI}

\maketitle

\begin{abstract}
Elastic full-waveform inversion (FWI) when successfully applied can provide accurate and high-resolution subsurface parameters. However, its high computational cost prevents the application of this method to large-scale field-data scenarios. To mitigate this limitation, we propose a target-oriented elastic FWI methodology based on a redatuming step that relies upon an extended least-squared migration process. In our approach, the surface-reflection data can be attributed to a given subsurface portion when mapped into the image space. This process allows us to reconstruct reflection data generated by a target area and recorded with a virtual acquisition geometry positioned directly above it. The redatuming step enables the application of an elastic FWI method within the target portion only. The entire workflow drastically diminishes the overall cost of the surface-data inversion and allows the retrieval of accurate elastic parameters of the area of interest. We demonstrate the effectiveness of our approach on a synthetic case based on the well-known Marmousi2 model and on a 3D ocean-bottom-node (OBN) pressure data recorded in the Gulf of Mexico. We first discuss the fundamental aspects of the methodology and apply the proposed workflow to the synthetic test case. We also employ the methodology on the field-data scenario and show its efficacy at correctly retrieving the elastic parameters and rock-physical properties of a gas-bearing sand reservoir positioned in proximity of a salt-dome flank. 
\end{abstract}

\section{Introduction}
Since its first definition by~\cite{Tarantola-FWI}, FWI has become a fundamental process in velocity-model-building workflows~\cite[]{virieux2009overview}. The study described by~\cite{sirgue2010thematic} represents one of the earliest successful applications of FWI to 3D field data. In their work, 3D data recorded in the Valhall field in the North Sea are acoustically inverted, and their FWI workflow produces significant image and velocity-model improvements compared to the initial subsurface information. Since then, FWI has continued to be a highly-relevant research topic in the seismic exploration field~\cite[]{guitton2013introduction}. This growing interested is due to the ability of FWI processes to retrieve high-resolution and accurate subsurface parameters with minimal data processing.

The technological advancements in the computer processing speed in recent years is allowing the application of computational intensive inversion and imaging methods~\cite[]{brandsberg2017high}. In fact, FWI methodologies are not limited to single-parameter and acoustic wave-equation inversion workflows, and multi-parameter estimation strategies applications for seismic exploration have been proposed and discussed by multiple authors~\cite[]{operto2013guided}. Specifically, elastic FWI is now widely considered to being able to accurately retrieve the subsurface material properties from seismic data~\cite[]{sears2010elastic,vigh2014elastic,pan2018elastic,zhang2020high,chen2022salt}. However, due to its high computational cost associated with the elastic wavefield modeling, elastic FWI approaches are usually applied to small 3D or only 2D field data.

To mitigate the intrinsic cost of elastic FWI and focus the inverse problem to only a limited portion of the subsurface, multiple target-oriented strategies have been proposed. These methods can be separated into two major categories. The first one is represented by approaches in which a redatuming step is performed before an elastic inversion processed is applied to data acquired with a virtual acquisition geometry positioned directly above any target area~\cite[]{wapenaar2014elastic,ravasi2017rayleigh,guo2019target,da2019target,garg2020surface,li2022target}. The second set of methods comprises workflows based on local solver strategies. Within them, the target-area wavefield is computed by changing the boundary conditions of the inversion domain~\cite[]{malcolm2016rapid,masson2017box,10.1093/gji/ggx118,kumar2019enabling,HuangTarget2020}. Both categories have been successfully applied to different geological scenarios, but none of these applications have shown a 3D field application.

Our method falls within the first category in which an extended least-squares migrated image is employed to perform the redatuming step before the target-oriented inversion process. Many authors employed this image domain to invert for a subsurface migration velocity based on optimal focusing of the image when the correct velocity is used during migration \cite[]{symes1994inversion,yang2009wave,alkhalifah2014scattering,biondi2014simultaneous,barnier2022full}. In other applications, a form of extended-image space based on ray parameter is used to generate angle gathers that preserve the amplitude information of the migrated events \cite[]{kuehl2002robust,wang2005high}. The ability to maintain the amplitude behavior of primary reflected events allows for the analysis of the amplitude-versus-angle (AVA) information and its subsequent inversion to retrieve the elastic subsurface parameters \cite[]{schleicher19933,albertin2004true,gray2009true,biondi2022true}. 

In our method, the extended image generated using the surface data is used to simulate data acquired using a virtual acquisition geometry placed in proximity of a target area. The redatum dataset is then inverted within an elastic FWI process whose modeling domain comprises of only a selected subsurface area. We first describe the theory behind the entire workflow and use synthetic tests to highlights the fundamental aspects of our elastic inversion process. We test the method on synthetic data generated using the Marmousi2 model and estimate the elastic properties of one of the reservoirs present in the subsurface. To demonstrate the efficacy of our method on 3D field data, we apply it to the pressure component of an OBN dataset recorded in the Gulf of Mexico by Shell Exploration and Production company. In this application, we invert the elastic property of a subsurface prospect. The estimated rock-physical parameters agree with field observations showing the possible presence of a gas-bearing reservoir trapped by the salt-dome flank and a high-shale-content formation.

\section*{Theory}
Our workflow can be separated into two main steps; namely, the surface-data redatuming and the target-oriented inversion of the virtual-acquisition data. The redatuming is based on the usage of an extended linearized waveform inversion, which is also referred to as extended least-squares reverse time migration (RTM). The generated image is then used to synthesize seismic data recorded by a virtual acquisition geometry placed in the proximity of a target area. The redatumed data are then used within an elastic FWI workflow to retrieve the subsurface elastic parameters. We start this discussion by reviewing the theory behind extended imaging and least-squares migration and then describe the elastic FWI framework we use to estimate the elastic properties of the target portion.

\subsection{Extended imaging}
The formation of a subsurface-offset extended image $\tilde{m}$ is performed using a source and a receiver wavefield, $p_0$ and $q$, as follows:
\begin{align}\label{eqn:extended-imaging}
\tilde{m}(\mathbf{x},\mathbf{h}) = \int \ddot{p}_0(\mathbf{x}-\mathbf{h},t) q(\mathbf{x}+\mathbf{h},t)dt,
\end{align}
where $\mathbf{x}$ represents the spatial coordinates, $\mathbf{h}$ is the subsurface offset variables, $t$ is the time axis, and the $\ddot{}$ denotes the second-order time derivative of a function.
The source wavefield is computed using an acoustic isotropic medium and thus it compliant with the following partial-differential equation (PDE):
\begin{align}\label{eqn:acoustic-fwd-wave-eq}
\left[\frac{1}{v^2(\mathbf{x})}\frac{\partial^2}{\partial t^2 } - \nabla^2\right]p_0(\mathbf{x},t) = s(\mathbf{x},t),
\end{align}
where $v$ represents the seismic P-wave speed or velocity and $s$ is the known seismic source. The receiver wavefield is obtained by solving the adjoint wave equation~\cite[]{fichtner2010full}:
\begin{align}\label{eqn:acoustic-adj-wave-eq}
\left[\frac{1}{v^2(\mathbf{x})}\frac{\partial^2}{\partial t^2 } - \nabla^2\right]^{*}q(\mathbf{x},t) = d(\mathbf{x},t),
\end{align}
where $^*$ denotes the adjoint sign and $d$ represents the recorded seismic data for a given shot. By discretizing the previous equations we can define the adjoint Born extended operator $\tilde{\mathbf{B}}^{*}$ as follows:
\begin{align}\label{eqn:ext-born-adj}
\tilde{\mathbf{m}}= \tilde{\mathbf{B}}(\mathbf{v})^{*} \mathbf{d},
\end{align}
where $\tilde{\mathbf{m}}$ represents the extended image, $\mathbf{d}$ is the data vector, and the extended Born operator depends non-linearly on the velocity vector $\mathbf{v}$. 
In the forward extended Born operator we employ the following scattering condition:
\begin{align}\label{eqn:extended-scattering}
s'(\mathbf{x},t) = \int \ddot{p}_0(\mathbf{x}-2\mathbf{h},t) \tilde{m}(\mathbf{x}-\mathbf{h},\mathbf{h})\mathbf{dh}.
\end{align}
The secondary source $s'$ is then propagated by solving the following PDE:
\begin{align}\label{eqn:acoustic-fwd-scat-wave-eq}
\left[\frac{1}{v^2(\mathbf{x})}\frac{\partial^2}{\partial t^2 } - \nabla^2\right]q'(\mathbf{x},t) = s'(\mathbf{x},t),
\end{align}
in which we can extract the Born model data from the scattered wavefield $q'$. The discretization of equations~\ref{eqn:acoustic-fwd-wave-eq},~\ref{eqn:extended-scattering}, and~\ref{eqn:acoustic-fwd-scat-wave-eq} is used to define the forward extended Born modeling operator as follows:
\begin{align}\label{eqn:ext-born-fwd}
\mathbf{d}'= \tilde{\mathbf{B}}(\mathbf{v})\tilde{\mathbf{m}},
\end{align}
where $\mathbf{d}'$ represents the scattered data vector. The extended Born forward and adjoint operators are employed within the extended linearized waveform inversion step.

\subsection{Extended linearized waveform inversion}
To form an optimal extended subsurface image, which is then used to synthesize data for a new acquisition, we minimize the following quadratic objective function:
\begin{eqnarray} \label{eqn:ext-lsrtm}
\phi(\tilde{\mathbf{m}})=\frac{1}{2}\left \|\tilde{\mathbf{B}}\tilde{\mathbf{m}} - \mathbf{d} \right \|_2^2,
\end{eqnarray}
in which we dropped the velocity dependency of the extended Born operator for brevity.  Given the dimensions of the extended Born operator and its computational cost when applied to a given vector, especially, for the 3D case, we solve the problem in equation~\ref{eqn:ext-lsrtm} using an iterative method, such as linear conjugate gradient (LCG)~\cite[]{Aster}. In certain scenarios, such as sparse acquisition geometry or complex overburden (e.g., subsalt targets), we instead solve the following regularized quadratic problem:
\begin{eqnarray} \label{eqn:ext-lsrtm-dso}
\phi(\tilde{\mathbf{m}})=\frac{1}{2}\left \|\tilde{\mathbf{B}}\tilde{\mathbf{m}} - \mathbf{d} \right \|_2^2 + \frac{\epsilon}{2} \left \|\mathbf{D}\tilde{\mathbf{m}} \right \|_2^2,
\end{eqnarray}
where $\mathbf{D}$ represents the regularization operator, and $\epsilon$ is a scalar weight associated with the regularization term. In this work, a differential semblance optimization (DSO) operator is employed within the regularization term \cite[]{symes1991velocity}. The effect of the added regularization term on the optimal image is to enhance its focusing, which in turn corresponds to an enhancement of  the coherency of the image across reflection angles~\cite[]{shen2005wave,biondi2021target}. Effectively, this process interpolates across reflection angles that have not been illuminated by the surface acquisition in the extended-image space~\cite[]{prucha2002subsalt}.

\subsection{Redatuming through extended least-squares migration}
The goal of any redatuming method is to transform the observed data acquired at a certain location (e.g., at the surface) into a new dataset as if they had been acquired at a different location in the subsurface \cite[]{wapenaar1992elastic,mulder2005rigorous}. Here, we seek to reconstruct the data generated from a target area that is recorded with sources and receivers placed directly above the target. This process enables the application of an FWI algorithm only within the target area. 

In our redatuming step, the optimal solution to the imaging problems of equations~\ref{eqn:ext-lsrtm} and~\ref{eqn:ext-lsrtm-dso}, $\tilde{\mathbf{m}}_{opt}$, is used to reconstruct the data $\mathbf{d}'$ corresponding to sources and receivers placed at a new subsurface acquisition level.  The reconstruction is performed by the following demigration process:
\begin{eqnarray} \label{eqn:demig-datum}
\mathbf{d}'=\tilde{\mathbf{B}}'\tilde{\mathbf{M}}\tilde{\mathbf{m}}_{opt},
\end{eqnarray}
where $\tilde{\mathbf{M}}$ is a restriction operator that limits the extended image to only the target area. The symbol $'$ denotes quantities related to the new acquisition geometry. The success of this reconstruction method depends on the knowledge of an accurate overburden, which is a common assumption within any redatuming technique. The advantage of this redatuming process compared to other methods resides in the usage of the image space to reconstruct the subsurface data. In fact, the regularization term employed during this step releases the common strict constraint of having dense source-receiver surface sampling.

To intuitively understand how this redatuming step works, let $d_{z_0}$ represent the subset of the observed data of interest (e.g., reflected events). Furthermore, we assume that $d_{z_0}$ is given by the following relation:
\begin{align}\label{eqn:surface-data-cont}
d_{z_0}(\mathbf{x}_r,\mathbf{x}_s,t) = \int g(\mathbf{x}_r, \mathbf{x},t) * g(\mathbf{x},\mathbf{x}_s,t)p(\mathbf{x})\mathbf{dx},
\end{align}
where $*$ denotes the time convolution, $p$ represents a scattering potential (also referred to as subsurface image), and $g$ is the Green's function for the acoustic wave equation (equation~\ref{eqn:acoustic-fwd-wave-eq}). The usage of wave-equation operators in this step allows us to take into account all the finite-frequency effects (e.g., Fresnel’s zone, scale sensitivity as a function of frequency). The usage of an extended image space should not be confused with a Born approximation since the extended Born approximation can be used to fit any kind of wave arrivals~\cite[]{barnier2022full1}. 

\begin{figure}[t]
    \centering
    \subfigure[]{\label{fig:redatuming1}\includegraphics[width=0.475\linewidth]{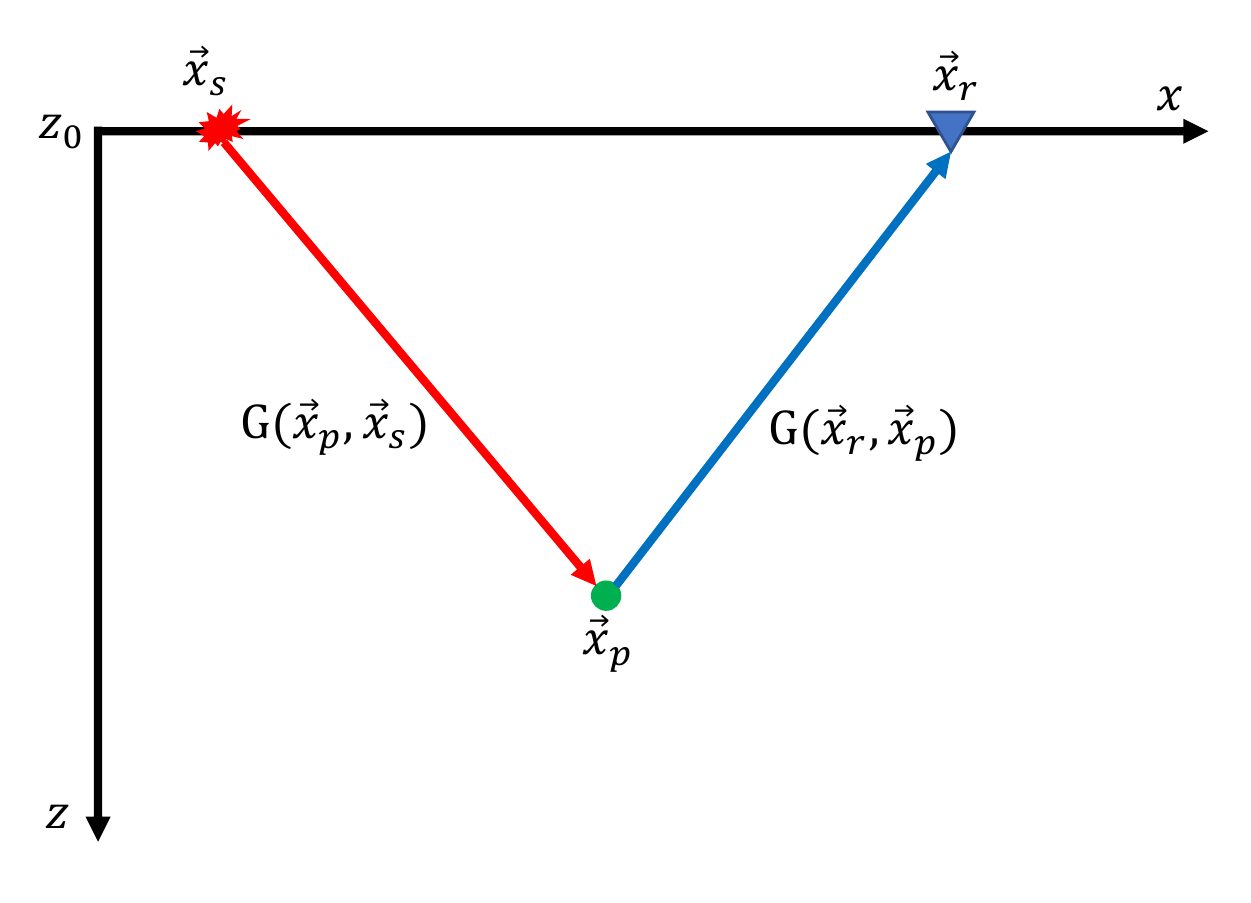}}
    \subfigure[]{\label{fig:redatuming2}\includegraphics[width=0.5\linewidth]{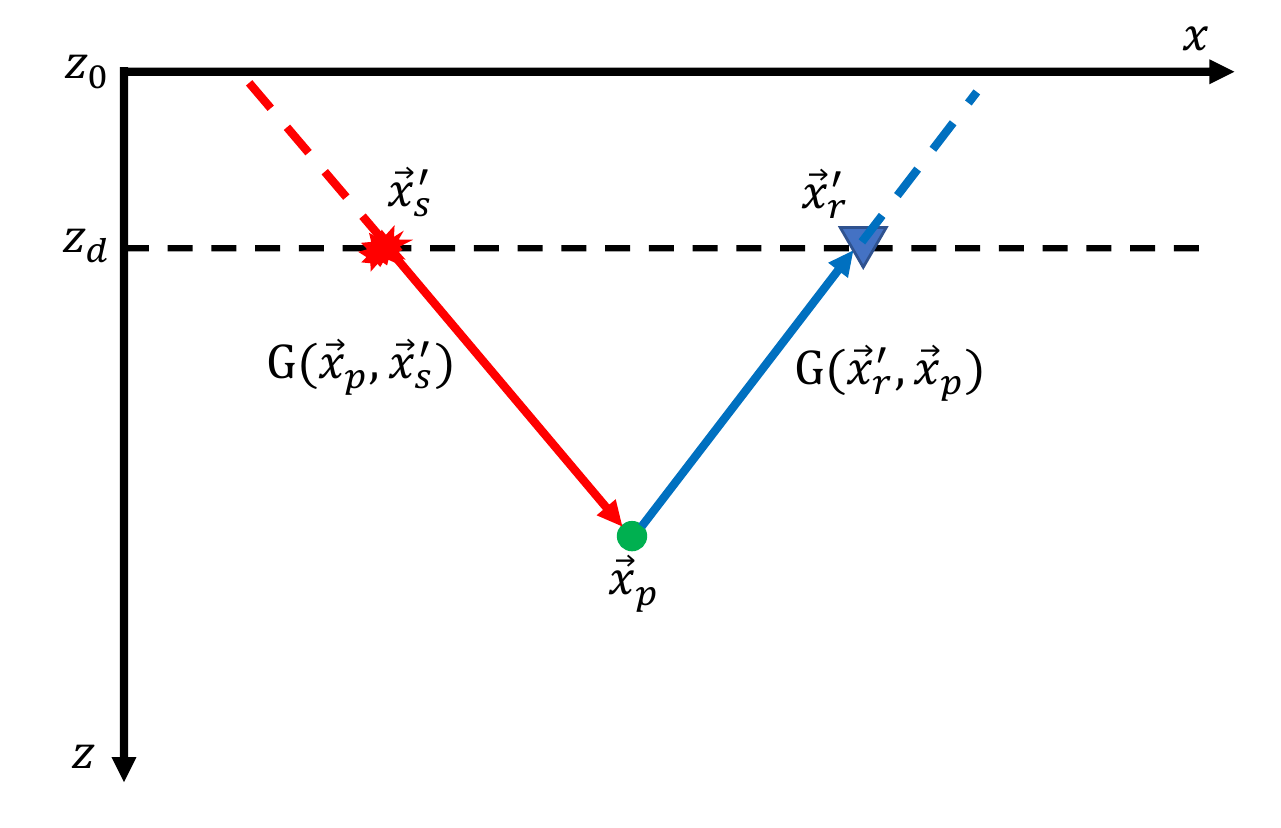}}
    \caption{(a) Schematic plot of equation~\ref{eqn:surface-data-cont}. The data observed at $\vec{x}_r$ are generated by the field propagating from $\vec{x}_s$ impinging on the scattering point at $\vec{x}_p$. (b) The same process can be used to generate the same data but with the source and receiver placed at $z_d$.}
    \label{fig:redatuming}
\end{figure}

Figure~\ref{fig:redatuming1} schematically illustrates the process of computing the data $d_{z_d}$ for a single point in the scattering potential and one source and receiver pair. The source wavefield $G(\vec{x}_p,\vec{x}_s)$ propagates from the source position $\vec{x}_s$ at the surface $z_0$ and it is then scattered by an image point placed at $\vec{x}_p$. This secondary source is then propagated by the receiver-side Green's function $G(\vec{x}_r,\vec{x}_p)$ and recorded from the device placed at $\vec{x}_r$. To obtain the same data but for the source-receiver pair placed at $z_d$, deeper than the surface level $z_0$, the same scattering potential can be used to generate the data for a source-receiver pair placed at $\vec{x}_s'$ and $\vec{x}_r'$ (Figure~\ref{fig:redatuming2}). Hence, the knowledge of the scattering potential $p$ enables the computation of the same event for a given source-receiver pair placed at two different depth levels. 

\begin{figure}[t]
    \centering
    \includegraphics[width=0.5\linewidth]{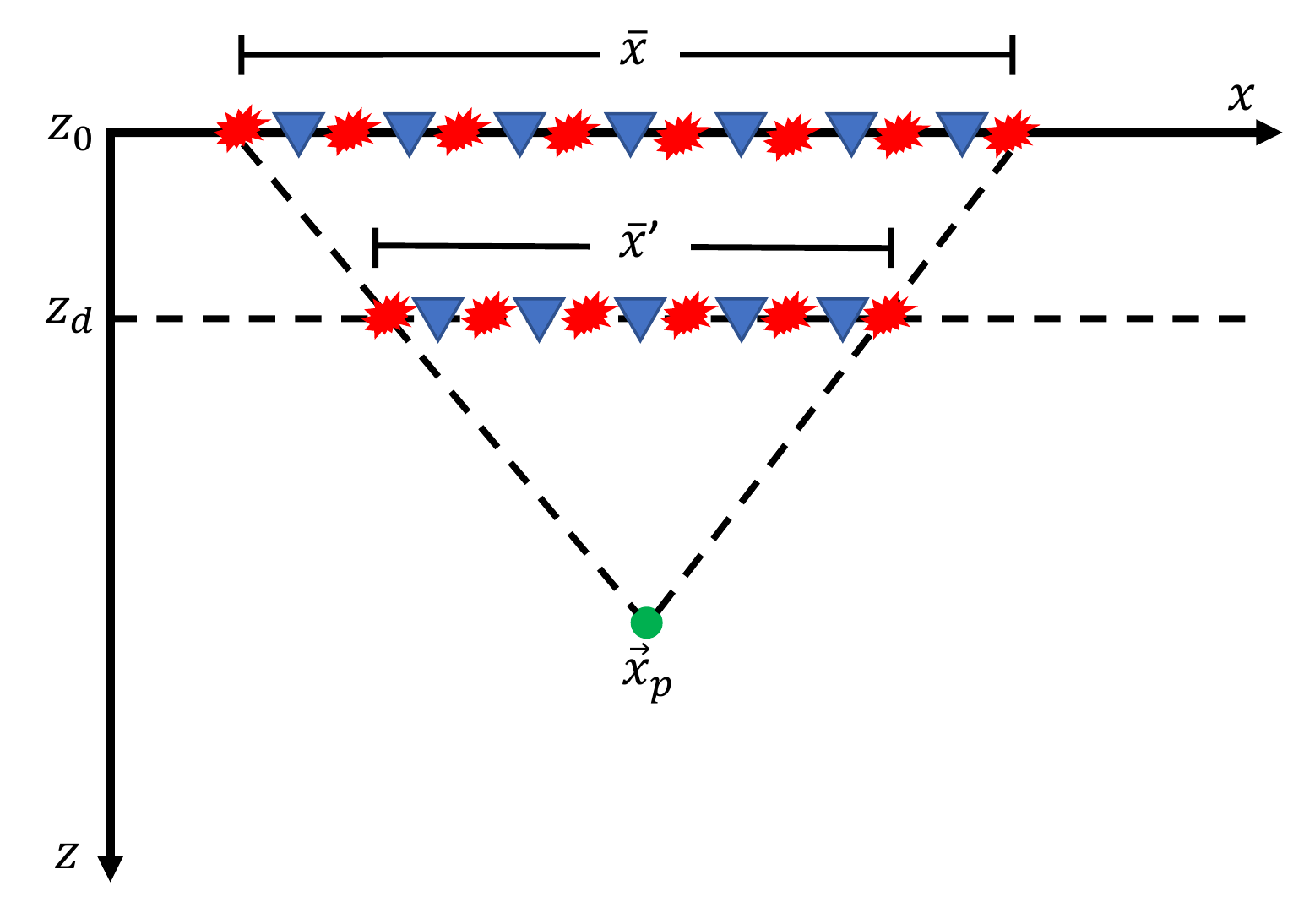}
    \caption{Schematic illustrating how the surface acquisition extent $\bar{x}$ at $z_0$ becomes $\bar{x}'$ when moved to $z_d$ assuming a constant velocity and identical illumination for the point $\vec{x}_p$.}
    \label{fig:redatumingGeom}
\end{figure}

Since this redatuming procedure is based on the formation of an image, the source-receiver distribution at the new datum depends on the maximum extent of the surface acquisition. Figure~\ref{fig:redatumingGeom} shows how the surface acquisition geometry extent changes when mapped to a deeper subsurface position $z_d$ assuming a constant velocity. The surface and datum acquisition extents $\bar{x}$ and $\bar{x}'$ identically illuminates the image point $\vec{x}_p$. Therefore, an image formed using the data acquired at $z_0$ with a source-receiver extent $\bar{x}$ can be employed to synthesize the data with an acquisition extent $\bar{x}'$ at $z_d$. The datumed acquisition geometry is reduced compared to the surface one. Thus, when generating the datumed dataset, a reduced source-receiver distribution must be used to avoid the introduction of data artifacts due to the limited illumination of the surface acquisition. Finally, the selection of the virtual geometry in complex geological scenarios can be performed using a ray-tracing algorithm for reflected events~\cite[]{vcerveny1987ray}.

In this simplified discussion, we assume that the recorded events are generated by a scattering or reflection process. Therefore, the virtual datumed geometry is kept horizontal and moved only along the vertical direction. However, as shown by~\cite{biondi2014simultaneous}, transmitted events, such as diving and head waves, can be reconstructed using an extended image. Thus, allowing for potentially positioning the virtual geometry around the target area. Despite this fact, this possibility is not explored in this work, which is focused on reflected events. We employ this method only to redatum single-component pressure data, but this process can be modified to compute the particle velocities associated with P-wave reflected events. In the shown examples, we employ an acoustic wave-equation operator during the image formation since transmission effects can be approximated by using an acoustic engine. When strong mode conversion or elastic transmission effects are present (e.g., basalt layers), then elastic wavefields should be considered during the extended migration step. This change would also allow a proper reconstruction of converted waves when an accurate overburden model is employed in this process. Finally, one could potentially employ the extended-image gathers to estimate the elastic properties of the reflectors using a Zoeppritz approximation~\cite[]{aki2002quantitative}. However, this approach would be hampered by the limitation of this approximation (e.g., locally flat interface, ray-based approximation), and thus would limit the application of an elastic inversion to simple geological scenarios.

\subsection{Elastic FWI}
Once the redatumed data have been reconstructed, they can be used within any elastic FWI framework. We assume the subsurface to be an elastic isotropic medium. Thus, to predict the observed data we employ the elastic isotropic wave equation in the velocity-stress form~\cite[]{virieux1986p}:
\begin{align}\label{eqn:elastic-wave-cont}
    \rho(\mathbf{x})\frac{\partial v_i(\mathbf{x},t)}{\partial t} &= \frac{\partial \sigma_{ik}(\mathbf{x},t)}{\partial x_k}+f_i(\mathbf{x},t),\\\nonumber
    \frac{\partial \sigma_{ij}(\mathbf{x},t)}{\partial t} &=\lambda(\mathbf{x})\frac{\partial v_k(\mathbf{x},t)}{\partial x_k}\delta_{ij}+\mu(\mathbf{x})\left[\frac{\partial v_i(\mathbf{x},t)}{\partial x_j}+\frac{\partial v_j(\mathbf{x},t)}{\partial x_i}\right]+M_{ij}(\mathbf{x},t),
\end{align}
where I employ the Einstein notation and $\delta_{ij}$ represents the Kronecker delta. The subsurface is fully characterized by the three elastic parameters: density $\rho$, first Lam\'e parameter $\lambda$, and shear modulus $\mu$. The wavefield variables in this equation are given by the particle velocities $v_i$ and the stress tensor components $\sigma_{ij}$. The wave propagation is due to the presence of the source terms $f_i$ and $M_{ij}$ that represent a volumetric force field and the time derivative of the moment tensor, respectively \cite[]{aki2002quantitative}. The pressure data can be obtained by averaging the normal-stress components. In our applications, we consider single component pressure data generated by explosive sources (i.e., $M_{ij} \neq 0$ for $i=j$). 

By discretizing the PDEs in equation~\ref{eqn:elastic-wave-cont}, we can define the elastic wave-equation modeling operator $\mathbf{f}$. We define the elastic FWI objective function as follows:
\begin{eqnarray} \label{eqn:elastic-fwi}
\psi(\mathbf{m}_{ela})=\frac{1}{2}\left\| \mathbf{f}(\mathbf{m}_{ela}) - \mathbf{d}\right\|_2^2,
\end{eqnarray}
where $\mathbf{m}_{ela}$ represents the elastic parameters of the subsurface. In our elastic FWI workflow, we parameterize the elastic isotropic medium using the P- and S-wave velocities $V_p$ and $V_s$, and density $\rho$. 

\section*{Synthetic tests}
In this section, we demonstrate using 2D numerical tests how the extended image space can synthesize elastic pressure data as if the acquisition geometry is sunk into the subsurface. First, we show how this image-based redatuming is applied to a flat-layer model. Then, we apply the redatuming technique on elastic pressure data generated on the Marmousi2~\cite[]{martin2006marmousi2}, to retrieve the elastic properties associated with a gas-bearing sand reservoir only within a target area.
We compare elastic FWI results of this target obtained using the surface and the reconstructed data and report the computational speed-up factor achieved by the target-oriented inversion approach.
All the numerical simulations are performed using finite-difference approximations of the wave-equation operators previously described.

\subsection{Redatuming of elastic pressure waves through extended linearized waveform inversion}
\begin{figure}[t]
    \centering
    \includegraphics[width=0.9\linewidth]{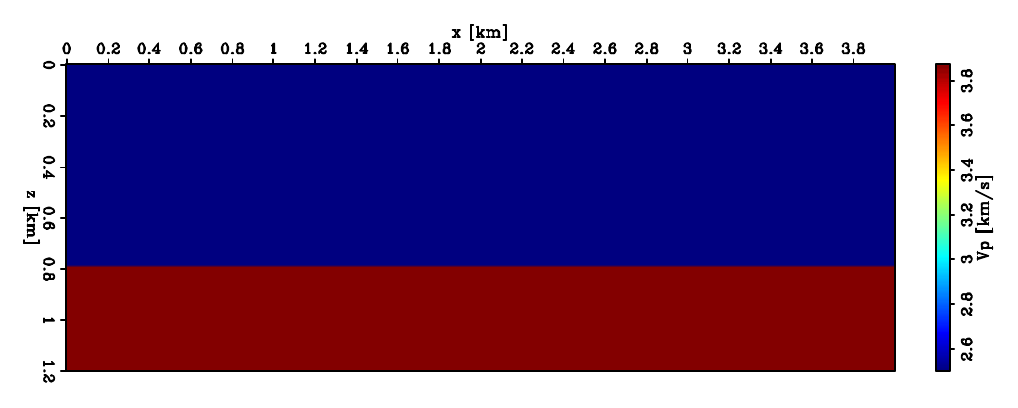}
    \caption{P-wave velocity model of the elastic single-interface model used in the first amplitude-preserving numerical test.}
    \label{fig:flatVp2D}
\end{figure}

\begin{figure}[t]
    \centering
    \subfigure[]{\label{fig:flatVpProfile}\includegraphics[width=0.25\columnwidth]{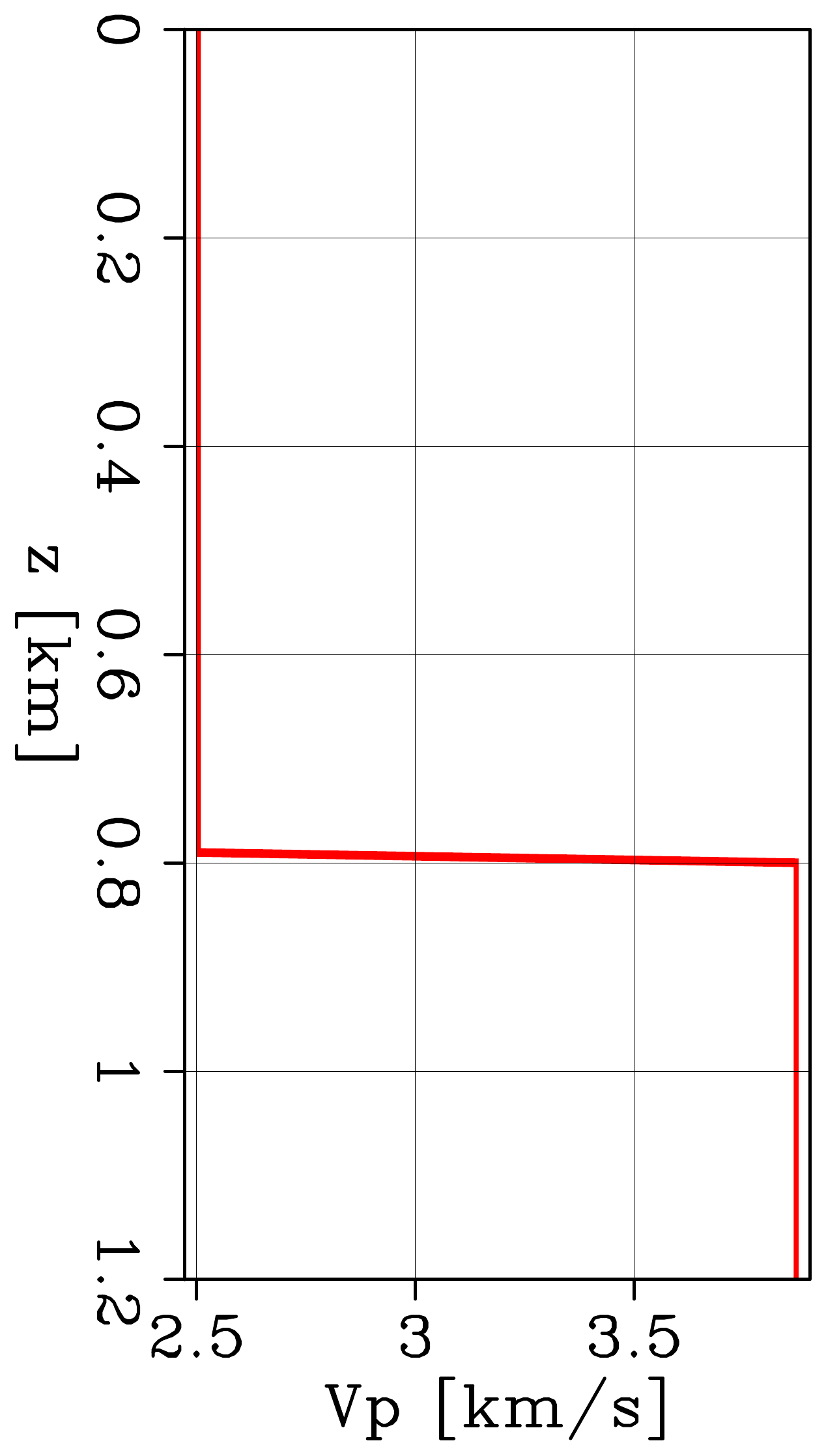}}
    \subfigure[]{\label{fig:flatVsProfile}\includegraphics[width=0.26\columnwidth]{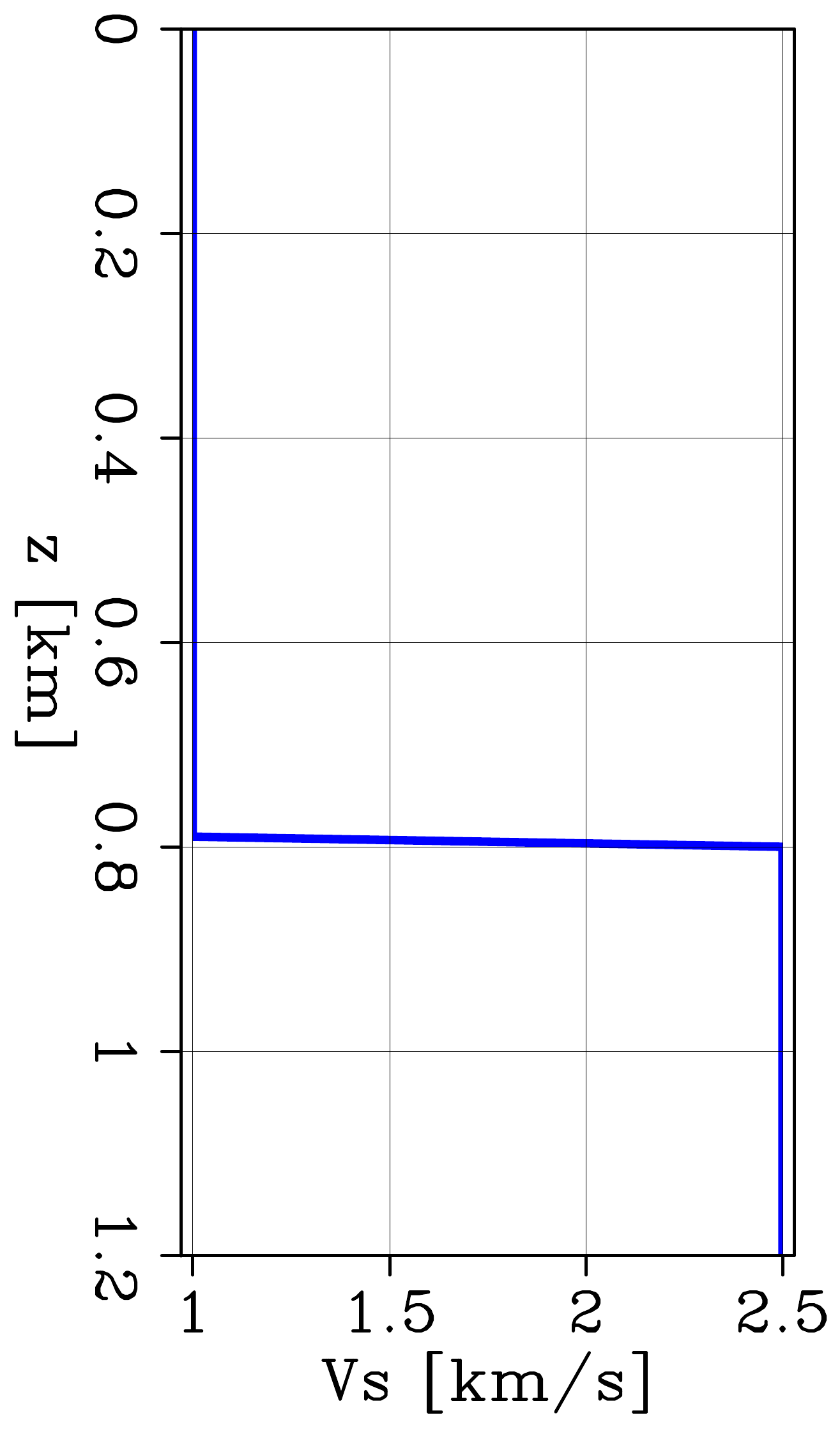}}
    \subfigure[]{\label{fig:flatRhoProfile}\includegraphics[width=0.26\columnwidth]{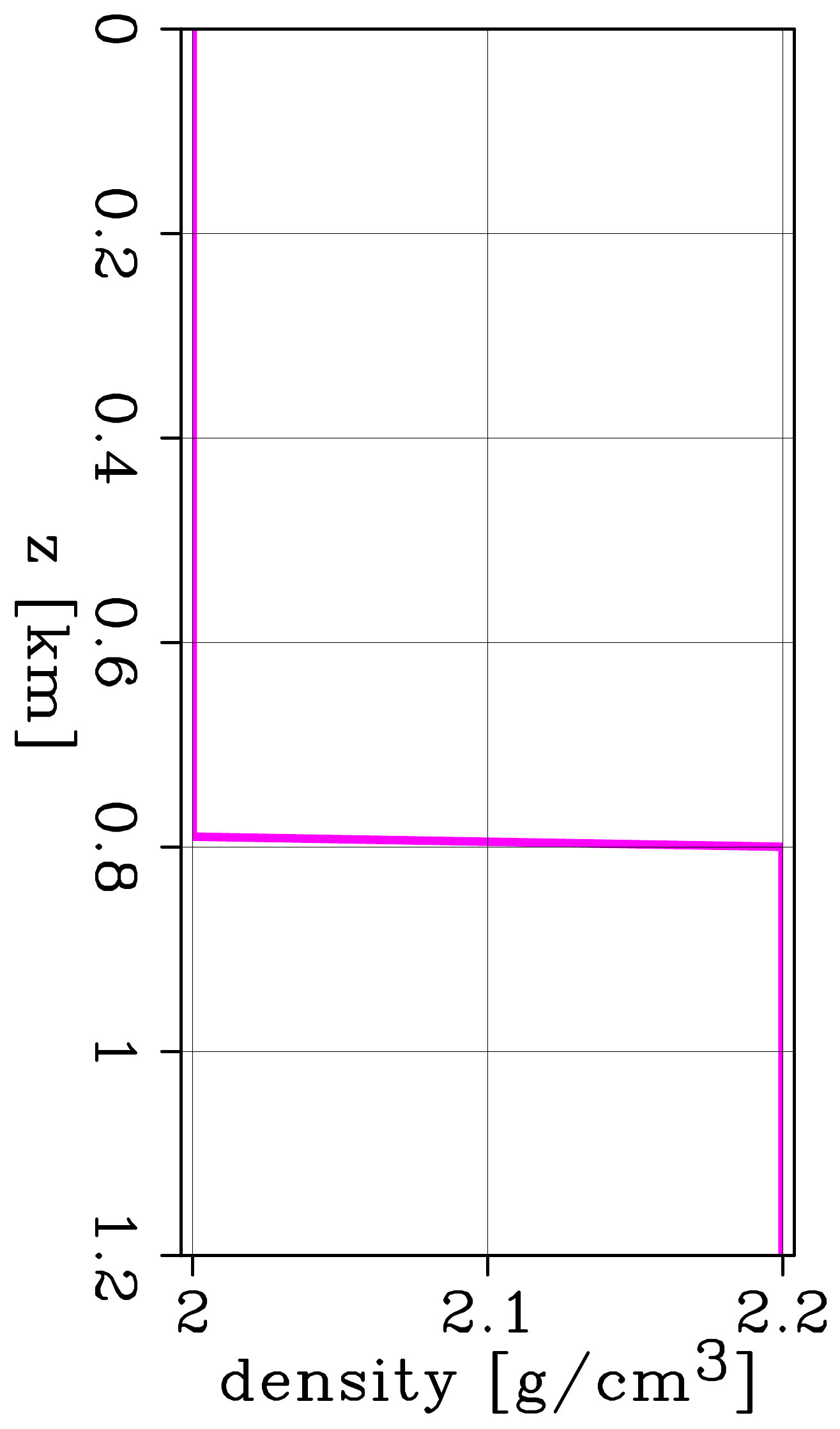}}
    \caption{Elastic parameter profiles of the single-interface model. The three panels show the (a) P-wave, (b) S-wave, and (c) density profiles, respectively.}
    \label{fig:flatProfiles}
\end{figure}

\begin{figure}[t]
    \centering
    \subfigure[]{\label{fig:flatWaveletTime}\includegraphics[width=0.45\columnwidth]{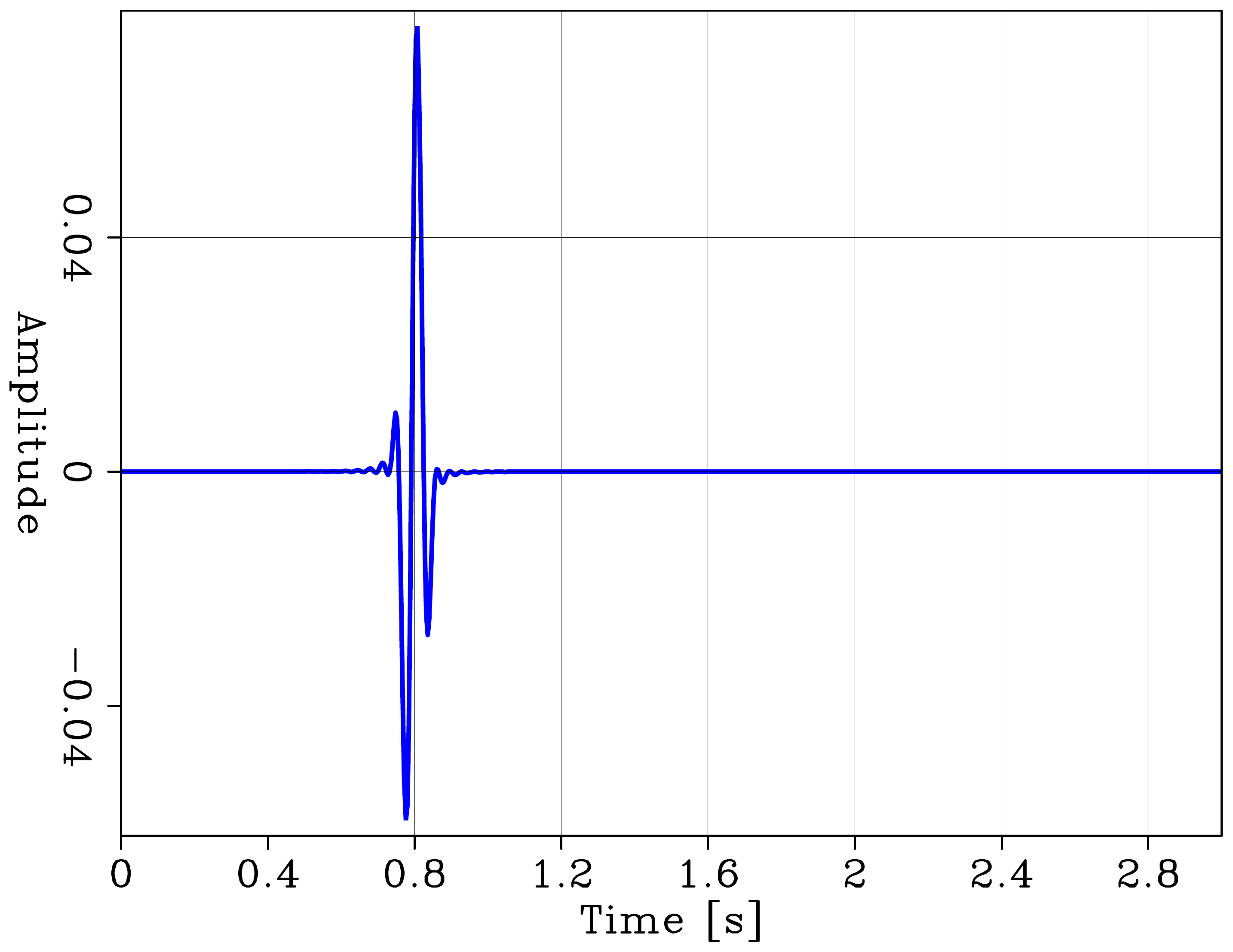}}
    \subfigure[]{\label{fig:flatWaveletSpectrum}\includegraphics[width=0.455\columnwidth]{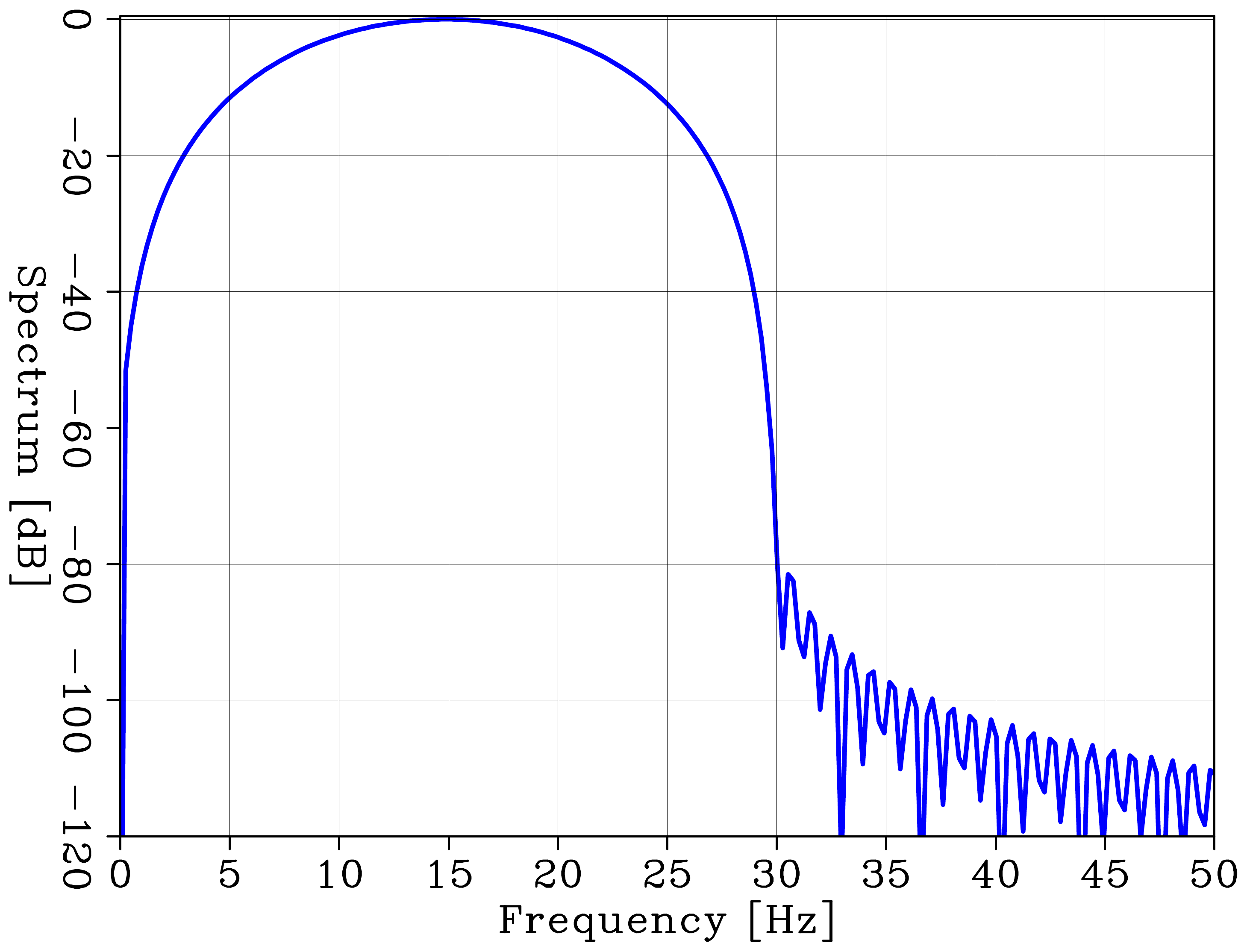}}
    \caption{Explosive source used to generate the elastic pressure data for the true-amplitude migration tests. Panels (a) and (b) show the time signature and frequency spectrum, respectively.}
    \label{fig:flatWavelet}
\end{figure}

Figure~\ref{fig:flatVp2D} shows the P-wave velocity model for the flat interface test. The change in the elastic parameters is depicted in the three vertical profiles shown in Figure~\ref{fig:flatProfiles}, where an increase of all the elastic parameters is occurring across the interface. we generate elastic pressure data using an explosive source, whose time signature and spectrum are displayed in Figure~\ref{fig:flatWavelet}. The goal is to use the elastic pressure data recorded at $z=0$ km to reconstruct a new dataset as if the sources and the receivers could have been placed at $400$ m below the surface.

\begin{figure}[t]
    \centering
    \subfigure[]{\label{fig:flatDataLeft}\includegraphics[width=0.45\columnwidth]{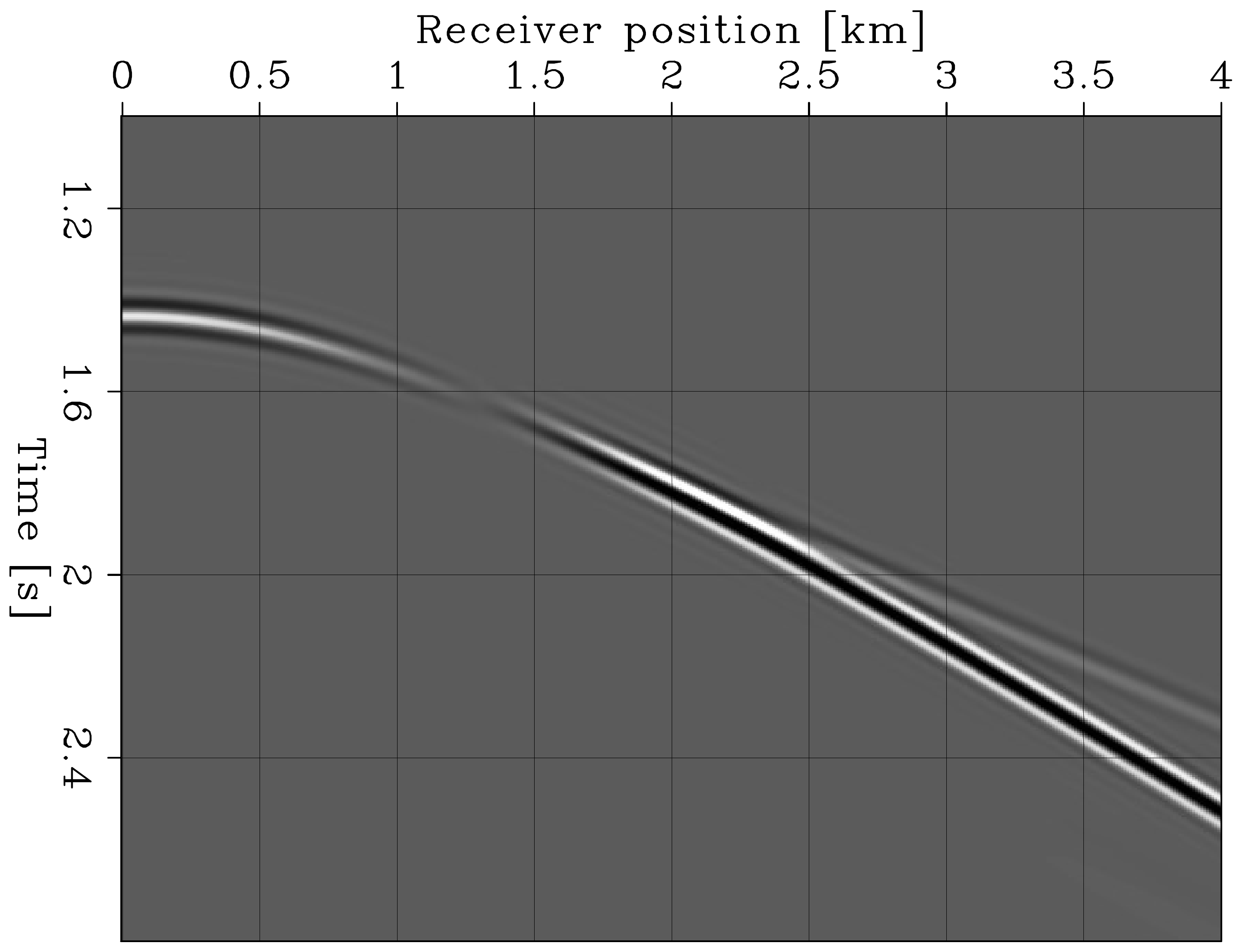}}
    \subfigure[]{\label{fig:flatDataMid}\includegraphics[width=0.45\columnwidth]{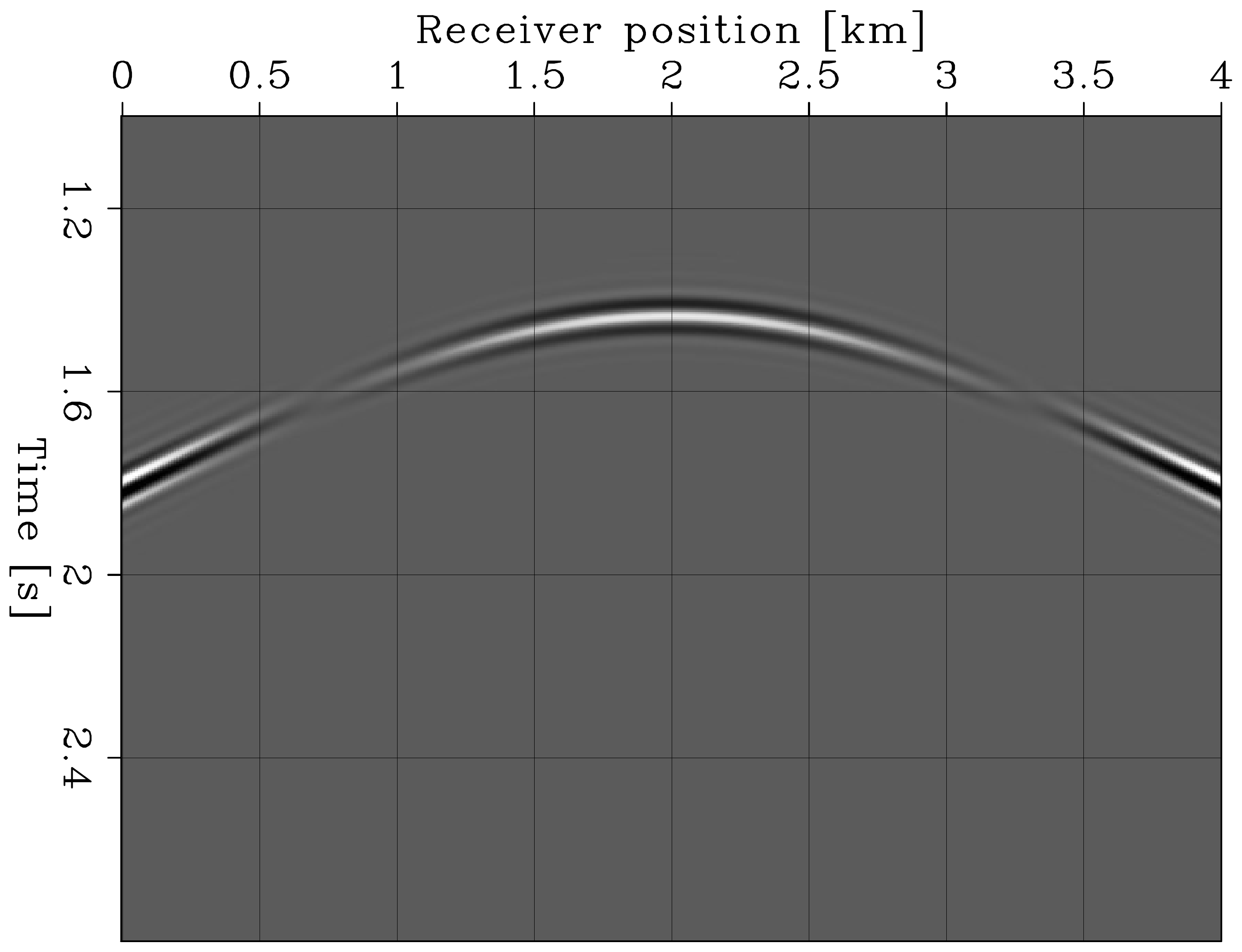}}
    \caption{Elastic pressure waves recorded at the surface for a shot placed at (a) $x=0$ km and (b) $x=2.0$ km, respectively. The direct arrival has been removed from the observed data.}
    \label{fig:flatData}
\end{figure}

We record the pressure using 81 sources and 401 receivers placed at the surface and spaced by 50 and 10 m, respectively. A single reflected event is recorded by the receivers for each experiment, where a clear phase rotation is present as the offset between the source and receiver pair increases (Figure~\ref{fig:flatData}).

To perform the imaging step, we solve the linearized waveform inversion problem using an acoustic extended Born modeling operator and minimize the objective function in equation~\ref{eqn:ext-lsrtm} using 500 iterations of the LCG algorithm. The migration velocity model is a constant speed set to 2.5 km/s, corresponding to the correct overburden wave speed. The extended-space inversion achieves a relative objective function decrease of an approximately numerical level of accuracy for single-precision operators (i.e., $10^{-6}$), showing the ability of the extended-image space to fully preserve all the elastic amplitude variations present in the recorded data.

\begin{figure}[t]
    \centering
    \subfigure[]{\label{fig:flatODCIGleft}\includegraphics[width=0.3\columnwidth]{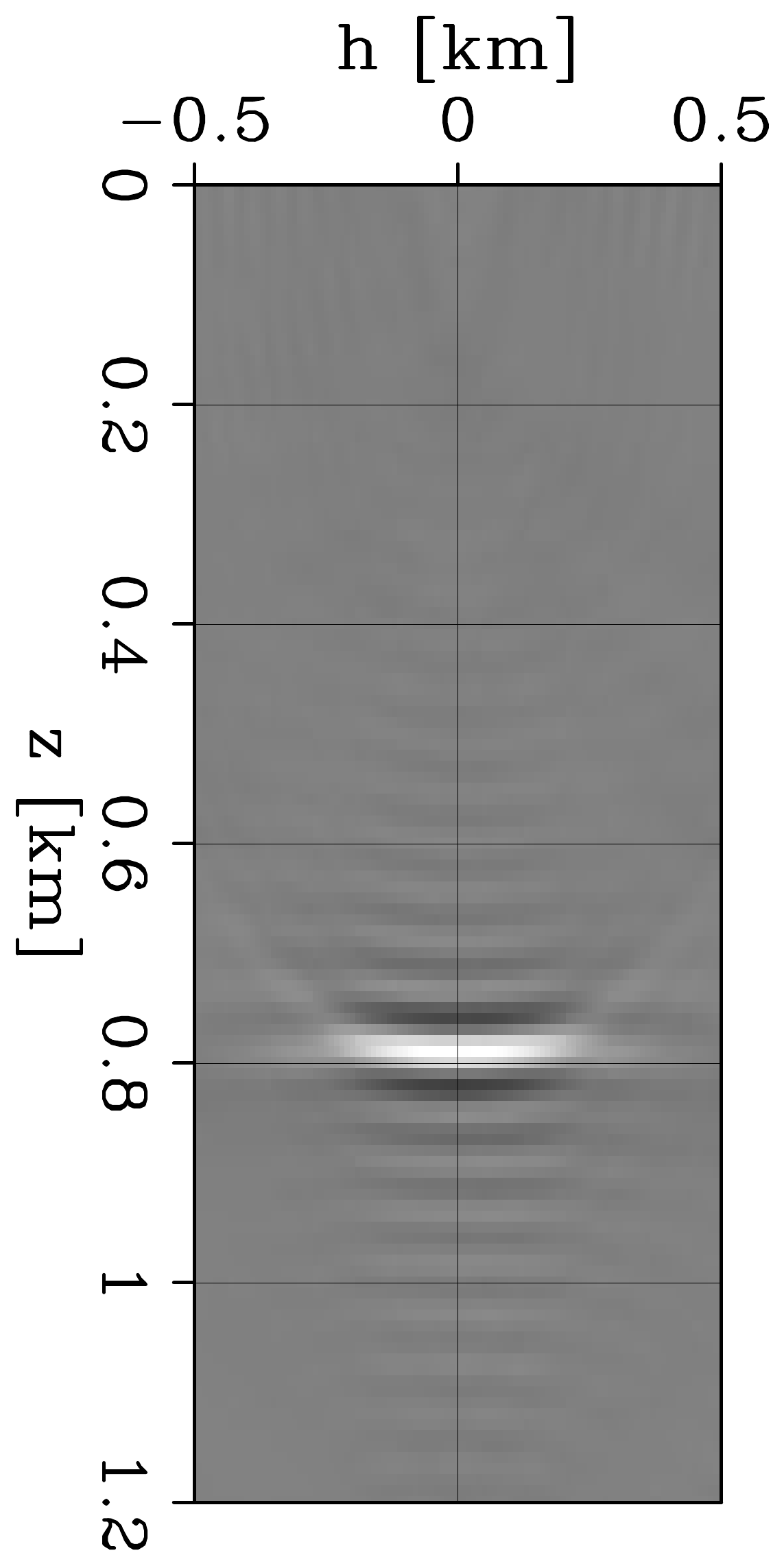}}
    \subfigure[]{\label{fig:flatODCIGmid}\includegraphics[width=0.3\columnwidth]{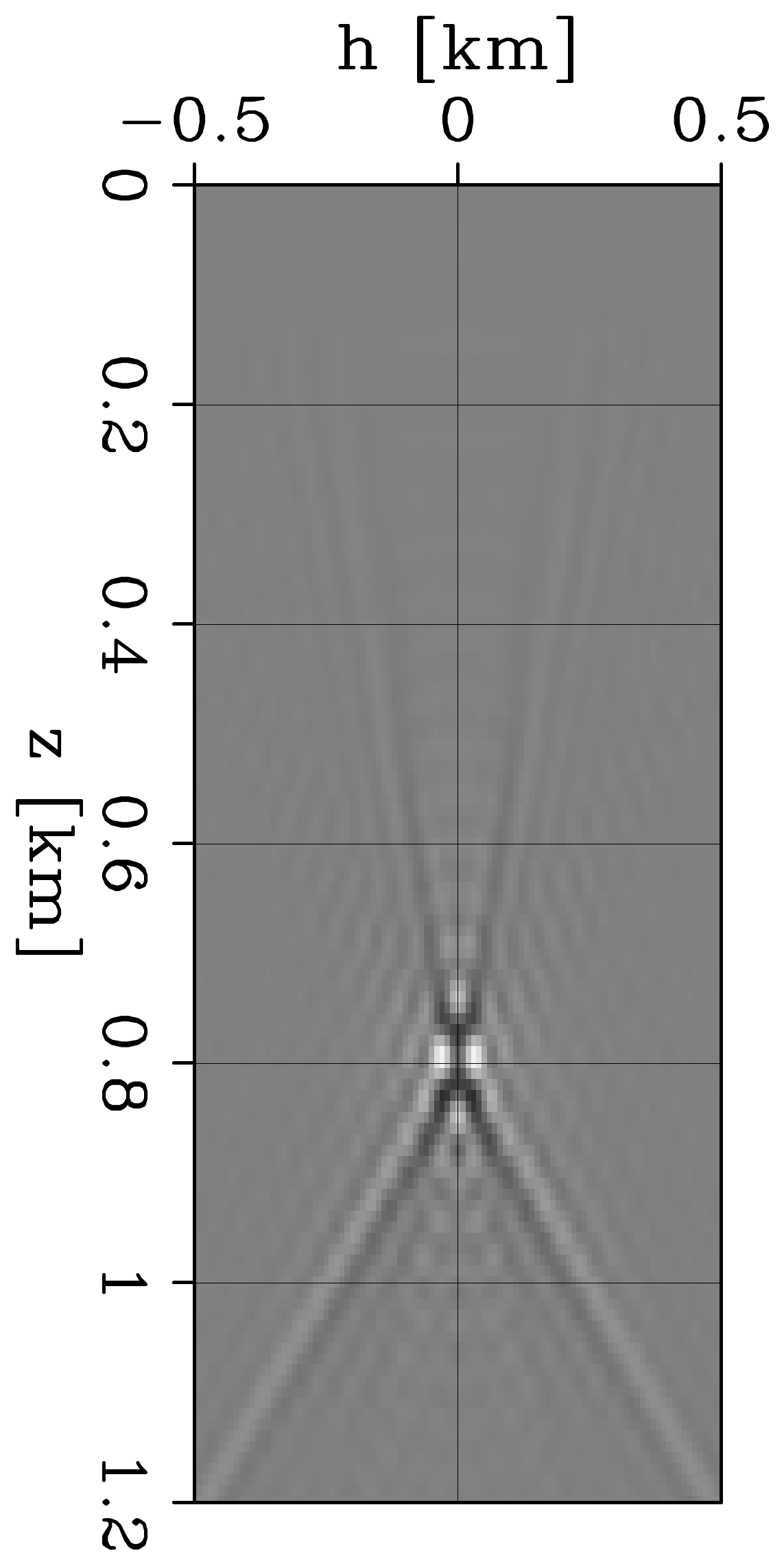}}
    \caption{Subsurface-offset common image gathers for the flat-interface example extracted at (a) $x=0$ km and (b) $x=2.0$ km, respectively.}
    \label{fig:flatODCIGs}
\end{figure}

The extended linearized waveform inversion problem's solution is a function of the spatial coordinates $x$ and $z$ and of the extended subsurface offset axis $h$. The shape of the image highly depends on the recorded events and the acquisition geometry employed. Figure~\ref{fig:flatODCIGs} shows the offset-domain common image gathers (ODCIGs) for two different $x$ coordinates. For $x=2.0$ km, the ODCIG appears to be focused around the zero-offset axis. This behavior is expected since the correct migration velocity has been used during the inversion process~\cite[]{biondi20043d}. The two linear features below $z=0.8$ km represent the head waves recorded in the longer offset shot gathers mapped into the image space. The other two faint linear features above $z=0.8$ km are caused by the limited acquisition aperture (i.e., the maximum source-receiver offset of $4$ km). On the other hand, when an ODCIG is extracted at $x=0.0$ km (Figure~\ref{fig:flatODCIGleft}), the image does not appear as focused as for the central-model position because fewer reflection angles have been illuminated from the surface acquisition.
 
\begin{figure}[t]
    \centering
    \subfigure[]{\label{fig:redatumFlatReflLeft}\includegraphics[width=0.4\linewidth]{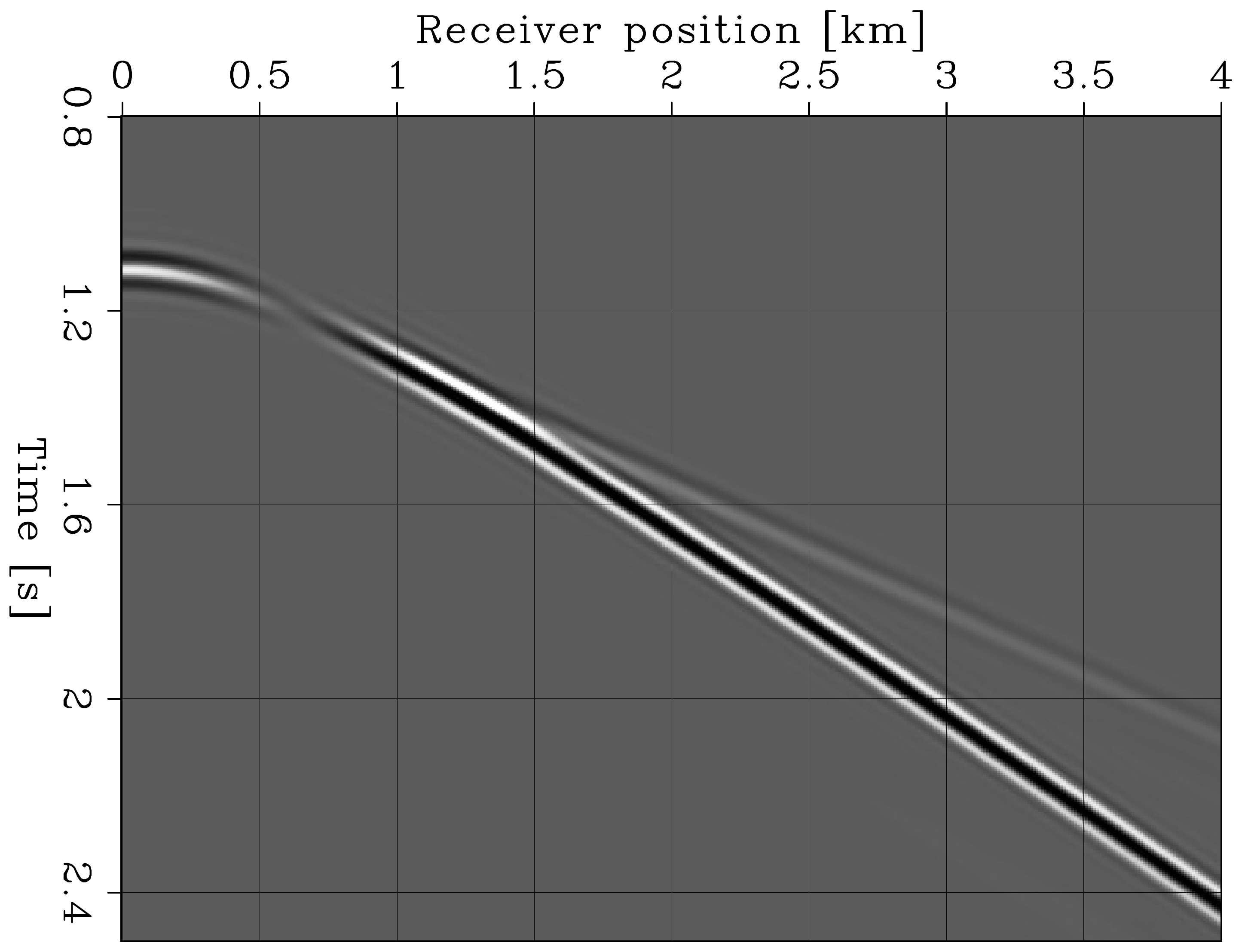}}
    \subfigure[]{\label{fig:redatumFlatReflMid}\includegraphics[width=0.4\linewidth]{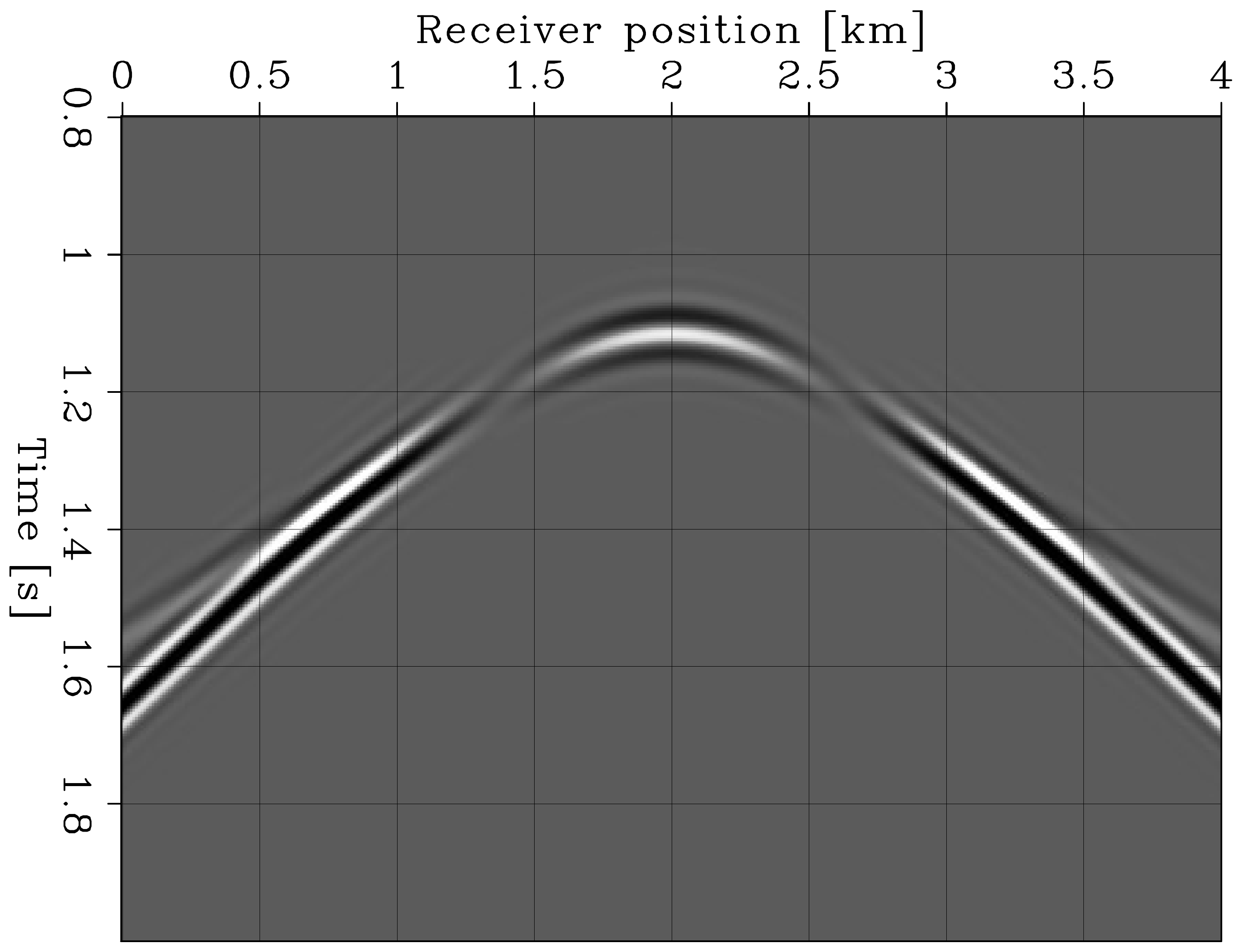}}
    
    \caption{Representative elastic pressure shot gathers for sources placed at (a) $x=0.0$ km and (b) $x=2.0$ km generated using the single flat interface model of Figure~\ref{fig:flatProfiles} and with an acquisition depth of $400$ m.}
    \label{fig:redatumFlatRefl}
\end{figure}

Figure~\ref{fig:redatumFlatRefl} displays two representative shot gathers obtained when the acquisition geometry is placed at $z=400$ m. The same amplitude-versus-offset behavior is observed as in the surface pressure data (Figure~\ref{fig:flatData}).

\begin{figure}[t]
    \centering
    \subfigure[]{\label{fig:redatumFlatSurfOdcig}\includegraphics[width=0.3\linewidth]{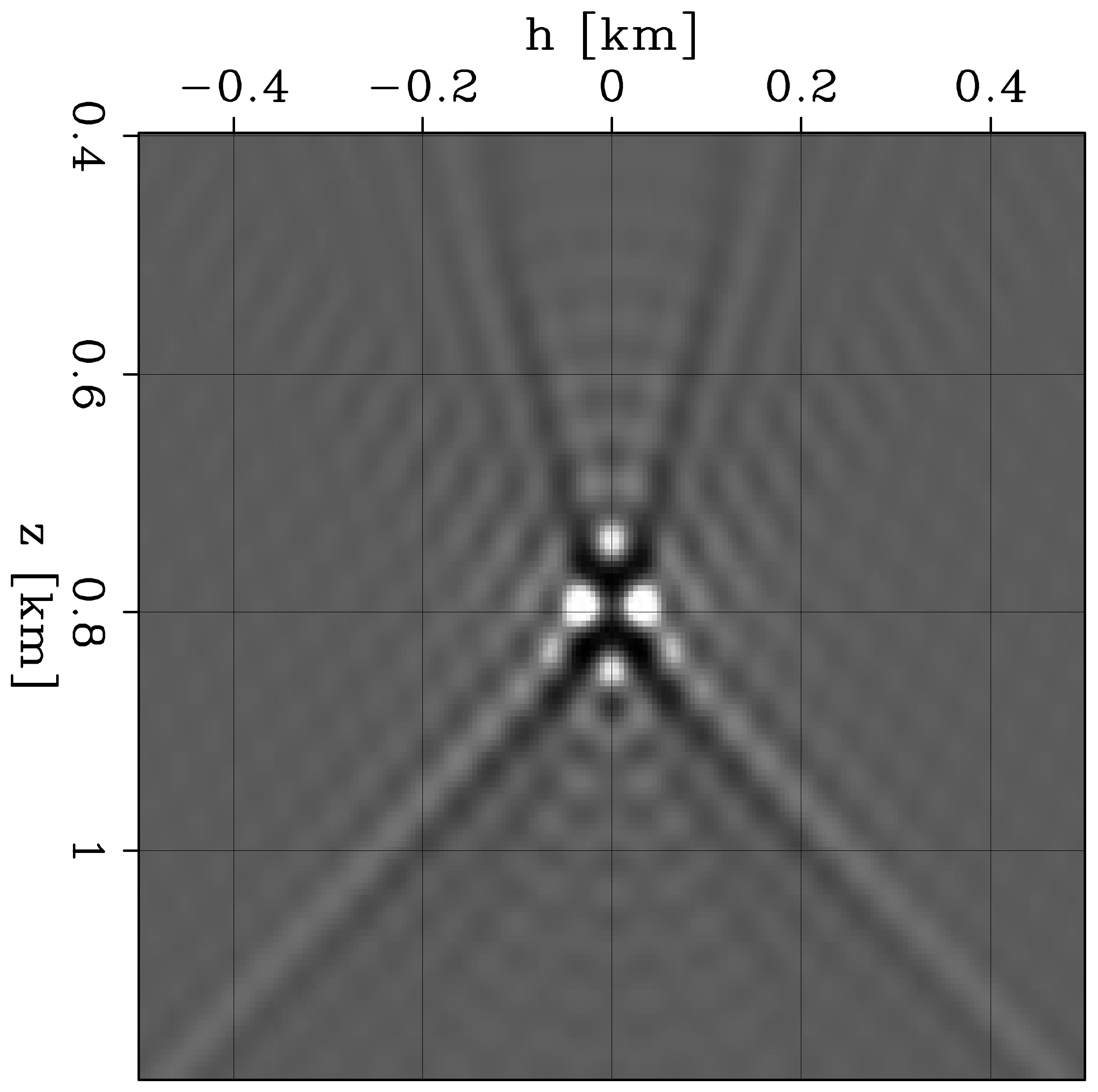}}
    \subfigure[]{\label{fig:redatumFlatDatOdcig}\includegraphics[width=0.3\linewidth]{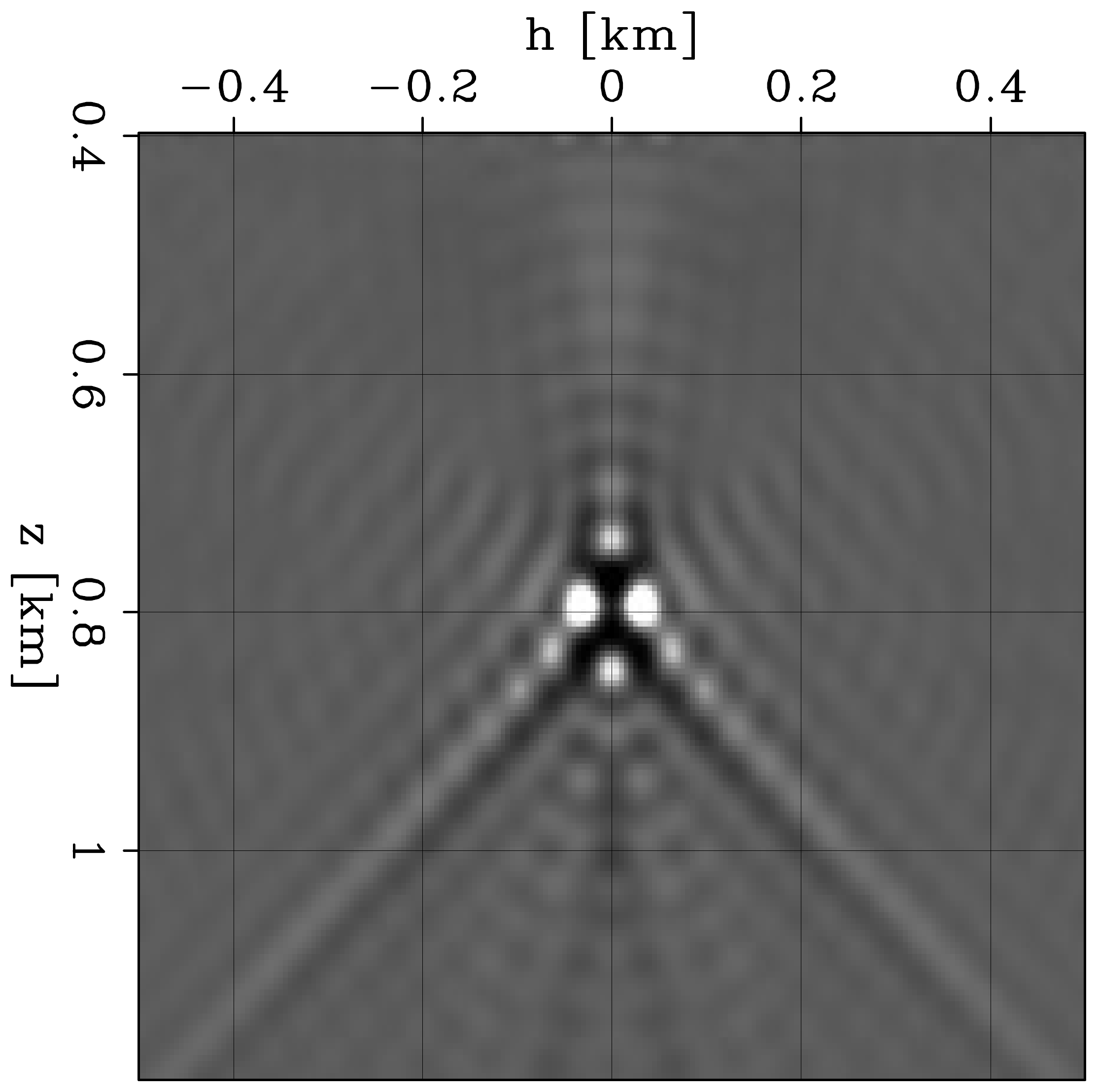}}
    \subfigure[]{\label{fig:redatumFlatDiffOdcig}\includegraphics[width=0.3\linewidth]{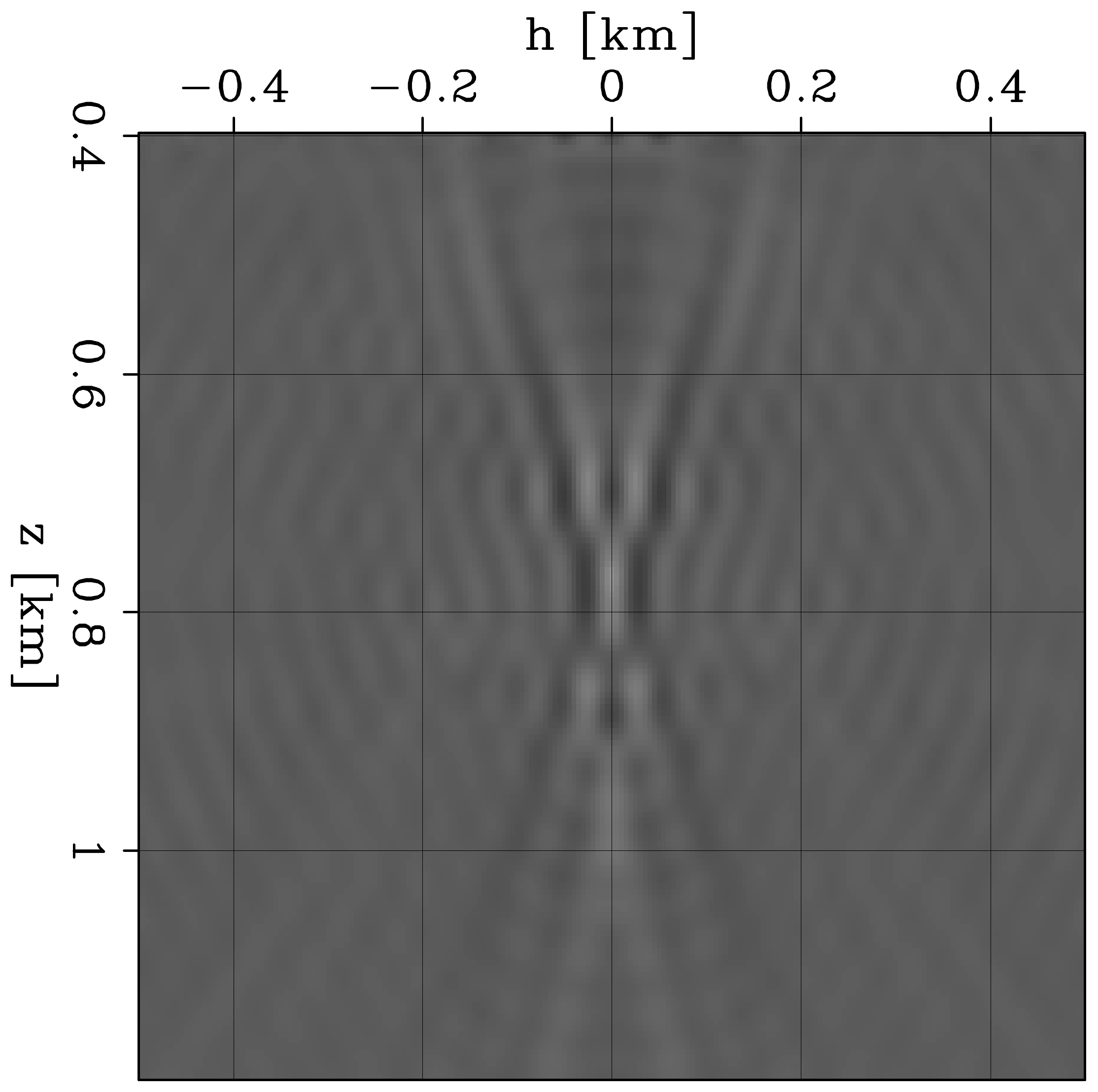}}
    
    \caption{(a) Close-up of the ODCIG of Figure~\ref{fig:flatODCIGmid}. (b) ODCIG generated by solving a linearized waveform inversion using the buried acquisition geometry. (c) Difference between panels (a) and (b). All panels are displayed using the same gain.}
    \label{fig:redatumFlatOdcig}
\end{figure}

The schematic of Figure~\ref{fig:redatuming} shows that a scattering point can be used to generate the surface and the sunk acquisition datasets. This observation also implies that the images obtained from two acquisitions, assuming infinite source-receiver extent, are identical. To demonstrate this statement, we compare the ODCIGs obtained by inverting the surface and sunk-acquisition data, respectively (Figures~\ref{fig:redatumFlatSurfOdcig} and~\ref{fig:redatumFlatDatOdcig}). Indeed, the only difference between the two ODCIGs is due to the limited acquisition aperture (Figure~\ref{fig:redatumFlatDiffOdcig}).

\begin{figure}[t]
    \centering
    \subfigure[]{\label{fig:redatumFlatRecMid}\includegraphics[width=0.4\linewidth]{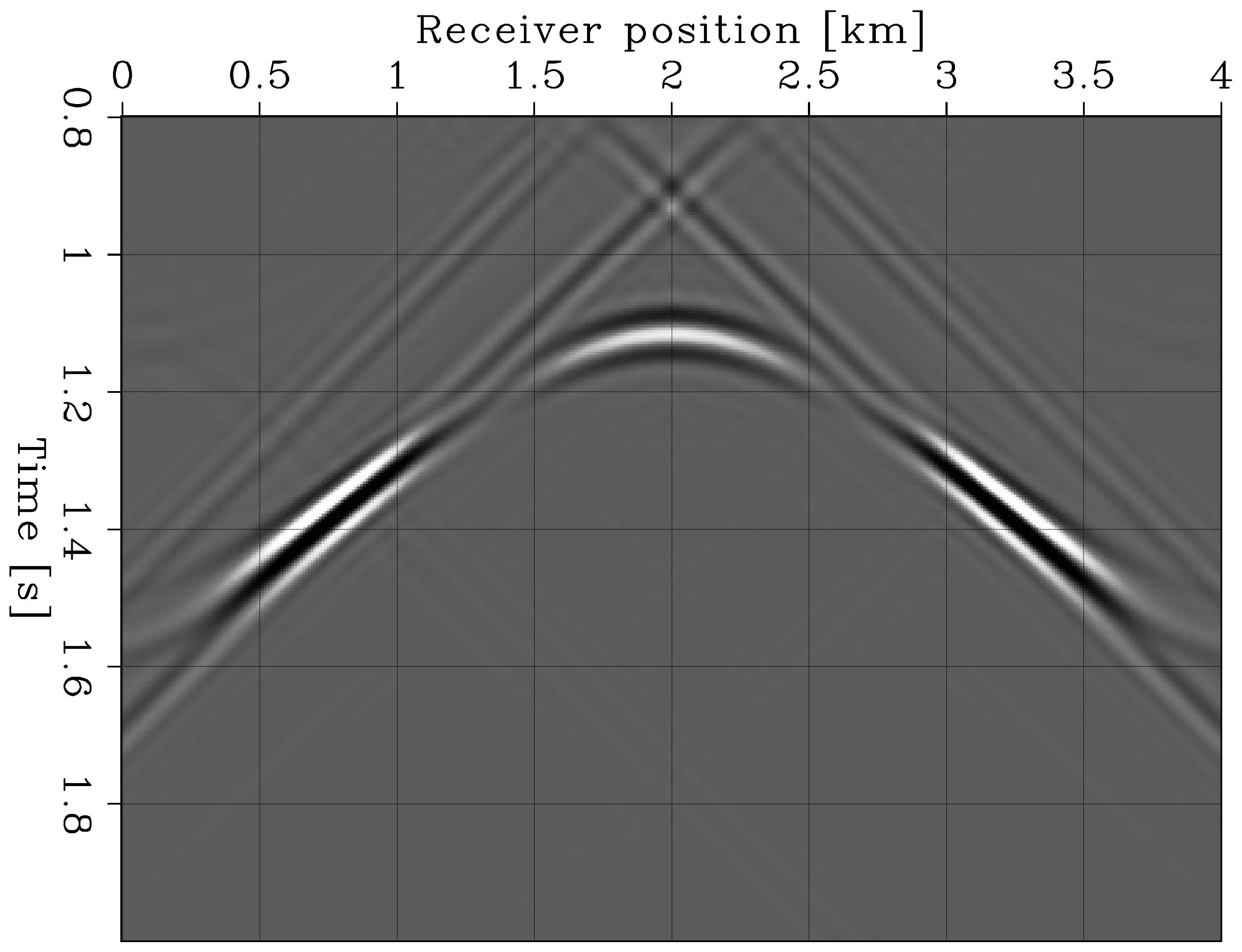}}
    \subfigure[]{\label{fig:redatumFlatRecMaskOdcig}\includegraphics[width=0.31\linewidth]{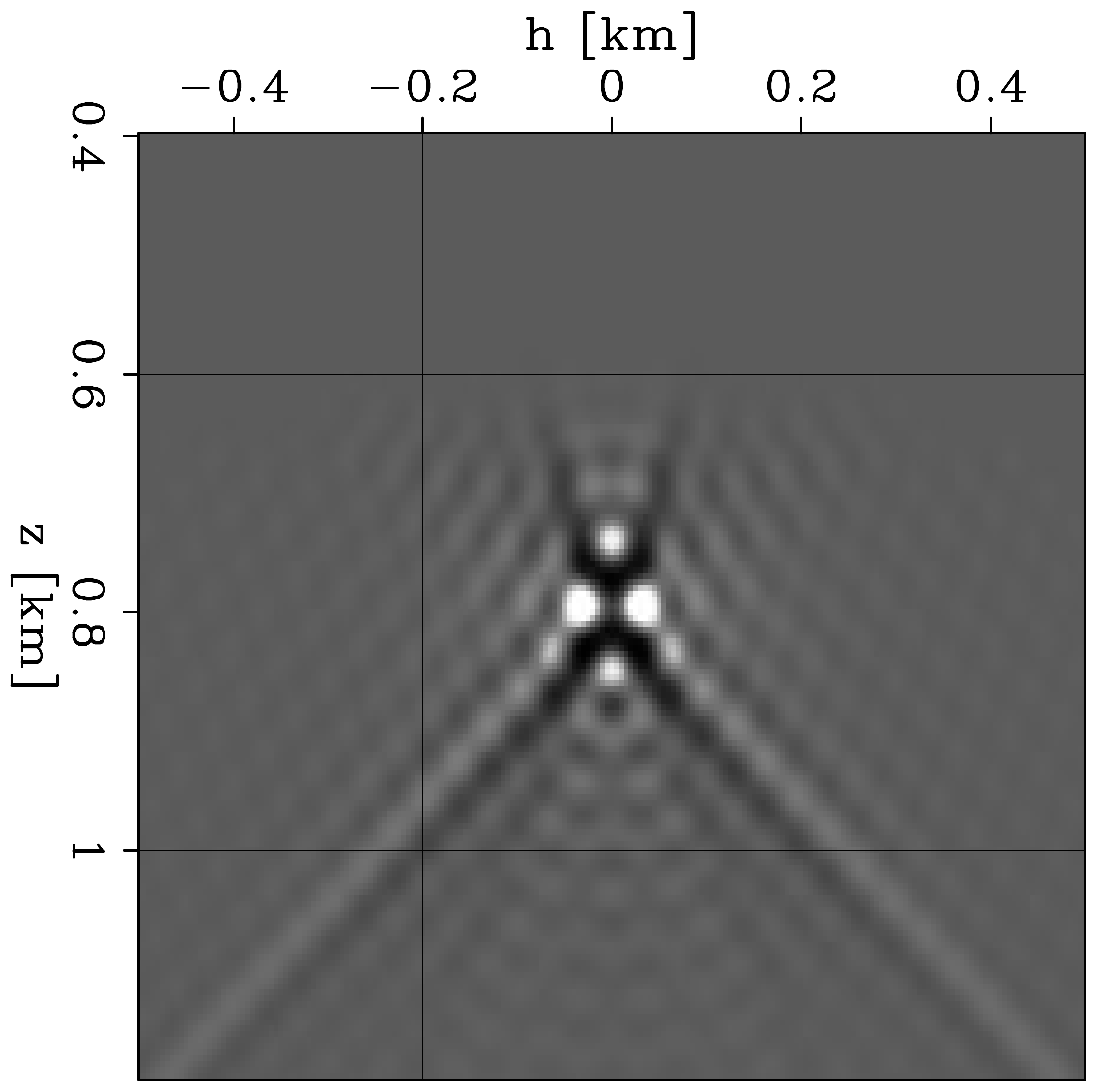}}
    
    \subfigure[]{\label{fig:redatumFlatRecMidMask}\includegraphics[width=0.4\linewidth]{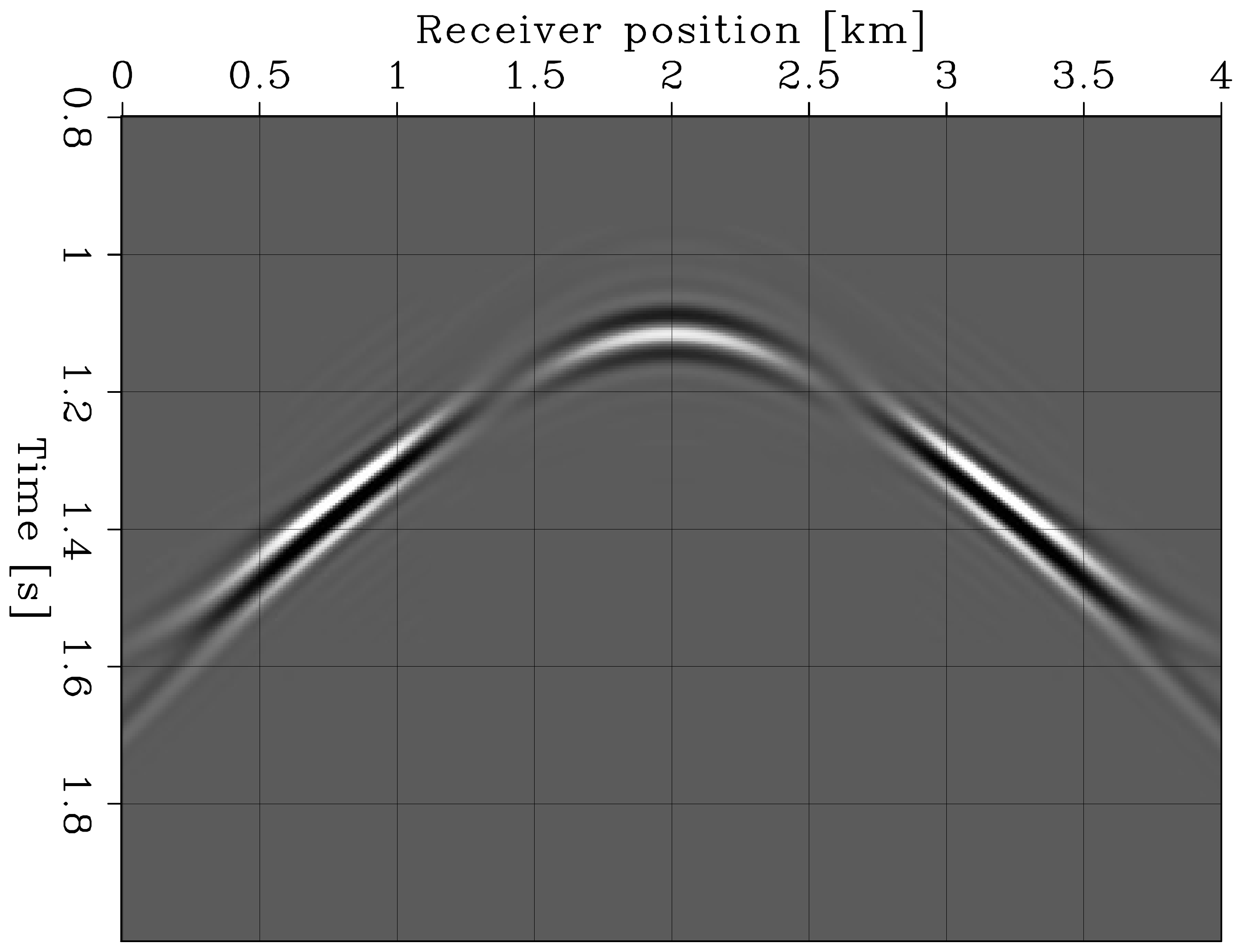}}
    \subfigure[]{\label{fig:redatumFlatRecMidMaskDiff}\includegraphics[width=0.4\linewidth]{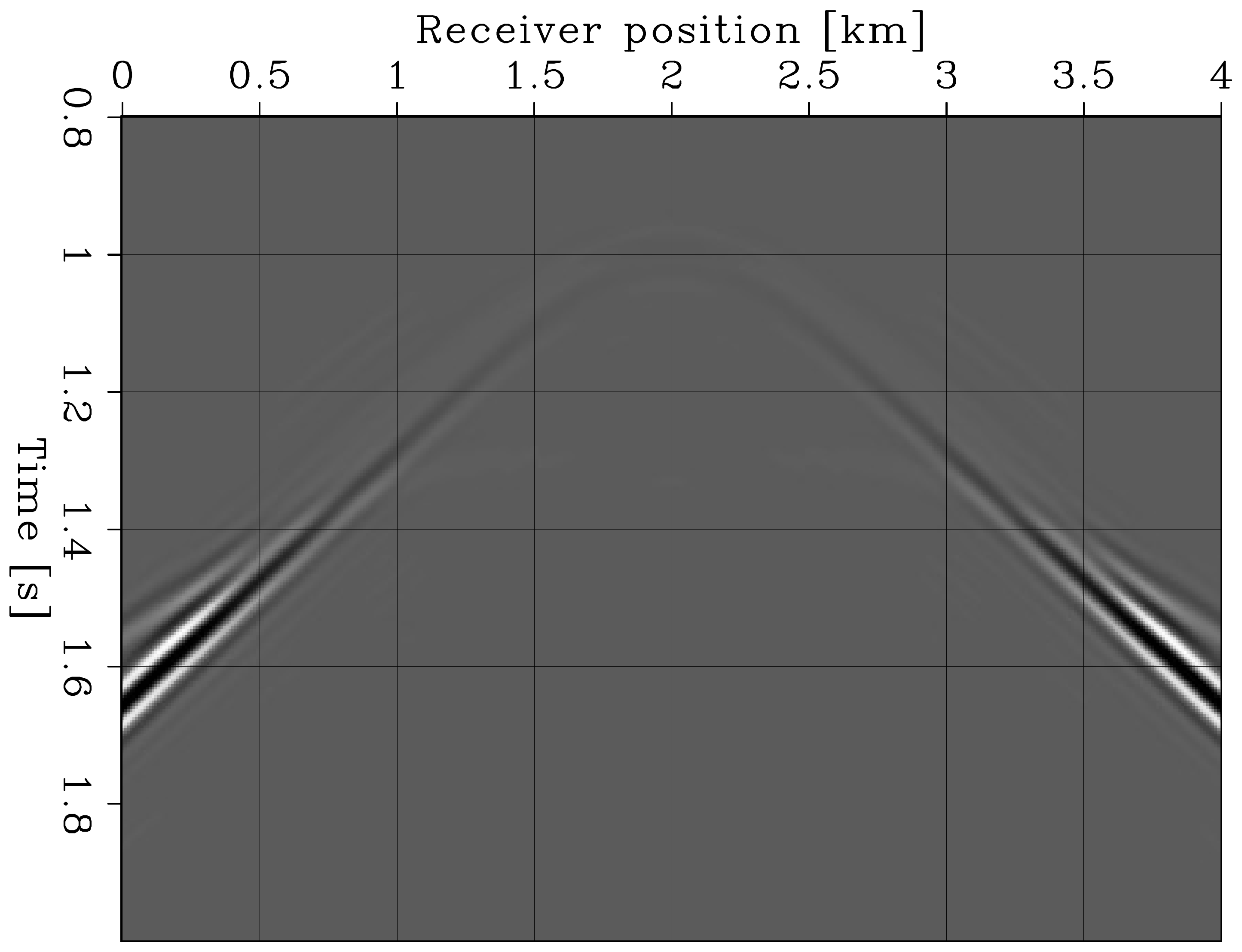}}
    
    \caption{(a) Elastic pressure reconstructed by demigrating the surface ODCIGs. (b) ODCIG of Figure~\ref{fig:redatumFlatSurfOdcig} where a muting mask has been applied to dampen the acquisition artifacts above $0.6$ km. (c) Reconstructed pressure data obtained by demigrating the ODCIGs where a image mask has been applied. (d) Difference between panel (c) and the one of Figure~\ref{fig:redatumFlatReflMid}.}
    \label{fig:redatumFlatRec}
\end{figure}

Since the data from the two acquisition geometries maps into the same extended image, we can use the ODCIGs obtained from the surface pressure to synthesize the events recorded by the sunk sources and receivers. Figure~\ref{fig:redatumFlatRecMid} shows the shot gather at $x=2.0$ km obtained by demigrating the ODCIGs of Figure~\ref{fig:redatumFlatSurfOdcig}. A similar amplitude behavior is present compared to the shot gather of Figure~\ref{fig:redatumFlatReflMid} up to an offset of $1$ km. The artifacts above the apex of the reflected event are due to the truncation of the surface acquisition geometry. In fact, when we demigrate the image where those truncation artifacts are masked (Figure~\ref{fig:redatumFlatRecMaskOdcig}), the reconstructed reflection does not present any spurious events (Figure~\ref{fig:redatumFlatRecMidMask}).

As we described using the schematic of Figure~\ref{fig:redatumingGeom}, not all the events associated with any source-receiver pairs can be reconstructed from an image obtained with surface data. In this case, the maximum illuminated reflection angle from the surface geometry is approximately $64^{\circ}$, which corresponds to a maximum half offset of $1$ km for the sunk acquisition geometry. Figure~\ref{fig:redatumFlatRecMidMaskDiff} shows the difference between the reference and the reconstructed data of Figures~\ref{fig:redatumFlatReflMid} and~\ref{fig:redatumFlatRecMidMask}, where only energy for an offset greater than $1$ km are present as expected.

\subsubsection{Sensitivity to assumed source wavelet}

\begin{figure}[t]
    \centering
    \subfigure[]{\label{fig:redatumWav90rot}\includegraphics[width=0.4\linewidth]{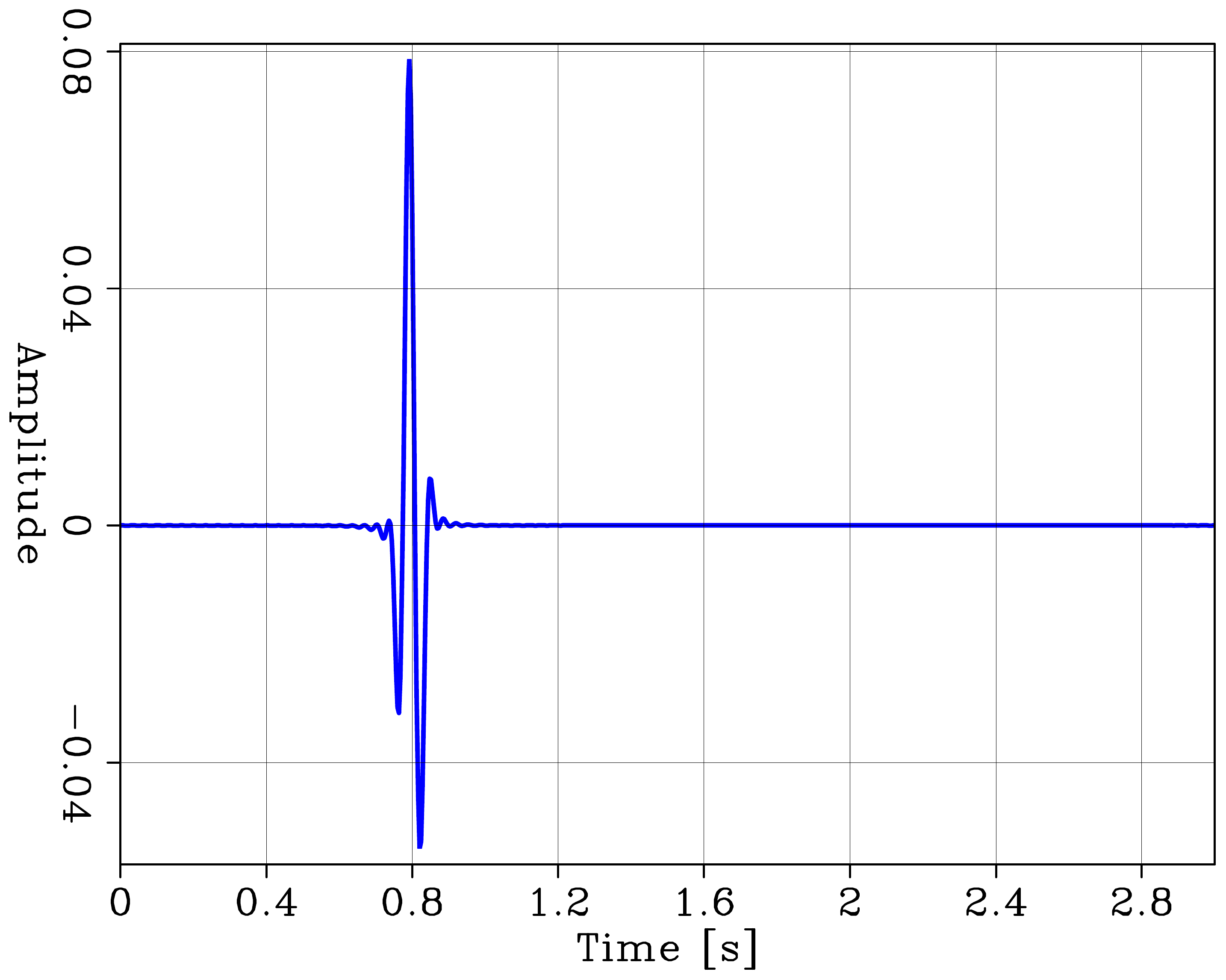}}
    \subfigure[]{\label{fig:redatumWavRick}\includegraphics[width=0.4\linewidth]{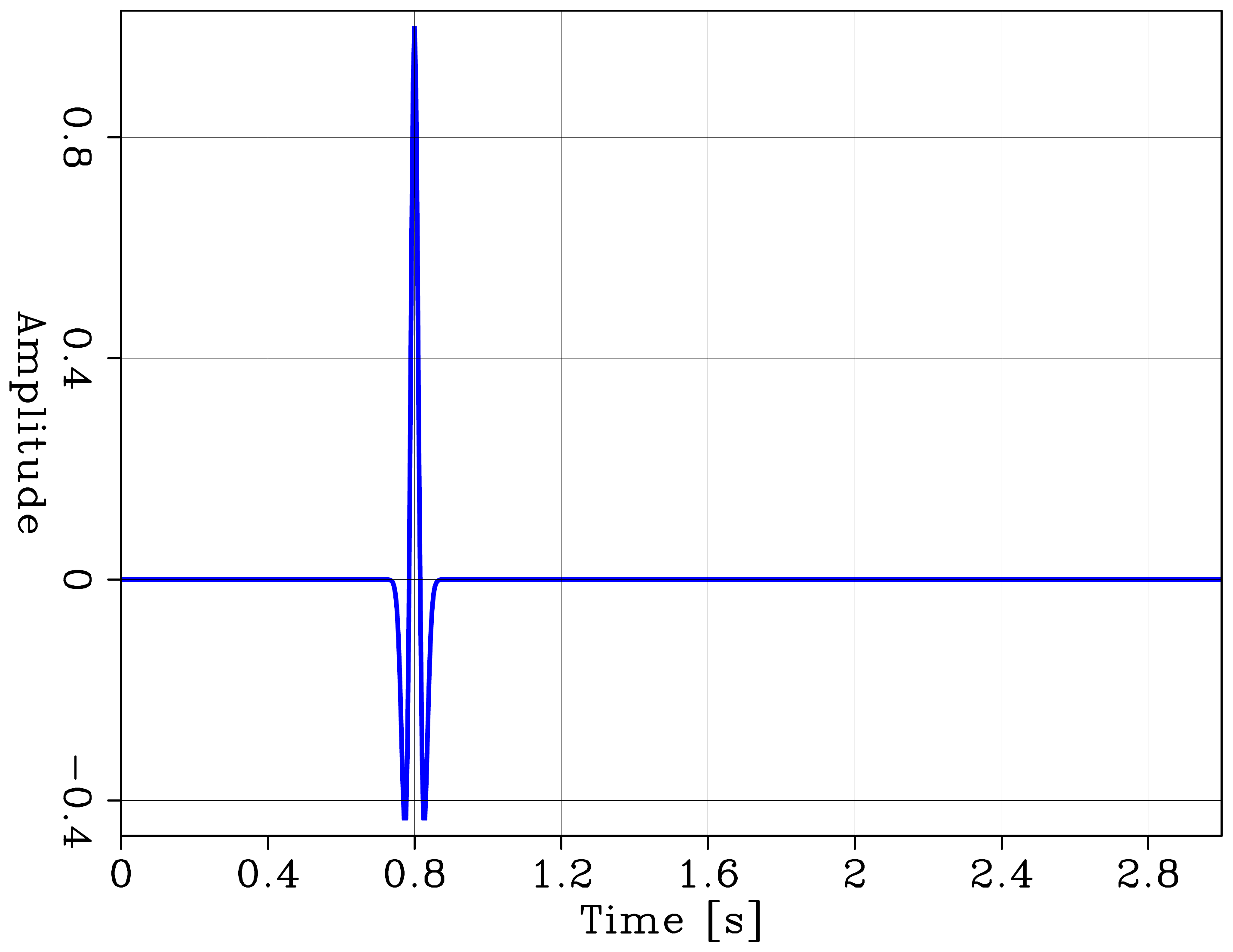}}
    
    \caption{Time-domain wavelet plots for testing invariability of the data reconstruction procedure. (a) Wavelet where a 90-degree phase rotation has been applied to the original signal. (b) Ricker wavelet with dominant frequency of $15$ Hz.}
    \label{fig:redatumWav}
\end{figure}

Since the redatuming technique is based on an imaging step, it is necessary to create a source wavelet signature. However, the data reconstruction is invariable to the choice of the source signature. To numerically verify this statement, we reconstruct the same events of Figure~\ref{fig:redatumFlatRecMidMask}, but employ different waveforms during the linearized waveform inversion and demigration steps. Figure~\ref{fig:redatumWav90rot} displays the same wavelet of Figure~\ref{fig:flatWaveletTime} on which a 90-degree phase rotation has been applied. The right panel in Figure~\ref{fig:redatumWavRick} shows a Ricker wavelet with a domain frequency of 15 Hz. These two waveforms are independently used to solve the extended linearized waveform inversion problem defined on the flat-interface model (Figure~\ref{fig:flatVp2D}). The extended gathers generated by this process are then used to reconstruct the elastic pressure events at the new datum (i.e., $400$ m).

\begin{figure}[t]
    \centering
    \subfigure[]{\label{fig:redatumWavRecRick}\includegraphics[width=0.4\linewidth]{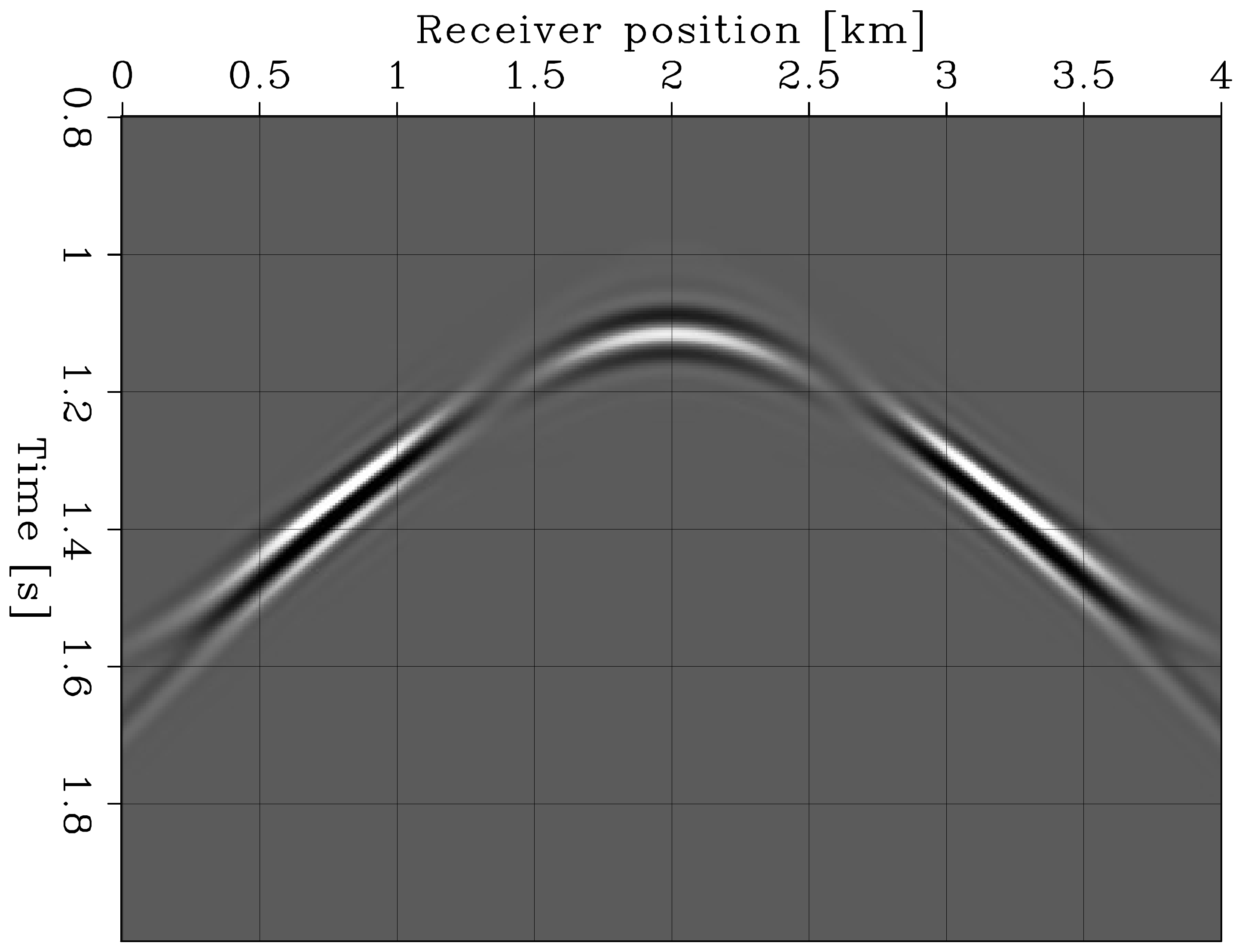}}
    \subfigure[]{\label{fig:redatumWavRec90rot}\includegraphics[width=0.4\linewidth]{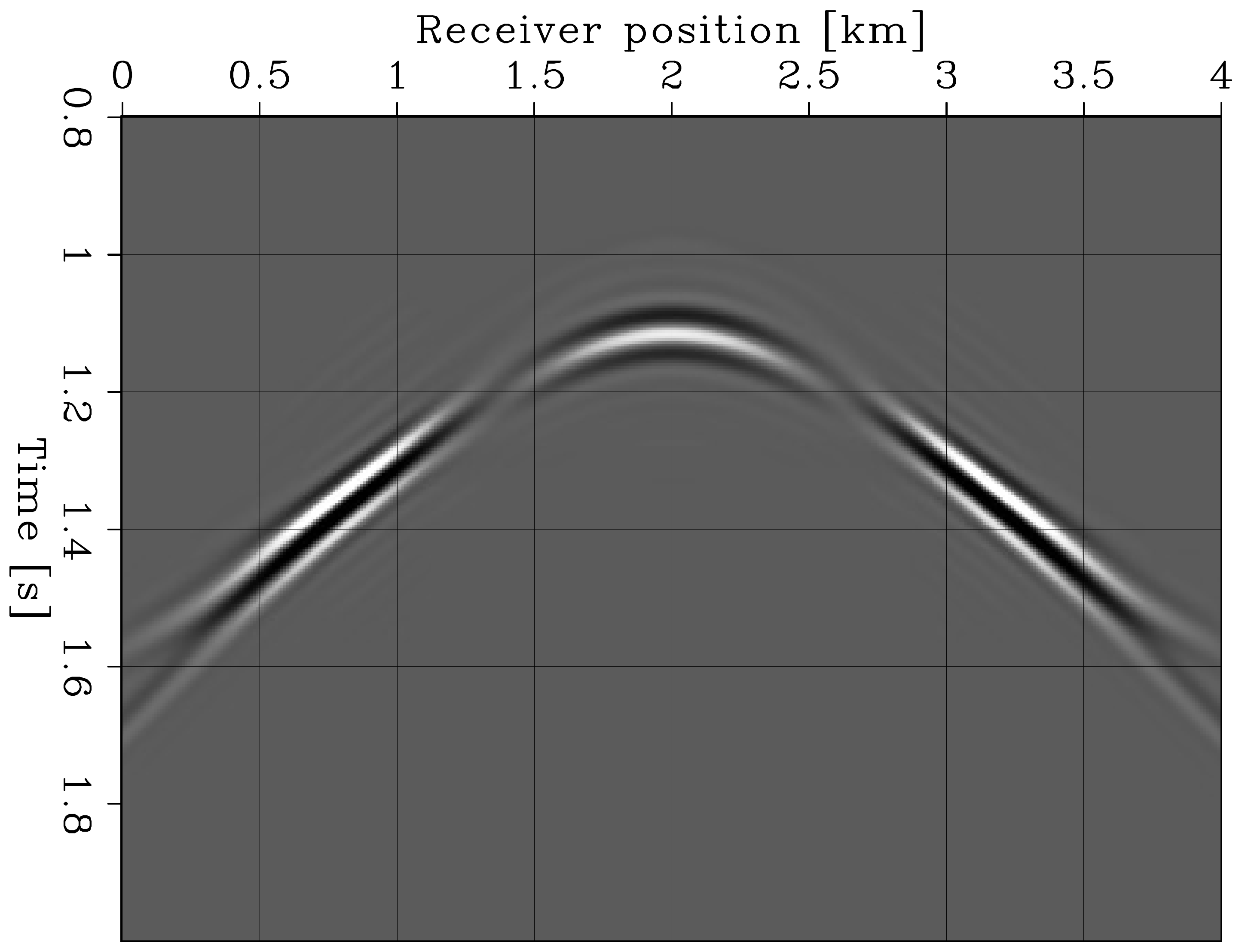}}
    
    \caption{Pressure data reconstructed from the extended images obtained using the wavelets signals of Figure~\ref{fig:redatumWav}}
    \label{fig:redatumWavRec}
\end{figure}

Figure~\ref{fig:redatumWavRec} shows the redatumed pressure when the 90-degree rotated waveform and the Ricket wavelet are employed during the demigration process, respectively. No evident difference is visible when these two panels are compared to the one displayed in Figure~\ref{fig:redatumFlatRecMidMask}. 

\begin{figure}[t]
    \centering
    \subfigure[]{\label{fig:redatumWavADCIGtrue}\includegraphics[width=0.25\linewidth]{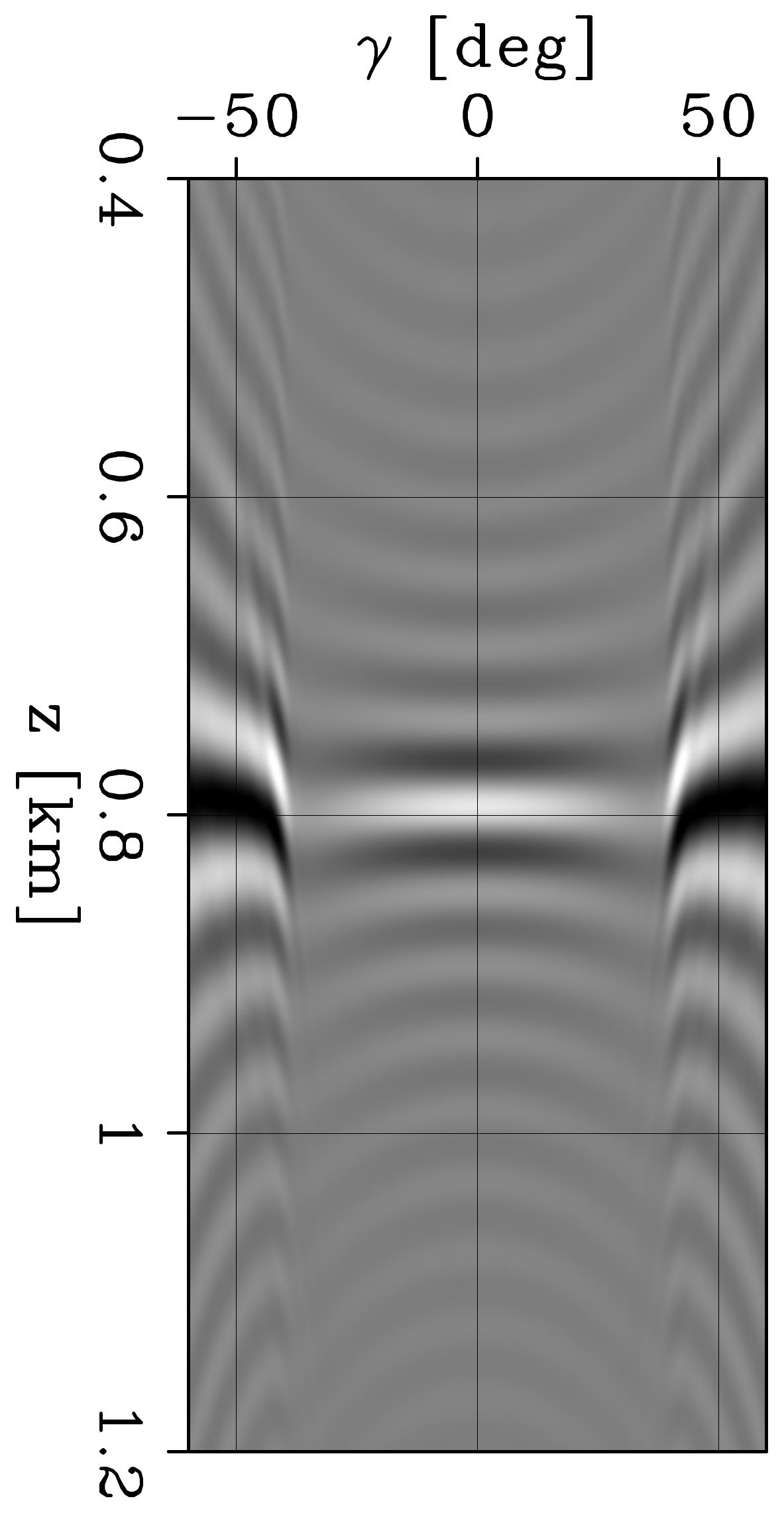}}
    \subfigure[]{\label{fig:redatumWavADCIGRick}\includegraphics[width=0.25\linewidth]{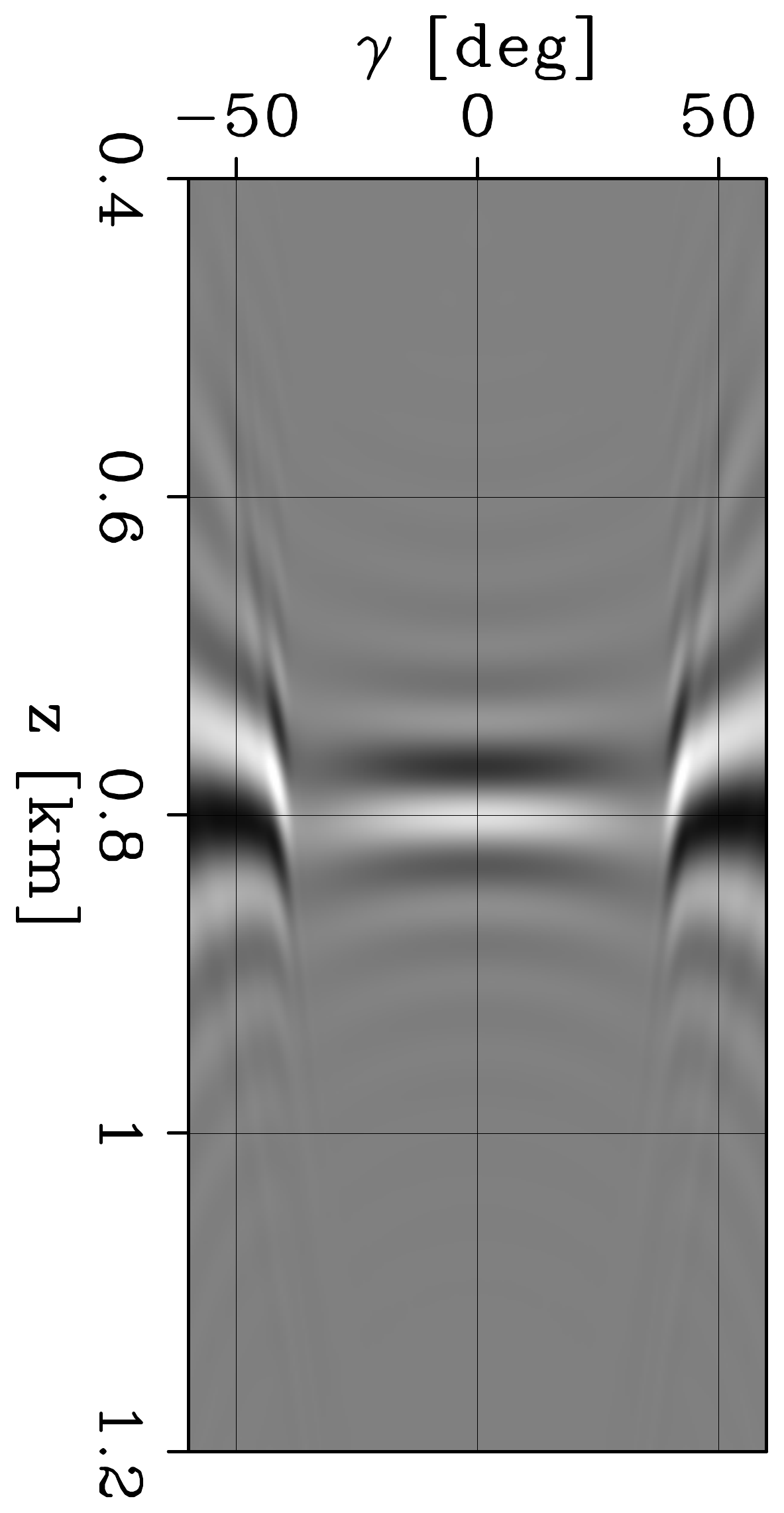}}
    \subfigure[]{\label{fig:redatumWavADCIG90rot}\includegraphics[width=0.25\linewidth]{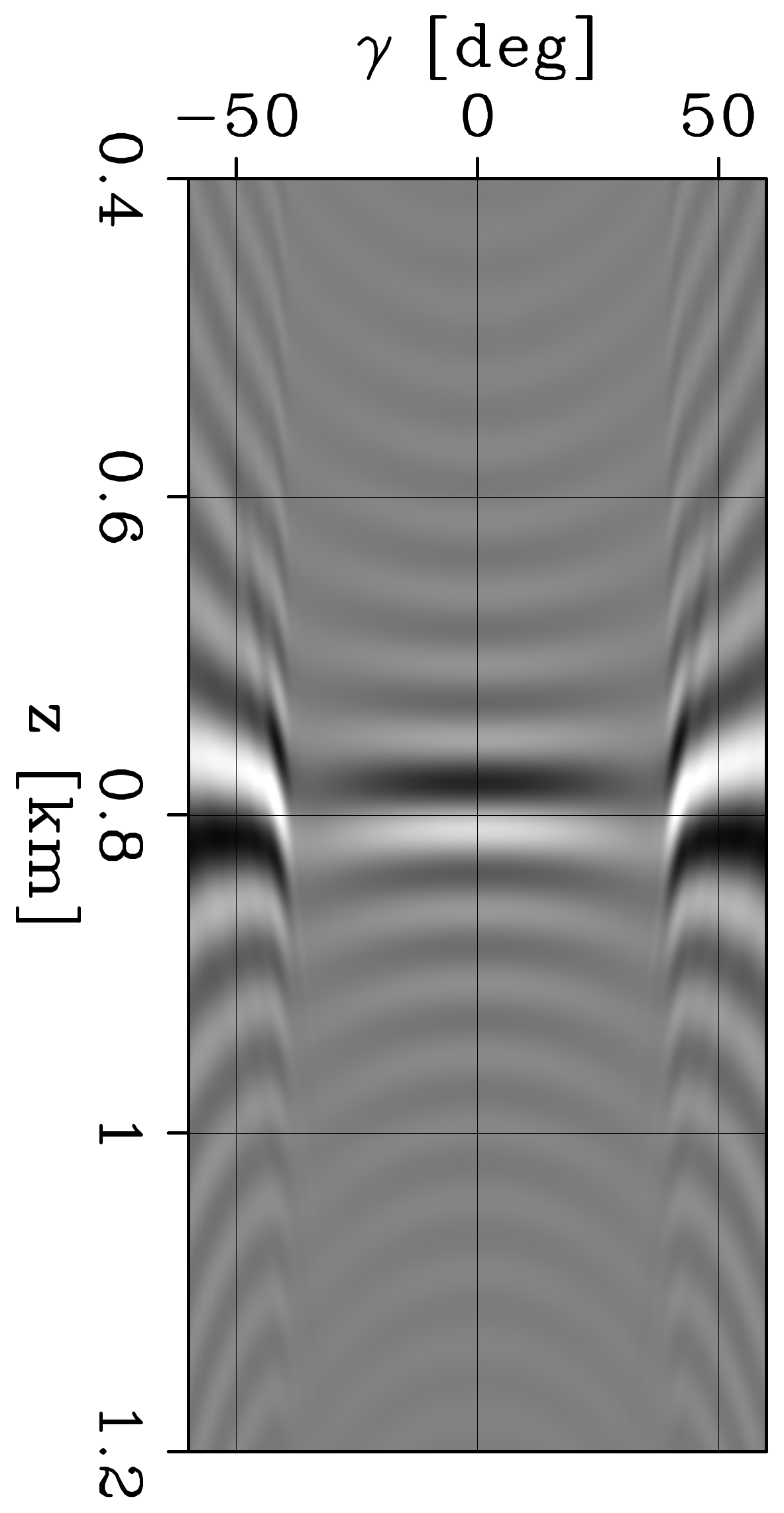}}
    
    \caption{ADCIGs extracted at $x=2.0$ km for the extended images obtained using: (a) the correct signature, (b) the 90-degreee rotated signal, and (c) the Ricker wavelet, respectively.}
    \label{fig:redatumWavADCIG}
\end{figure}

This invariability can also be seen by analyzing the amplitude behavior of the ADCIGs generated by the extended linearized waveform inversion when different wavelets are employed. The same AVA pattern is visible in the three panels of Figure~\ref{fig:redatumWavADCIG} showing the ADCIGs generated with three source signatures described in this section.

\subsubsection{Sensitivity to migration velocity}

\begin{figure}[t]
    \centering
    \subfigure[]{\label{fig:redatumRecWrongOdcig}\includegraphics[width=0.4\linewidth]{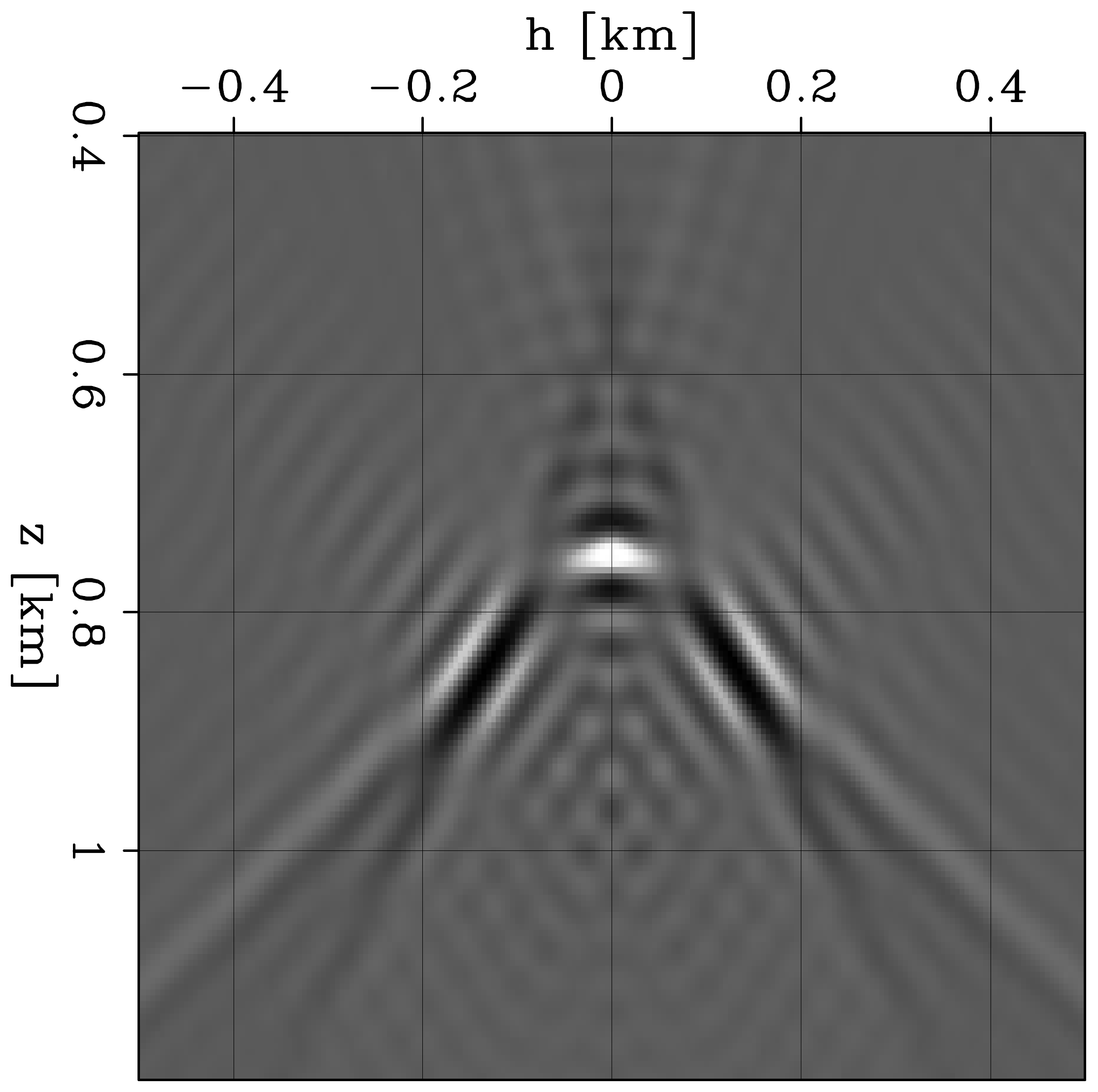}}
    
    \subfigure[]{\label{fig:redatumRecWrongAdcig}\includegraphics[width=0.25\linewidth]{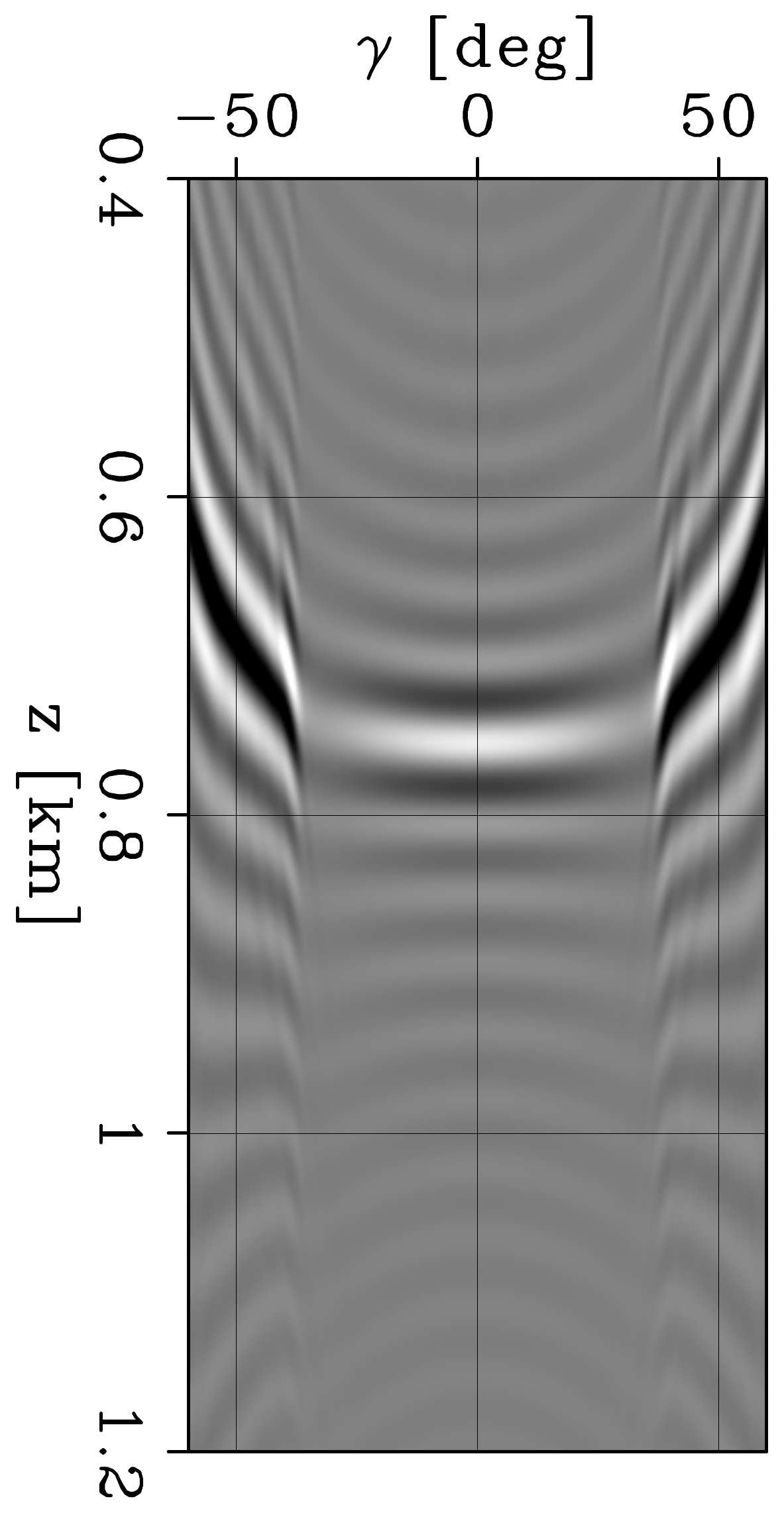}}
 
    \caption{(a) ODCIG and (b) ADCIG extracted at $x=2.0$ km on the extended image obtained from the surface pressure migrated employing a constant velocity model of $2375$ m/s.}
    \label{fig:redatumCigWrong}
\end{figure}

Finally, we analyze the sensitivity of the reconstruction process with respect to the migration velocity map used during the linearized waveform inversion step. To this end, we perform the same redatuming steps previously described for the flat-interface case but in which a 5\% slower velocity is used compared to the correct one (i.e., $2375$ m/s). As expected the ODCIG obtained during the migration process is not as focused as when the correct velocity is employed (compare Figures~\ref{fig:redatumRecWrongOdcig} and~\ref{fig:redatumFlatSurfOdcig}). Moreover, the typical curving effect within the angle gather is observed when analyzing Figure~\ref{fig:redatumRecWrongAdcig} \cite[]{biondi2004angle}. The successful application of any target-oriented method, whether is based on local solvers or on a redatuming step, is dependent on the accuracy of the overburden velocity model. Here, we show how our redatuming method is affected by overburden inaccuracies.

\begin{figure}[t]
    \centering
    \subfigure[]{\label{fig:redatumRecWrongRec}\includegraphics[width=0.4\linewidth]{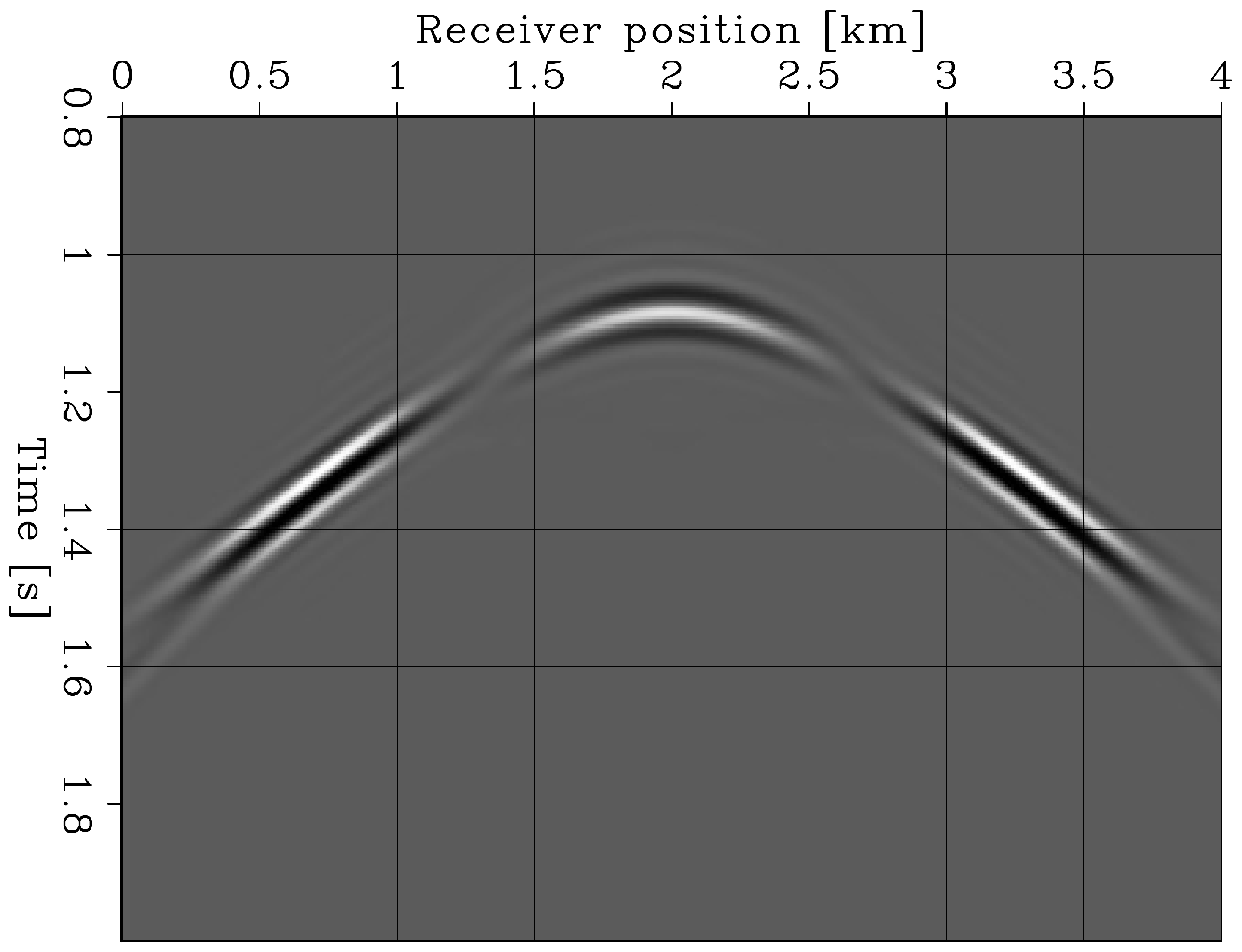}}
    \subfigure[]{\label{fig:redatumRecWrongRecDiff}\includegraphics[width=0.4\linewidth]{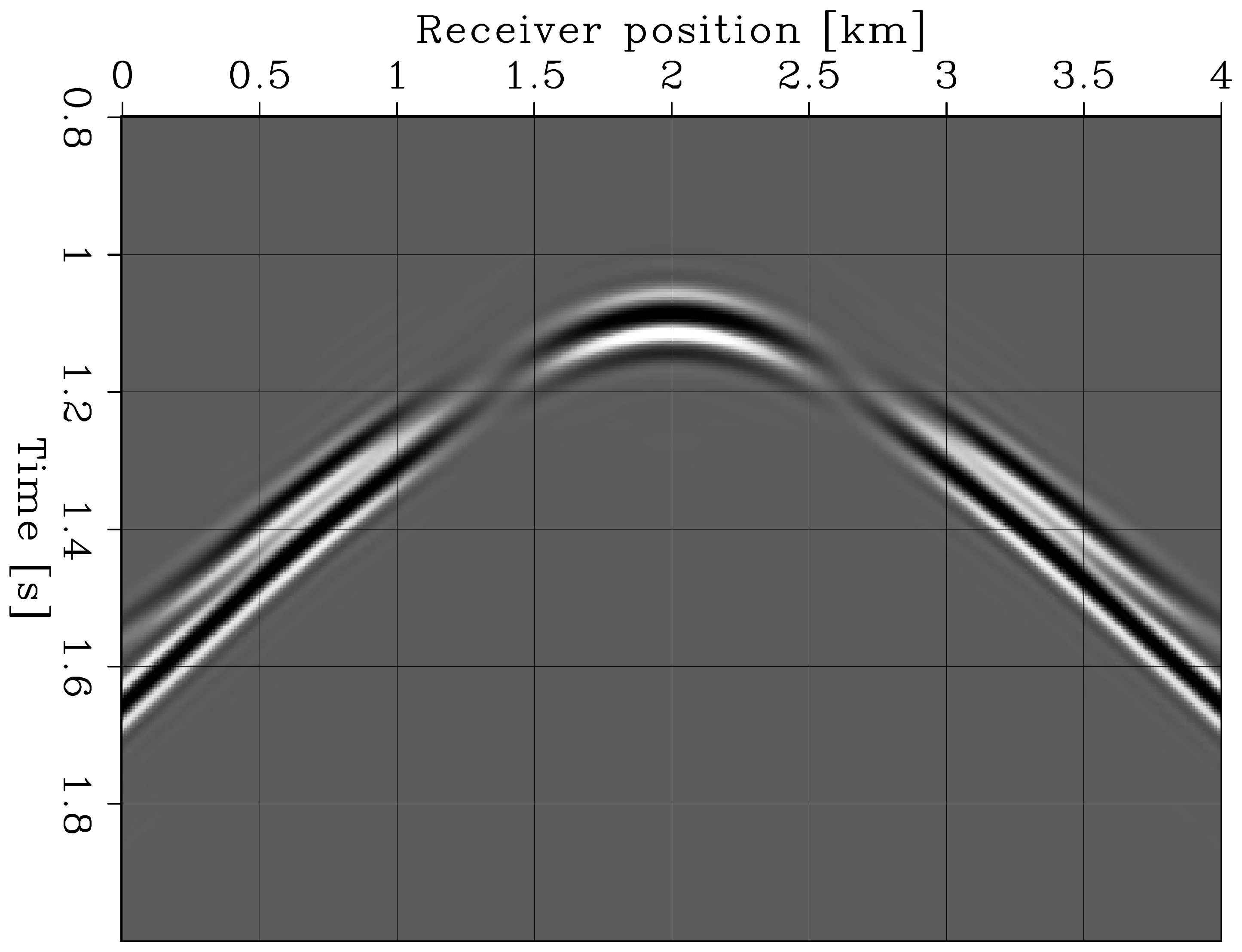}}
    
    \caption{(a) Reconstructed pressure using the extended image obtained using an incorrect velocity model. (b) Difference between panel (a) and the shot gather of Figure~\ref{fig:redatumFlatReflMid}.}
    \label{fig:redatumRecWrong}
\end{figure}

When the ODCIGs obtained using the slower migration velocity are demigrated to reconstruct the datumed elastic pressure, the AVO pattern is reconstructed but the kinematics of the events result incorrect (Figure~\ref{fig:redatumRecWrong}). However, when the data are reconstructed at the original acquisition depth, then both kinematics and amplitude effects are perfectly reconstructed.

This test demonstrates the importance of obtaining an accurate migration velocity model of the overburden before performing the redatuming step. This observation is generally true for any other redatuming technique. However, the proposed technique, since it is based on an imaging step, provides a quality control step thanks to the kinematic behavior of the generated ODCIGs and ADCIGs with respect to the migration velocity model.

\subsection{Elastic target-oriented inversion applied to the Marmousi2 model}
\begin{figure}[t]
    \centering
    \subfigure[]{\label{fig:MarmElaVpTrue}\includegraphics[width=0.8\linewidth]{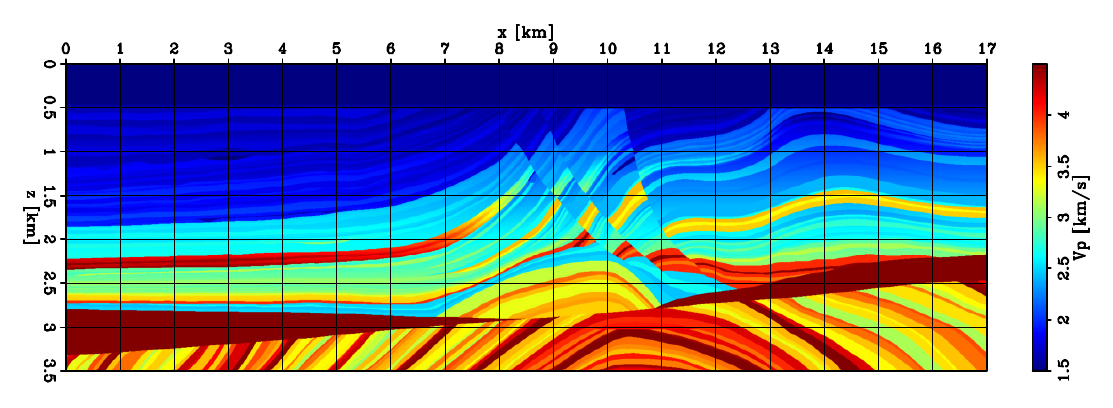}}
    
    \subfigure[]{\label{fig:MarmElaVsTrue}\includegraphics[width=0.8\linewidth]{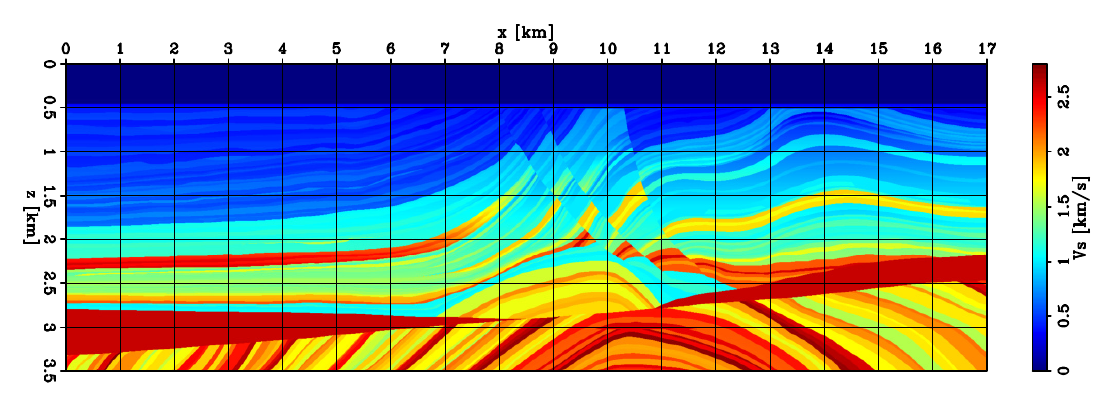}}
    
    \subfigure[]{\label{fig:MarmElaRhoTrue}\includegraphics[width=0.8\linewidth]{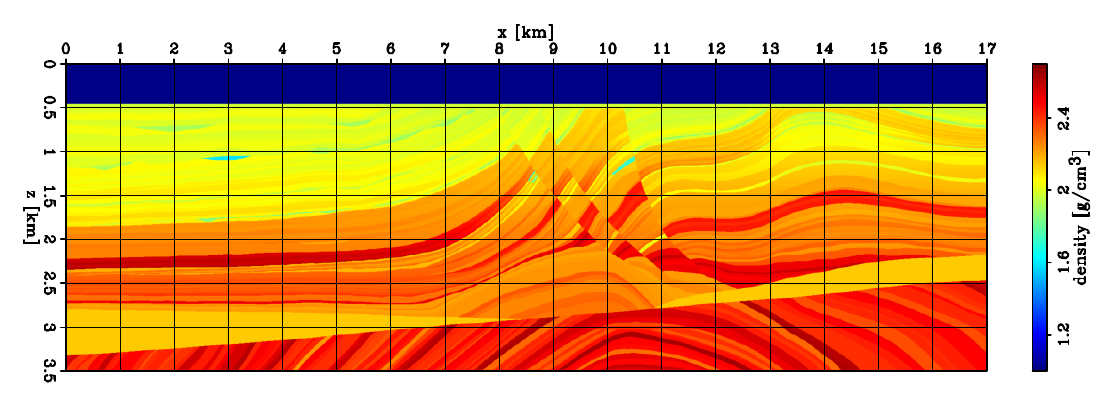}}
    
    \caption{Elastic parameters of the Marmousi2 model. From top to bottom: (a) P-wave velocity, (b) S-wave velocity, (c) density.}
    \label{fig:MarmElaTrue}
\end{figure}

We apply the described redatuming and elastic inversion techniques to the Marmousi2 model to estimate the elastic parameters associated with a gas-bearing reservoir located within a faulted anticline structure. The true subsurface elastic parameters are displayed in Figure~\ref{fig:MarmElaTrue}. This gas reservoir is located at a depth of $1.1$ km and spans approximately $500$ m in the horizontal direction, starting from $x=10$ km.

First, we apply an elastic FWI workflow to a surface dataset to retrieve the entire model's subsurface parameters starting from a smoothed version of the true model. Then, we solve an extended linearized waveform inversion to synthesize the reflected events generated by the gas reservoir recorded with an acquisition located in its vicinity. The redatumed dataset is then used within the same elastic FWI workflow to estimate the reservoir's elastic properties. Finally, we compare the target-oriented results with the elastic FWI applied to the entire surface dataset. 

\begin{figure}[t]
    \centering
    \subfigure[]{\label{fig:MarmWaveletTime}\includegraphics[width=0.45\columnwidth]{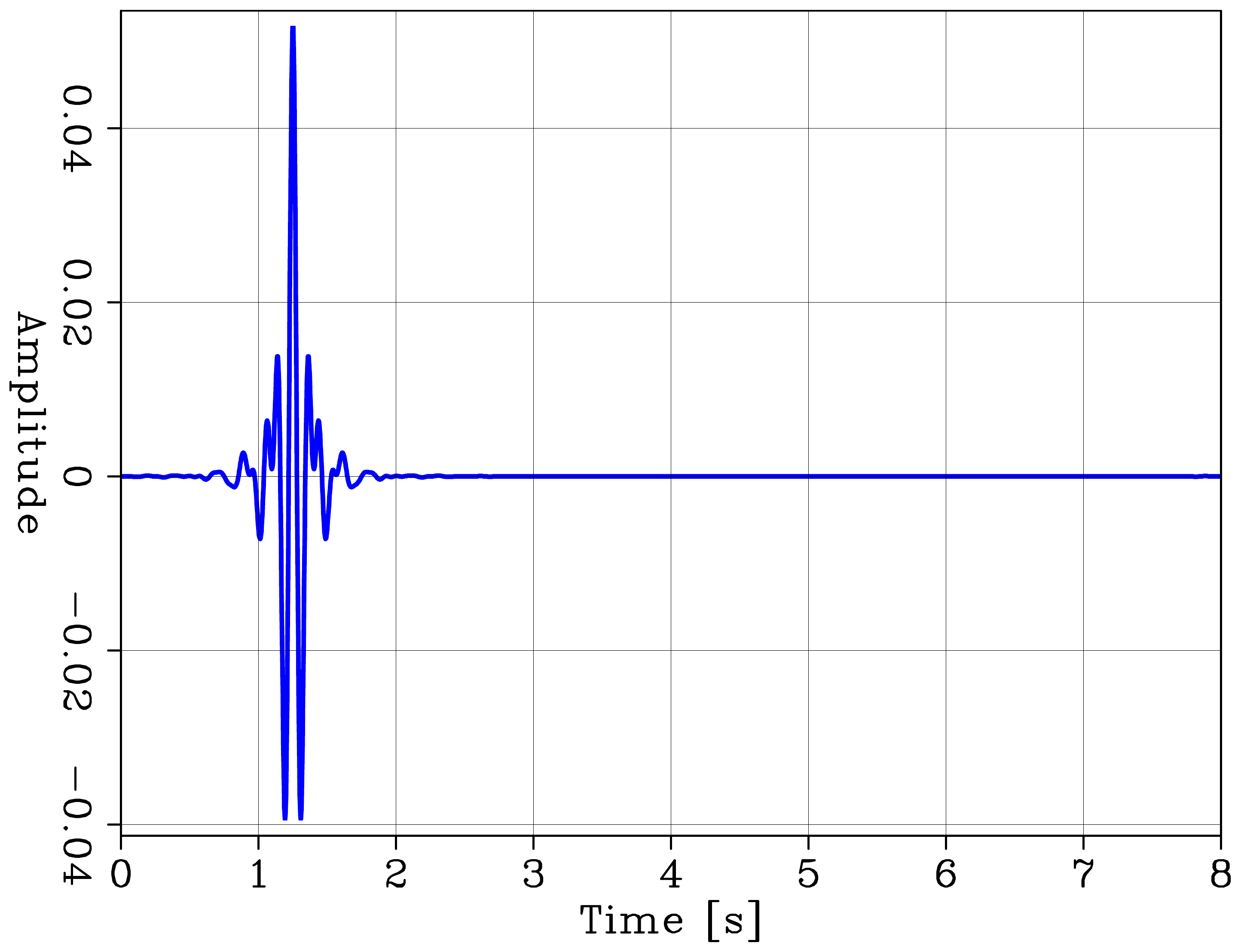}}
    \subfigure[]{\label{fig:MarmWaveletSpectrum}\includegraphics[width=0.455\columnwidth]{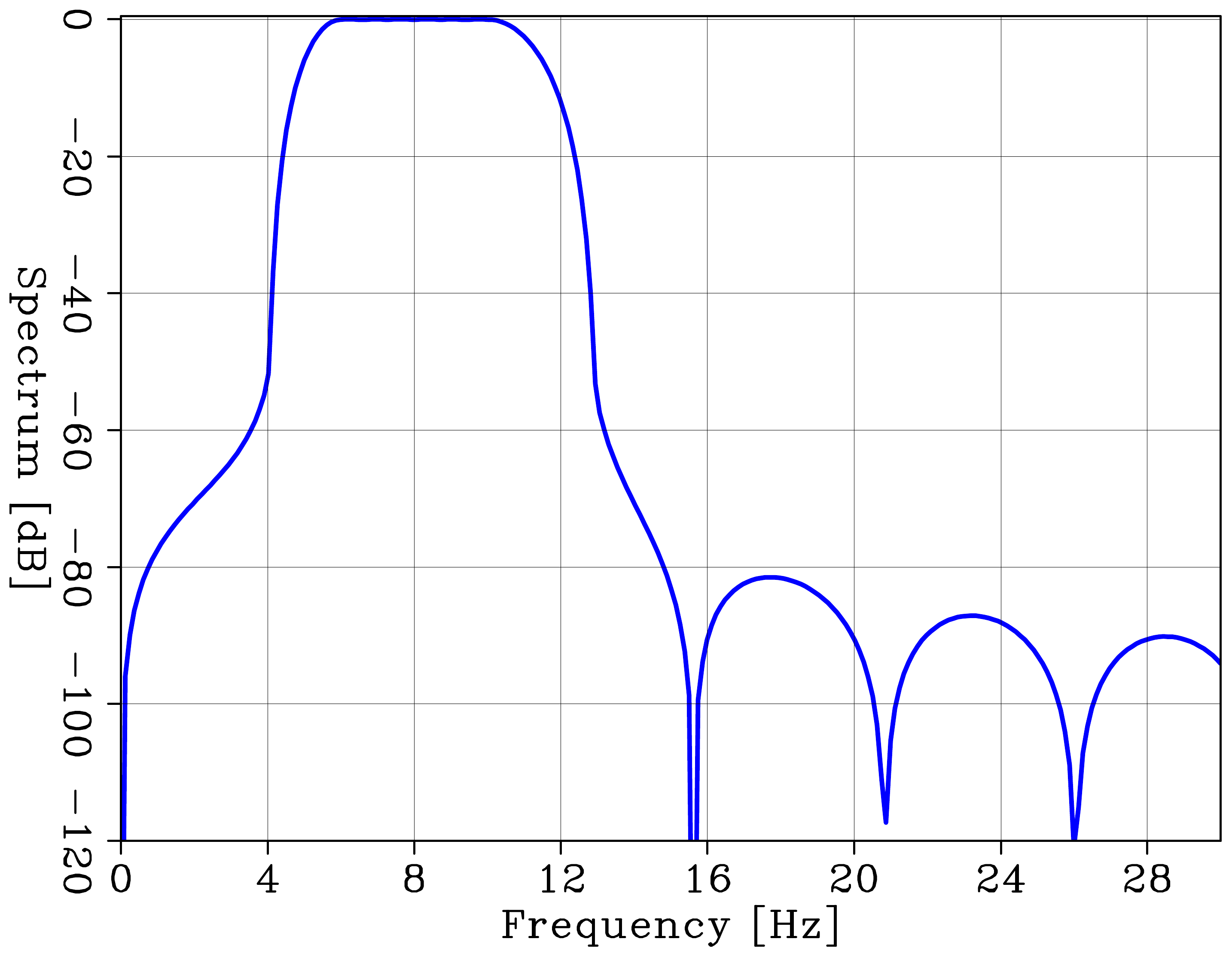}}
    \caption{Explosive source used to generate the elastic pressure data on the elastic Marmousi2 model. Panel (a) and (b) show the time signature and frequency spectrum, respectively.}
    \label{fig:MarmWavelet}
\end{figure}

The observed elastic pressure data is generated from a surface acquisition composed of 140 sources and 567 receivers spaced by $120$ m and $30$ m along the x-axis, respectively. The modeling is performed using absorbing boundaries around all the four edges of the simulation domain~\cite[]{israeli1981approximation}. Figure~\ref{fig:MarmWaveletTime} shows the time signature of the explosive source employed in this synthetic experiment. This wavelet's frequency content is effectively contained between $4$ and $13$ Hz with a flat response between $6$ and $10$ Hz (Figure~\ref{fig:MarmWaveletSpectrum}). The choice of the lowest frequency wants to simulate a field scenario in which the low-frequency content is commonly removed given its low signal-noise ratio (SNR).

\begin{figure}[t]
    \centering
    \subfigure[]{\label{fig:MarmShotlLeft}\includegraphics[width=0.45\linewidth]{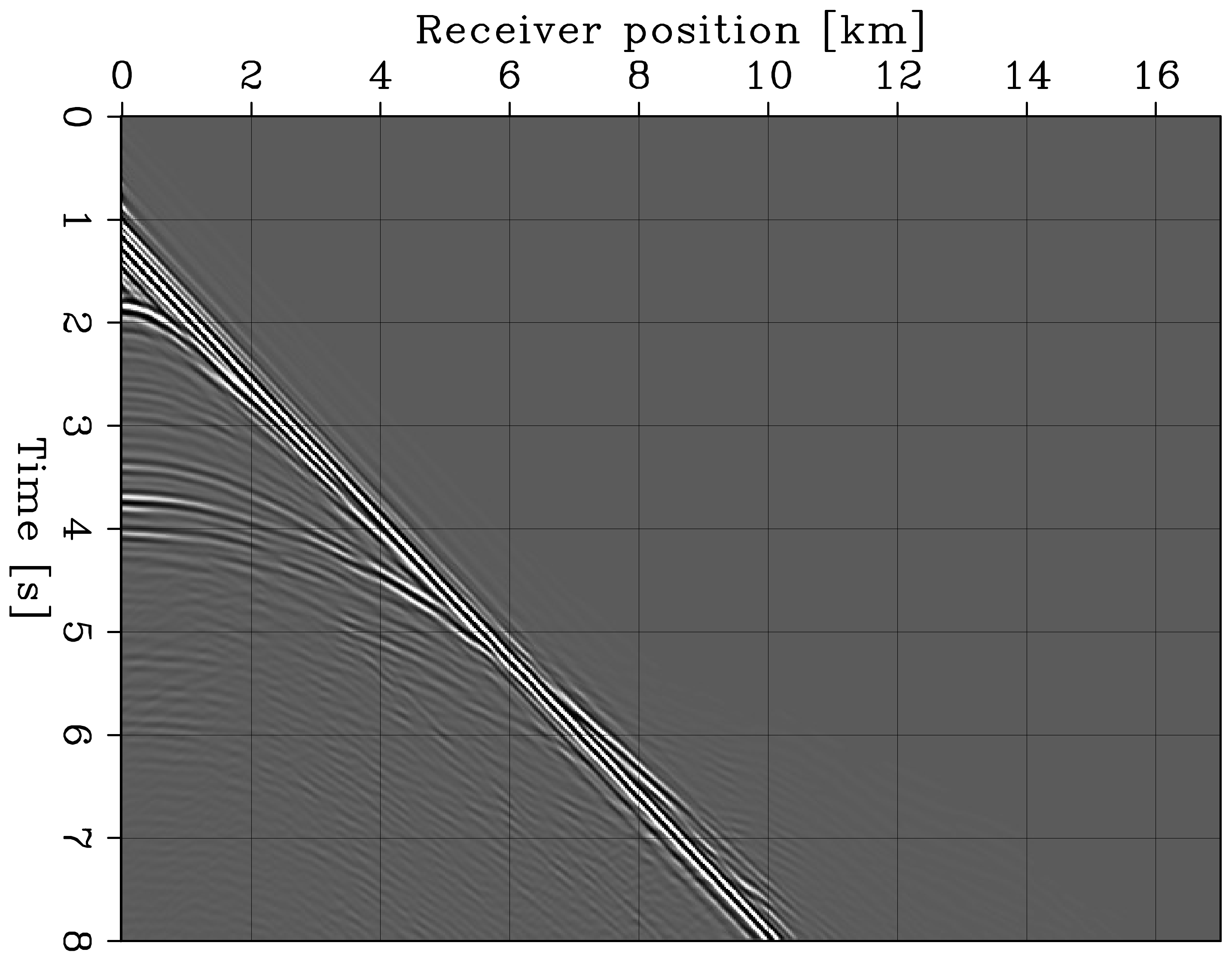}}
    \subfigure[]{\label{fig:MarmShotlMid}\includegraphics[width=0.45\linewidth]{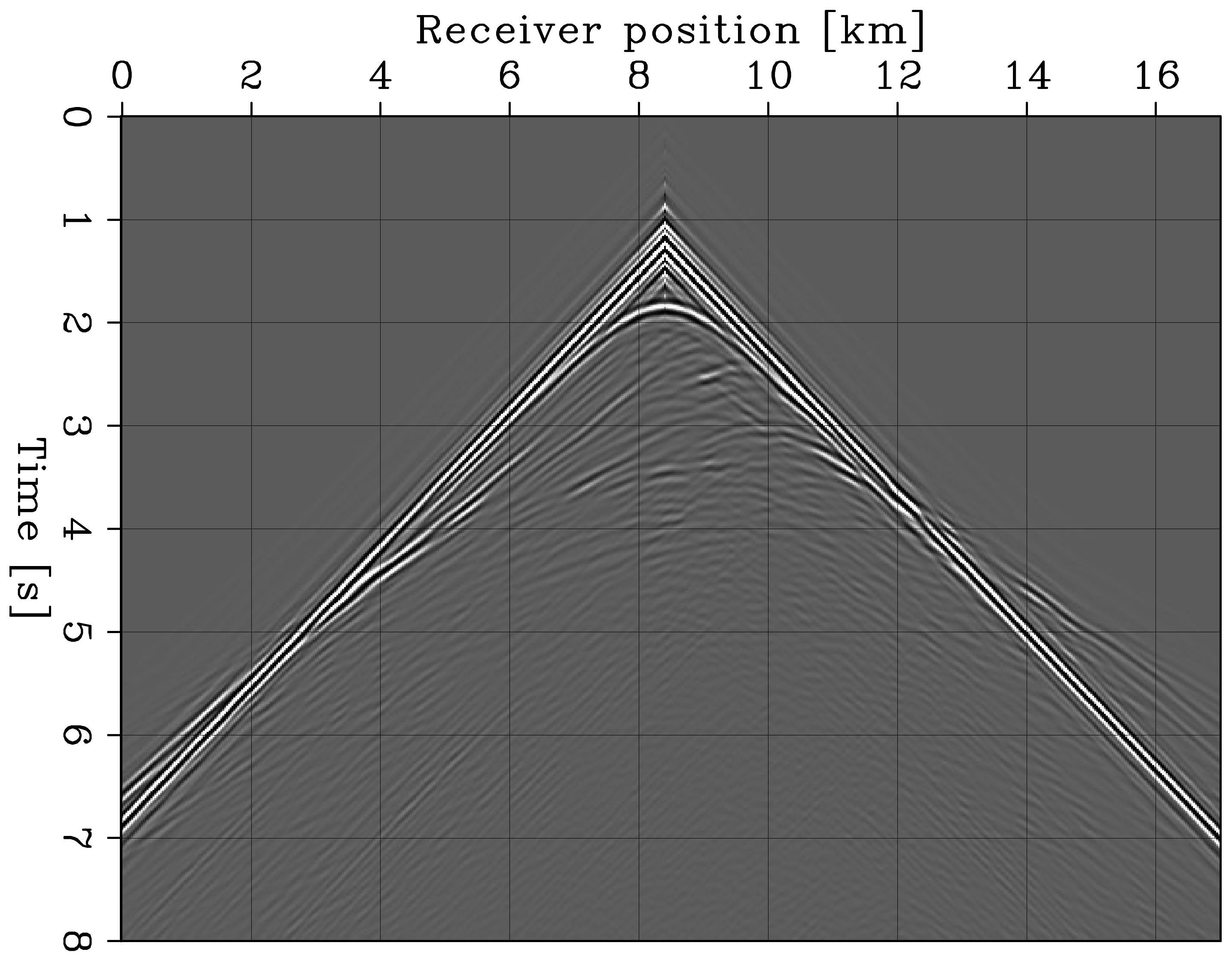}}
    
    \caption{Representative elastic pressure shot gathers for sources placed at (a) $x=0.0$ km and (b) $x=8.5$ km on the Marmousi2 model.}
    \label{fig:MarmShots}
\end{figure}

Given the amplitude response of the reflected events from the subsurface interfaces, the dataset is mostly dominated by reflections. In fact, by analyzing the two representative shot gathers displayed in Figure~\ref{fig:MarmShots}, the mentioned reflections present greater amplitudes than the transmitted waves. This presence of these reflected events represents the ideal application scenario for the redatuming technique. The linearized waveform inversion process can map the AVO of the reflected events within the extended image space.

\begin{figure}[tbhp]
    \centering
    \subfigure[]{}[]{\label{fig:MarmElaVpInit}\includegraphics[width=0.8\linewidth]{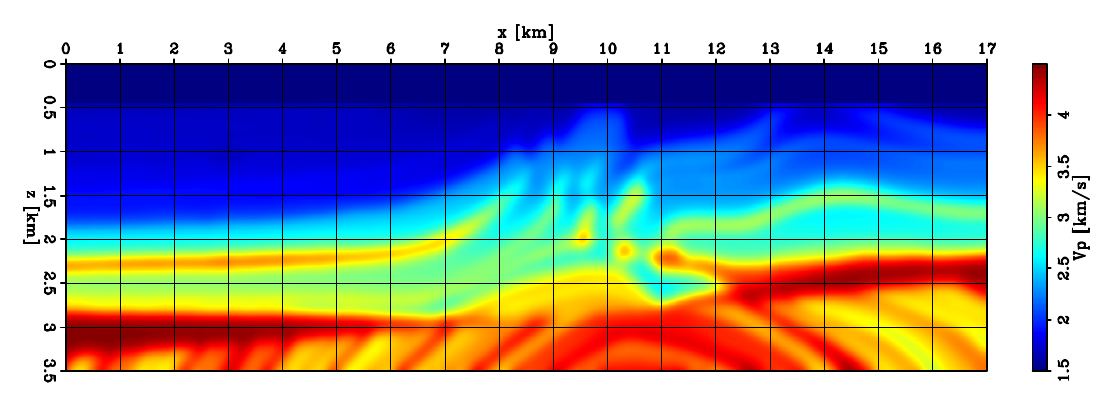}}
    
    \subfigure[]{\label{fig:MarmElaVsInit}\includegraphics[width=0.8\linewidth]{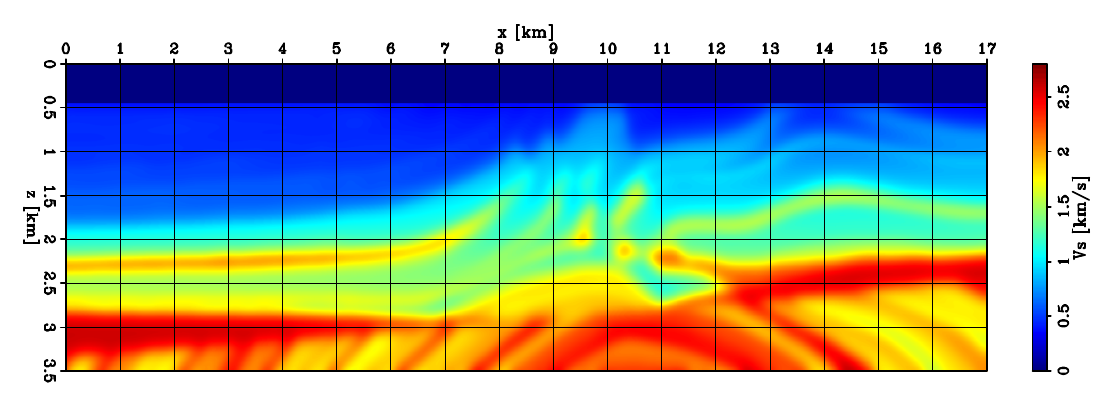}}
    
    \subfigure[]{\label{fig:MarmElaRhoInit}\includegraphics[width=0.8\linewidth]{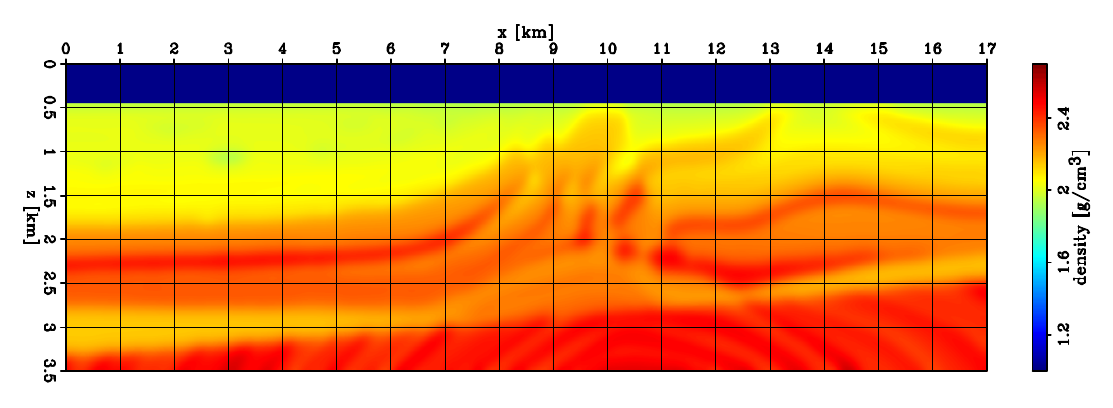}}
    
    \caption{Initial elastic parameters obtained by smoothing the sediments composing the Marmousi2 model. From top to bottom: (a) P-wave velocity, (b) S-wave velocity, (c) density.}
    \label{fig:MarmElaInit}
\end{figure}

The initial elastic parameters are obtained by applying a moving average filter to the true model (Figure~\ref{fig:MarmElaInit}). This process produces an accurate initial elastic model and mitigates the possibility of falling into a non-useful local minimum given the chosen frequency content. Additionally, all the short-wavelength features of all the reservoirs present within the subsurface are entirely missing from the initial guess.

\begin{figure}[t]
    \centering
    \includegraphics[width=0.7\linewidth]{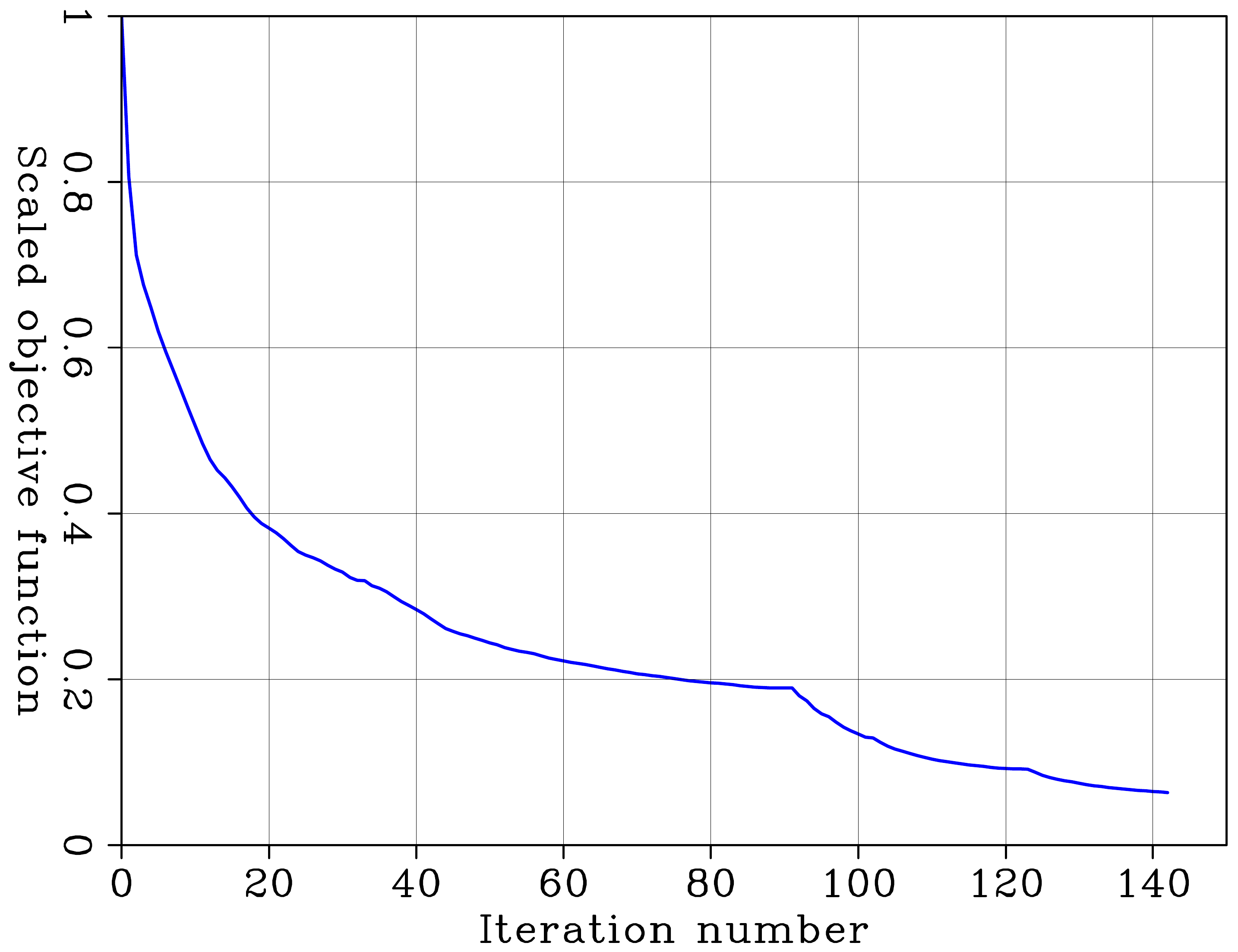}
    \caption{Convergence curve of the Marmousi2 elastic FWI problem. The two changes in convexity of the curve at 90 and 125 iterations are due to the change in spline grid of the elastic parameters.}
    \label{fig:MarmElaObj}
\end{figure}

The full bandwidth of the data is simultaneously injected, and the three elastic parameters are jointly inverted within an elastic FWI procedure. Moreover, we apply the model-space multi-scale approach described in the previous chapter to mitigate the presence of local minima and mitigate any spatial artifacts that may arise during the inversion process. Three sequentially refined spline grids are employed, namely, $100$ m, $50$ m, and $25$ m spacing, while the propagation is performed with a $5$ m sampling in both directions. For each spline grid, the elastic FWI process employs the L-BFGS optimization method, and the inversion is stopped when an appropriate step-length value cannot be found. The convergence curve obtained using the described elastic FWI workflow is shown in Figure~\ref{fig:MarmElaObj}. The first spline grid reaches the closest local minimum after 90 iterations and achieves a relative objective function decrease of more than 80\%. The final elastic model is then projected onto a finer spline grid, and then other 35 iterations are employed to further diminish the objective function. The spline refinement is performed again to obtain an additional objective function decrease.

\begin{figure}[t]
    \centering
    \subfigure[]{}[]{\label{fig:MarmElaVpInv}\includegraphics[width=0.8\linewidth]{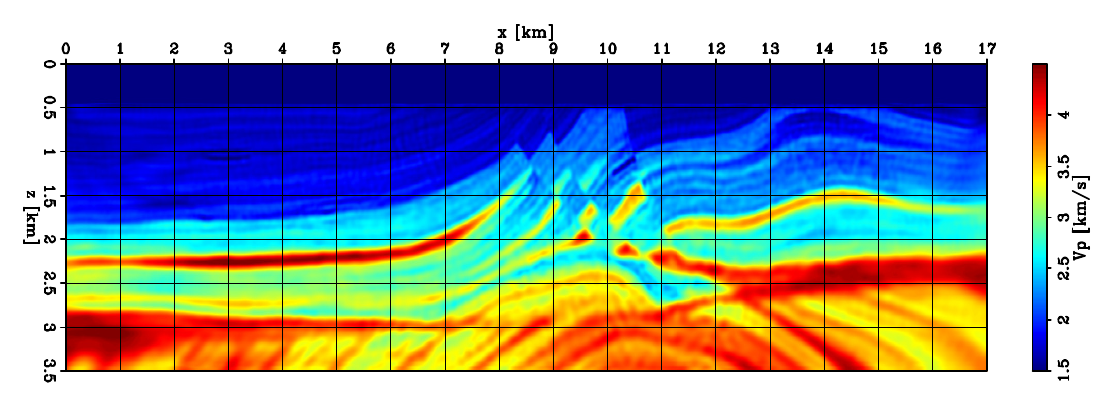}}
    
    \subfigure[]{\label{fig:MarmElaVsInv}\includegraphics[width=0.8\linewidth]{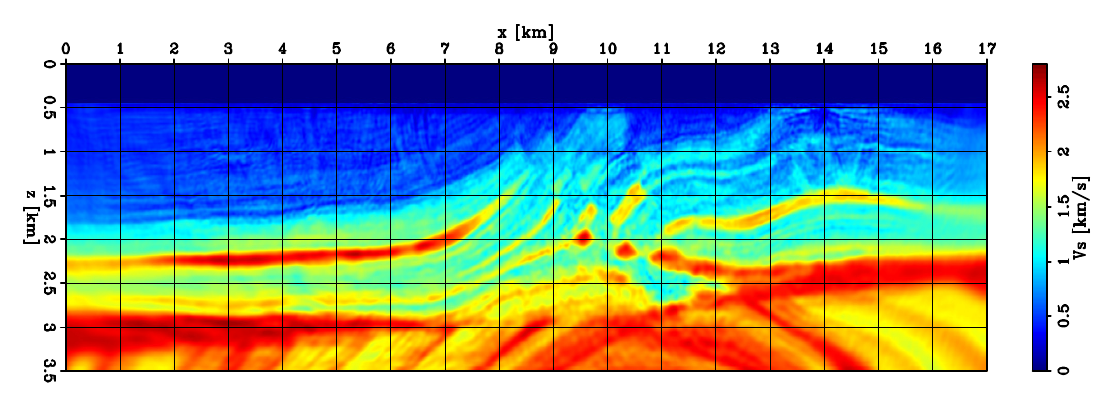}}
    
    \subfigure[]{\label{fig:MarmElaRhoInv}\includegraphics[width=0.8\linewidth]{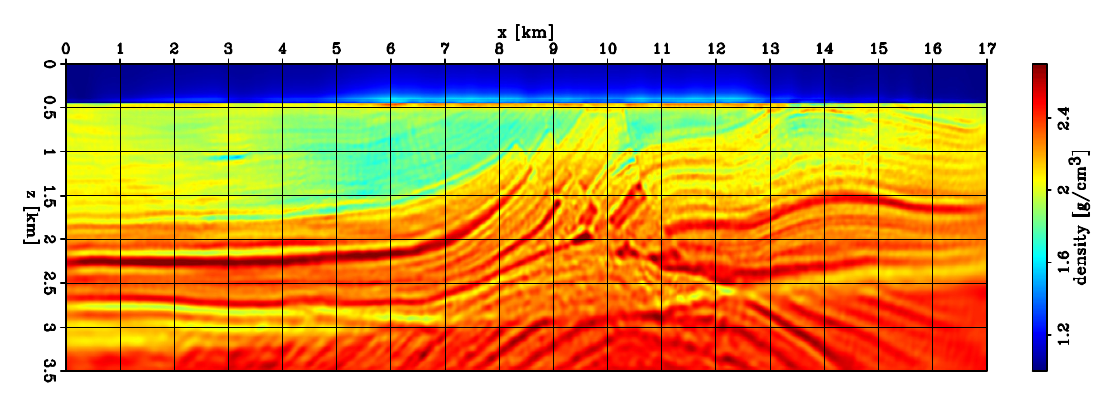}}
    
    \caption{Inverted elastic parameters obtained by solving the elastic FWI defined on the Marmousi2 model. From top to bottom: (a) P-wave velocity, (b) S-wave velocity, (c) density.}
    \label{fig:MarmElaInv}
\end{figure}

The panels in Figure~\ref{fig:MarmElaInv} show the final elastic parameters obtained at the end of the described elastic FWI workflow. The P-wave velocity is the parameter accurately retrieved and does not present any evident artifacts. On the other hand, the S-wave velocity is affected by some inversion artifacts and a potential cross-talk positioned at $x=3.0$ km and $z=1.0$ km. However, overall this parameter is in agreement with the true one shown in Figure~\ref{fig:MarmElaVsTrue}. The density parameter is also in good agreement with the true one, and the anomaly associated with the gas reservoir is correctly retrieved.

\begin{figure}[t]
    \centering
    \subfigure[]{\label{fig:MarmElaModObsDataInit}\includegraphics[width=0.65\linewidth]{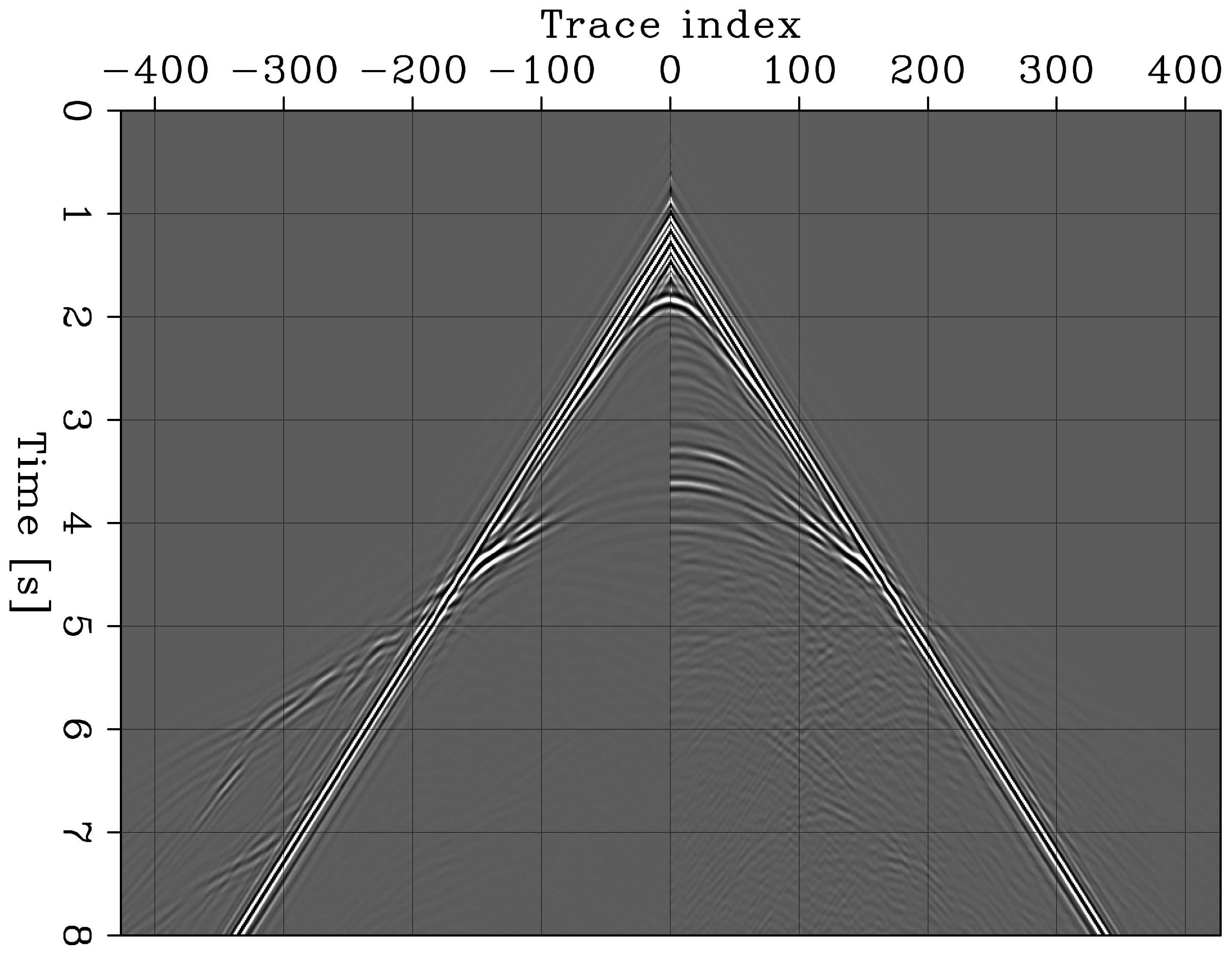}}
    
    \subfigure[]{\label{fig:MarmElaModObsDataInv}\includegraphics[width=0.65\linewidth]{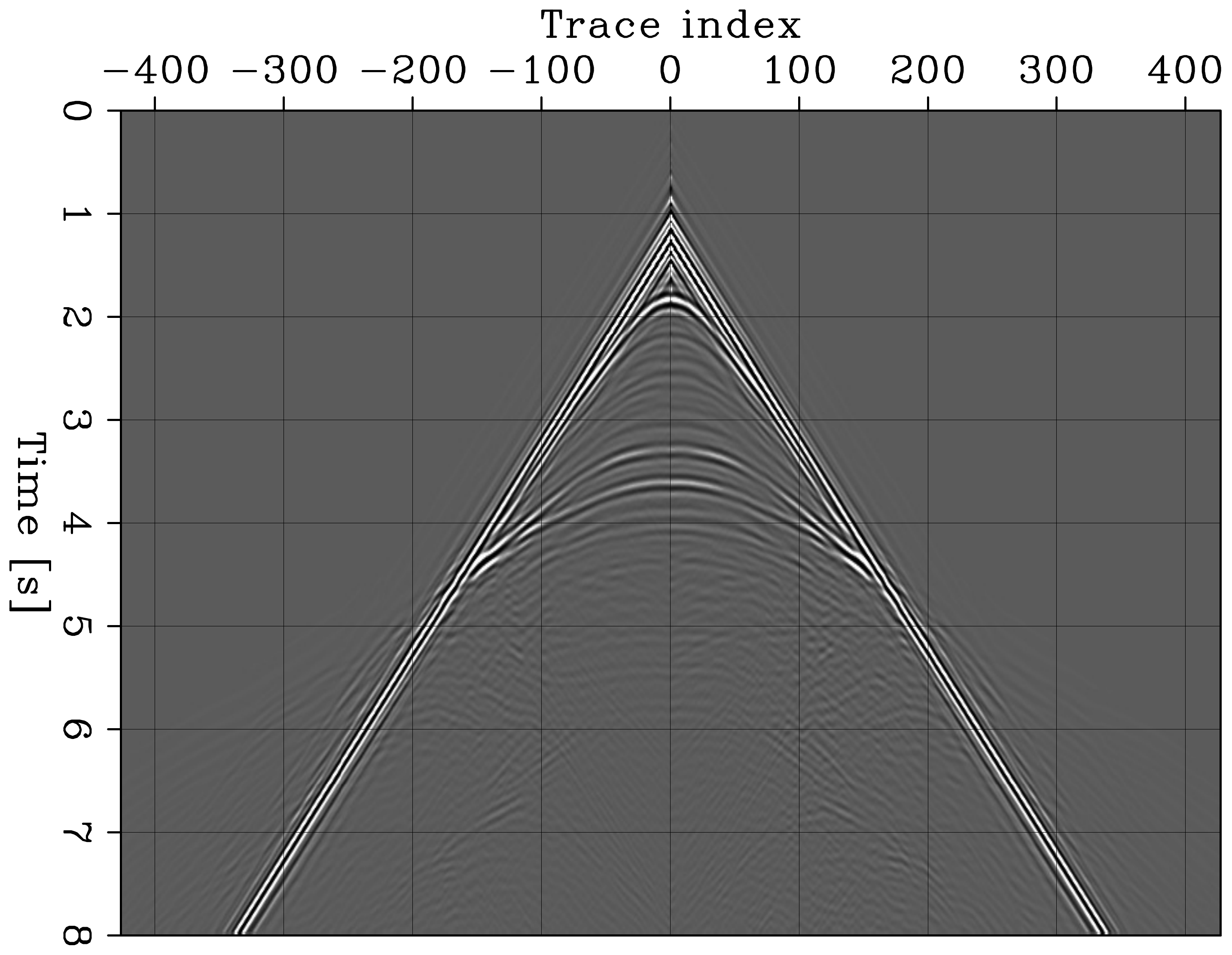}}
    
    \caption{Comparison between the predicted and observed elastic pressure data on the initial (a) and inverted (b) models, respectively. The negative trace indices indicate the predicted data, while the positive ones denote the observed data. Both panel are shown using the same gain.}
    \label{fig:MarmElaModObsData}
\end{figure}

\begin{figure}[t]
    \centering
    \subfigure[]{\label{fig:MarmElaResInit}\includegraphics[width=0.65\linewidth]{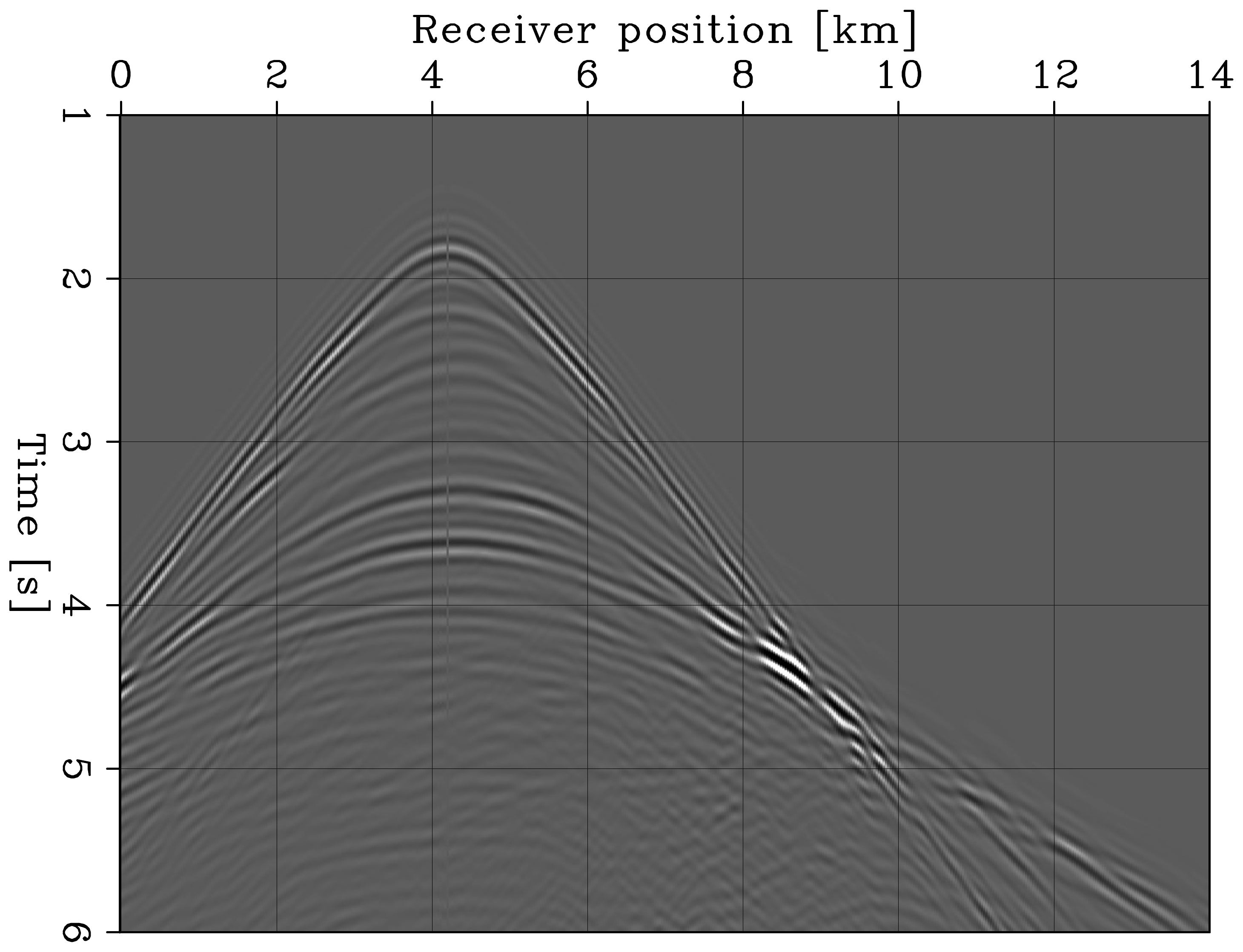}}
    
    \subfigure[]{\label{fig:MarmElaResInv}\includegraphics[width=0.65\linewidth]{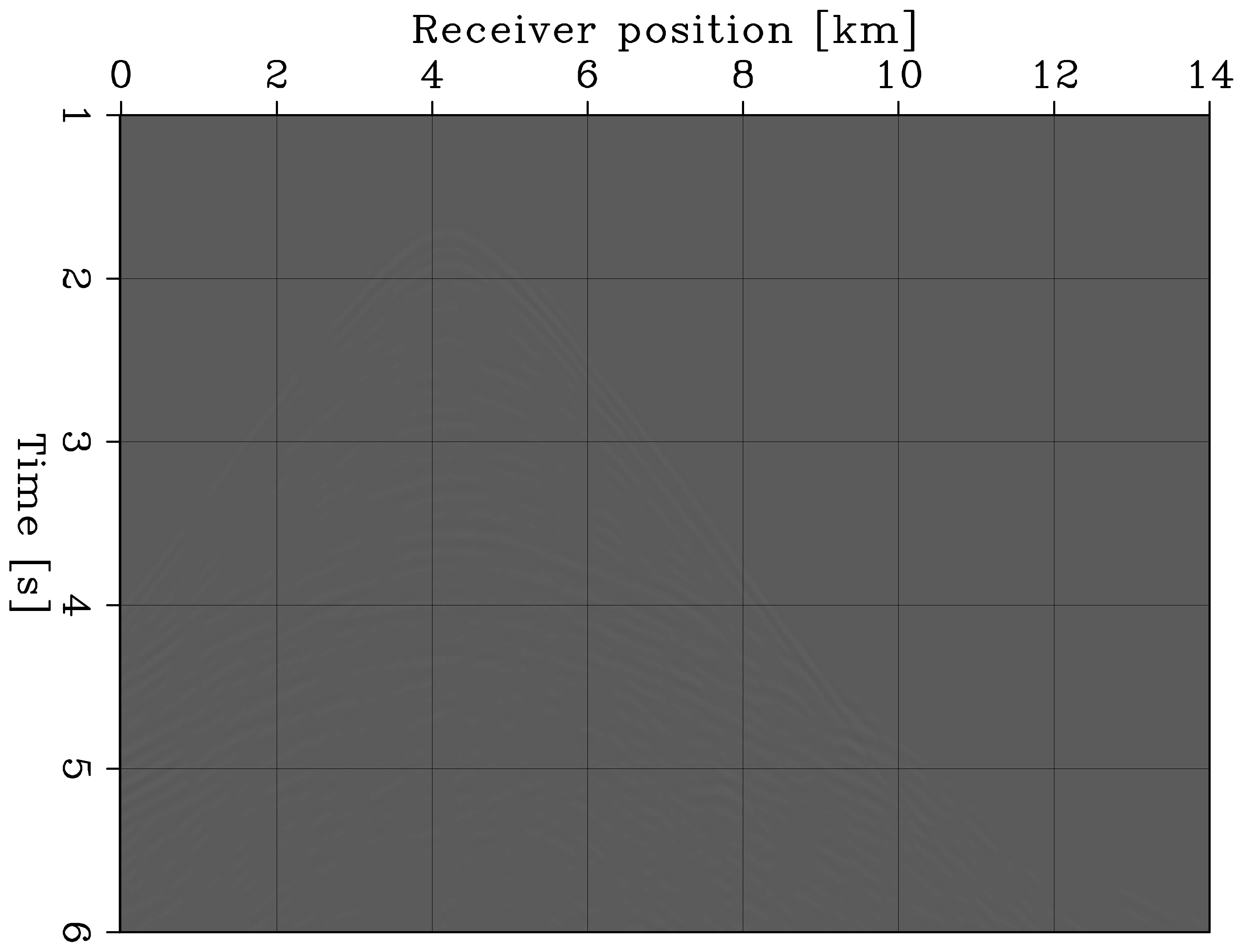}}
    
    \caption{Representative source gather showing the initial (a) and final (b) residual pressure data for the surface-acquisition inversion. Both panels are displayed using the same gain.}
    \label{fig:MarmElaRes}
\end{figure}

To evaluate the quality of the inverted elastic model, we display the predicted and the observed data within the same plot to compare the amplitude and timing of the reflected event. The representative shot gather is located at $x=4.190$ km, and only the positive offsets are compared. Figure~\ref{fig:MarmElaModObsDataInit} shows this comparison when the observed data are plotted along with the predicted pressure obtained using the initial elastic model. All the reflected events are not modeled from the initial guess, and a clear mismatch in the long offset events is evident. On the contrary, after applying the FWI workflow (Figure~\ref{fig:MarmElaModObsDataInv}), the predicted data using the inverted elastic parameters are in excellent agreement with the observed events. This observation is also evident when comparing the initial and final data residuals for a representative shot gather (Figure~\ref{fig:MarmElaRes}).

Despite the elastic FWI workflow's ability to retrieve accurate elastic subsurface parameters from surface data, the overall computational cost makes the method hardly applicable to 3D field datasets. In fact, in this 2D synthetic example, each model point evaluation, which comprised of an objective function and gradient evaluations, took approximately 3 hours on an Intel(R) Xeon(R) Gold 6126 CPU @ 2.60GHz connected to 4 Nvidia Tesla V100-PCIe-16GB graphics processing units (GPUs). In the reported example, 180 model points have been tested, making the total elapsed time approximately 540 hours, corresponding to almost 23 days of computation. 

This test shows the high computational cost associated with solving an elastic FWI problem. In field applications, higher frequencies than those used in this test may contain valuable information on the subsurface. However, the increase of the computational cost as the fourth power of the maximum frequency highly limits elastic FWI methodologies' applicability at high frequency. The proposed target-oriented technique has the potential of overcoming this limitation. High-resolution elastic parameters are usually necessary to be estimated only within potential areas of interest or hazard (e.g., over-pressured zones, gas pockets, and natural resource reservoirs). These subsurface targets are recognizable from images generated from surface data, making the image-space redatuming and target-oriented elastic FWI subsequent steps of an exploration project.

\begin{figure}[t]
    \centering
    \subfigure[]{\label{fig:MarmTargTrueVp}\includegraphics[width=0.45\linewidth]{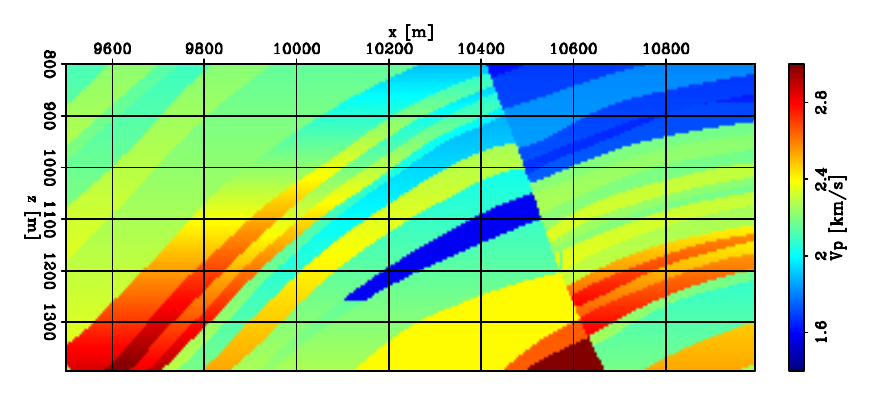}}
    \subfigure[]{\label{fig:MarmTargInitVp}\includegraphics[width=0.45\linewidth]{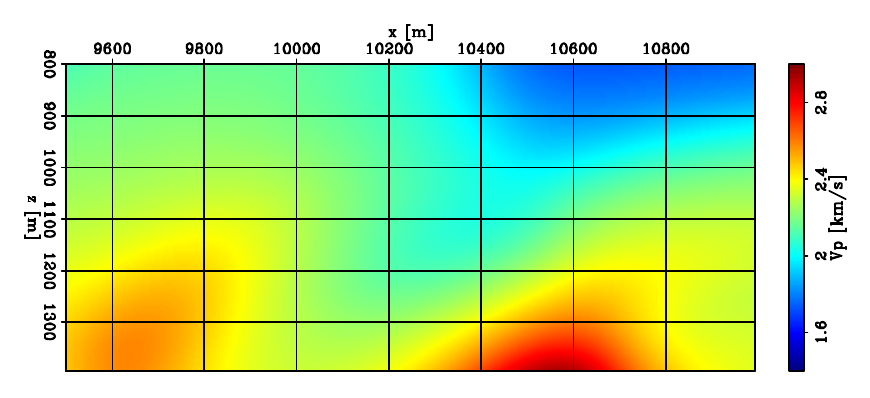}}
    
    \subfigure[]{\label{MarmTargTrueVs}\includegraphics[width=0.45\linewidth]{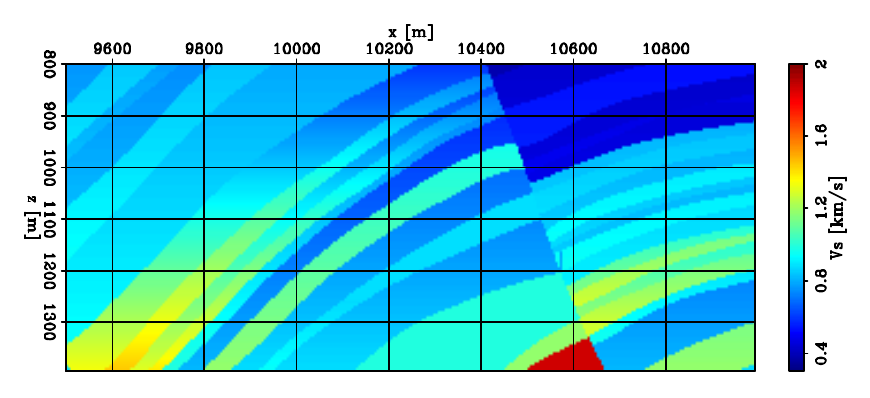}}
    \subfigure[]{\label{fig:MarmTargInitVs}\includegraphics[width=0.45\linewidth]{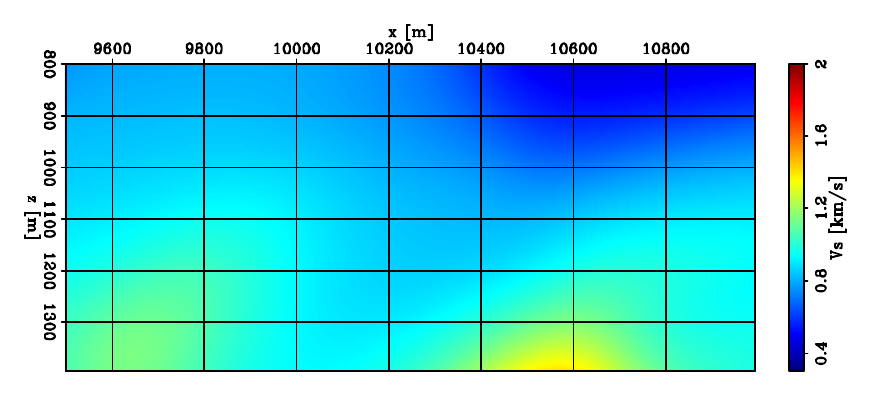}}
    
    \subfigure[]{\label{fig:MarmTargTrueRho}\includegraphics[width=0.45\linewidth]{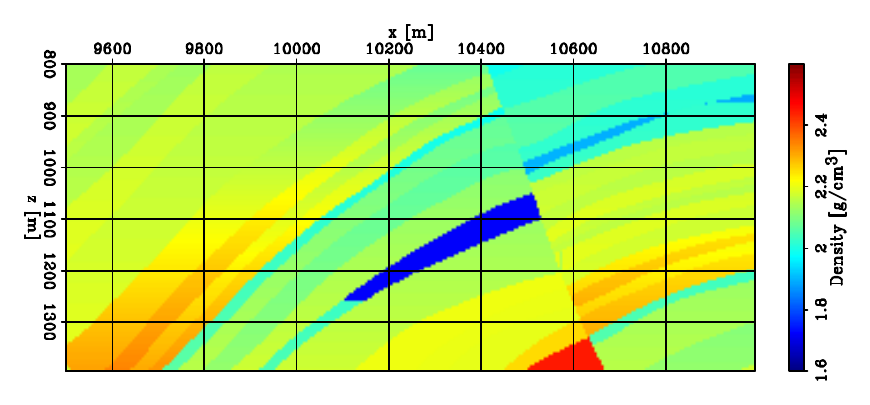}}
    \subfigure[]{\label{fig:MarmTargInitRho}\includegraphics[width=0.45\linewidth]{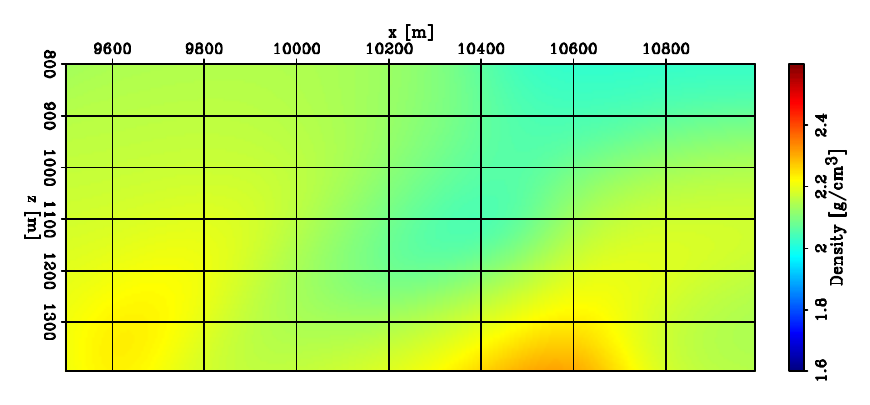}}
    
    \caption{True and initial elastic model parameters of the target area plotted on the left and right columns, respectively: (a-b) P-wave velocity, (c-d) S-wave velocity, and (e-f) density.}
    \label{fig:MarmTargInit}
\end{figure}

As previously mentioned, the goal of this test is to characterize the elastic properties associated with a gas-bearing reservoir placed within the faulted anticline structure. The panels on the left column of Figure~\ref{fig:MarmTargInit} display the elastic parameters of the target structure located at $z = 1100$ m and $x = 10300$ m. A clear decrease in the  P-wave velocity and density parameters is noticeable, while the S-wave velocity does not present such variation. The initial elastic model is obtained by extracting the same parameters from the ones shown in Figure~\ref{fig:MarmElaInit} and are shown in the panels on the right panels of Figure~\ref{fig:MarmTargInit}.

\begin{figure}[t]
    \centering
    \subfigure[]{\label{fig:MarmTargLSRTMObjLog}\includegraphics[width=0.35\linewidth]{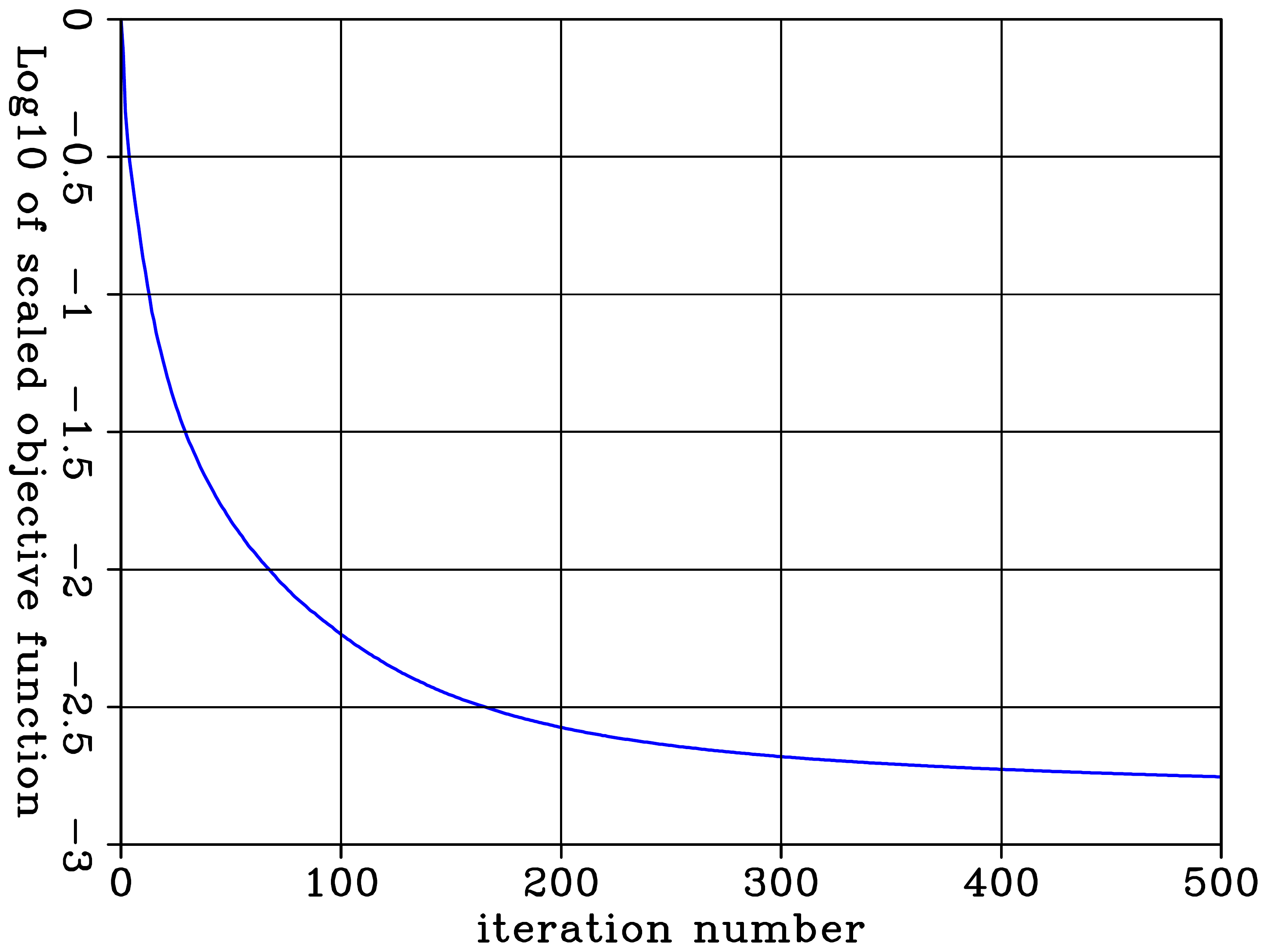}}
    \subfigure[]{\label{fig:MarmTargLSRTMZeroOff}\includegraphics[width=0.5\linewidth]{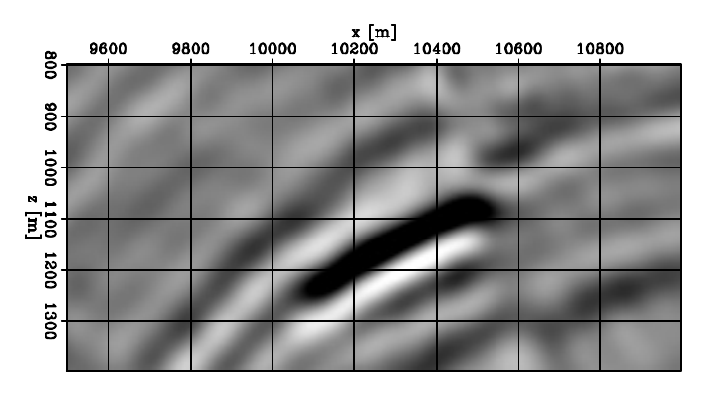}}
    
    \subfigure[]{\label{fig:MarmTargLSRTMADCIG}\includegraphics[width=0.35\linewidth]{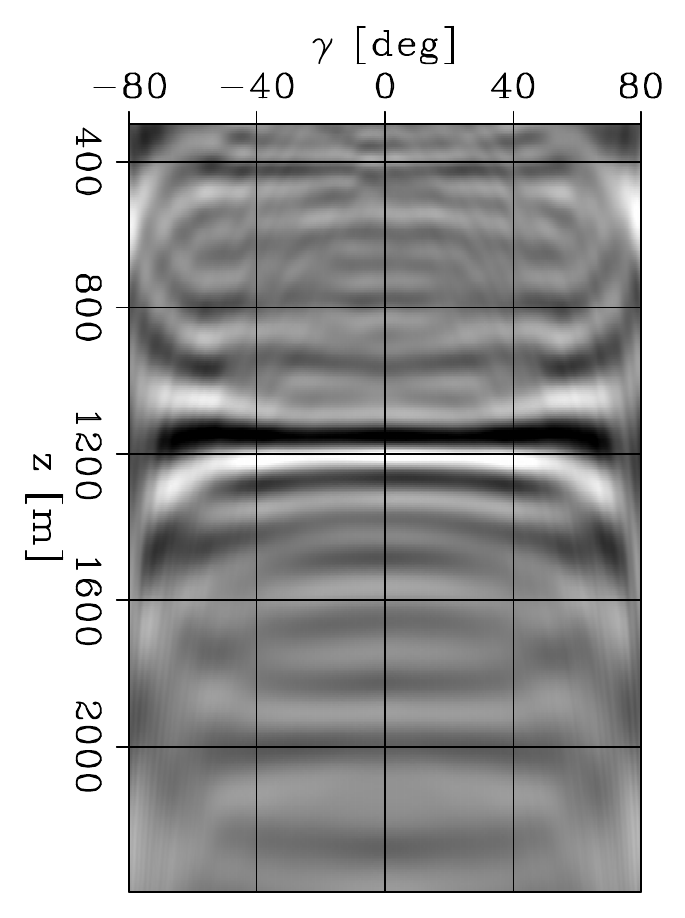}}
    
    \caption{(a) Convergence curve of the extended linearized waveform inversion of the elastic data generated on the Marmousi2 model. (b) Zero-subsurface offset image of the target area. (c) ADCIG extracted at $x = 10.3$ km highlighting a high-amplitude event at $z=1.2$ km associated with the gas reservoir.}
    \label{fig:MarmTargLSRTM}
\end{figure}

The smoothed P-wave velocity parameter of Figure~\ref{fig:MarmElaVpInit} is employed to solve an acoustic extended linearized waveform inversion of the surface elastic pressure data. The same absorbing boundary conditions have been used during the imaging step as those employed in the elastic surface-data computation and inversion. we apply 500 iterations of the linear conjugate-gradient method to reach the numerical minimum of the problem (Figure~\ref{fig:MarmTargLSRTMObjLog}). Within the zero-subsurface offset image of the target area, a high-amplitude response is associated with the reservoir (Figure~\ref{fig:MarmTargLSRTMZeroOff}). Additionally, the subsurface structures are correctly imaged since an accurate velocity model has been employed. This observation is also supported by the flat response of ADCIG extracted at $x=10.3$ km (Figure~\ref{fig:MarmTargLSRTMADCIG}).

\begin{figure}[t!]
    \centering
    \subfigure[]{\label{fig:MarmTargSurfInvVp}\includegraphics[width=0.45\linewidth]{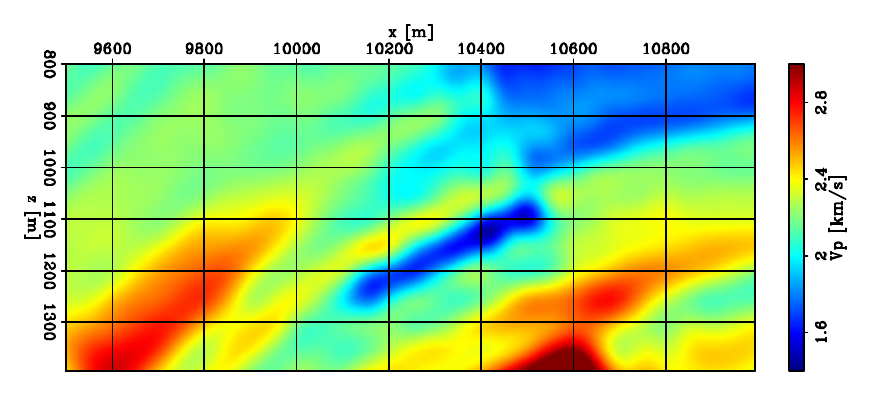}}
    \subfigure[]{\label{fig:MarmTargInvVp}\includegraphics[width=0.45\linewidth]{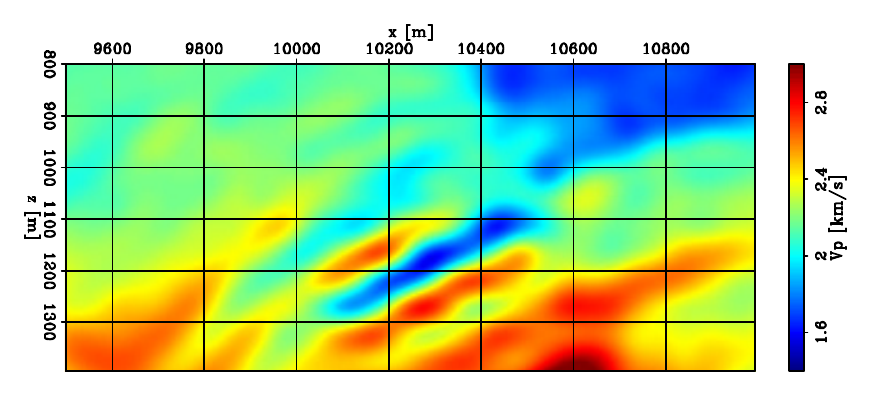}}
    
    \subfigure[]{\label{fig:MarmTargSurfInvVs}\includegraphics[width=0.45\linewidth]{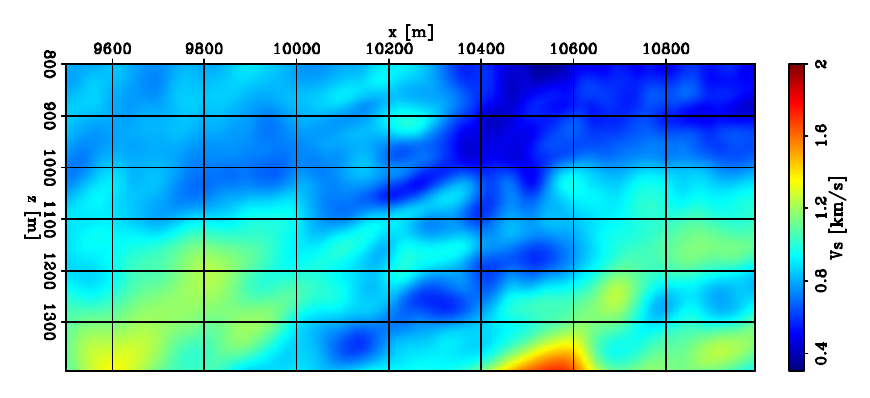}}
    \subfigure[]{\label{fig:MarmTargInvVs}\includegraphics[width=0.45\linewidth]{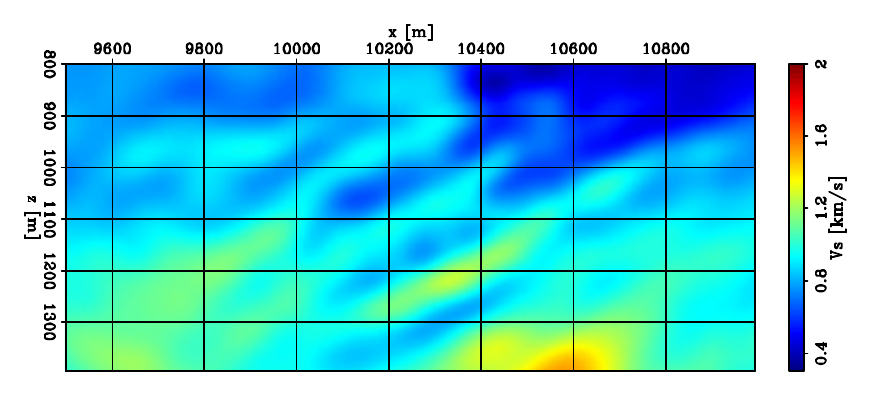}}
    
    \subfigure[]{\label{fig:MarmTargSurfInvRho}\includegraphics[width=0.45\linewidth]{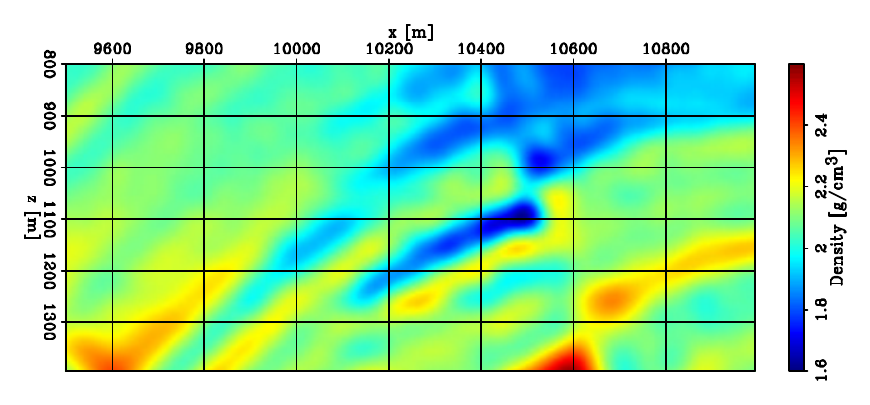}}
    \subfigure[]{\label{fig:MarmTargInvRho}\includegraphics[width=0.45\linewidth]{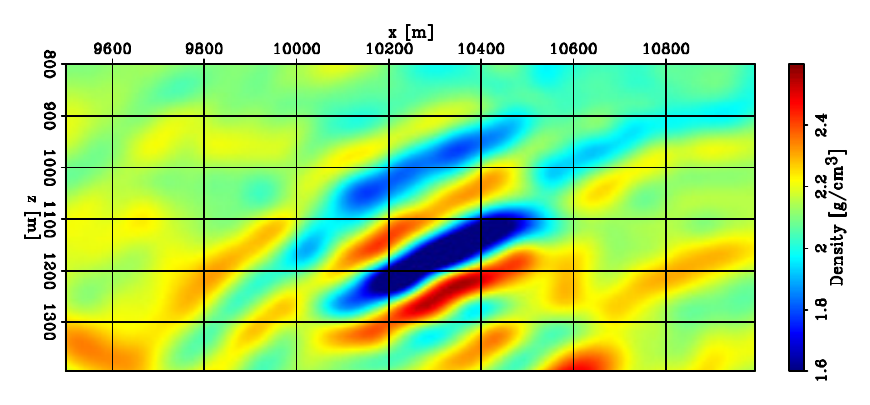}}
    
    \caption{Inverted elastic parameters of the target area obtained from the surface data (left column) and the target-oriented approach (right column): (a-b) P-wave velocity, (c-d) S-wave velocity, and (e-f) density.}
    \label{fig:MarmTargInv}
\end{figure}

\begin{figure}[t]
    \centering
    \includegraphics[width=0.5\linewidth]{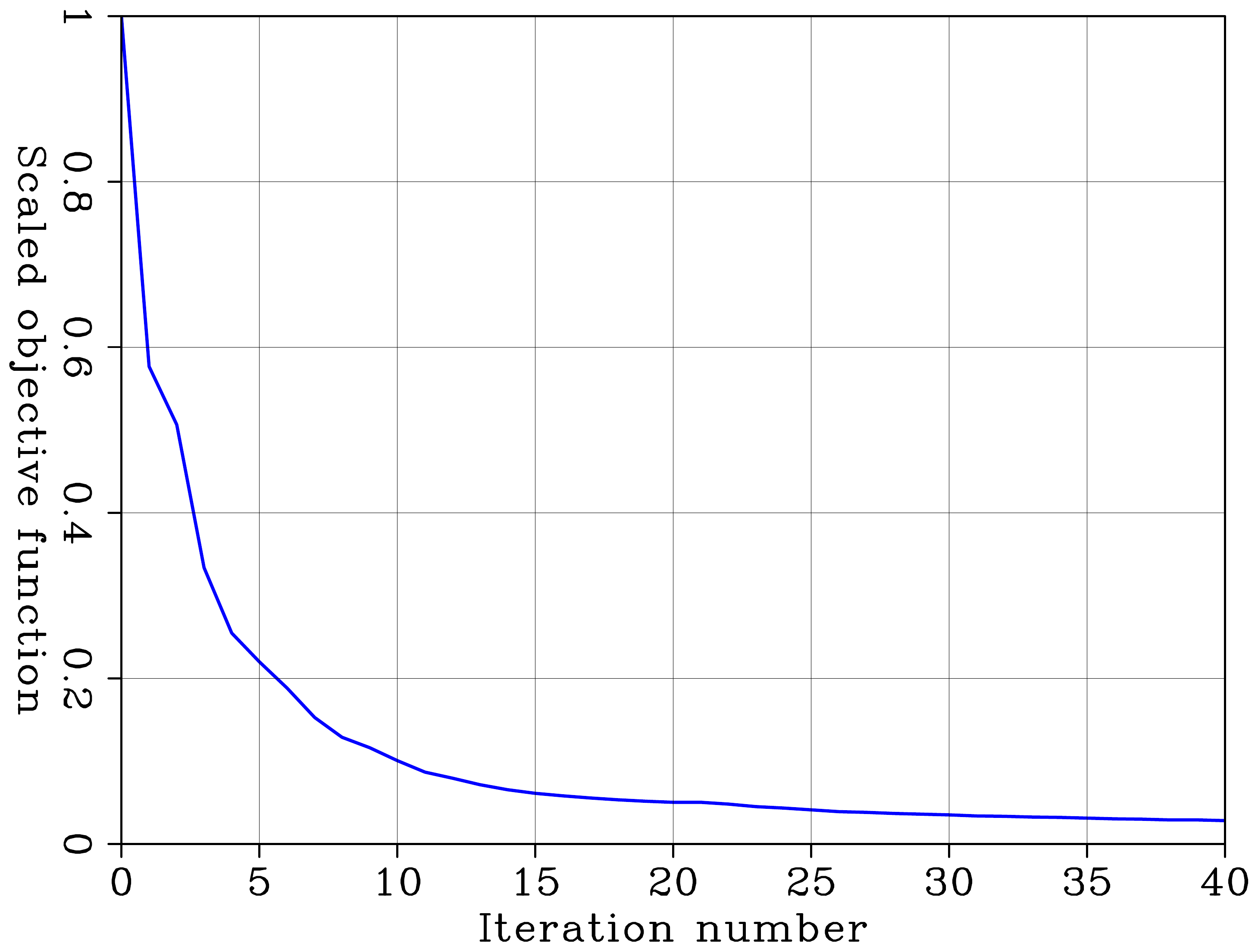}
    \caption{Convergence curve of the target-oriented elastic FWI problem.}
    \label{fig:MarmElaObjTarget}
\end{figure}

\begin{figure}[t]
    \centering
    \subfigure[]{\label{fig:MarmElaResTargetInit}\includegraphics[width=0.65\linewidth]{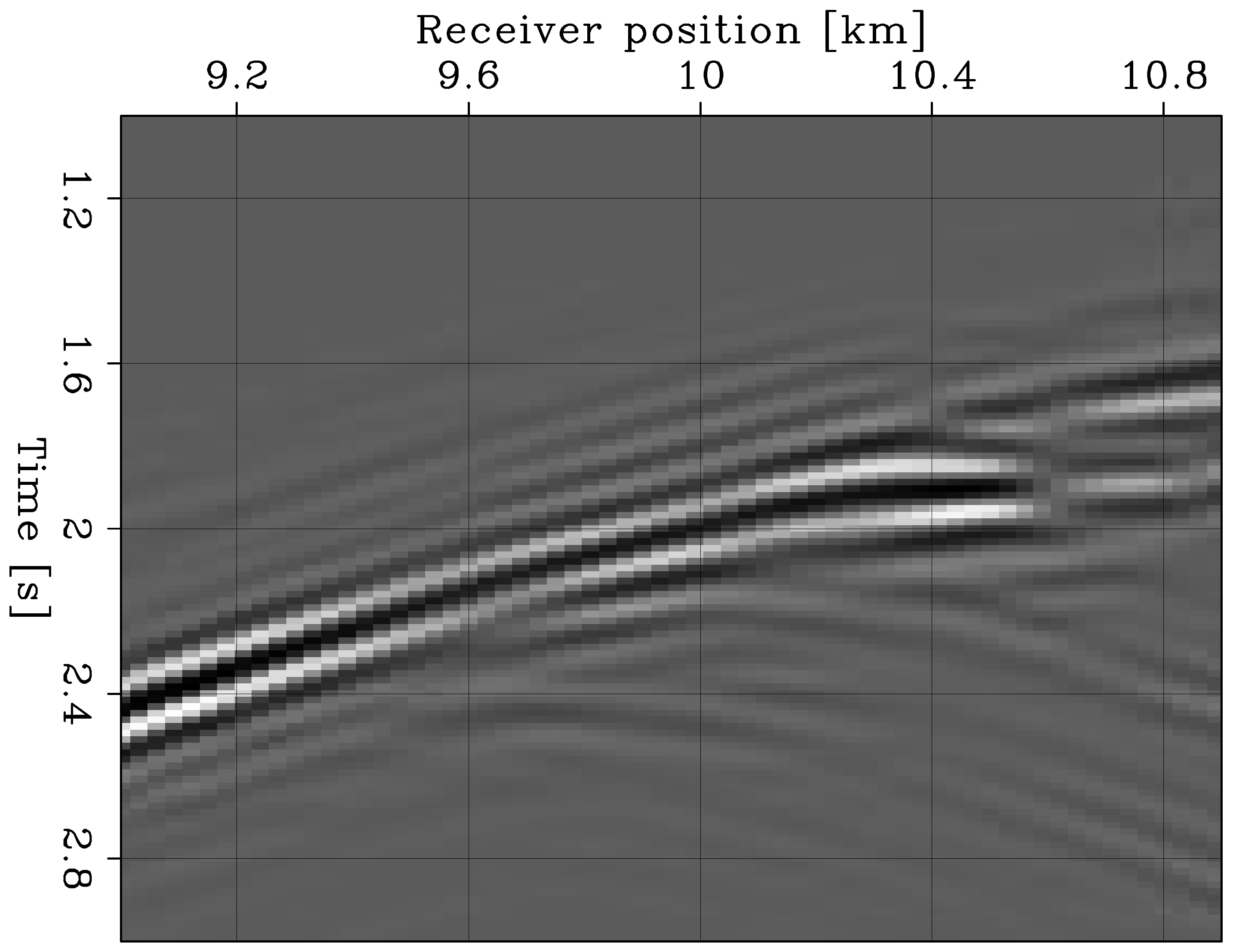}}
    
    \subfigure[]{\label{fig:MarmElaResTargetInv}\includegraphics[width=0.65\linewidth]{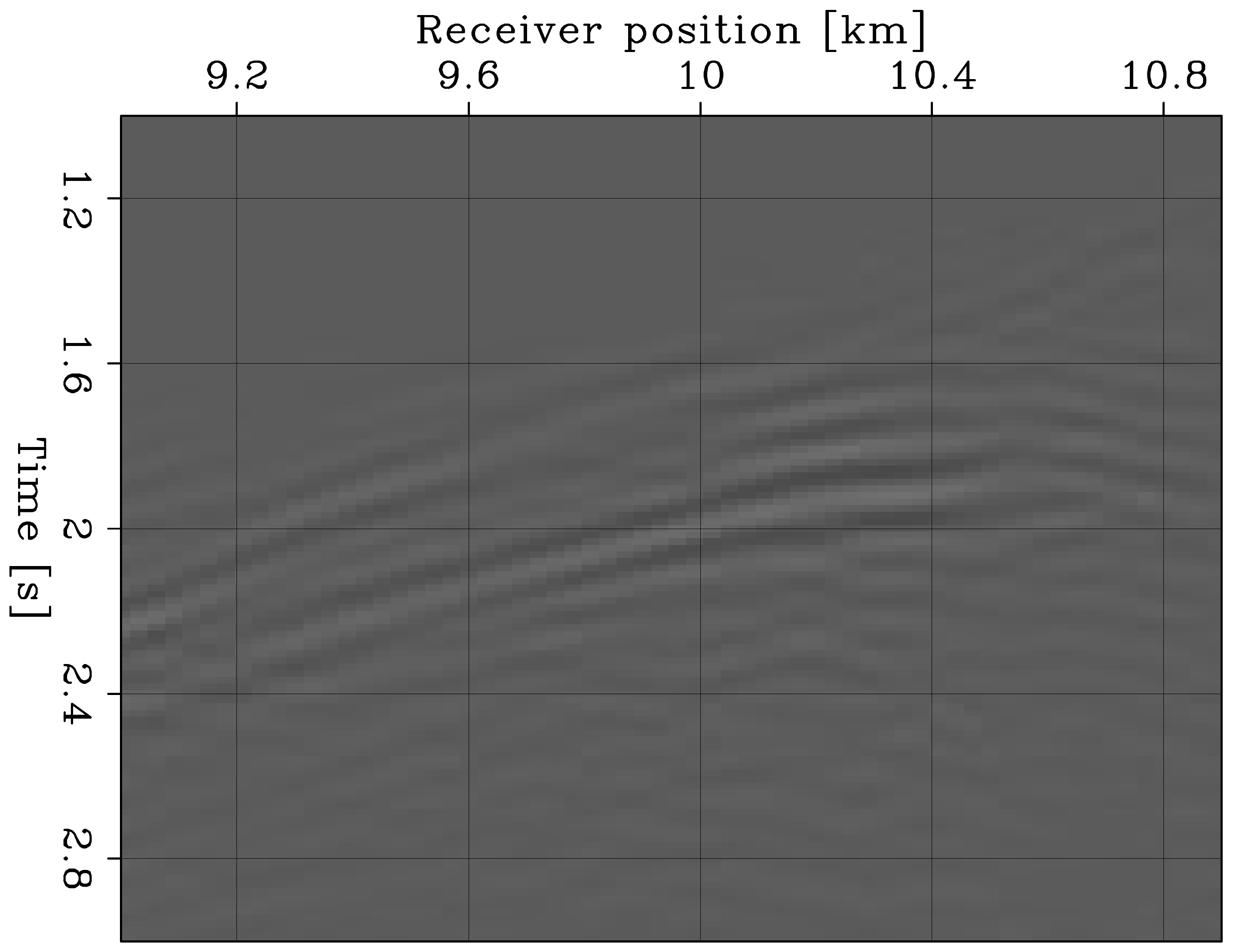}}
    
    \caption{Representative source gather showing the initial (a) and final (b) residual pressure data for the target-oriented inversion. Both panels are displayed using the same gain.}
    \label{fig:MarmElaResTarget}
\end{figure}

The extended image of the target area is demigrated to synthesize the elastic data as if the acquisition geometry was placed in the reservoir's proximity. The elastic pressure is reconstructed assuming 33 sources and 67 fixed receivers spaced every $60$ m and $30$ m, respectively, and recorded for $4$ s. we employ four absorbing boundary conditions to reconstruct and invert the redatumed data because we assume that most of the energy scattered from the target leaves the area of interest, and it is not reflected back from top interfaces. To retrieve the target's elastic parameters, we employ a similar elastic FWI workflow as for the surface data (Figure~\ref{fig:MarmTargInv}). In this case, we only use two spline grid refinements; namely, $50$ and $25$ m sampling. Overall, 20 iterations of BFGS have been applied to invert the elastic pressure on each spline grid (Figure~\ref{fig:MarmElaObjTarget}). A decrease of 98\% is reached after only $40$ iterations instead of the $145$ needed for the surface-data inversion to achieve the same data fitting level (Figure~\ref{fig:MarmElaResTarget}). The P-wave and density parameters of the reservoir gas anomaly are correctly retrieved. No leakage of the gas anomaly is observed in the inverted S-wave parameter. An increase in all three parameters is present right below the reservoir. This artifact is related to the limited frequency range and the regularization parameter used during the imaging step. Moreover, different regularization weights and image masks applied during the linearized waveform inversion step can diminish this artifact's impact. This artifact seems to affect only the first reconstructed event, thus a simple mask as described here can solve this issue. This test demonstrates the target-oriented approach's ability to retrieve the gas anomaly's elastic parameters and their spatial extent. 
Compared to the elastic FWI applied to the surface data, the target-oriented inversion is approximately 200 times computationally cheaper, including the migration process, leading to a memory usage decrease of 25 folds. The main computation speed-up is due to the target-oriented inversion workflow's ability to significantly diminish the simulation domain's size compared to the one where the data have been acquired. The imaging step is not as intensive as the elastic inversion. In fact, in the 2D case, the computational cost of elastic Green's function is approximately $12$ times higher than the one of acoustic wavefields. This observation is also valid for the 3D case, where this factor can be $30$. Moreover, the decreased domain size greatly simplified the implementation of inversion methods because the elastic wavefields can be stored within the computer memory, avoiding the need for applying checkpointing techniques~\cite[]{anderson2012time}. Finally, the computational and memory cost-saving factors can allow the application of elastic FWI methodologies to high-frequency data with reasonable processing time.

\section*{Field-data application}
\subsection*{Workflow of the methodology for field data}
We summarize the key steps necessary to apply our target-oriented elastic FWI approach to field data as follows:
\begin{itemize}
    \item Data selection and preprocessing
    \item Acoustic velocity model building
    \item Acoustic extended linearized waveform inversion
    \item Target area/s identification from subsurface image
    \item Target data redatuming/reconstruction
    \item Elastic FWI of the target data
\end{itemize}

The first step is to select the relevant portion of the data given any project or computational constraint. The preprocessing of the selected data is a crucial part of a successful application of the entire workflow. This step includes source wavelet estimation, frequency band selection for the subsequent FWI application, and data denoising or shaping.
Because our method relies on knowing an accurate overburden velocity model above any identified targets, we include a velocity model-building process before applying the following steps. In our case, we employ an acoustic FWI method given the complexity of the exploration area considered in which salt domes are present. However, any other velocity model-building approach can be used as long as an accurate P-wave speed can be retrieved. 

The next step is to obtain an extended image of the subsurface. In our method, we adopt an extended space based on subsurface offsets~\cite[]{rickett2002offset}. The length of each subsurface-offset axis during this imaging step depends on the azimuthal coverage of the surface acquisition. For instance, if narrow-azimuth streamer data are recorded, then a single axis subsurface-offset extension along the in-line direction could be employed. On the other hand, when a full-azimuth or an OBN setting is used to record a dataset, then the image space should be extended in both horizontal directions to retain the azimuthal behavior of the amplitudes~\cite[]{biondi20043d}. Moreover, to improve the coherency of the subsurface image when sparse acquisitions are employed, a regularized inversion is used to obtain the extended-image volume~\cite[]{prucha2002subsalt,biondi2019amplitude}. Once the image volume is retrieved, we use it to identify the potential subsurface prospects for which the elastic parameters must be retrieved.

Once a target has been determined, the image portion around it is used to synthesize the data for which the acquisition geometry is placed in the proximity of this area. This target-related acquisition geometry depends on the surface acquisition geometry. When full-azimuth data are employed, a similar geometry can be employed for the reconstructed or redatumed data. Conversely, if the surface acquisition is a narrow-azimuth one, then the target geometry should have a similar azimuthal coverage to the subsurface reflector. This redatuming or reconstruction process allows us to limit the elastic FWI modeling extent to only the target area. In addition, the number of receivers and sources within the target acquisition is drastically reduced compared to the surface-related dataset.

The final step of our algorithm is to retrieve the subsurface parameters of the target by employing an elastic FWI methodology. The choice of parameterization used during this optimization step is crucial~\cite[]{pan2019interparameter}. In our case, we parameterize the elastic model by P-wave speed, S-wave speed, and density using the following units: m/s, m/s, and kg/m$^3$. A sufficient number of iterations is necessary to mitigate the well-known problem of inter-parameter cross-talks~\cite[]{operto2013guided}. To this end, we let the optimization algorithm progress until a relative change of the objective function between iterations is below a certain threshold set based on trial and error. The retrieved parameters can then be employed to perform rock-physical analysis to estimate other fundamental rock properties~\cite[]{grana2012quantitative}.
\subsection*{OBN field-data application}
We demonstrate the performance of target-oriented elastic FWI workflow to retrieve geophysical parameters of the subsurface using the pressure component of an OBN dataset. First, we describe the dataset, and all pre-processing steps followed to apply the proposed workflow. Then, we use an acoustic FWI procedure to improve the initial velocity for the proposed method's necessary imaging step. Finally, we synthesize the elastic pressure data generated by a potential prospect positioned on the salt-diapir flank and retrieve estimates of its elastic properties by applying an elastic 3D FWI method. 

\subsection{Dataset overview}
\begin{figure}[t]
    \centering
    \includegraphics[width=0.8\linewidth]{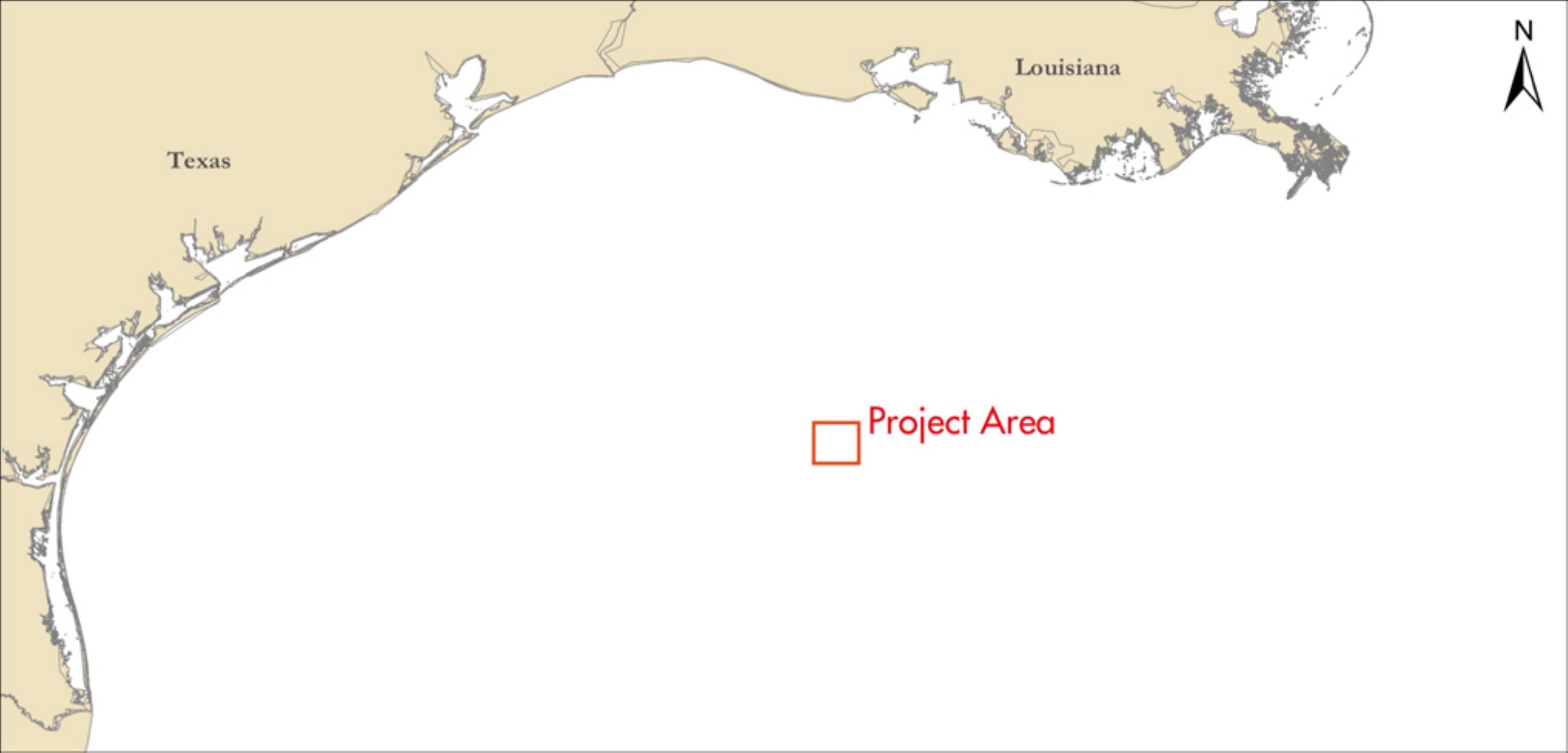}
    \caption{Geographic map highlighting the portion of the Gulf Mexico in which the OBN dataset was acquired.}
    \label{fig:CardamomOverview}
\end{figure}

The field dataset used in this example was acquired within the Gulf of Mexico by Shell Exploration and Production Company in 2010. The area sits in the Garden Banks region, about 362km southwest of New Orleans, Louisiana (Figure~\ref{fig:CardamomOverview}). This data aims to illuminate prospects belonging to a producing field to improve the subsurface images around the diffuse salt bodies by leveraging the full wide-azimuth capability of an OBN geometry. The area has been subjected to diffuse diapirism and presents multiple salt bodies making its exploration possibly challenging from a seismic perspective \cite[]{murray1966salt,thompson2011salt}.

\begin{figure}[t]
    \centering
    \subfigure[]{\label{fig:CardamoSouGeo}\includegraphics[width=0.55\columnwidth]{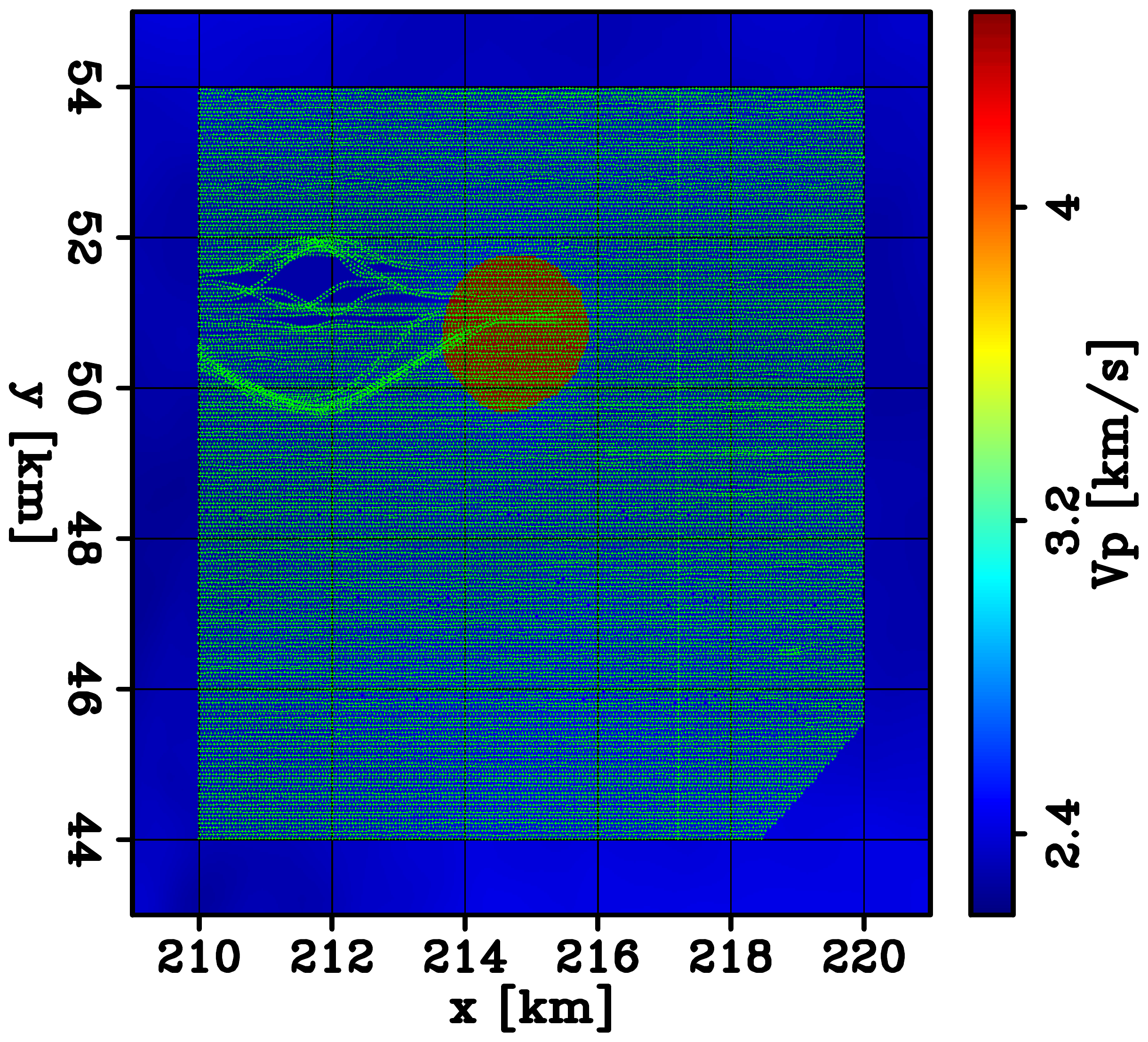}}
    
    \subfigure[]{\label{fig:CardamoRecGeo}\includegraphics[width=0.55\columnwidth]{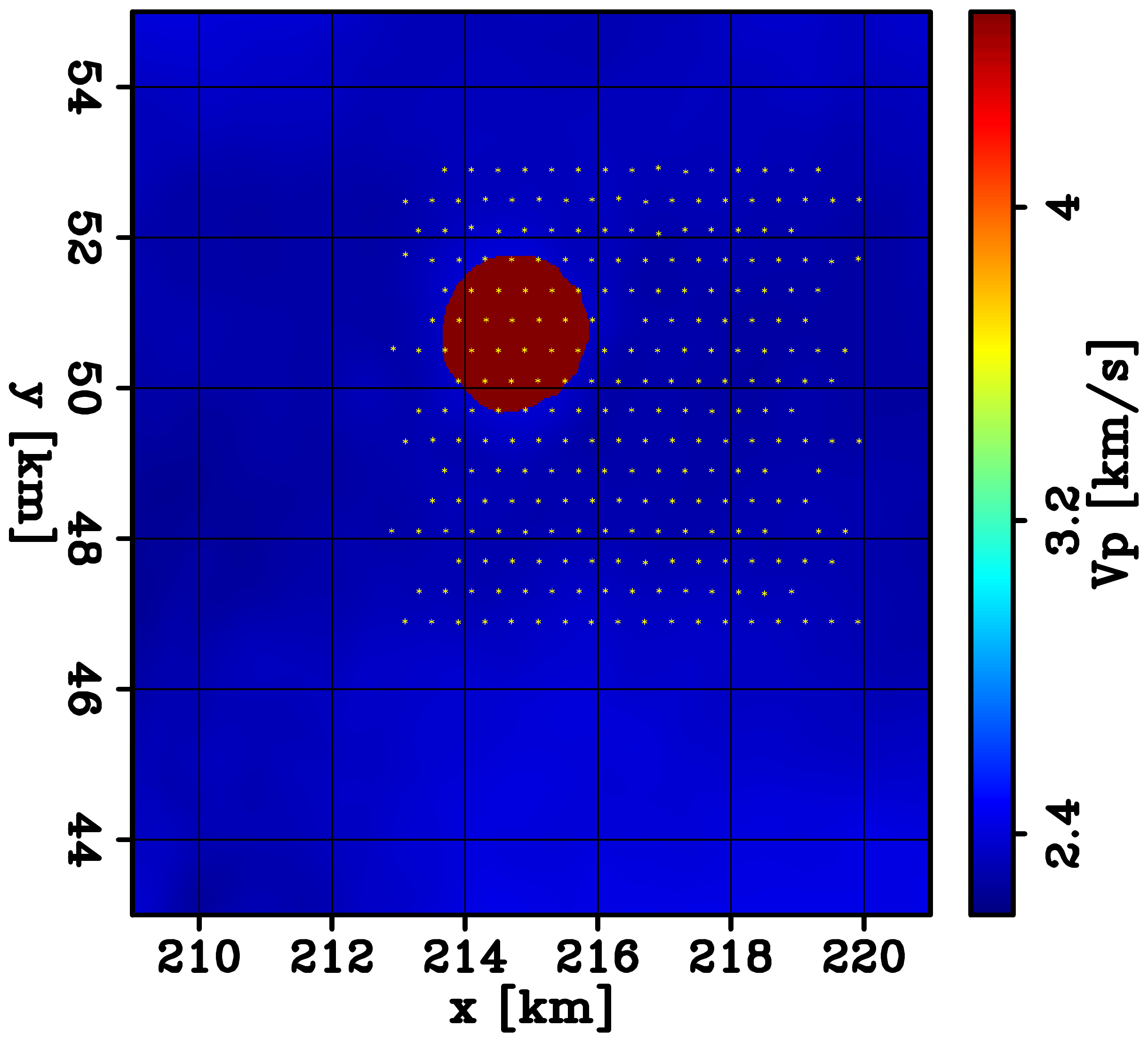}}
    \caption{(a) Sources' and (b) receivers' x-y positions overlaid on the initial velocity model depth slice extracted at $z=2.5$ km. The high-velocity portion is associated to the presence of a salt diapir.}
    \label{fig:CardamoGeo}
\end{figure}

From the entire dataset, we select $255$ nodes that recorded multi-component seismic data generated by $41000$ airgun sources covering an area of $100$ $\text{km}^2$. Figure~\ref{fig:CardamoGeo} shows the sources' and nodes' x-y positions overlaid on a depth slice of the initial velocity model depicting a salt diapir (i.e., high-velocity circular portion). The sources have been acquired using a flip-flop acquisition geometry with a source interval of approximately $25$ m, whose depth is $9.8$ m. The sail lines are aligned with the x-axis, and their interval on the cross-line or y-axis is approximately $100$ m. From Figure~\ref{fig:CardamoSouGeo}, it is clear where the source vessel had to divert its trajectory to abide by the acquisition restrictions in the proximity of production platforms. The multi-component nodes are placed at the seabed, and their depth varies between $0.83$ and $1.0$ km. Their spatial x-y interval is approximately $250$ m in both directions.

\begin{figure}[t]
    \centering
    \subfigure[]{\label{fig:CardamomInitX212}\includegraphics[width=0.48\columnwidth]{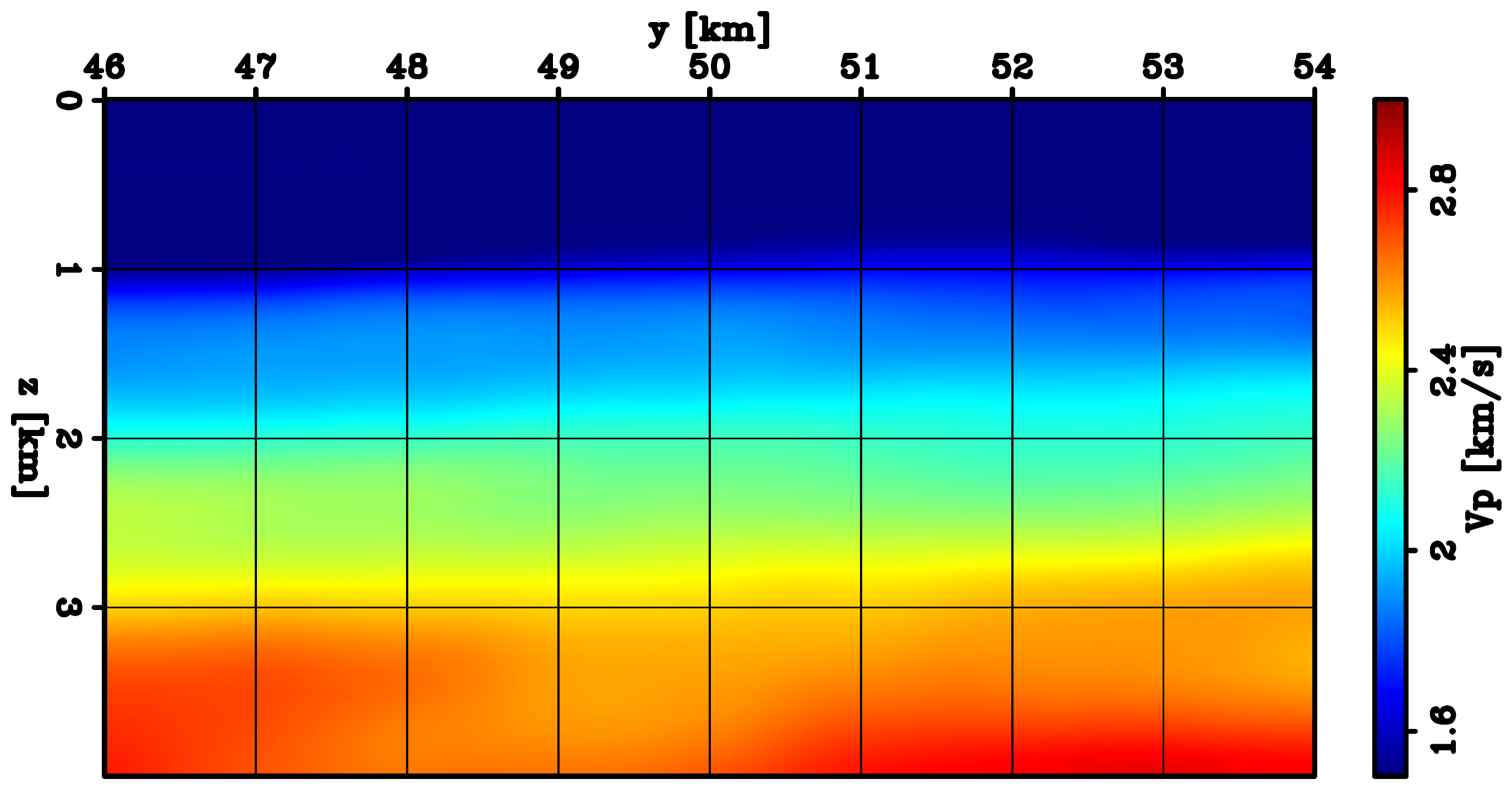}}
    \subfigure[]{\label{fig:CardamomInitY49}\includegraphics[width=0.48\columnwidth]{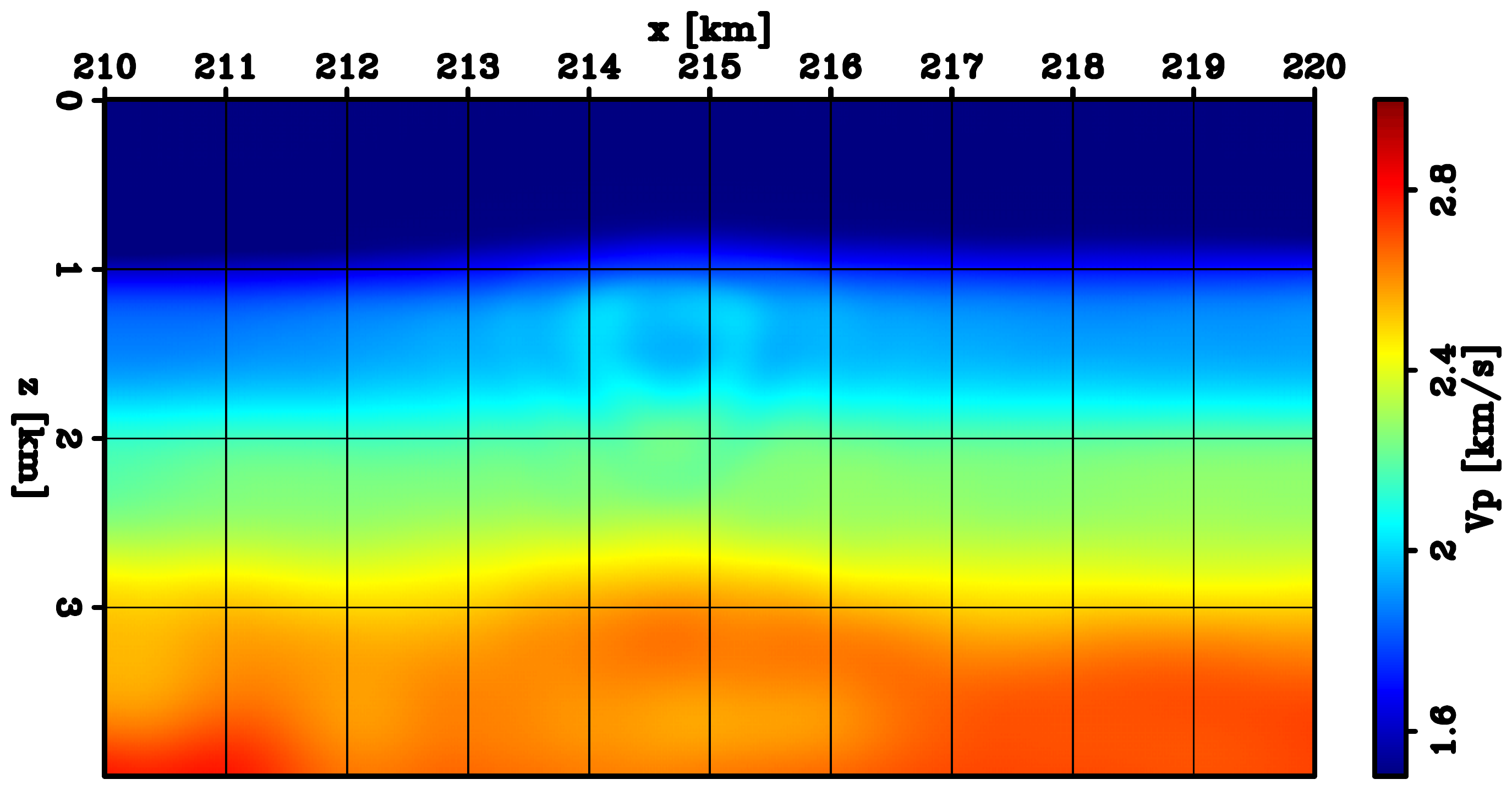}}
    
    \subfigure[]{\label{fig:CardamomInitX214}\includegraphics[width=0.48\columnwidth]{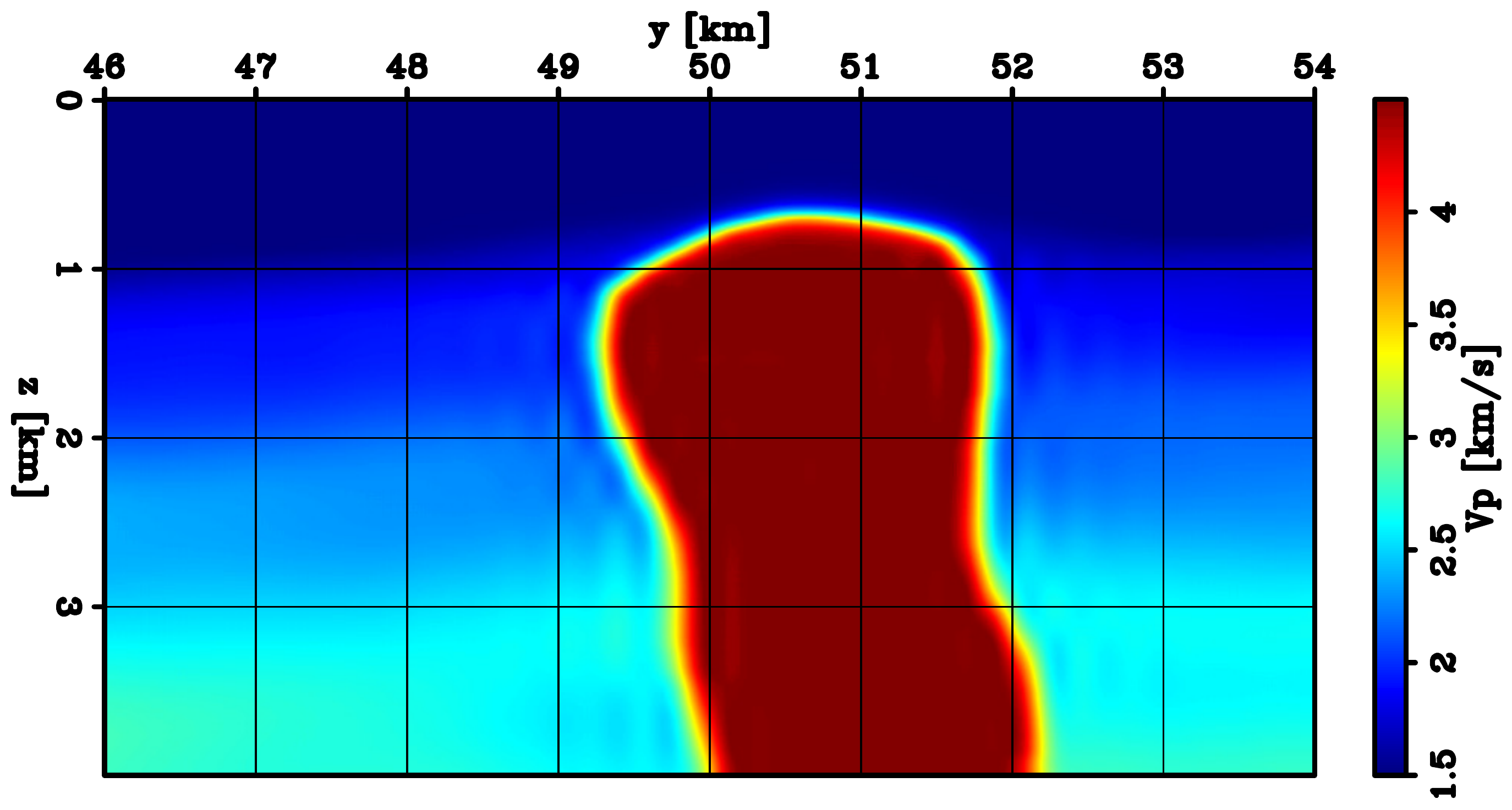}}
    \subfigure[]{\label{fig:CardamomInitY51}\includegraphics[width=0.48\columnwidth]{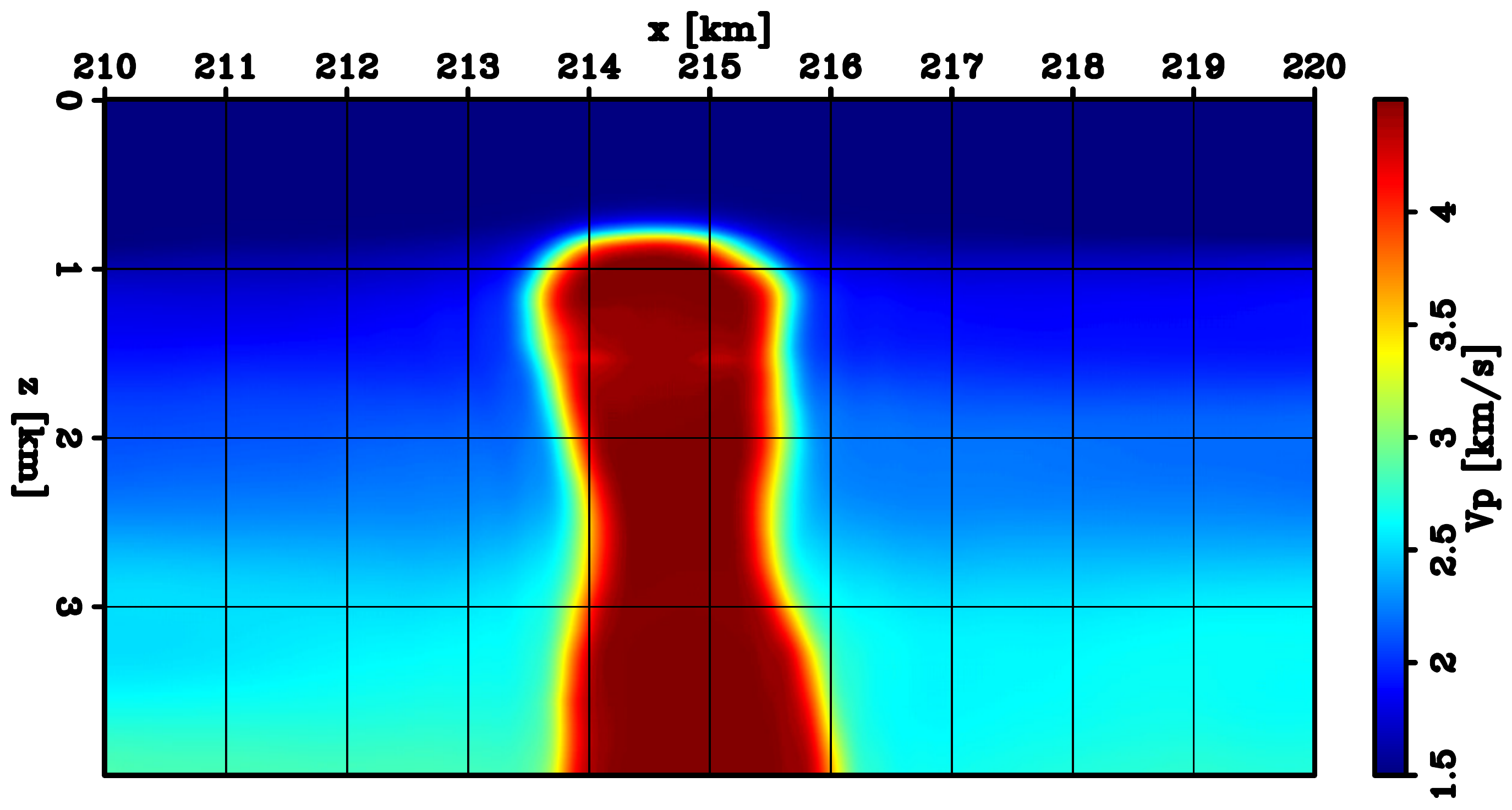}}
    
    \subfigure[]{\label{fig:CardamomInitX217}\includegraphics[width=0.48\columnwidth]{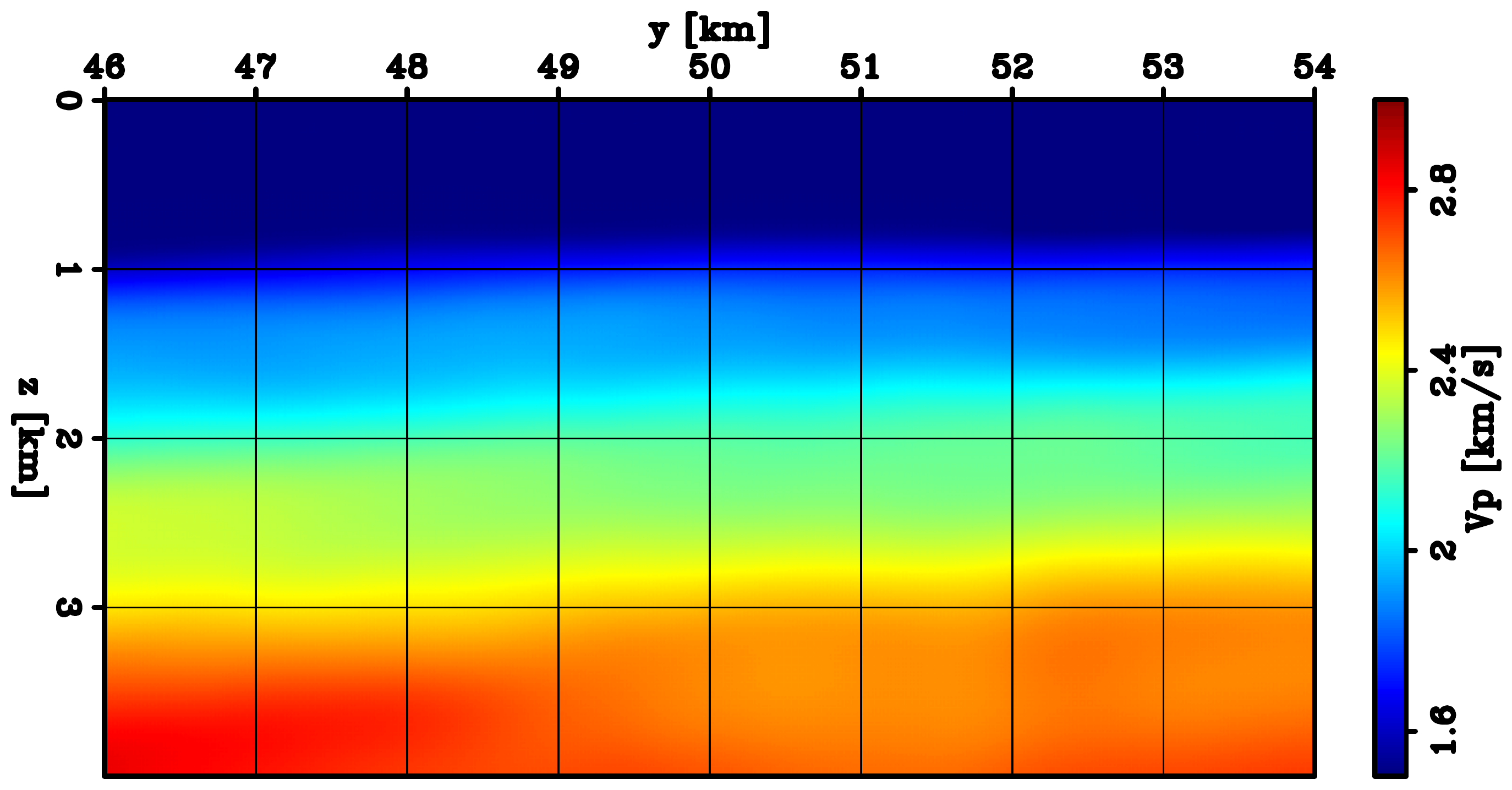}}
    \subfigure[]{\label{fig:CardamomInitY52}\includegraphics[width=0.48\columnwidth]{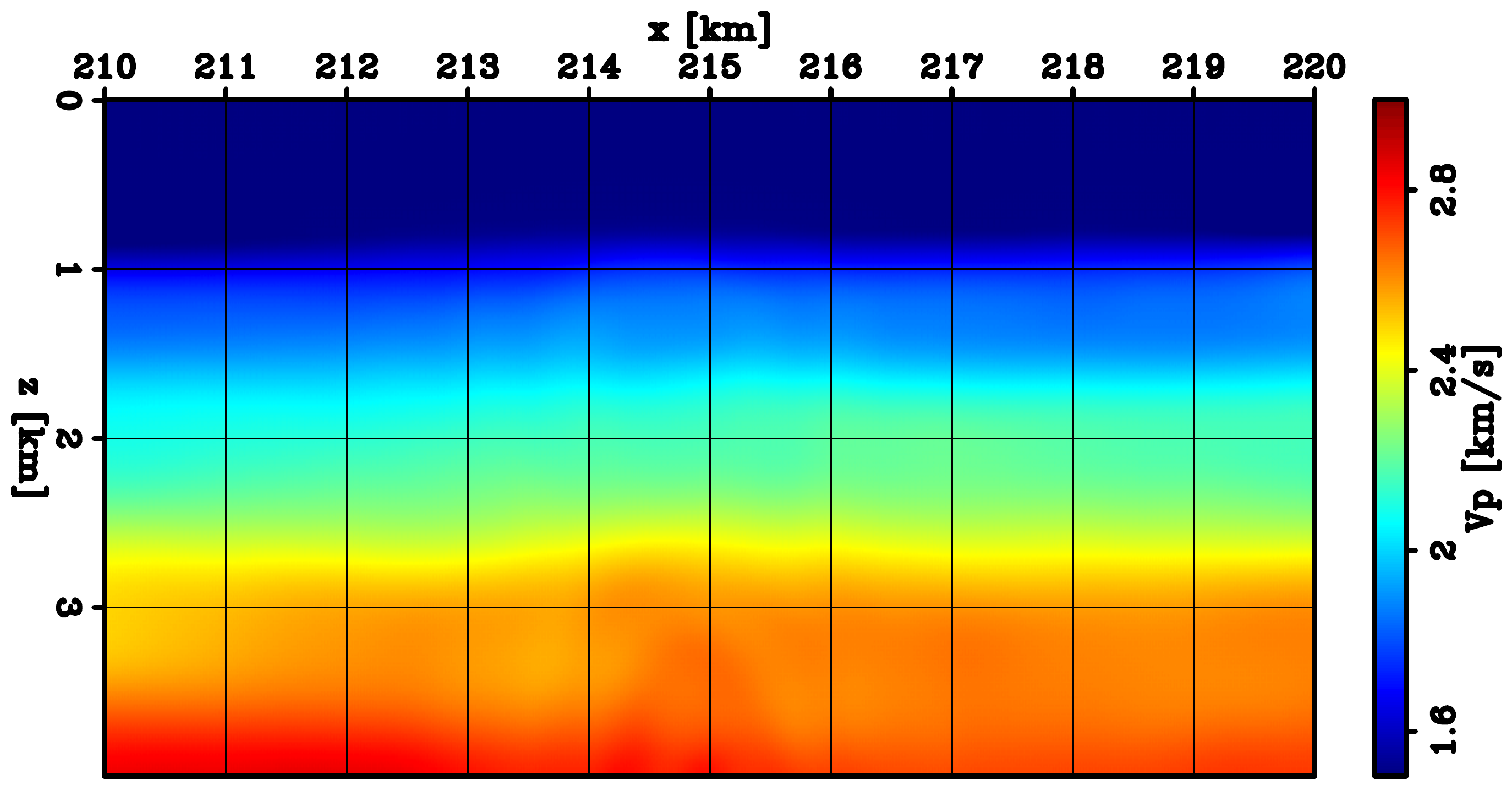}}

    \caption{Initial interpreted P-wave velocity model of the GOM field application: Cross-line slices at $x=212$ km (a), $x=214.8$ km (c), and $x=217$ km (e). In-line slices at $y=49$ km (b), $y=51.5$ km (d), and $y=52.5$ km (f).}
    \label{fig:CardamomInit}
\end{figure}

Besides the observed data, Shell also provided the elastic stiffness components $C_{11}$, $C_{33}$, $C_{13}$, $C_{44}$, and $C_{66}$ and a constant density value of the sediment layers. In addition to this information, they included the salt-body edges' positions obtained by interpreting subsurface images of the area. From the stiffness components, we construct an initial P-wave velocity model based on the provided inputs assuming an isotropic medium~\cite[]{mah2003determination}. Figure~\ref{fig:CardamomInit} shows representative sections of the 3D volume of the velocity model. The initial sediment velocity does not present any distinguishable geological feature. A salt diapir, reaching the seabed floor depth, is located close to the center of the area of interest, whose P-wave velocity is set to $4.5$ km/s and is assumed to be homogeneous, a common assumption based on field observations~\cite[]{zong2015elastic}.

\subsection{Data preprocessing}
\begin{figure}[t]
    \centering
    \subfigure[]{\label{fig:CardamomRec1hz}\includegraphics[width=0.45\columnwidth]{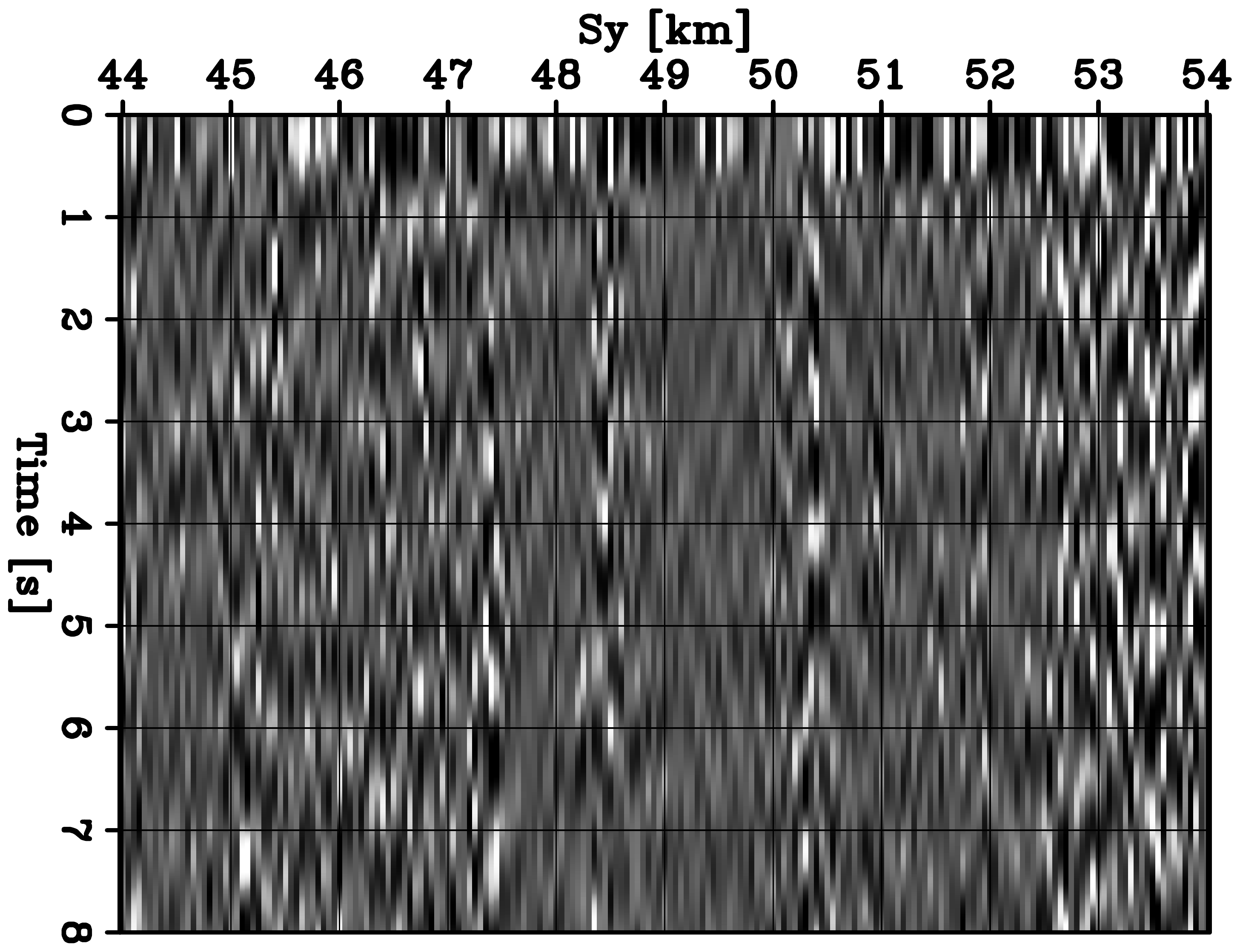}}
    \subfigure[]{\label{fig:CardamomRec2hz}\includegraphics[width=0.45\columnwidth]{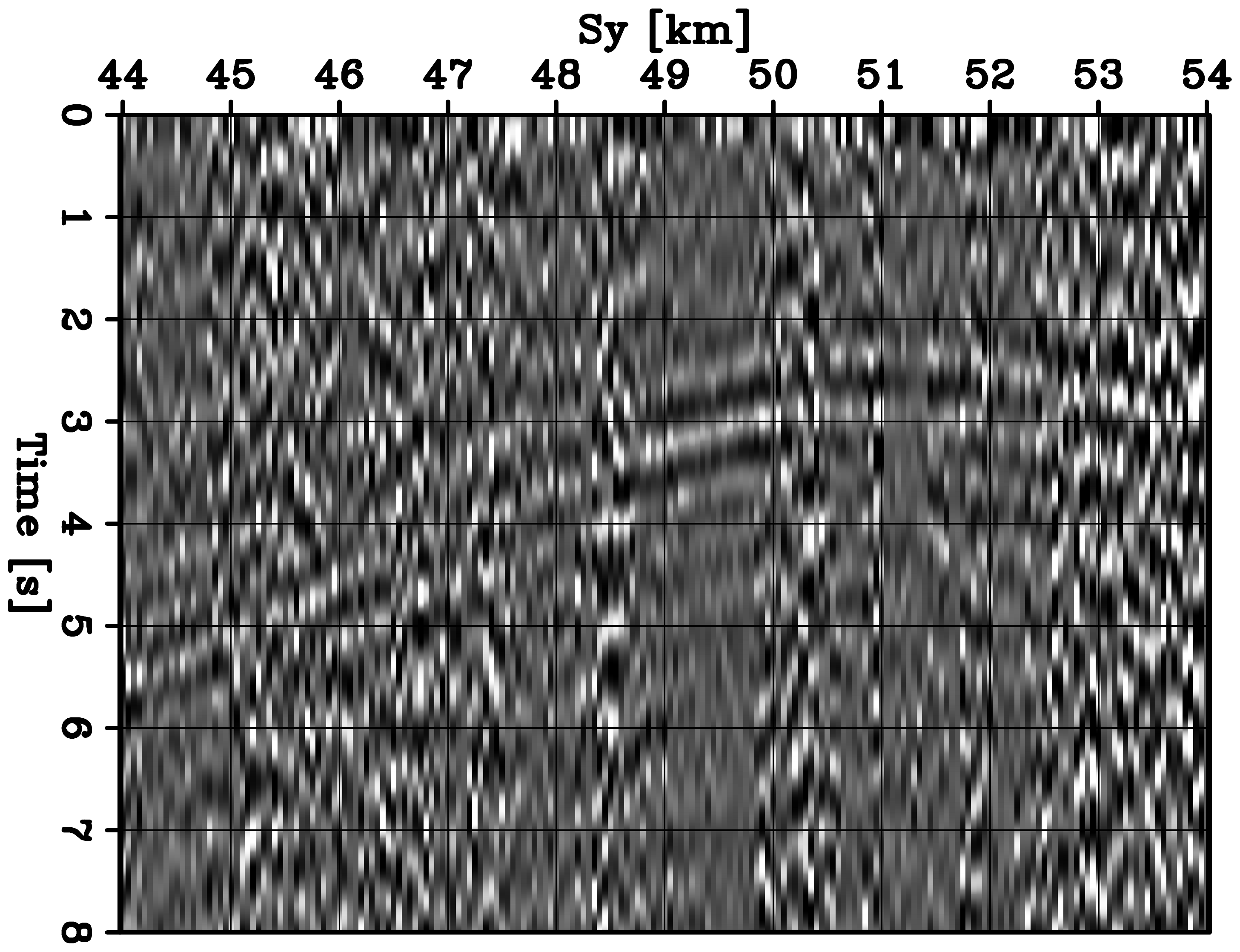}}
    
    \subfigure[]{\label{fig:CardamomRec3hz}\includegraphics[width=0.45\columnwidth]{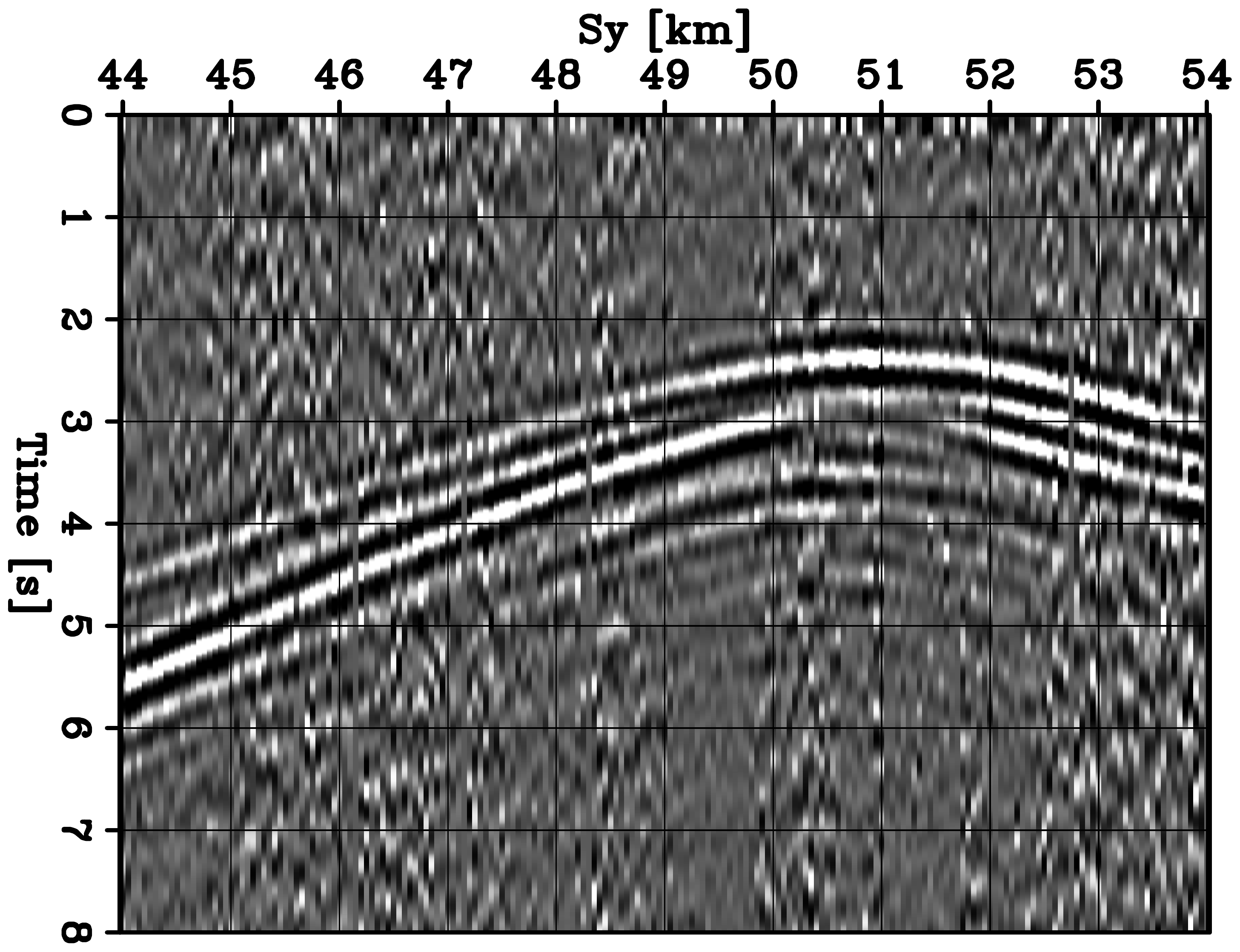}}
    \subfigure[]{\label{fig:CardamomRec4hz}\includegraphics[width=0.45\columnwidth]{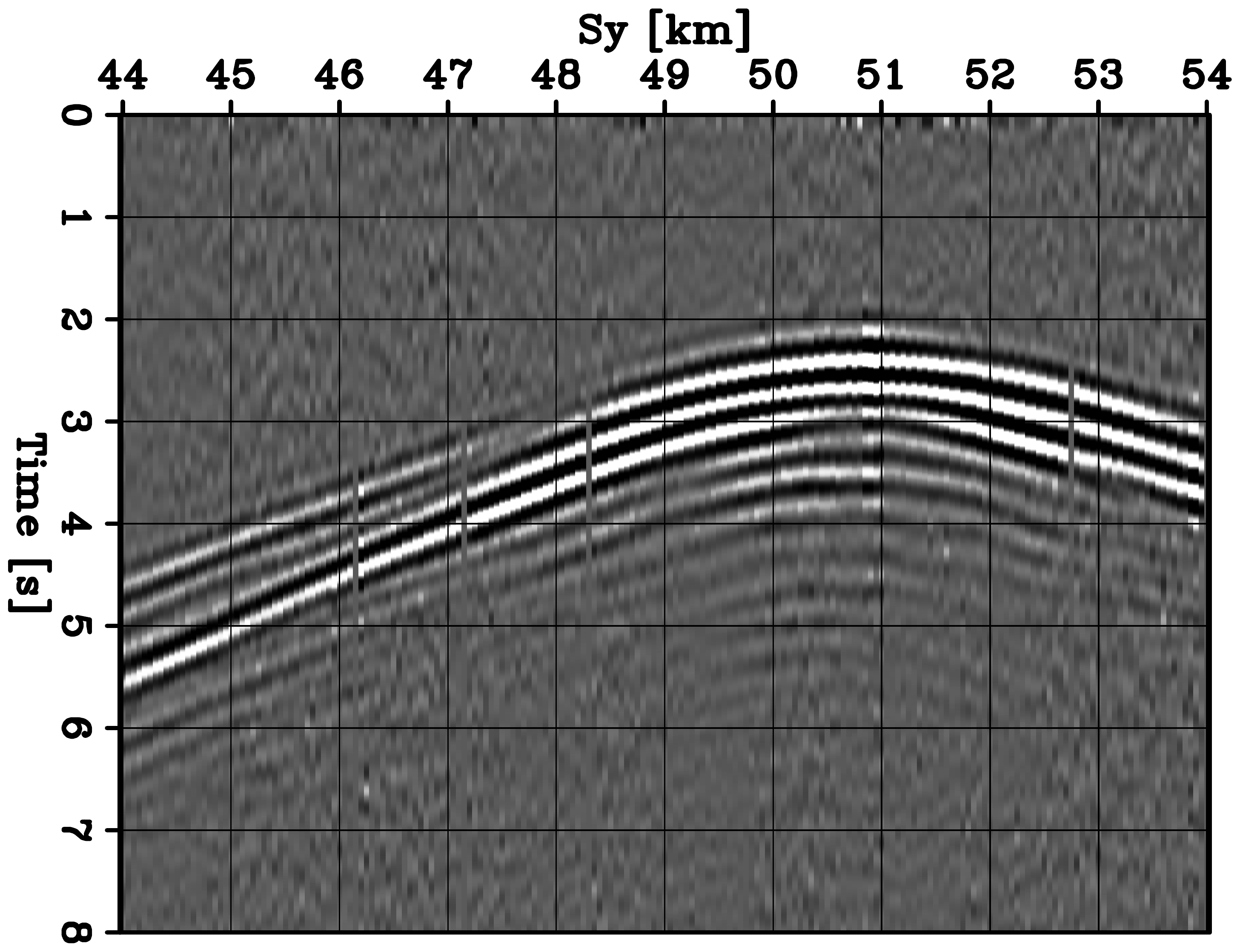}}
    \caption{Representative shot-binned common-receiver gather for $S_x=214.8$ km on which a low-pass filter has been applied with high-cut frequency of (a) $1$, (b) $2$, (c) $3$, and (d) $4$ Hz.}
    \label{fig:CardamoMinFreq}
\end{figure}

We describe in detail the pre-processing steps we followed to apply our target-oriented workflow. Since we rely on FWI approaches to estimate the migration velocity model and the elastic parameters of the subsurface, we first identify the minimum frequency for which the active source signal has a sufficient signal-to-noise ratio (SNR). To this end, we apply different high-cut filters to a representative shot-binned common-receiver gather and visually identify the frequency for which the direct arrival signal is distinguishable from the natural background noise (Figure~\ref{fig:CardamoMinFreq}). The top two panels clearly show that no useable active energy is recorded below $2$ Hz. Conversely, from the bottom two panels, we notice that the SNR increases between $3$ and $4$ Hz. Thus, we set the lowest frequency used in this field example to $3$ Hz. We observe a good signal-to-noise ratio (SNR) for the signal up to 40 Hz on the higher end of the spectrum. However, to reduce the modeling computational cost for this study, we apply a 30 Hz high-cut filter to the dataset.

\begin{figure}[t]
    \centering
    \subfigure[]{\label{fig:CardamomDirectTime}\includegraphics[width=0.45\columnwidth]{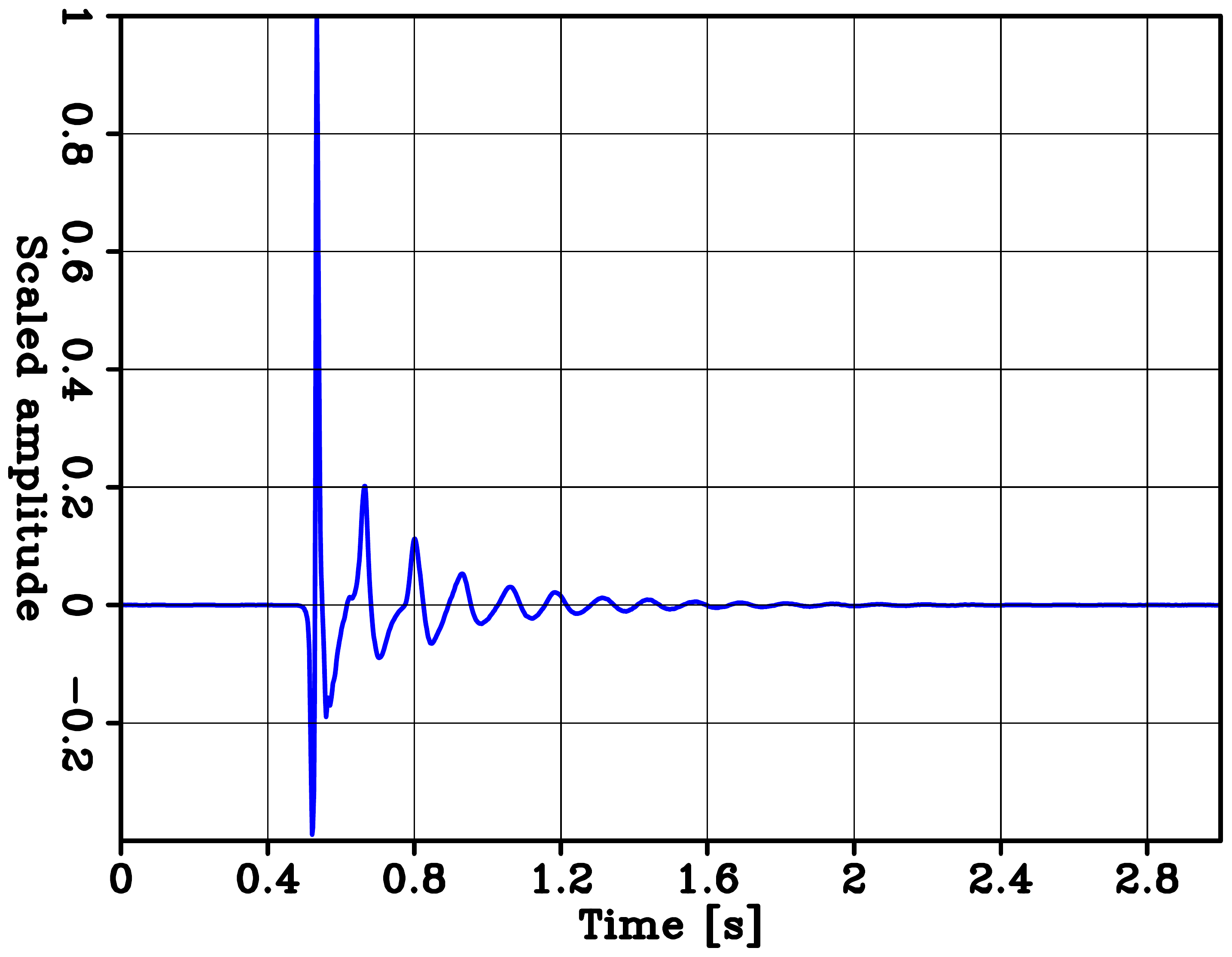}}
    \subfigure[]{\label{fig:CardamomDirectFreq}\includegraphics[width=0.46\columnwidth]{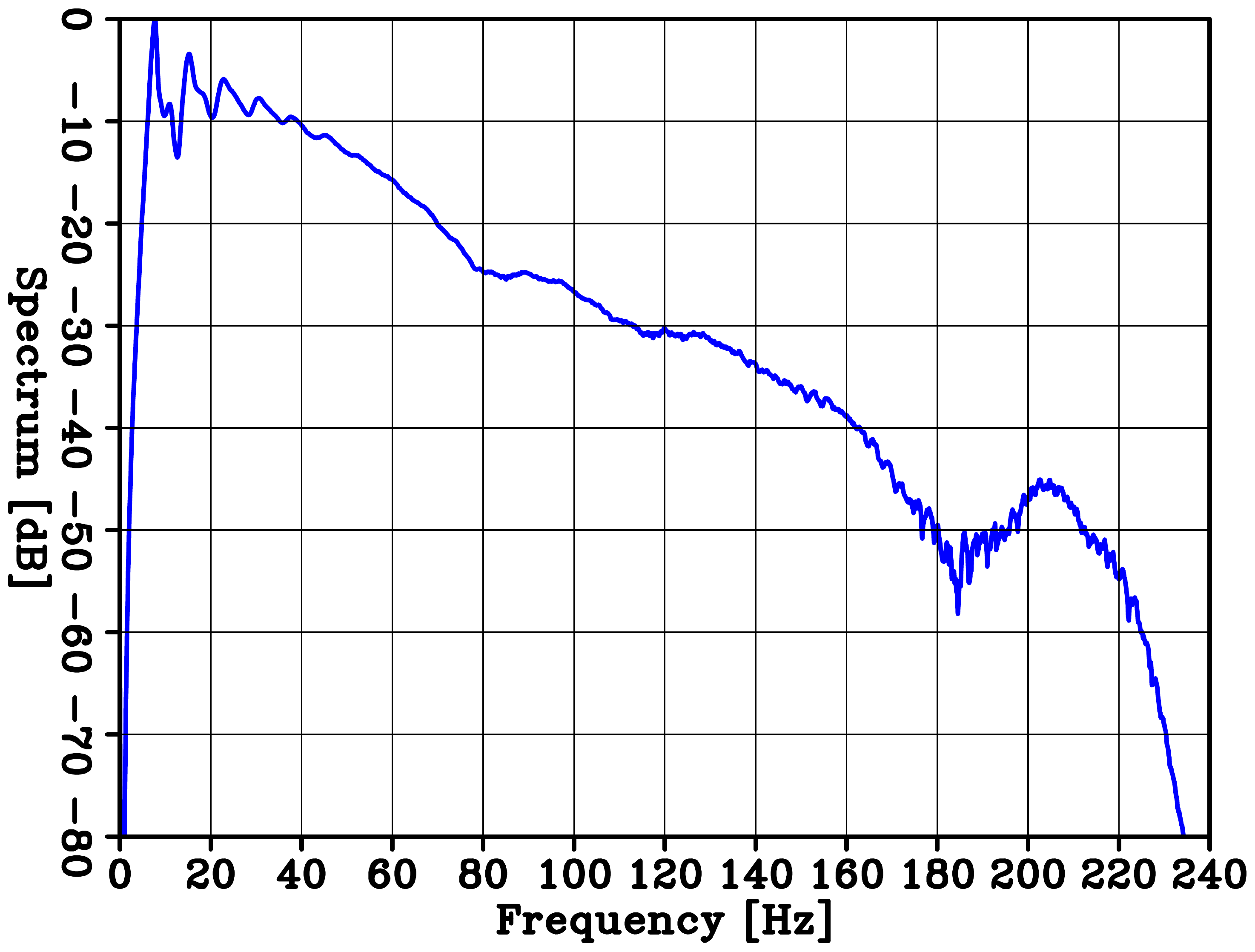}}
    \caption{(a) Time- and (b) frequency-domain signature for the observed direct-arrival signature for a representative receiver gather obtained after applying the described HMO correction and stacking procedures.}
    \label{fig:CardamomDirect}
\end{figure}

The estimation of the source signature is one of the fundamental steps for the successful application of any FWI workflow~\cite[]{rickett2013variable,sun2014source,skopintseva2016importance}. Theoretically, it is possible to compute such a signature from first principles by knowing the airgun experimental setup~\cite[]{ziolkowski1982signature}. However, all the necessary parameters are unknown, or other first-order experiment effects could prevent the theory from retrieving the correct source impulse response (e.g., faulty airguns in the battery, incorrect source parameters). For this reason, we follow a data-driven approach. To compute a proxy of the direct arrival waveform for each node, we apply a hyperbolic moveout (HMO) correction to all the nodes using a constant velocity of $1.5$ km/s~\cite[]{yilmaz2001seismic}. By stacking all the traces belonging to a representative common-receiver gather, we obtain the signal depicted in Figure~\ref{fig:CardamomDirectTime}. A strong peak is present on the signal's onset, which is then followed by the typical bubble response commonly generated by airgun sources~\cite[]{watson2019controls}. The frequency spectrum presents multiple notches due to the source-side ghost and the bubble response (Figure~\ref{fig:CardamomDirectFreq}). Additionally, the frequency content rapidly decreases after $40$ Hz, and minimal energy is present above $160$ Hz. 

Besides retrieving the P-wave speed using an FWI workflow, we also use the recorded data to produce a reverse time migration (RTM) image volume of the subsurface. This volume allows us to understand the area's geological setting and imaging challenges (i.e., position and shape of the salt dome). The usage of the time signature displayed in Figure~\ref{fig:CardamomDirectTime} for imaging would result in an image volume presenting multiple side lobes due to the bubble reverberations unless specific imaging conditions or procedures are used to form the subsurface image~\cite[]{valenciano2002deconvolution}. One standard practice to reduce the bubble reverberations is to apply prediction-error filters to the observed data to dampen these oscillations~\cite[]{wood1978debubbling}. However, in this application, we follow a different approach by combining acoustic wave-equation modeling and shaping filters. 

\begin{figure}[t]
    \centering
    \subfigure[]{\label{fig:CardamomObsRec}\includegraphics[width=0.48\columnwidth]{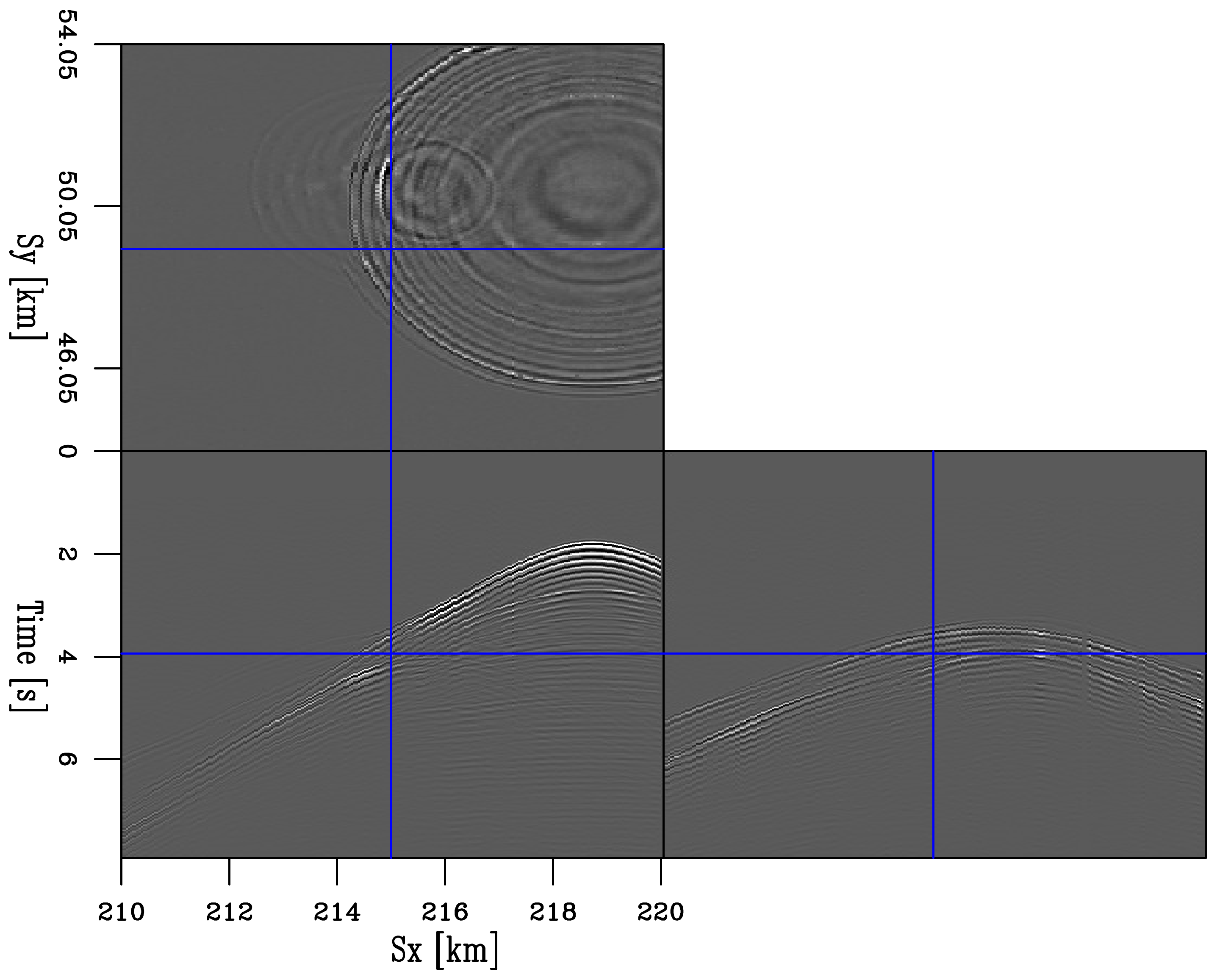}}
    \subfigure[]{\label{fig:CardamomPreRec}\includegraphics[width=0.48\columnwidth]{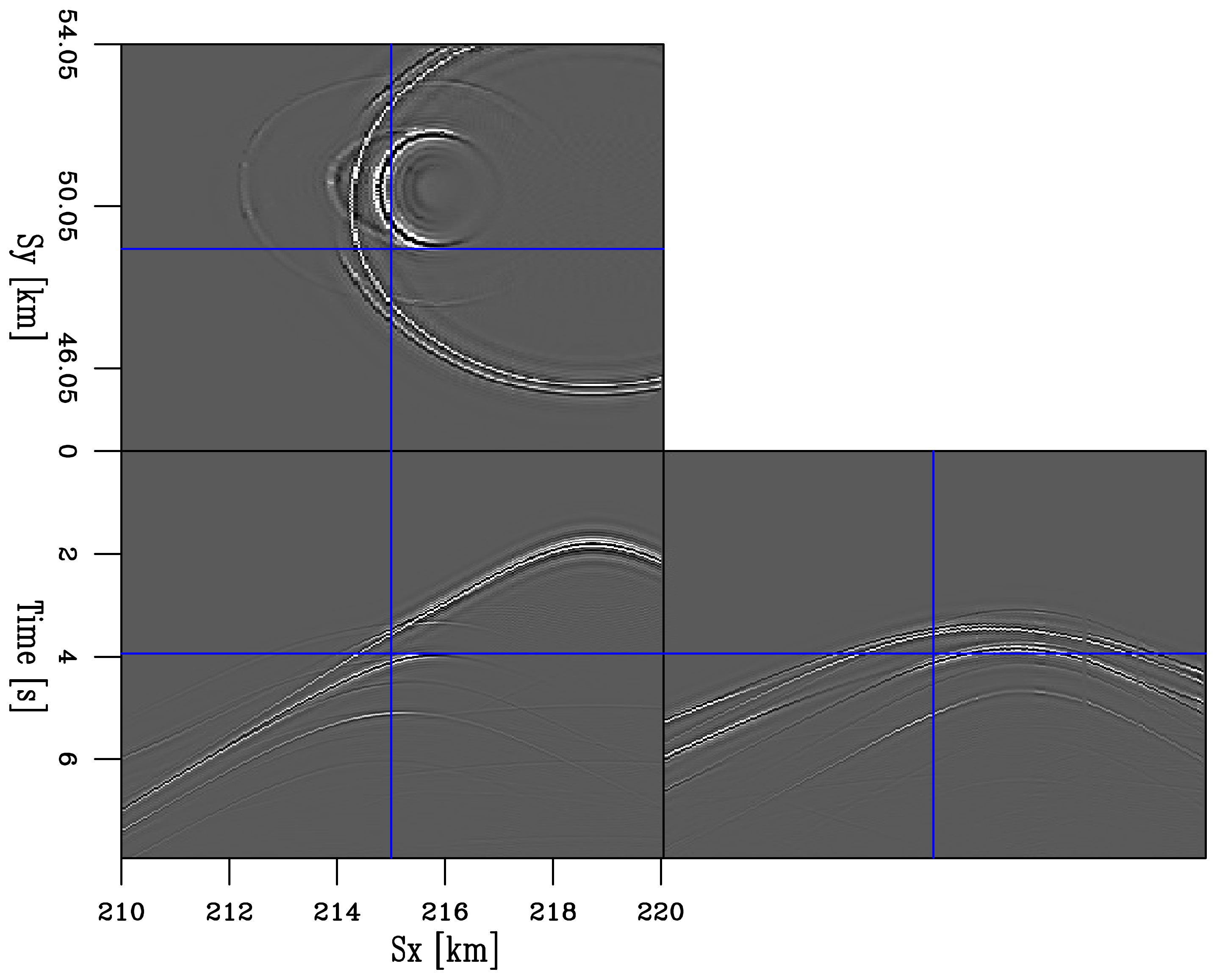}}
    
    \subfigure[]{\label{fig:CardamomShapeRec}\includegraphics[width=0.48\columnwidth]{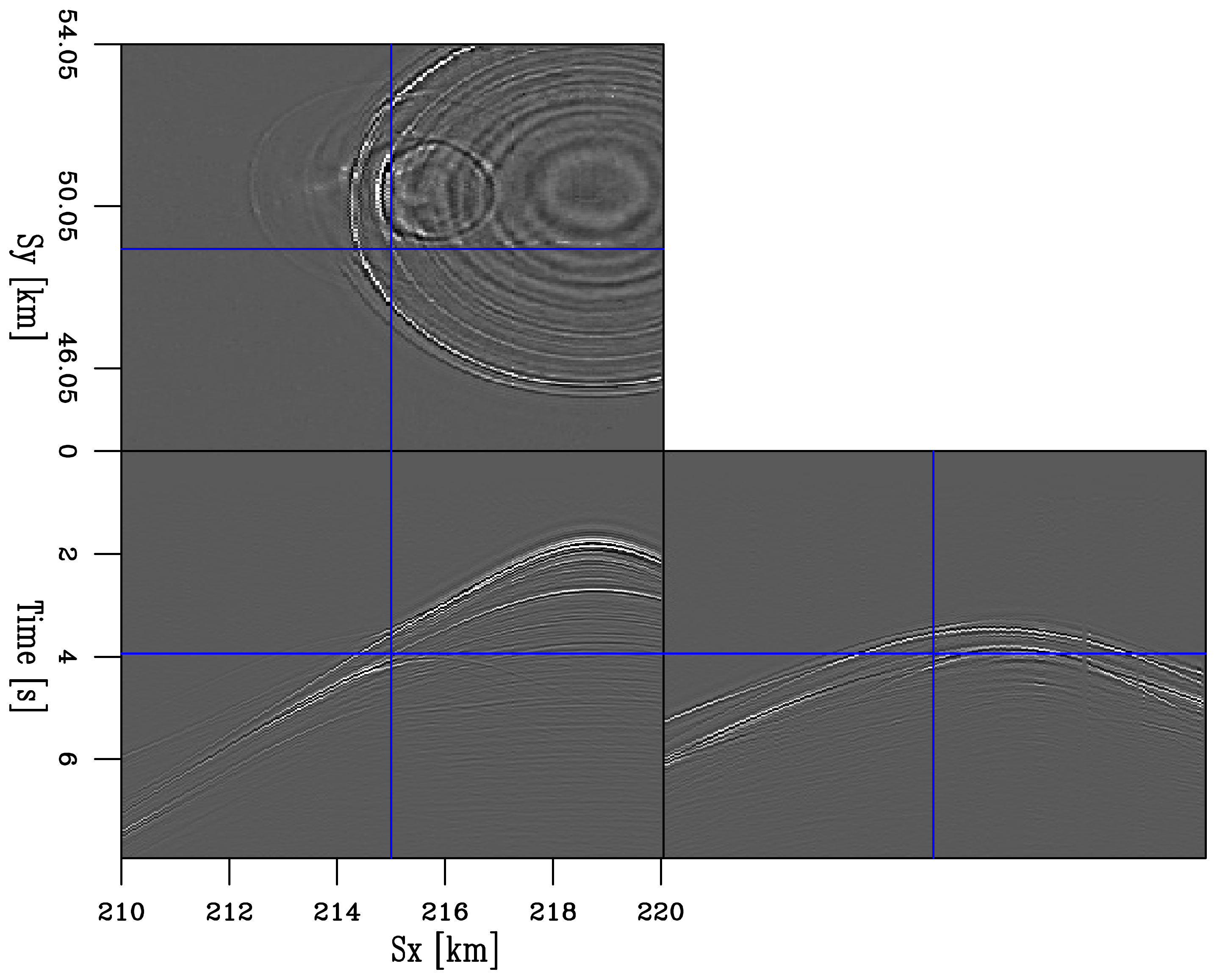}}
    
    \caption{Representative common-receiver gather from the (a) observed pressure data, (b) initial prediction, and (c) after the described waveform shaping filtering procedure. The maximum frequency of the plotted data is $20$ Hz.}
    \label{fig:CardamomShaping}
\end{figure}

The initial model of Figure~\ref{fig:CardamomInit} is used to compute an estimate of the observed data using an acoustic isotropic 3D approximation. We use the predicted pressure to calibrate the modeling operator on the datasets' observed amplitudes and reshape the data response. For each common-receiver gather, we compute the direct arrival proxy following the HMO-stacking procedure and shape the observed signal into the predicted one using a frequency-domain Wiener filtering operation. The proxy event is then used to estimate the filter by solving the following equation in the frequency domain,

\begin{eqnarray}
    \label{eqn:wiener-filter}
    d^{pre}_i(t) &=& \big ( f_i * d^{obs}_i \big) (t), 
\end{eqnarray}

where * denotes the time-convolution operator, $d^{obs}_i$ and $d^{pre}_i$ are observed and predicted direct arrival proxies for the $i$-th node, respectively. $f_i$ is the unknown filter for the $i$-th node. The panels of Figure~\ref{fig:CardamomShaping} show plots of the observed, initially predicted, and shaped data. The shaping procedure removes the bubble response from the observed data and makes the direct arrival waveforms consistent between the initial predicted pressure and the observed data. Compared to other techniques, this shaping procedure automatically removes the instrument response, which can be different for each node, makes the observed data consistent with the acoustic modeling operator, and gives complete freedom in choosing the source signature employed within the modeling and imaging operators.

\subsection{Initial RTM images and geological scenario description}

The shaped data are employed to compute RTM images using the initial P-wave velocity model. These images provide information on the geological subsurface structures present in this area. To simultaneously utilize the up- and down-going energy recorded by the nodes, we perform the acoustic Green's function computation using a free-surface condition at the water surface during the imaging procedure~\cite[]{robertsson1996numerical}. This choice avoids the necessity of performing an up-down separation step~\cite[]{schalkwijk1999application}. Moreover, we apply a source-side illumination compensation to diminish any acquisition artifacts within the subsurface image volume~\cite[]{kaelin2006imaging}. 

\begin{figure}[t]
    \centering
    \subfigure[]{\label{fig:CardamomInitRTMZ1}\includegraphics[width=0.49\columnwidth]{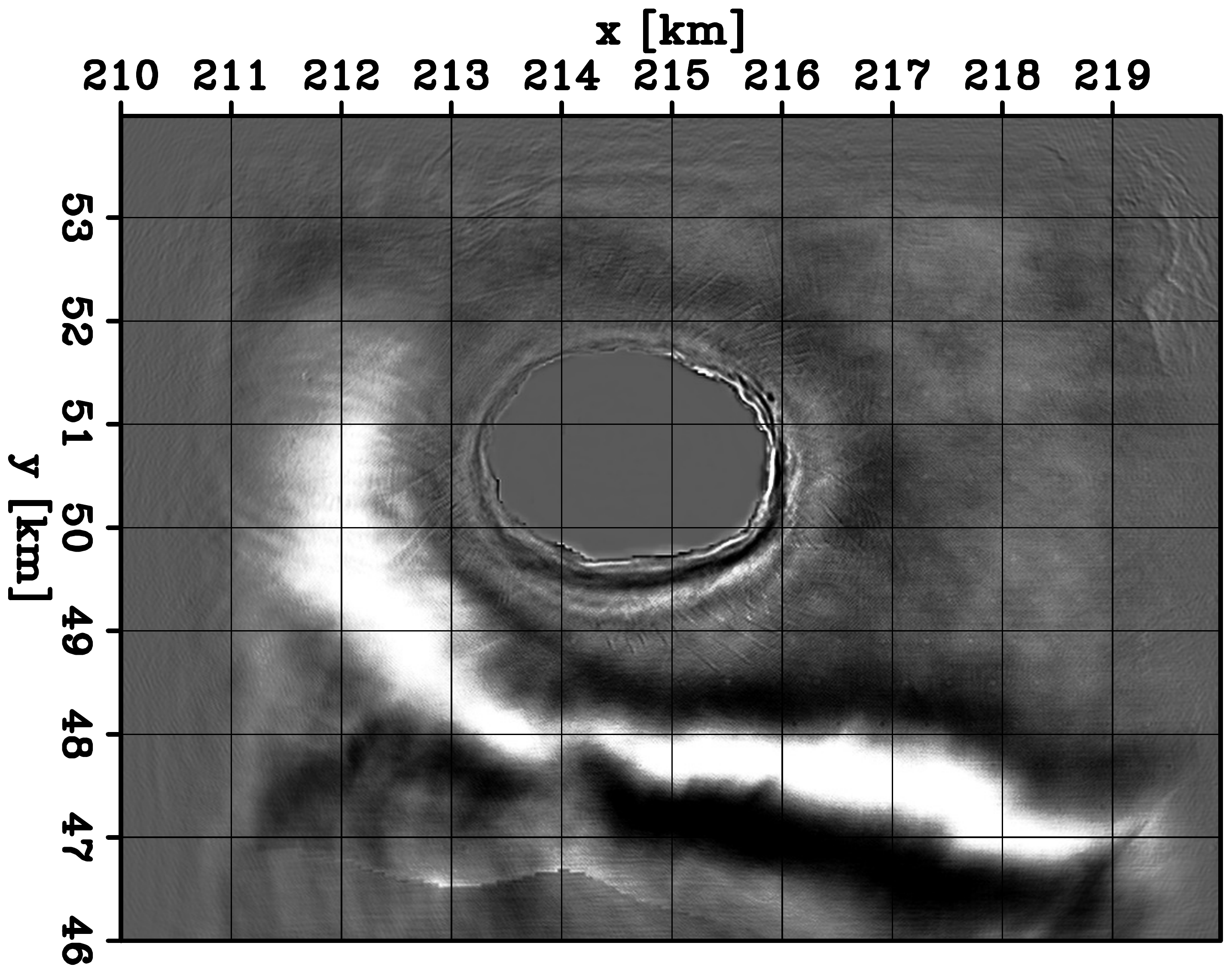}}
    \subfigure[]{\label{fig:CardamomInitRTMZ2}\includegraphics[width=0.49\columnwidth]{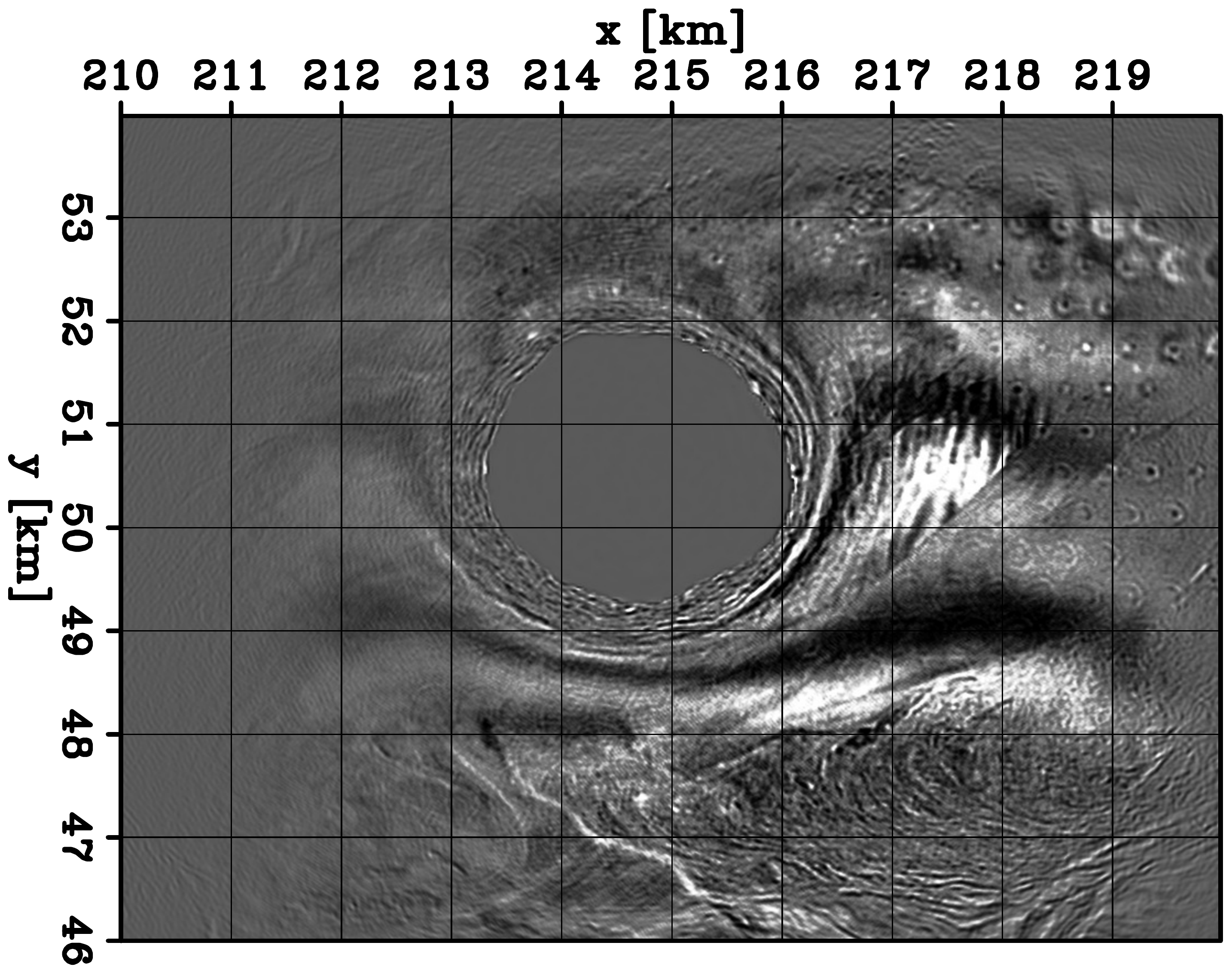}}
    
    \subfigure[]{\label{fig:CardamomInitRTMZ3}\includegraphics[width=0.49\columnwidth]{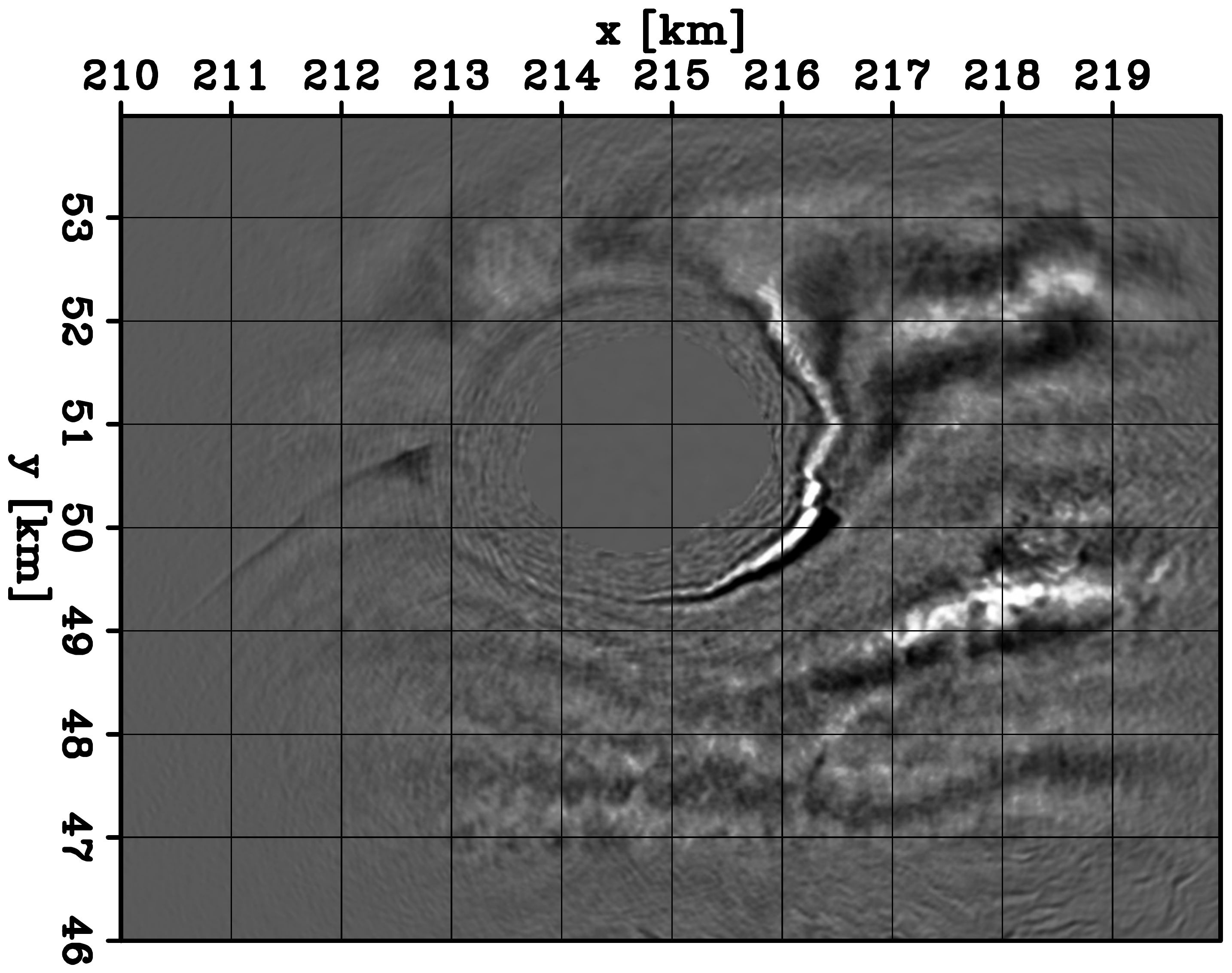}}

    \caption{Depth slices from the 30Hz RTM obtained using the initial provided velocity model. The slices are extracted at: $z=0.9$ km (a), $z=1.69$ km (b), and $z=2.865$ km (c).}
    \label{fig:CardamomInitRTM}
\end{figure}

Figure~\ref{fig:CardamomInitRTM} shows depth slices extracted from the RTM image obtained using the initial velocity model. From the depth section extracted at $z=0.9$ km (Figure~\ref{fig:CardamomInitRTMZ1}), the top edges of the diapir are recognizable, as well as radially distributed features around it. These structures are commonly observed in areas where diapirism is present and are related to the presence of radially distributed faults caused by the rising of the salt diapir~\cite[]{stewart2006implications,coleman2018and}. Furthermore, within deeper sections, multiple structures associated with turbidite sequences can be observed~\cite[]{berg1982seismic}. The panel of Figure~\ref{fig:CardamomInitRTMZ2} shows an example of such structures on the bottom-right corner of the depth slice. Mild acquisition-footprint artifacts are present despite the application of source-side compensation during the migration process. Finally, high-amplitude features are visible close to the diapir flanks in even deeper portions of the image volume. For instance, in the depth section extracted at $z=2.865$ km a noticeable faulted structure is present (Figure~\ref{fig:CardamomInitRTMZ3}). Such image features are potentially associated with hydrocarbon prospects~\cite[]{harding1979structural,tiapkina2008imaging}.

\subsection{Acoustic FWI}
The redatuming process we employ relies on the knowledge of an accurate overburden velocity model. Thus, we apply an acoustic isotropic constant-density FWI process to the shaped pressure data to improve the initial velocity model. The acoustic wave-equation operators employ a free-surface boundary condition at the top edge and absorbing boundaries on the other ones. To mitigate the inaccuracy of the modeling operator in correctly predicting the observed event amplitudes, we employ the objective function proposed by \cite{shen2010near}, where a trace-by-trace normalization is applied to both modeled and observed data vectors before computing their difference. It can be easily shown that the minimization of such an objective function corresponds to the maximization of the zero-lag cross-correlation between the predicted and observed data~\cite[]{shen2014early}. Furthermore, we invert the data using a data-space multiscale approach~\cite[]{bunks1995multiscale}, where progressively wider frequency bands undergo the inversion procedure. The chosen frequency bands are the following: $3-6$, $3-9$, $3-12$, and $3-18$ Hz. For the first three bands, the modeling is performed with an FD grid size of $35$ m in the three dimensions, while, for the last band, the modeling is performed with a grid size of $25$ m. To mitigate the introduction of any inversion artifacts, we parameterize the model on spline grids with an x-y sampling of $175$, $105$, and $50$ m for each band, respectively. The spline grid in the z-axis is as fine as the FD sampling. The usage of a spline parameterization during the inversion also has the additional advantage of limiting the convergence to local minima~\cite[]{barnier2019waveform}. Finally, we apply the acoustic reciprocity theorem so that the $255$ nodes act as sources and the $41000$ sources as receivers~\cite[]{aki2002quantitative}.

\begin{figure}[t]
    \centering
    \includegraphics[width=0.5\linewidth]{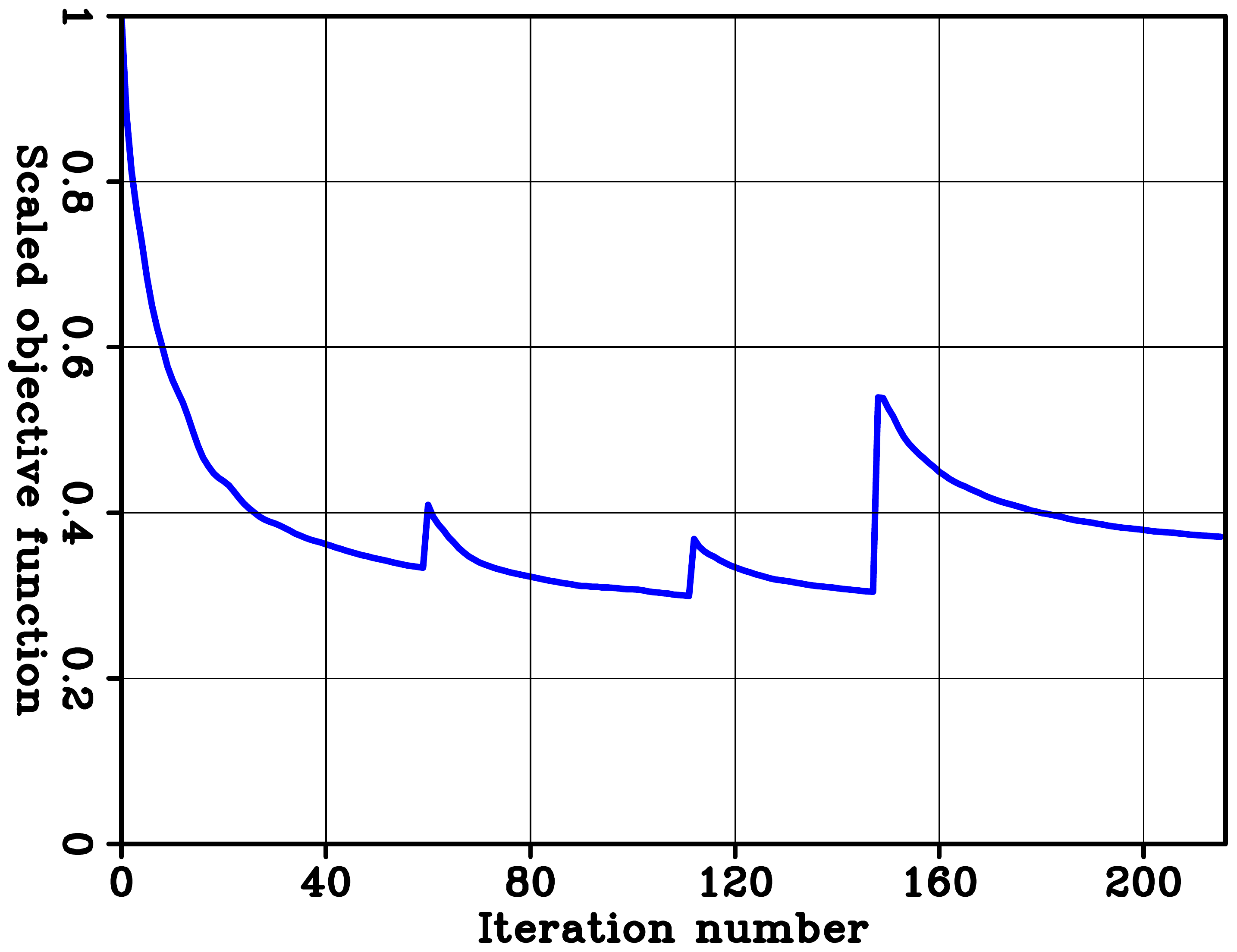}
    \caption{Normalized objective function convergence curve. The discontinuities in the curve corresponds to changes in the frequency content of the inverted data.}
    \label{fig:CardamomAcoFWIObj}
\end{figure}

The acoustic FWI objective function is minimized using a BFGS algorithm~\cite[]{liu1989limited}. Overall, $216$ iterations are performed to invert the data up to $18$ Hz. Figure~\ref{fig:CardamomAcoFWIObj} displays the convergence curve of the acoustic FWI problem. The discontinuities in the curve correspond to the changes in the frequency band during the inversion. For the two central bands, a decrease of approximately 70\% is achieved by the minimization algorithm.

\begin{figure}[t]
    \centering
    \subfigure[]{\label{fig:CardamomFinalX1}\includegraphics[width=0.49\columnwidth]{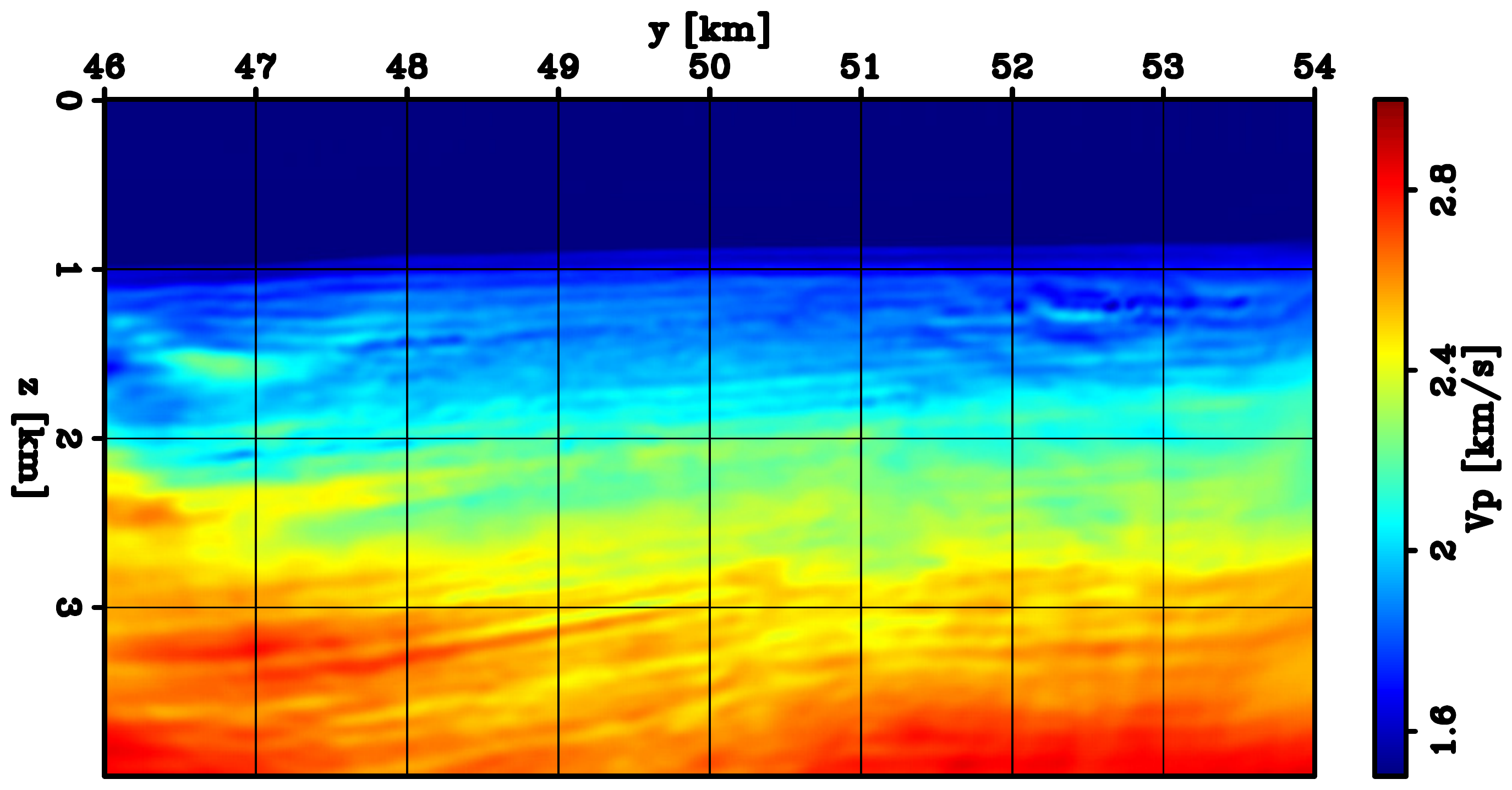}}
    \subfigure[]{\label{fig:CardamomFinalY1}\includegraphics[width=0.49\columnwidth]{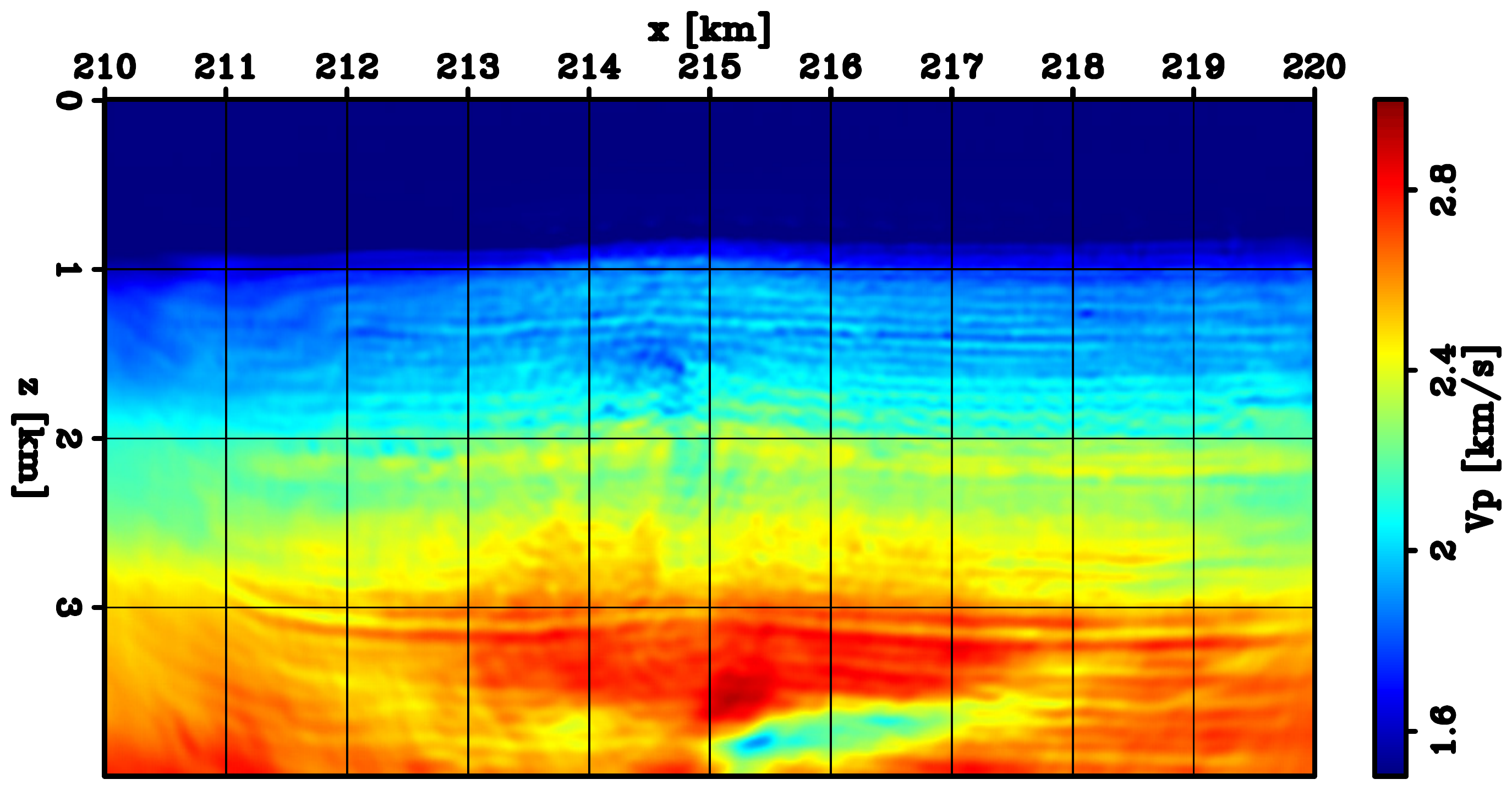}}
    
    \subfigure[]{\label{fig:CardamomFinalX2}\includegraphics[width=0.49\columnwidth]{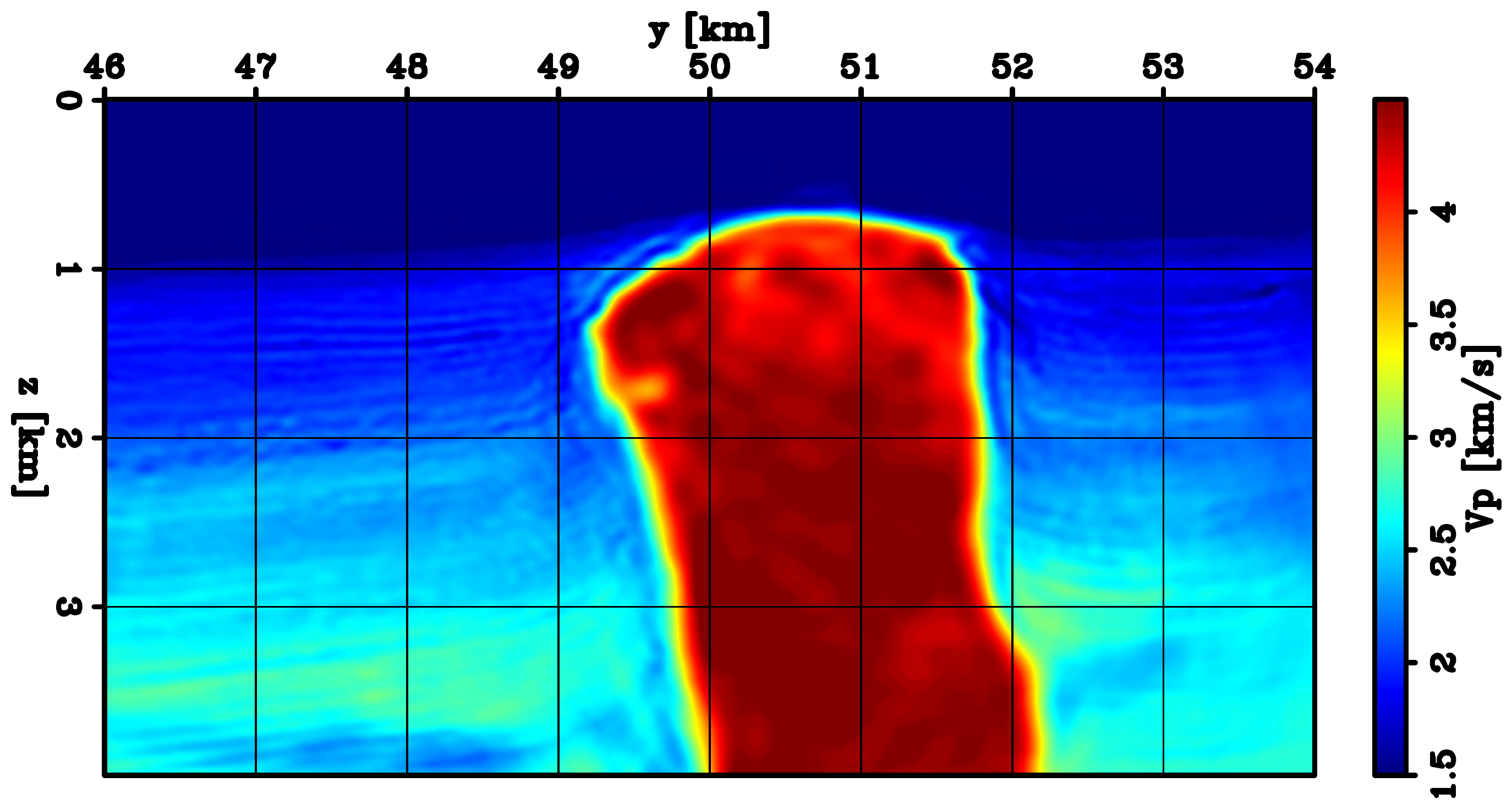}}
    \subfigure[]{\label{fig:CardamomFinalY2}\includegraphics[width=0.49\columnwidth]{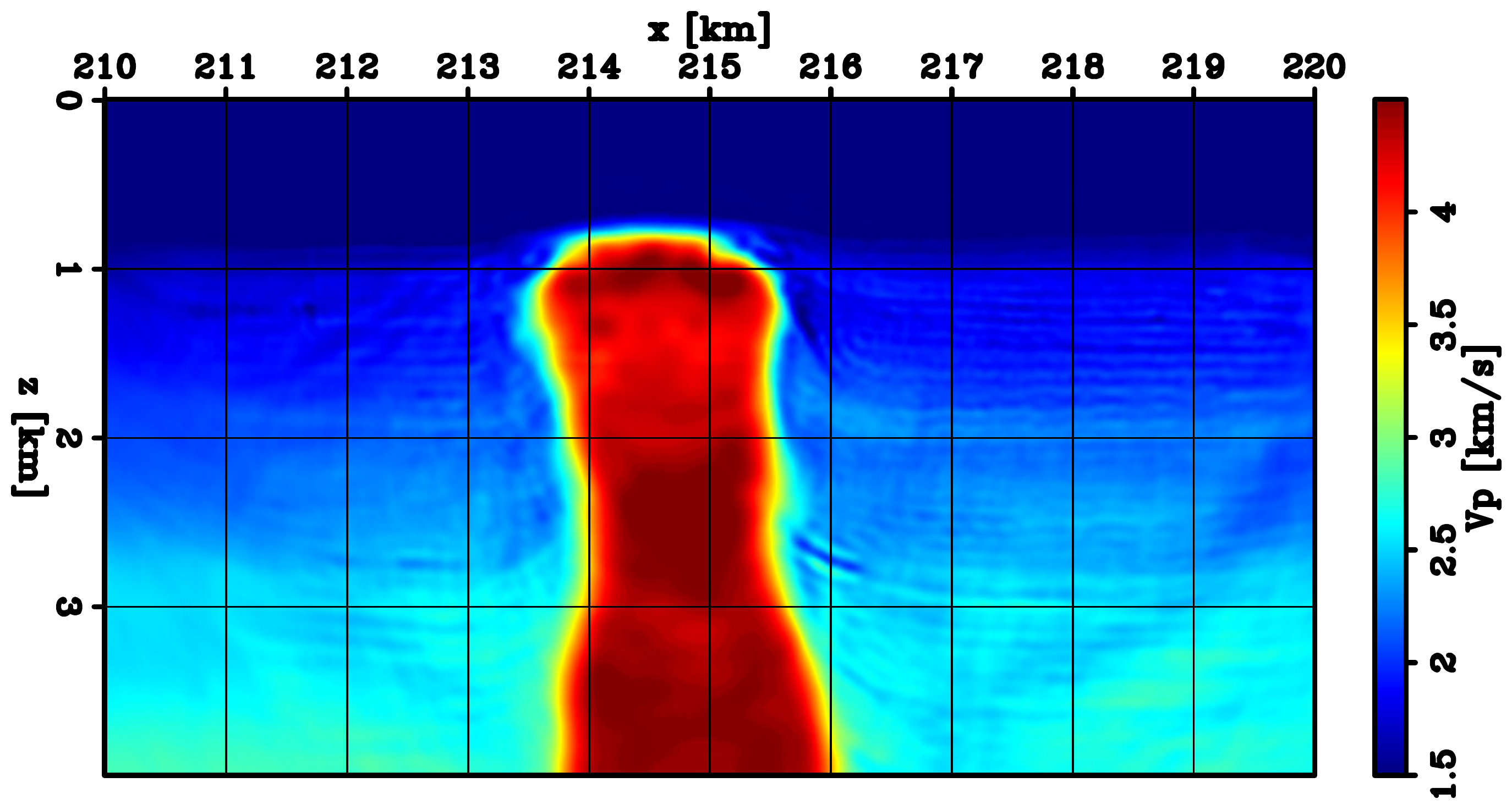}}
    
    \subfigure[]{\label{fig:CardamomFinalX3}\includegraphics[width=0.49\columnwidth]{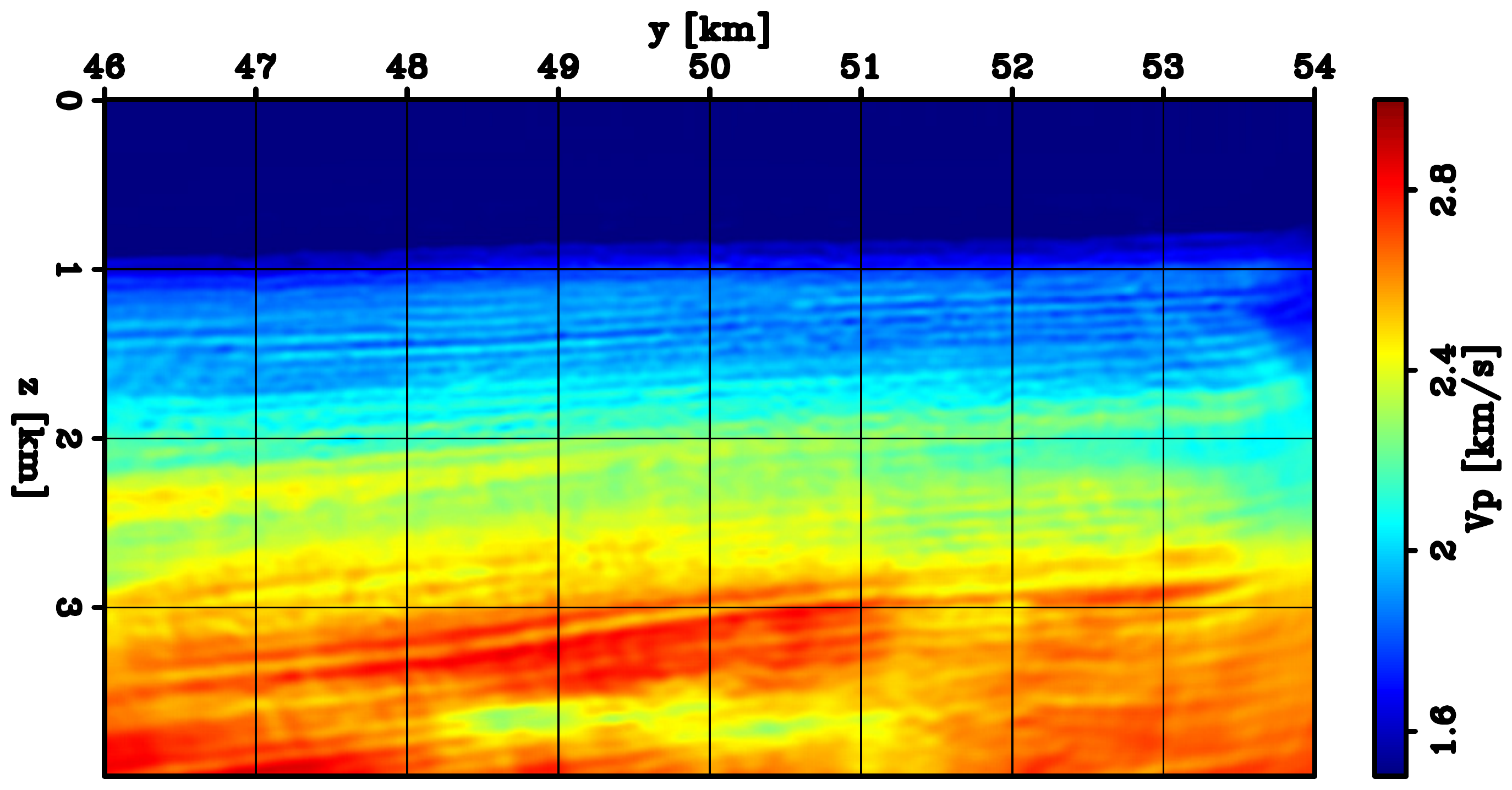}}
    \subfigure[]{\label{fig:CardamomFinalY3}\includegraphics[width=0.49\columnwidth]{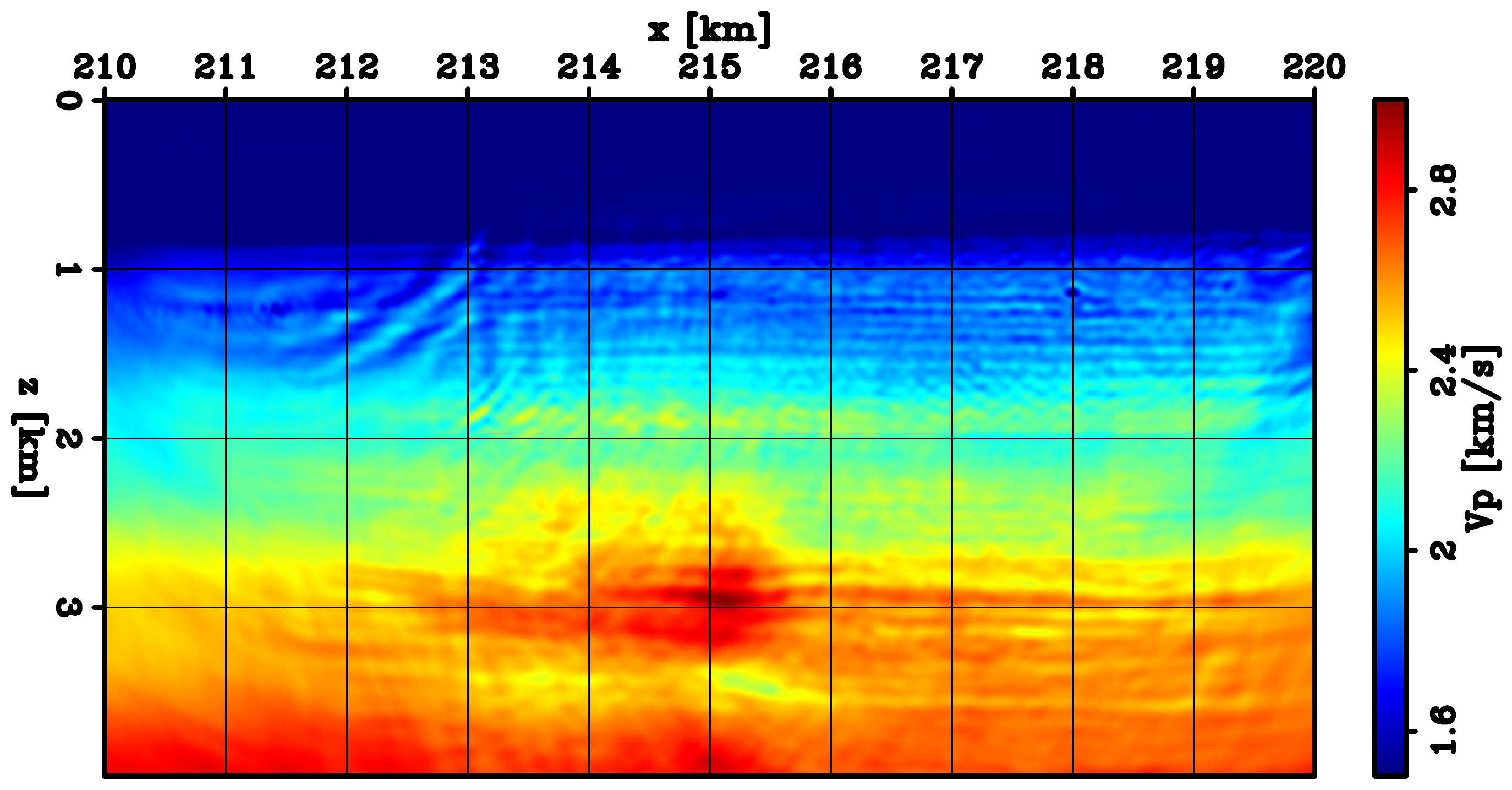}}
  
    \caption{Cross- (left column) and in-line (right column) slices extracted from the inverted acoustic FWI velocity model at (a) $x=212$ km, $y=49$ km (b), $x=214.8$ km (c), $y=51.5$ km (d), $x=217$ km (e), and $y=52.5$ km (f).}
    \label{fig:CardamomFinalCrossIn}
\end{figure}

From the final P-wave velocity we extract the same cross- and in-line sections of the 3D volume as the ones in Figure~\ref{fig:CardamomInit} (Figure~\ref{fig:CardamomFinalCrossIn}). The inverted model by the FWI scheme shows geologically consistent features. For instance, the panel in Figure~\ref{fig:CardamomFinalX1} presents a discontinuity potentially indicating the presence of a fault at $y=50.2$ km and $z=3$ km. Moreover, a clear low-velocity anomaly is placed on the salt flank at $z=2.8$ km (Figure~\ref{fig:CardamomFinalY2}). This decrease in velocity could be related to gas accumulation at the top of a hydrocarbon reservoir sealed by the salt body. Finally, the same inclusion reported by~\cite{dahlke2020applied} is retrieved by the acoustic FWI workflow (Figure~\ref{fig:CardamomFinalX2}). In addition, other velocity variations are present at the top of the diapir.

\begin{figure}[t]
    \centering
    \subfigure[]{\label{fig:CardamomInitZComp}\includegraphics[width=0.365\columnwidth]{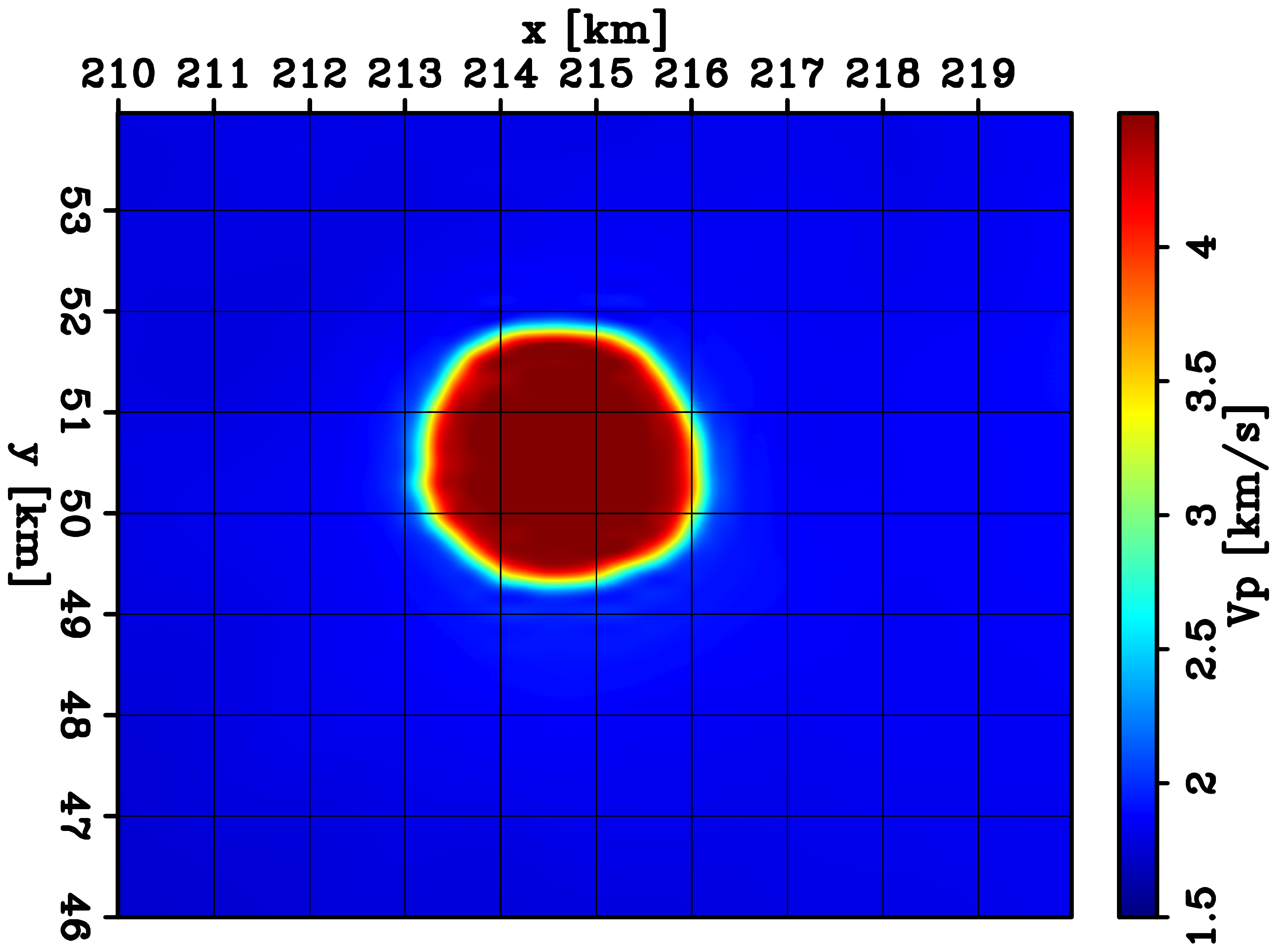}}
    \subfigure[]{\label{fig:CardamomInitXComp}\includegraphics[width=0.495\columnwidth]{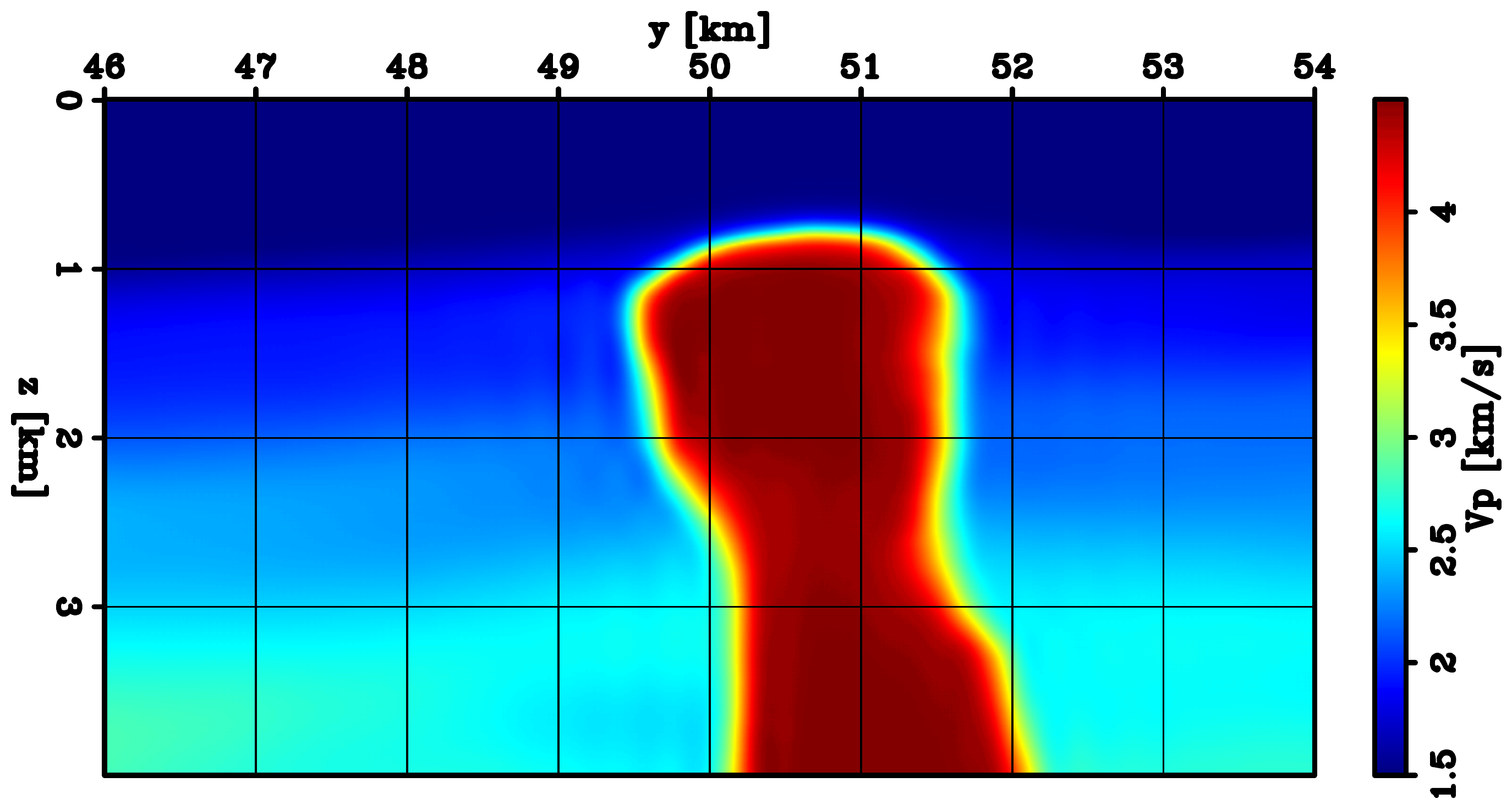}}
    
    \subfigure[]{\label{fig:CardamomFinalZComp}\includegraphics[width=0.365\columnwidth]{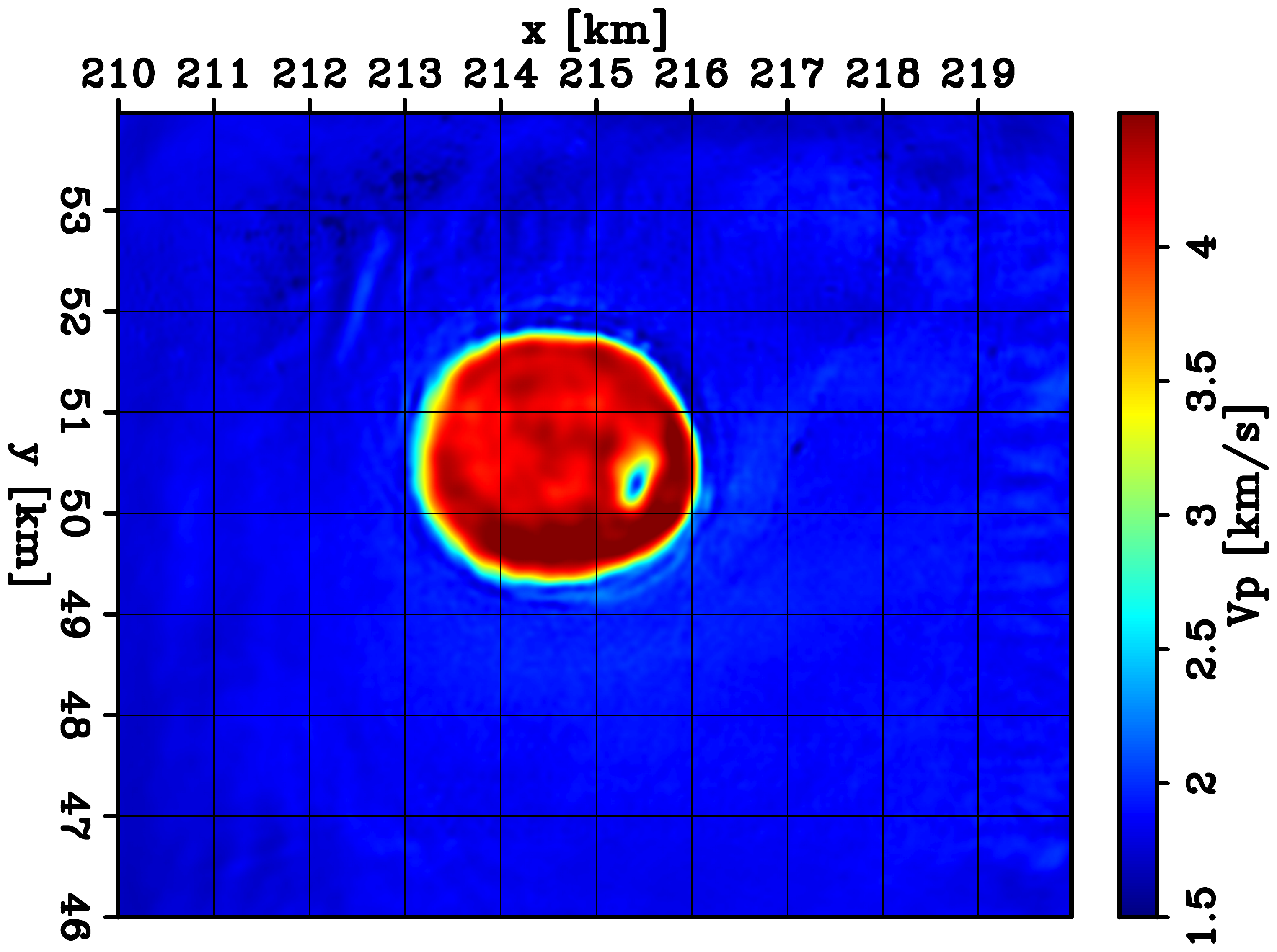}}
    \subfigure[]{\label{fig:CardamomFinalXComp}\includegraphics[width=0.495\columnwidth]{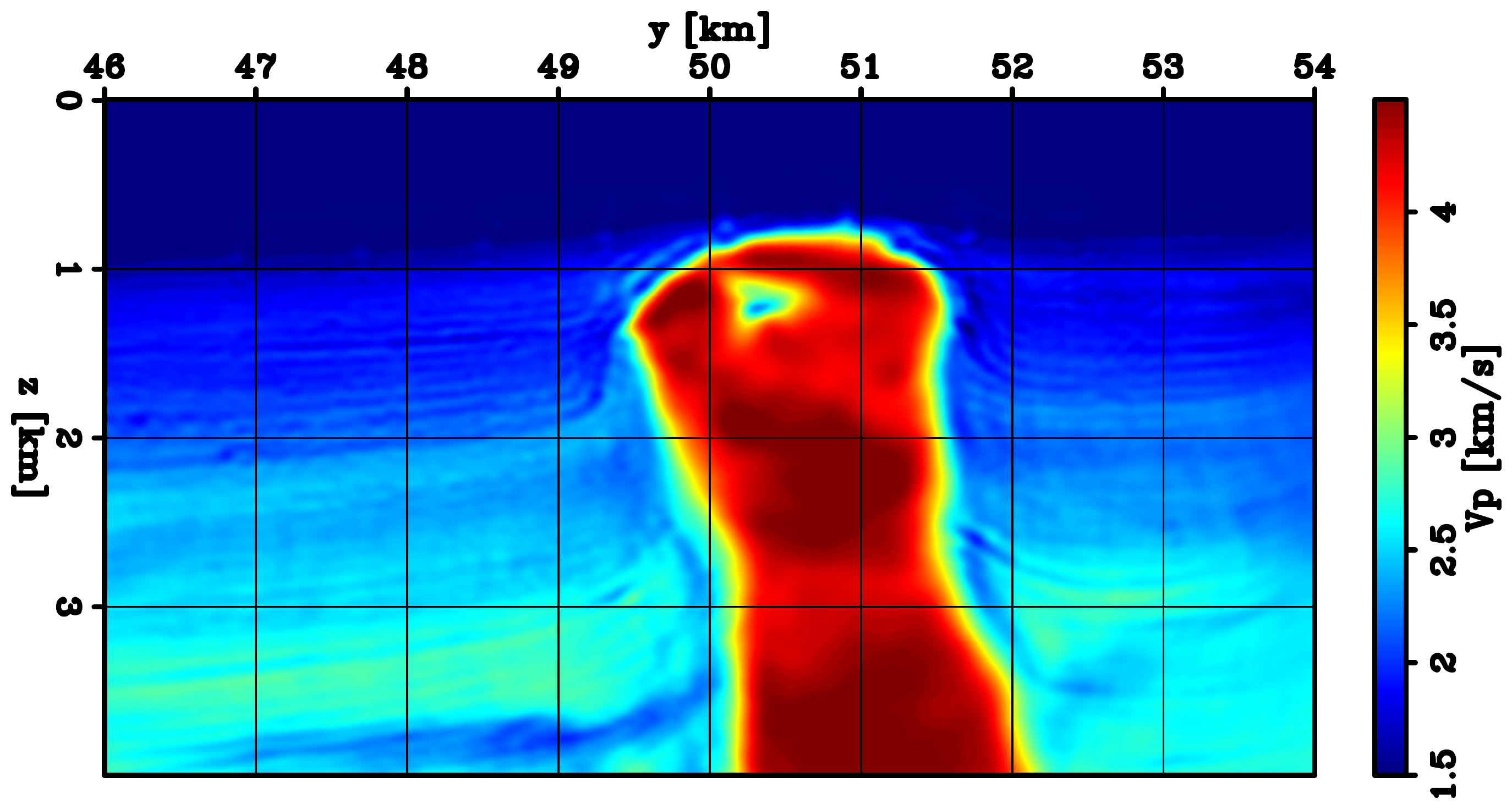}}

    \caption{Comparison between the initial (top panels) and the inverted (bottom panels) acoustic velocity models for (a-c) $z=1.2$ km and (b-d) $x=215.5$ km.}
    \label{fig:CardamomInitComp}
\end{figure}

The inversion is affected by artifacts due to the limited acquisition geometry (left sides of Figure~\ref{fig:CardamomFinalX1} and~\ref{fig:CardamomFinalY3}). Moreover, low-velocity anomalies are placed at depths higher than $3$ km. These features are probably due to the convergence to a local minimum of the optimization algorithm. Thus, we limit our area of search for any potential target to $3$ km of depth. In addition to the inclusion shown in Figure~\ref{fig:CardamomFinalX2}, the FWI workflow placed a low-velocity anomaly close to the top of the diapir (Figure~\ref{fig:CardamomInitComp}). This velocity decrease could be associated with salt-encased sediment packages included during the diapir formation~\cite[]{fernandez2017origin}.

\begin{figure}[t]
    \centering
    \subfigure[]{\label{fig:CardamomDataCompInitY}\includegraphics[width=0.65\columnwidth]{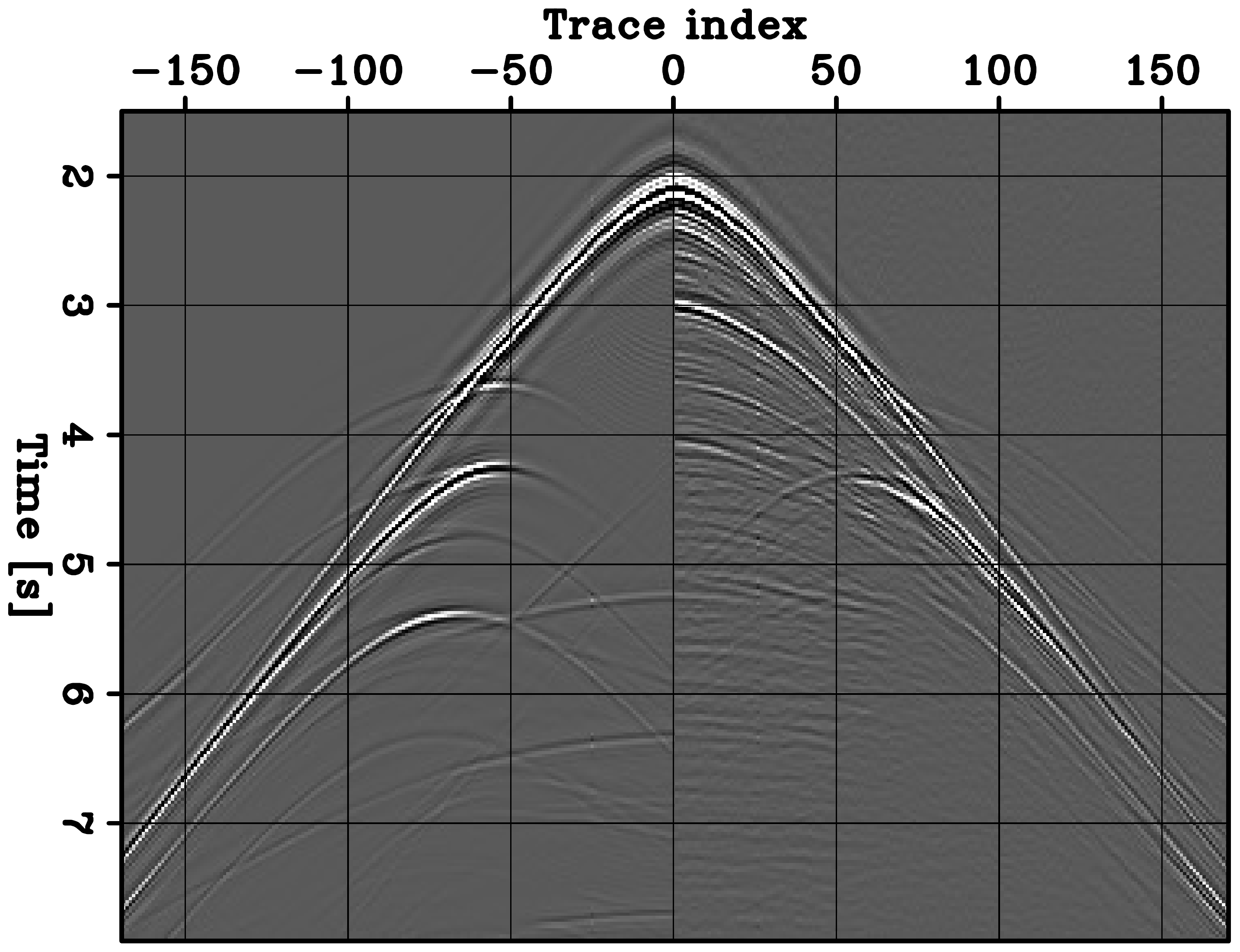}}
    
    \subfigure[]{\label{fig:CardamomDataCompFWIY}\includegraphics[width=0.65\columnwidth]{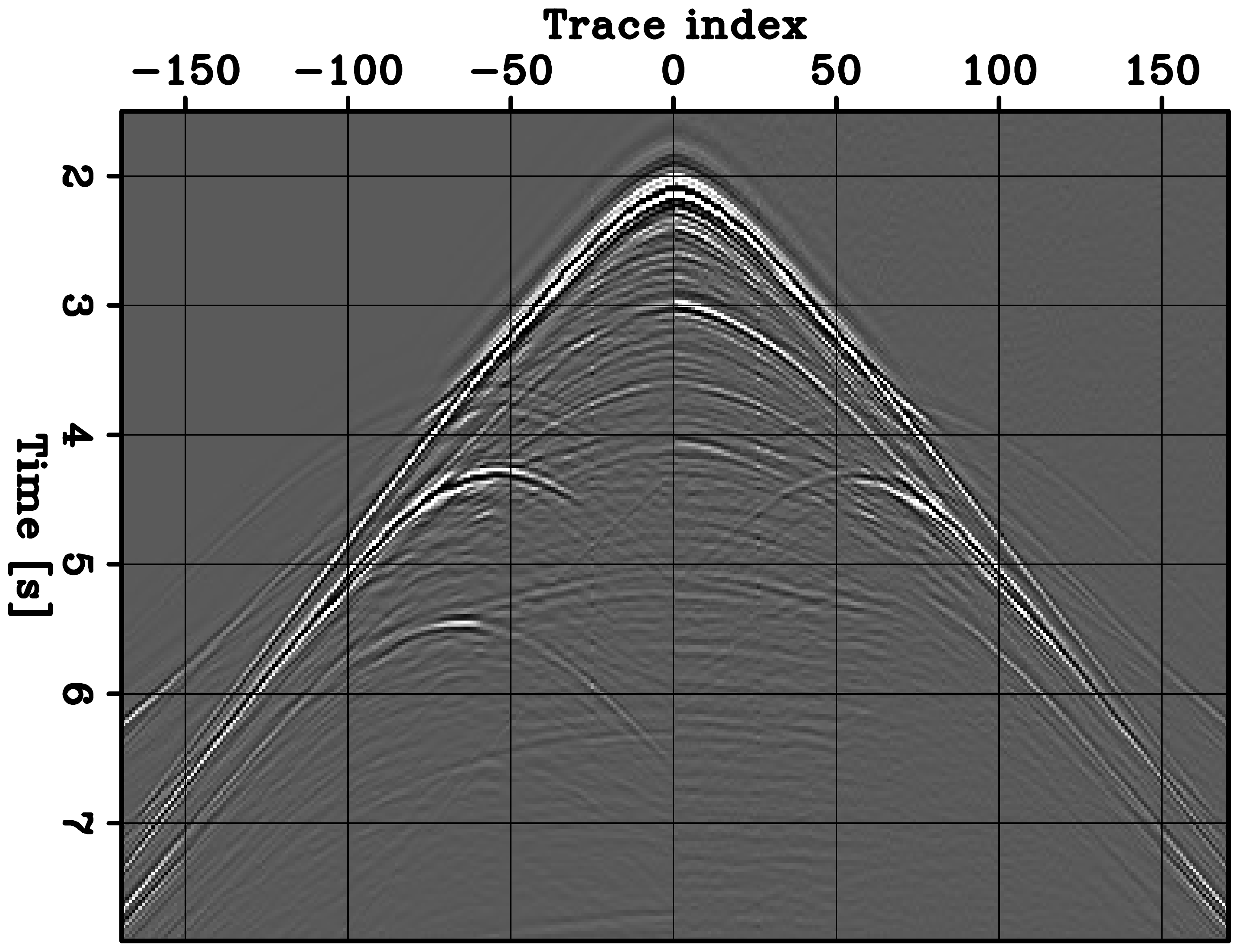}}

    \caption{Comparison between the predicted and observed pressure data on the initial (a) and inverted (b) models, respectively. The negative trace indices indicate the predicted data, while the positive ones denote the observed data. The noisy traces are due to the shot-binning process. Only the traces for $S_x=49.0$ km are plotted.}
    \label{fig:CardamomDataCompY}
\end{figure}

\begin{figure}[t]
    \centering
    \subfigure[]{\label{fig:CardamomDataCompInitT}\includegraphics[width=0.65\columnwidth]{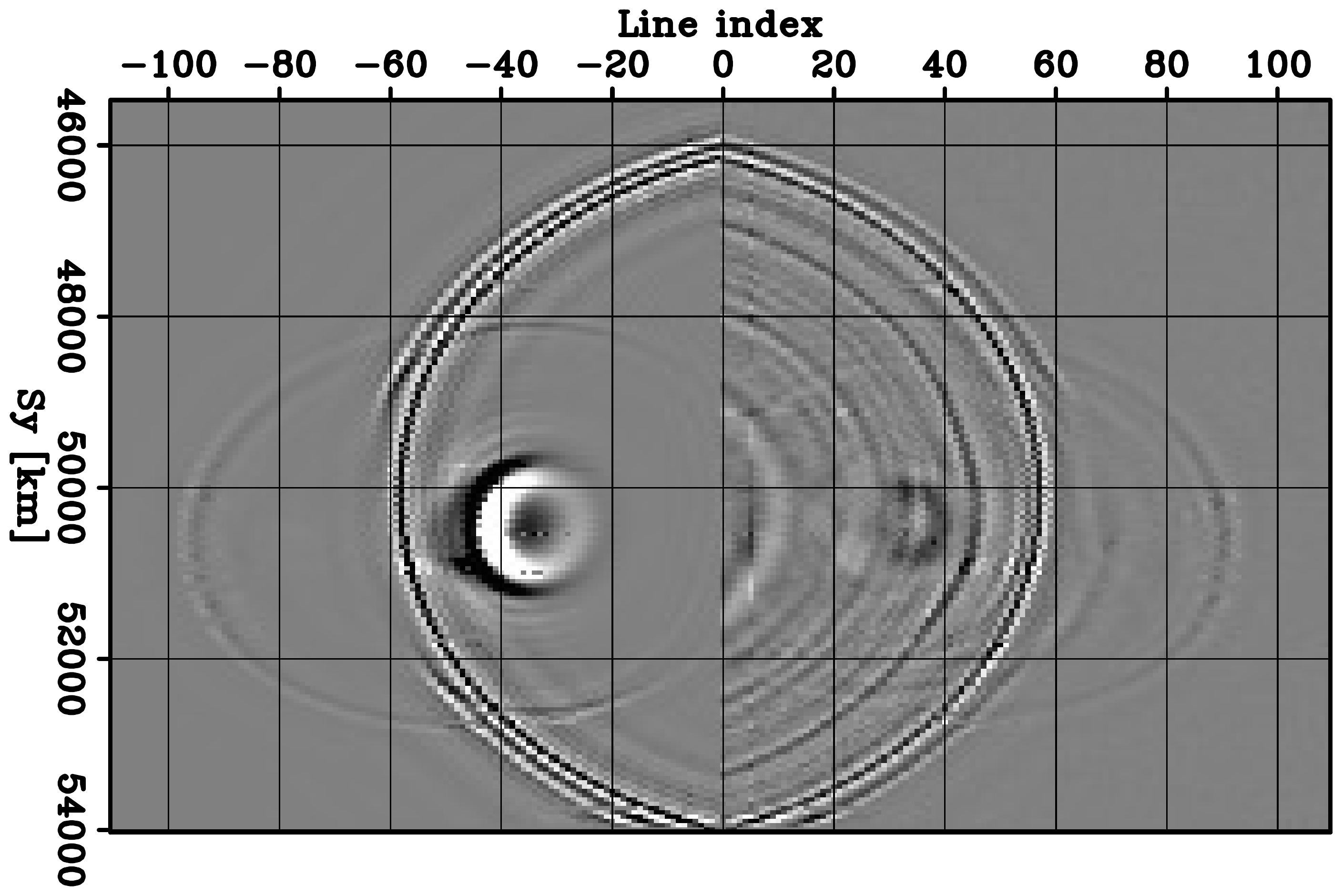}}
    
    \subfigure[]{\label{fig:CardamomDataCompFWIT}\includegraphics[width=0.65\columnwidth]{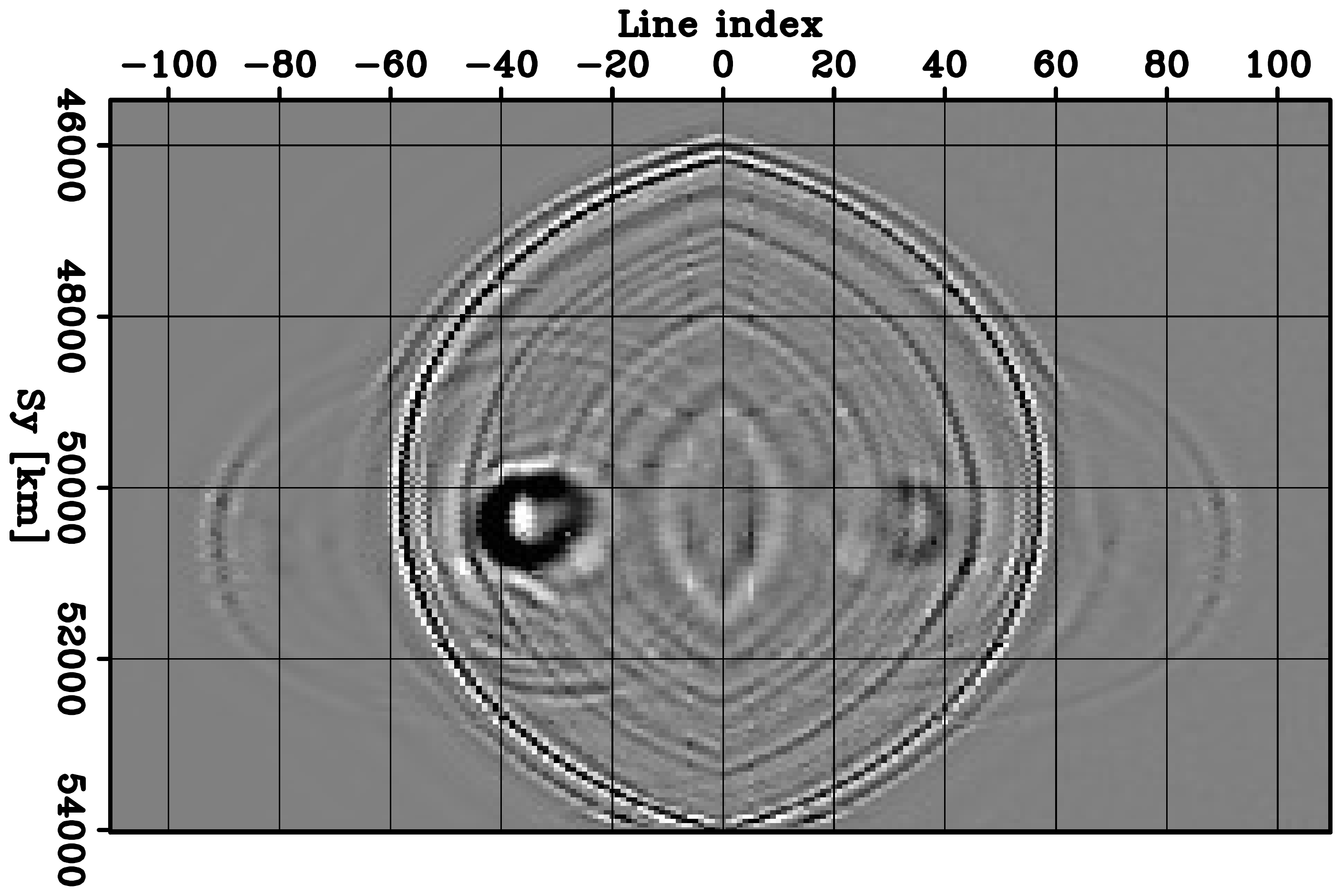}}

    \caption{Comparison between the predicted and observed pressure data on the initial (a) and inverted (b) models, respectively. The negative trace indices indicate the predicted data, while the positive ones denote the observed data. The noisy traces are due to the shot-binning process. The time slices are extracted at $t=4.0$ s.}
    \label{fig:CardamomDataCompT}
\end{figure}

As a quality control (QC) step, we compare the phase matching between the predicted and the observed pressure data before and after the inversion process is applied. Figure~\ref{fig:CardamomDataCompY} shows the phase matching when one of the source-spatial positions is fixed. The phase matching between modeled and observed data for both long- and short-offset traces improves after applying the FWI workflow. When a time slice is extracted from the modeled and the observed data, and we compare the phase agreement between the two (Figure~\ref{fig:CardamomDataCompT}), an excellent match is found using the final FWI acoustic model to generate the pressure data. On the other hand, from Figure~\ref{fig:CardamomDataCompY}, we notice that the accuracy of the matching diminishes for recording time greater than $5$ s for the mid- and short-offset ranges. This mismatch could explain the spurious low-velocity anomalies in the FWI acoustic model previously described.

\begin{figure}[t]
    \centering
    \subfigure[]{\label{fig:CardamomRTMInitx1}\includegraphics[width=0.49\columnwidth]{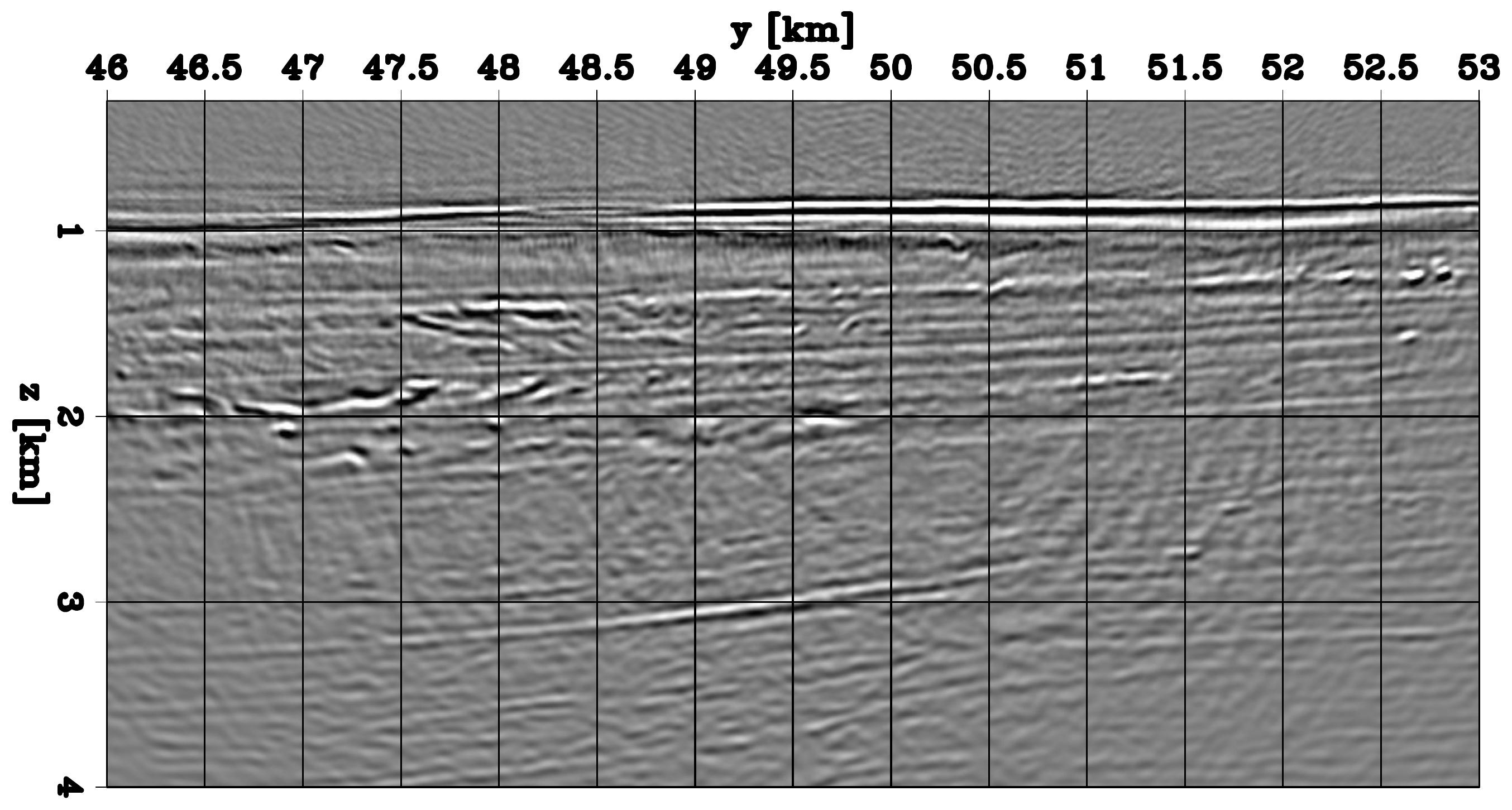}}
    \subfigure[]{\label{fig:CardamomRTMInitx2}\includegraphics[width=0.49\columnwidth]{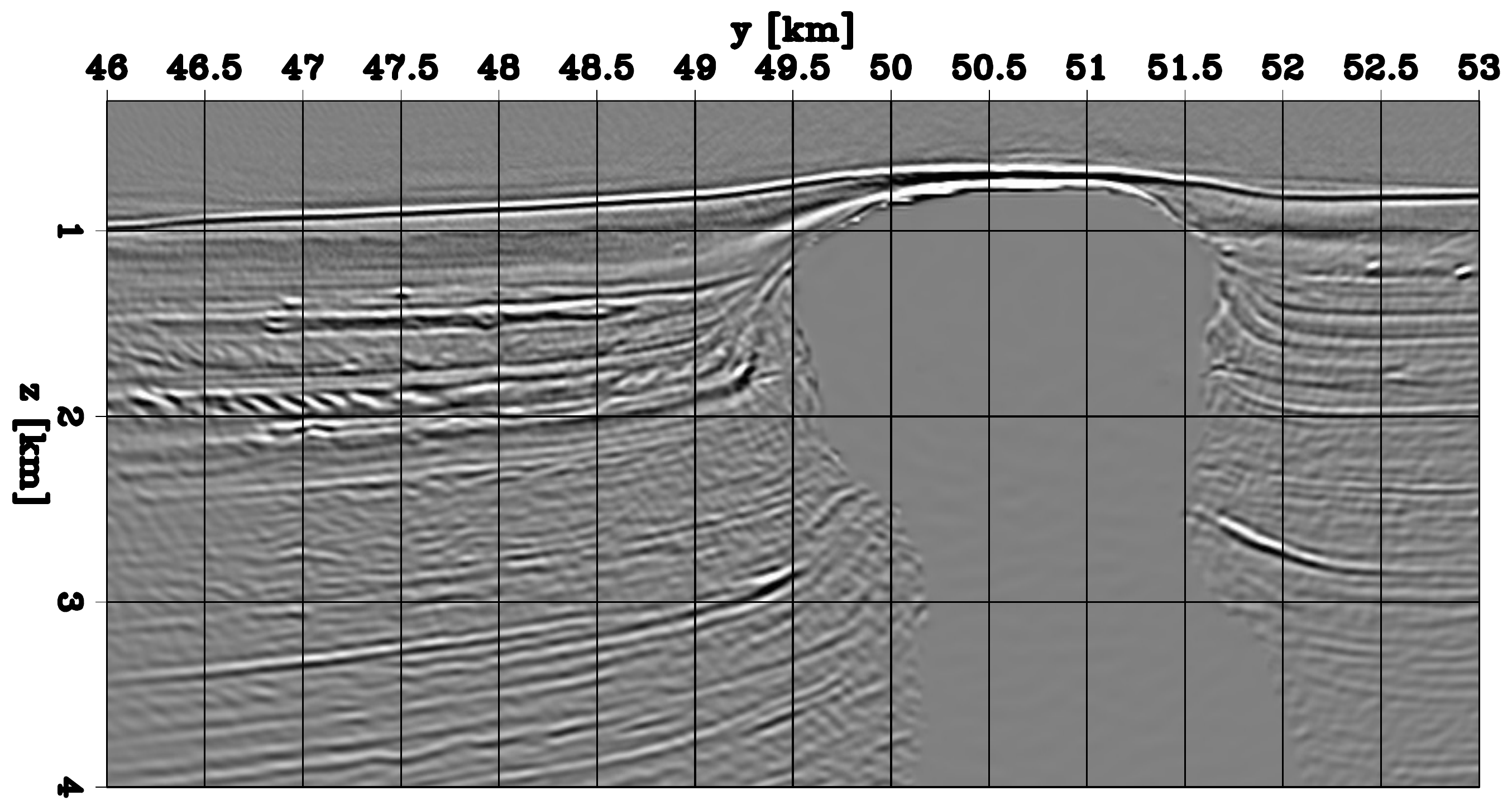}}
    
    \subfigure[]{\label{fig:CardamomRTMFWIx1}\includegraphics[width=0.49\columnwidth]{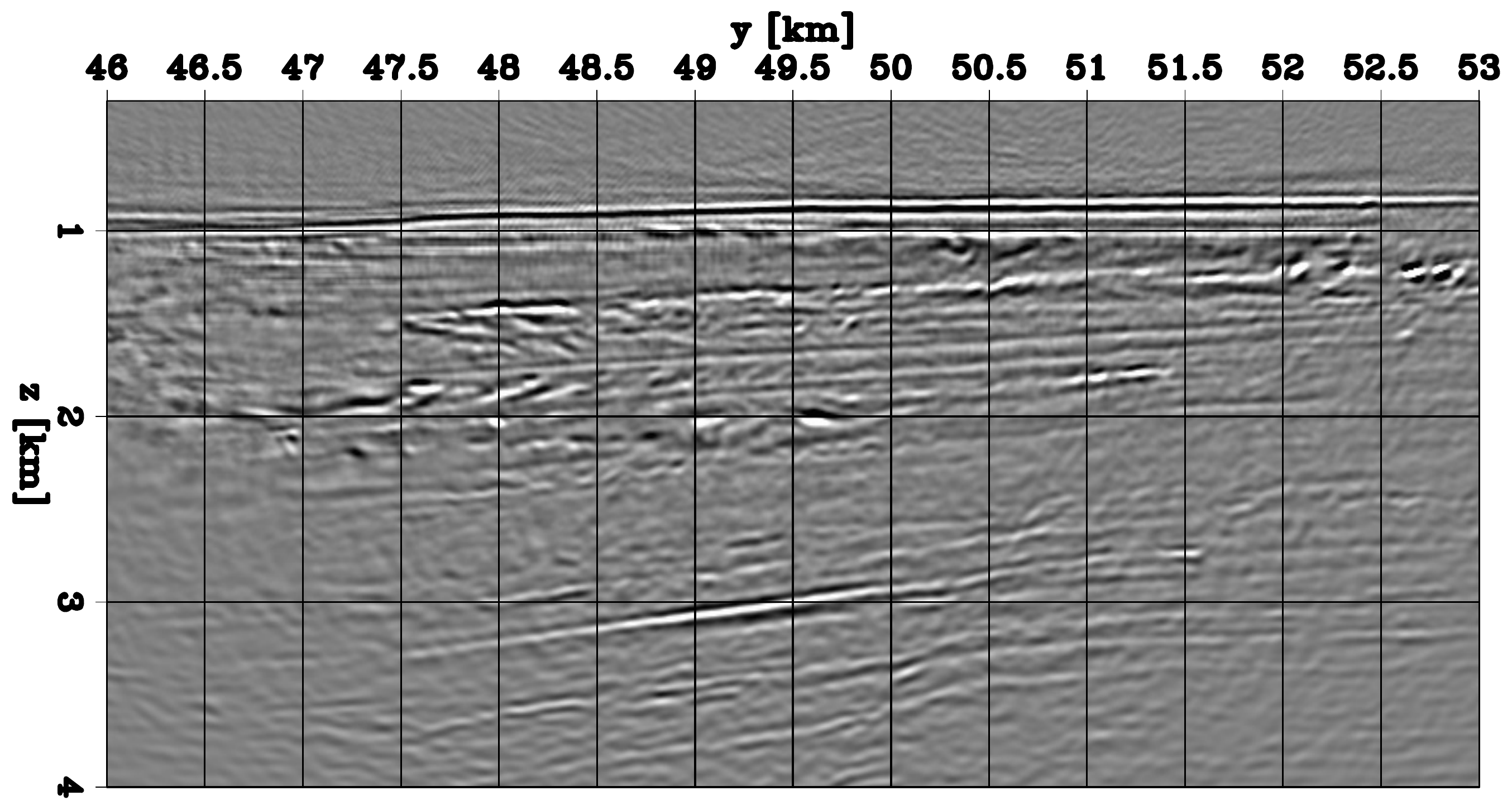}}
    \subfigure[]{\label{fig:CardamomRTMFWIx2}\includegraphics[width=0.49\columnwidth]{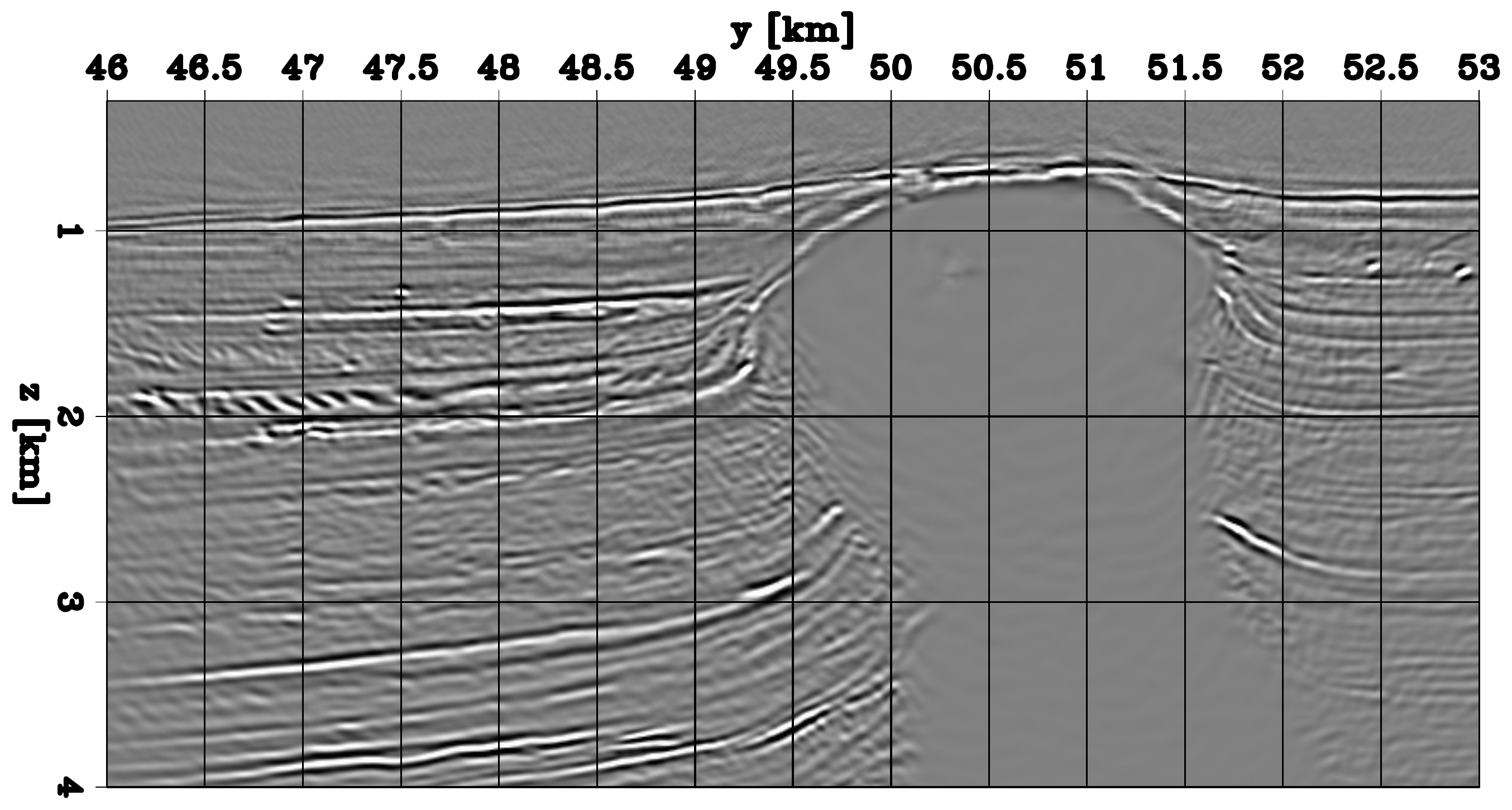}}
    
    \caption{Comparison between the $30$ Hz RTM images obtained using the initial (a-b) and the final FWI (c-d) models. The left and right panels are the cross-line sections extracted at $x=212$ km and $215.5$ km, respectively.}
    \label{fig:CardamomRTMcompX}
\end{figure}

Besides the satisfactory phase matching between the predicted and observed data, the quality of the RTM image greatly improves thanks to the more accurate velocity retrieved by the FWI process (Figure~\ref{fig:CardamomRTMcompX}). In fact, by comparing Figures~\ref{fig:CardamomRTMInitx1} and~\ref{fig:CardamomRTMFWIx1}, the fault planes between $y=49.8$ and $y=51.5$ km are more visible within the RTM image obtained on the FWI velocity model. For the sections passing through the salt diapir (Figures~\ref{fig:CardamomRTMInitx2} and~\ref{fig:CardamomRTMFWIx2}), the overall reflectors' continuity is improved for the RTM image generated on the FWI model; especially, for the reflectors close to the top of the salt body. Moreover, the high amplitude reflectors present a more consistent contact point with the salt flanks within the FWI-related RTM image. One interesting geological feature present on the left side in both sections of Figures~\ref{fig:CardamomRTMInitx2} and~\ref{fig:CardamomRTMFWIx2} is the sigmoidal shaped reflectors at $z=2.0$ km. These events are due to the presence of the turbidite deposits previously described.

\subsection{Target-oriented elastic FWI of a potential prospect}
\begin{figure}[t]
    \centering
    \includegraphics[width=1.0\linewidth]{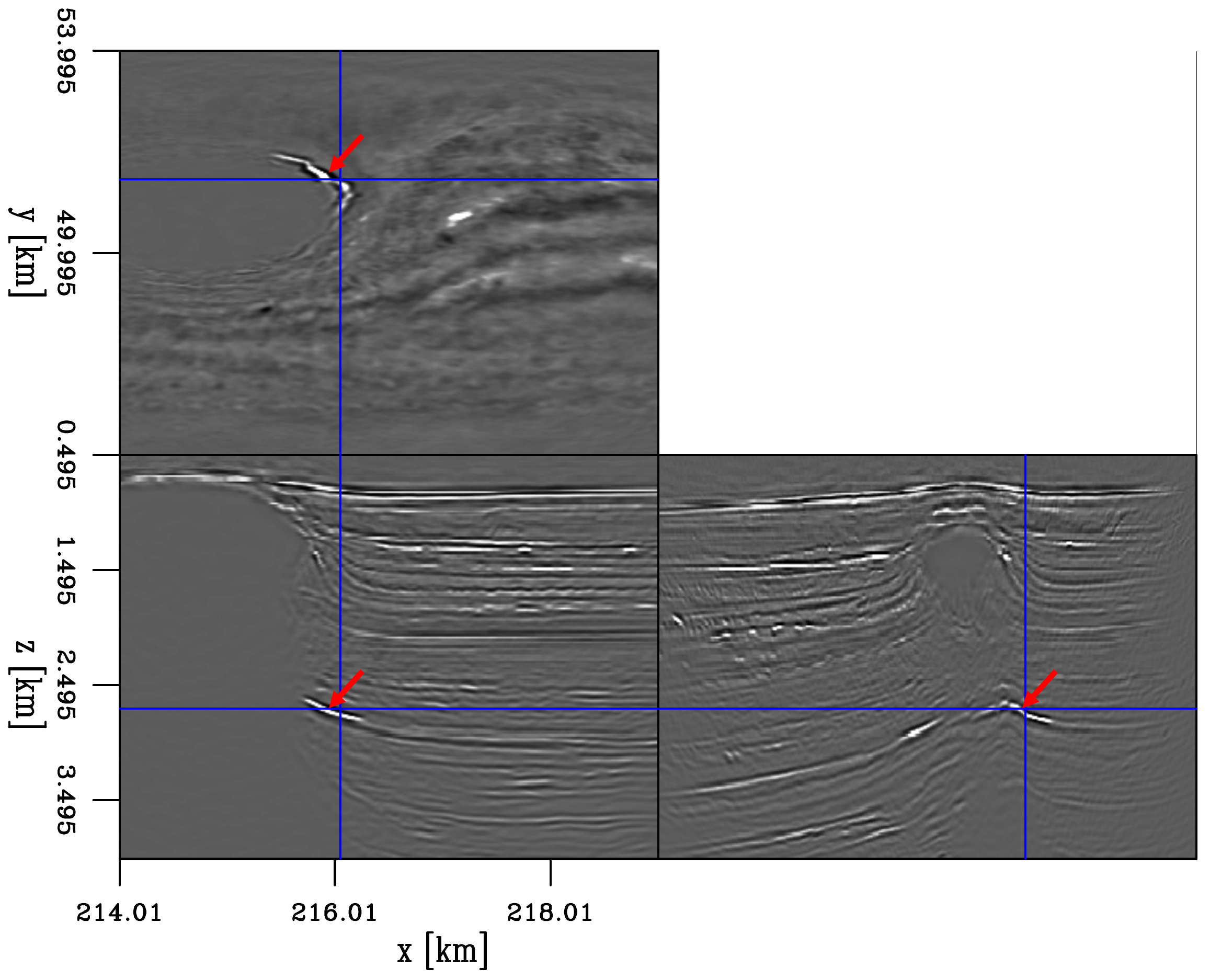}
    \caption{$30$-Hz RTM image volume obtained using the FWI model showing sections passing through a potential reservoir indicated by the red arrow.}
    \label{fig:CardamomRTMFWICube}
\end{figure}

By analyzing the RTM image volume obtained using the final FWI velocity model, we identify a clear high-amplitude reflector in the proximity of the salt flank (Figure~\ref{fig:CardamomRTMFWICube}). This amplitude response could be related to gas accumulation at the top of a hydrocarbon reservoir~\cite[]{mazzotti1990prestack}. Therefore, we apply the redatuming technique, followed by an elastic FWI workflow to retrieve this potential prospect's elastic properties.

\begin{figure}[t]
    \centering
    \subfigure[]{\label{fig:CardamomExtLSRTMObj}\includegraphics[width=0.49\columnwidth]{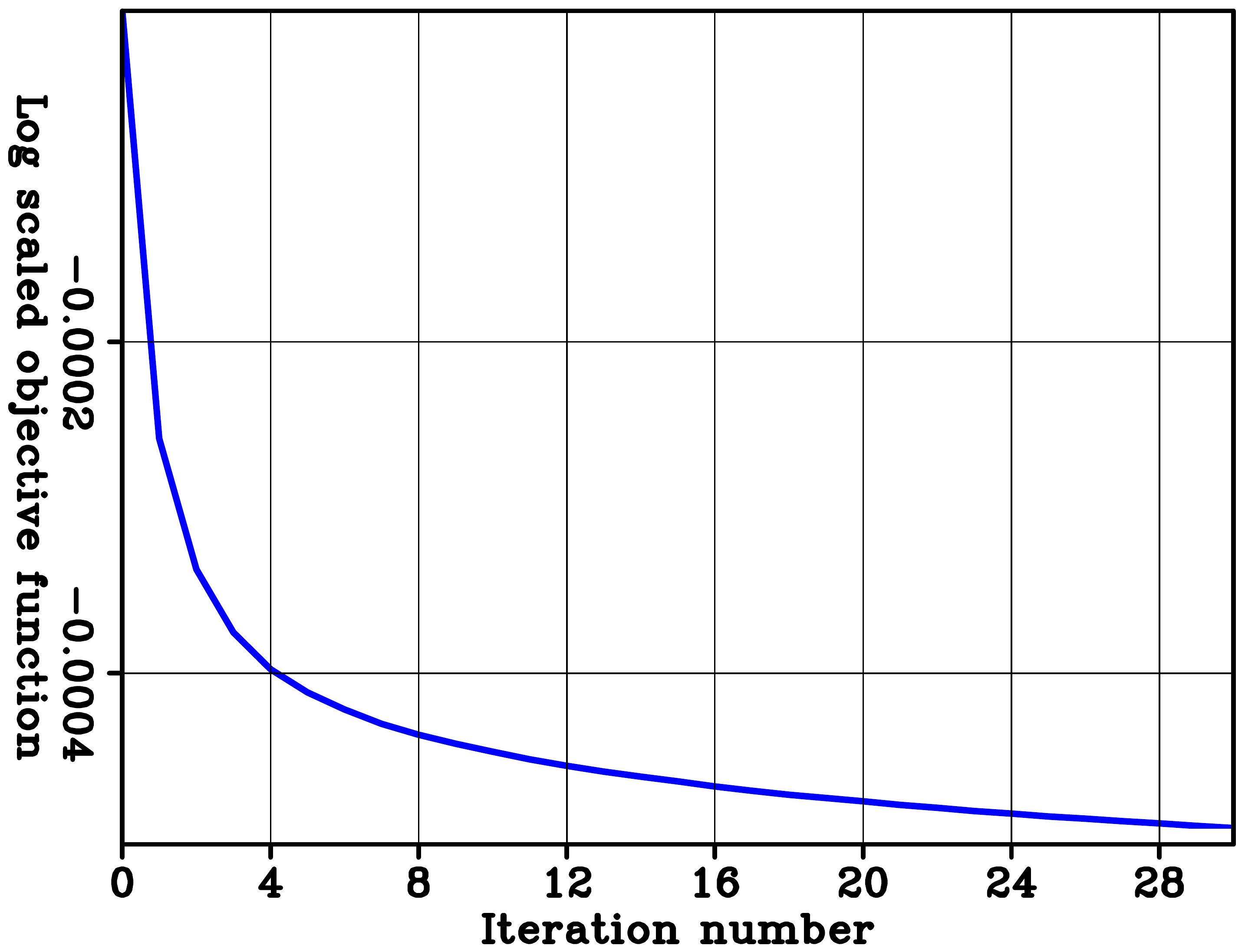}}
    
    \subfigure[]{\label{fig:CardamomExtLSRTMTarg}\includegraphics[width=0.8\columnwidth]{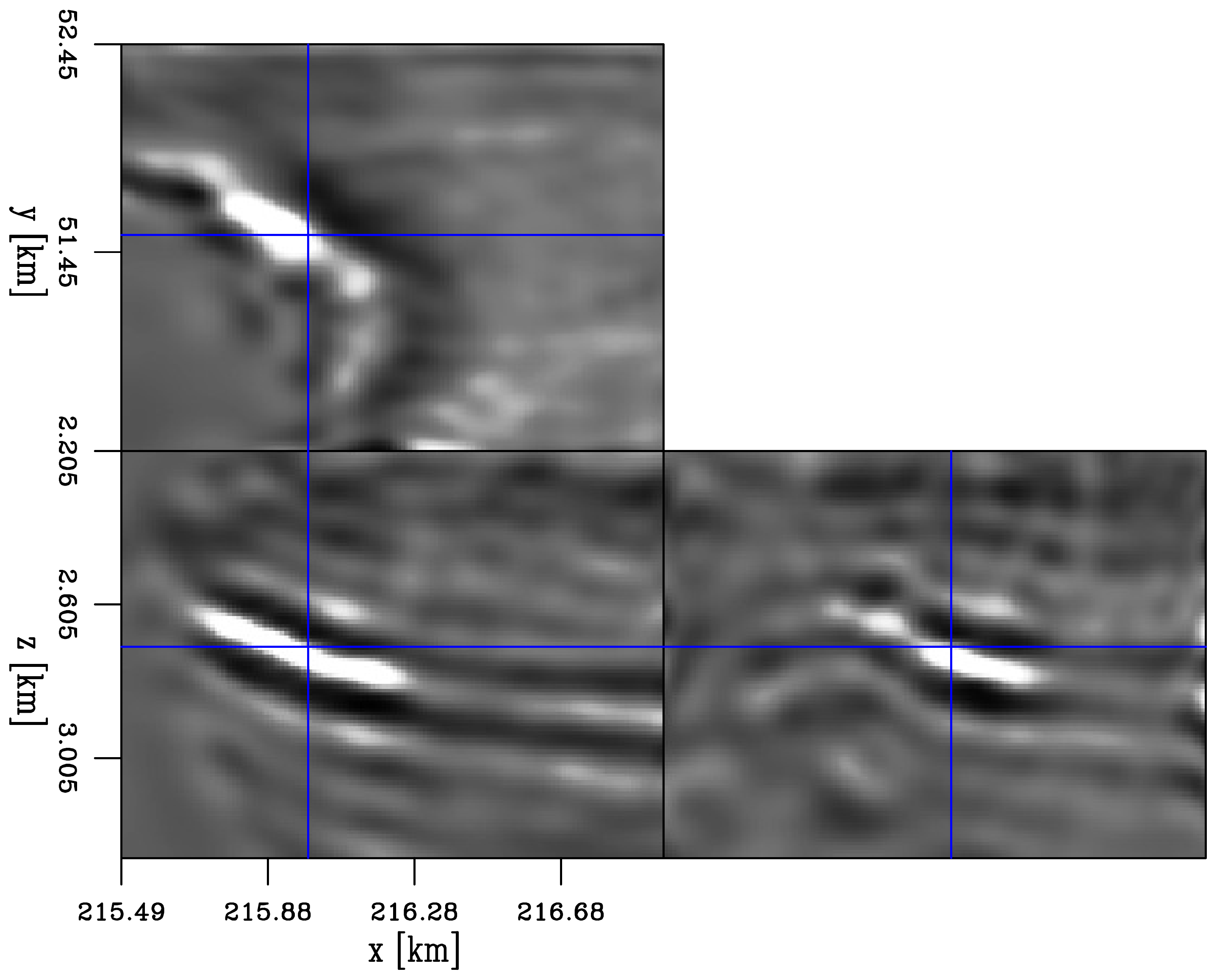}}
    
    \caption{(a) Convergence curve of the extended linearized waveform inversion problem of the GOM dataset with maximum frequency of $12$ Hz. (b) Closeup of the target extracted at $h_x=h_y=0.0$ km.}
    \label{fig:CardamomExtLSRTM}
\end{figure}

The first step is to solve an extended acoustic linearized inversion of the observed data to obtain an extended 3D image volume. We limit the observed data's maximum frequency to $12$ Hz to make the least-squares process feasible with the available computational resources. Higher frequencies can be employed as long as computer memory or disks can hold the migration wavefield and the extended images. Random-boundary conditions could be used to alleviate the memory requirements during the imaging step~\cite[]{clapp2009reverse}.  The FD grid is set to $35$ m in each direction. Moreover, given the acquisition's full-azimuth nature, we employ $h_x$ and $h_y$ subsurface-offset extensions of $9$ points in each direction, resulting in a maximum absolute subsurface offset of $140$ m.  Finally, a differential-semblance-optimization (DSO) regularization term is added to improve the image focusing~\cite{symes1994inversion}, and its weight is chosen on a heuristic basis. We focus the iterative process on inverting the data component stemming from the target area by employing a mask tailored for the potential prospect.  After 30 iterations of a linear CG algorithm, an acceptable numerical minimum of the objective function is reached given the selected parameters (Figure~\ref{fig:CardamomExtLSRTMObj}). Each linear iteration took approximately 1 hour and 15 minutes on the same machine used for the Marmousi2 experiment. When extracting the zero-subsurface offset image within the target, an evident high-amplitude reflector is visible (Figure~\ref{fig:CardamomExtLSRTMTarg}). Moreover, fault planes are clearly affecting the horizon of interest.

\begin{figure}[t]
    \centering
    \subfigure[]{\label{fig:CardamomExtODCIG}\includegraphics[width=0.49\columnwidth]{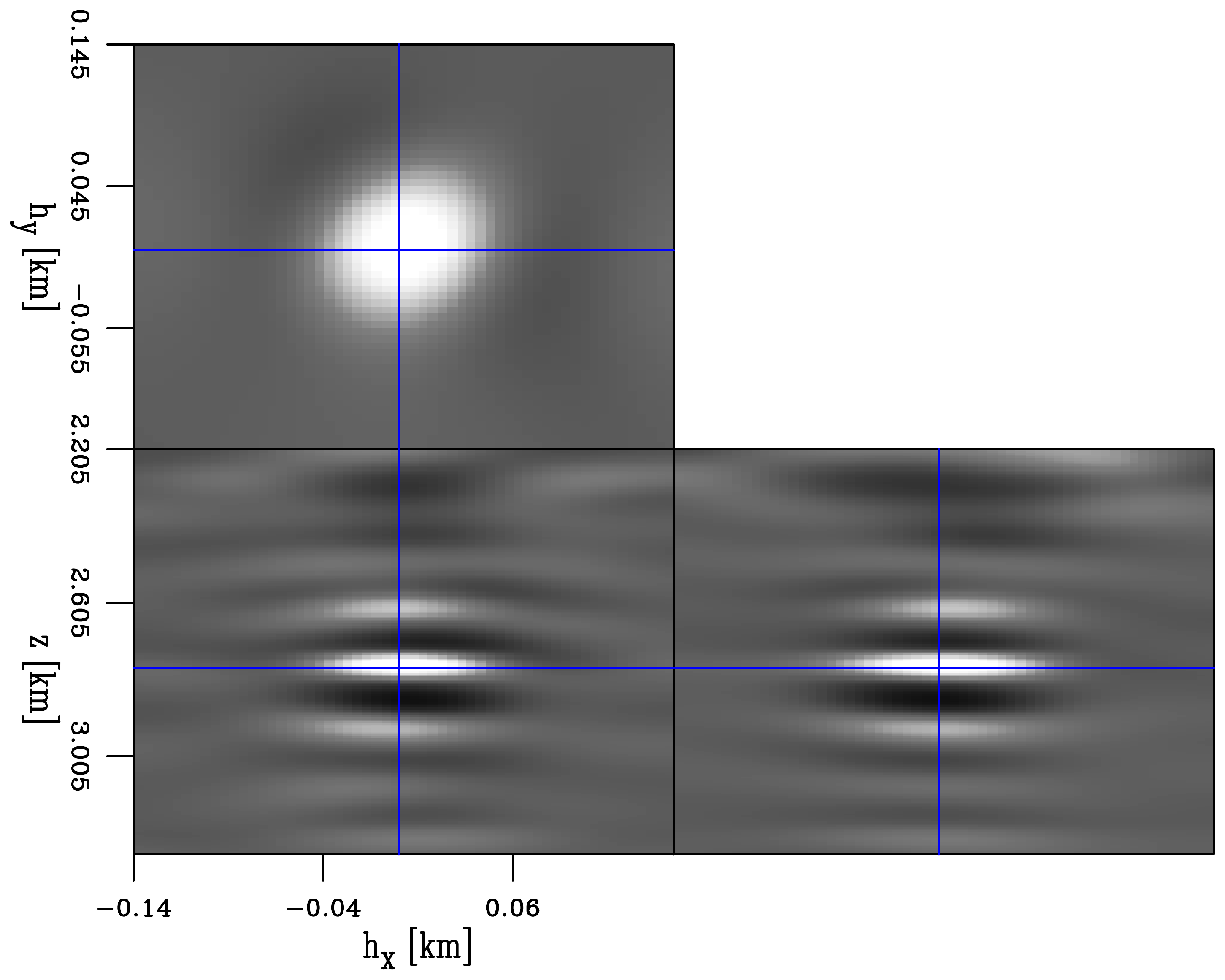}}
    \subfigure[]{\label{fig:CardamomExtADCIG}\includegraphics[width=0.49\columnwidth]{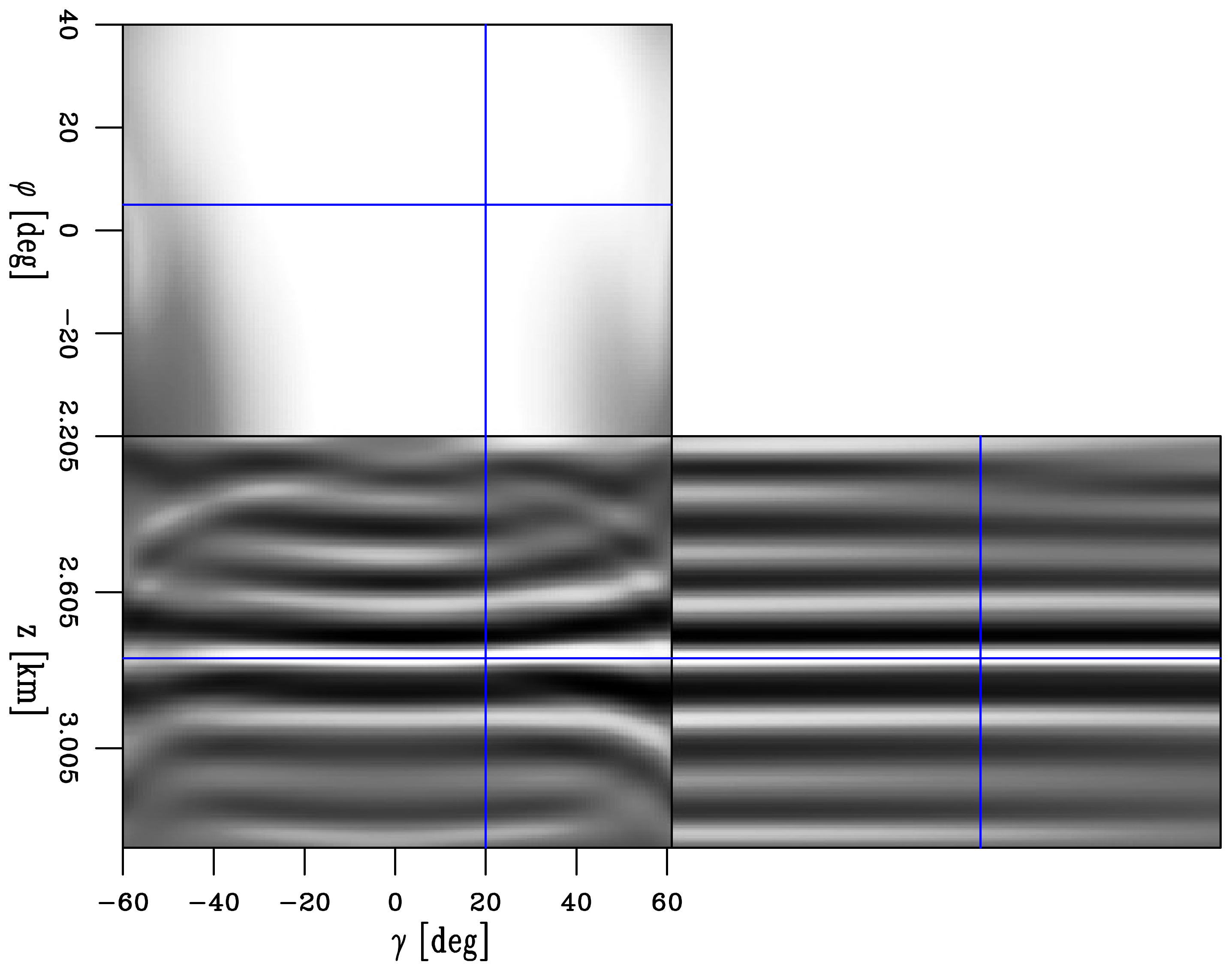}}
    
    \subfigure[]{\label{fig:CardamomExtADCIG-AVA}\includegraphics[width=0.49\columnwidth]{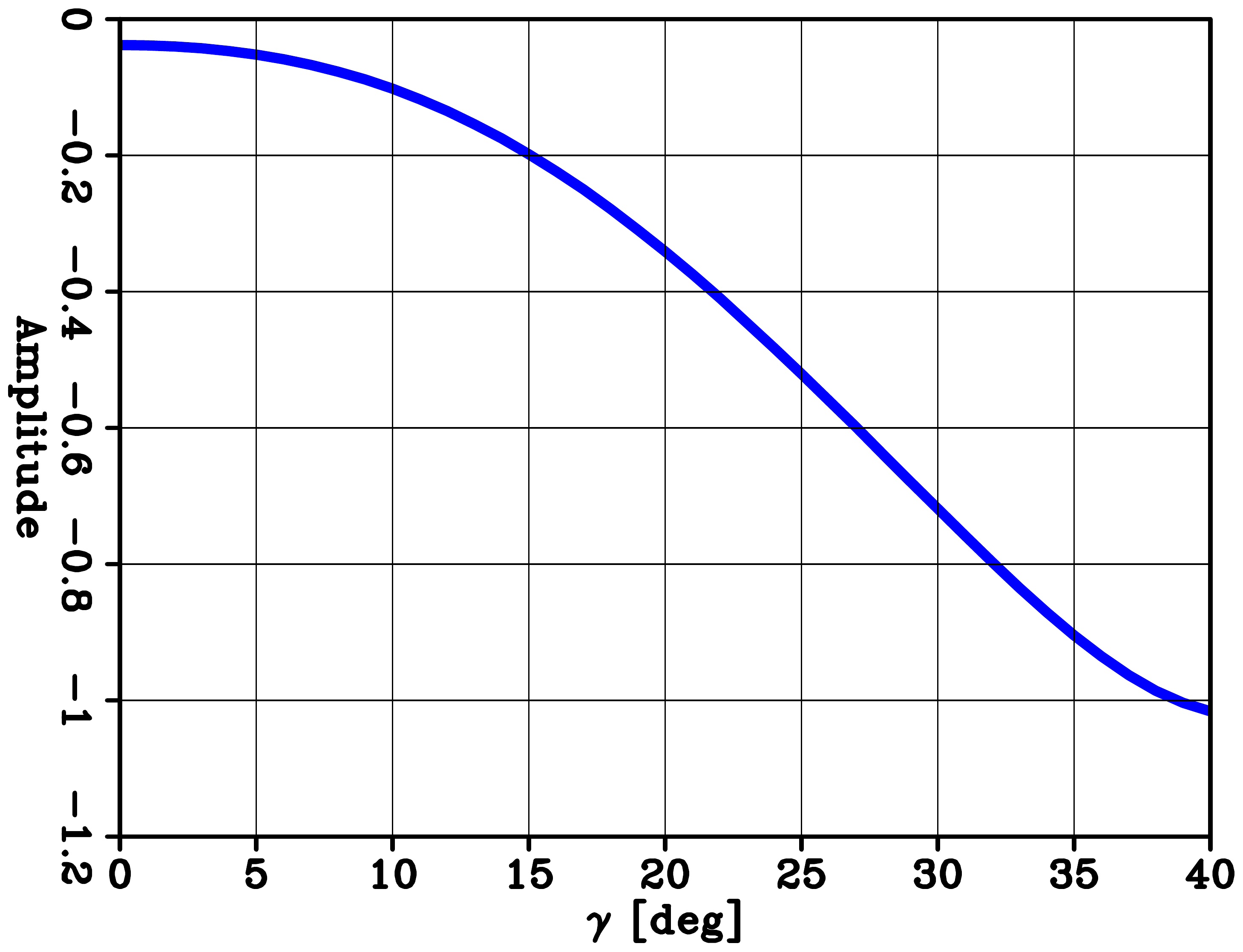}}
    
    \caption{(a) ODCIG of the potential prospect extracted at $x=216.1$ and $y=51.5$ km. (b) ADCIG extracted at $x=216.1$ and $y=51.5$ km. (c) Amplitude response of the ADCIG of panel (b) extracted at $z=2.7$ km and $\varphi=45^{\circ}$}
    \label{fig:CardamomExtCIG}
\end{figure}

The focusing of the offset-domain common-image gathers (ODCIGs) of the target area provides an additional QC step for assessing the migration velocity model's accuracy during the linear inversion process. A representative ODCIG extracted from the target volume is displayed in Figure~\ref{fig:CardamomExtODCIG}, which presents a clear focus around the zero-subsurface offset axes. Furthermore, we convert this ODCIG into the angle-azimuth domain~\cite[]{biondi20043d}, and extract the angle-domain common-image gather (ADCIG) (Figure~\ref{fig:CardamomExtADCIG}). This angle gather presents a flat response across reflection angles. These two observations suggest that the acoustic FWI process can retrieve an accurate overburden velocity. Finally, Figure~\ref{fig:CardamomExtADCIG-AVA} shows the amplitude response the the ADCIG at $z=2.7$ km and $\varphi=45^{\circ}$ following an class-3 AVO signature~\cite[]{castagna1993avo}. This amplitude response in addition to the bright reflectivity displayed in Figure~\ref{fig:CardamomRTMFWICube} is compelling evidence of the presence of a gas-bearing sand reservoir in that subsurface area. 

\begin{figure}[t]
    \centering
    \subfigure[]{\label{fig:CardamomTargetGeoSou}\includegraphics[width=0.49\columnwidth]{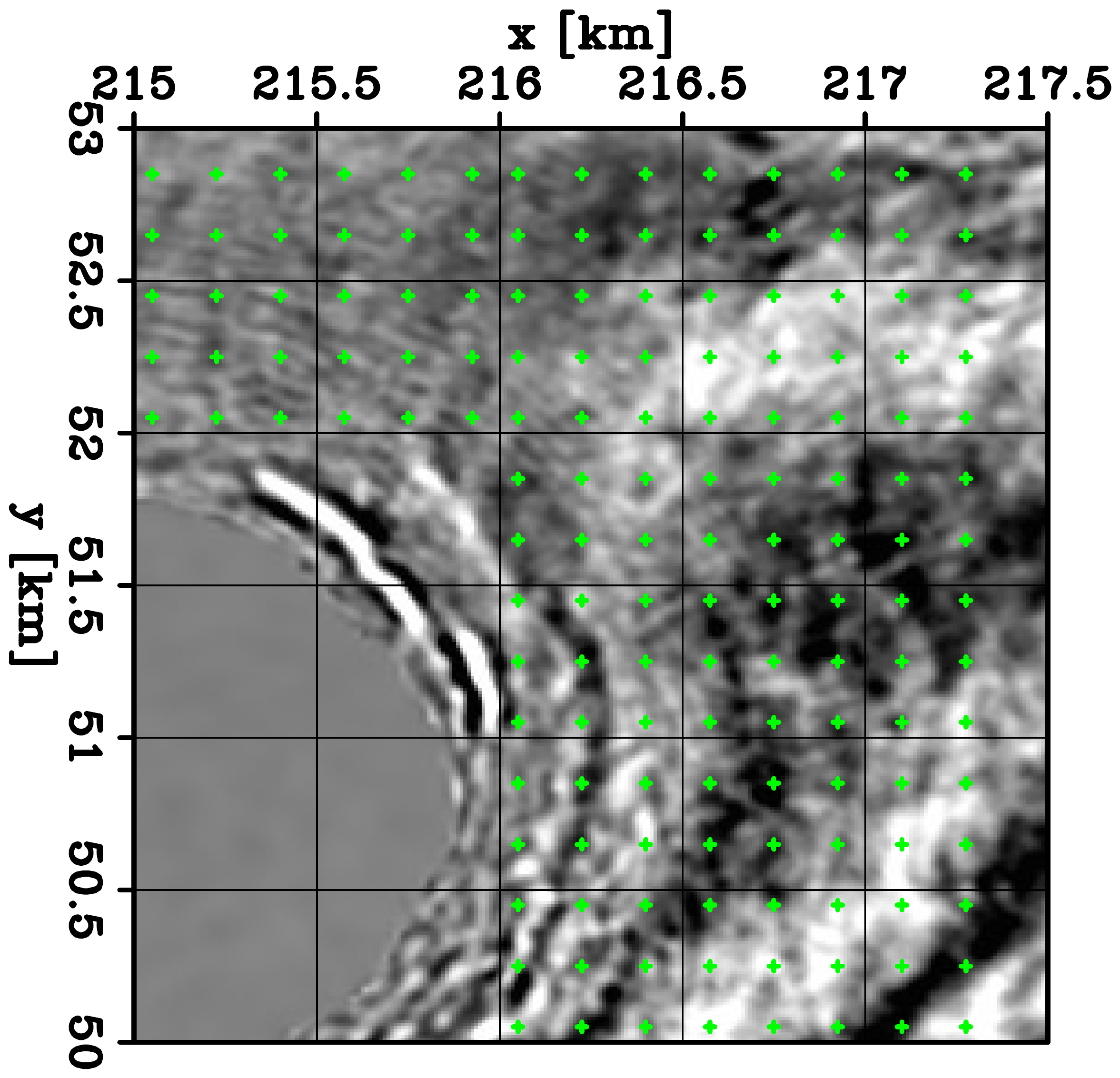}}
    
    \subfigure[]{\label{fig:CardamomTargetGeoRec}\includegraphics[width=0.49\columnwidth]{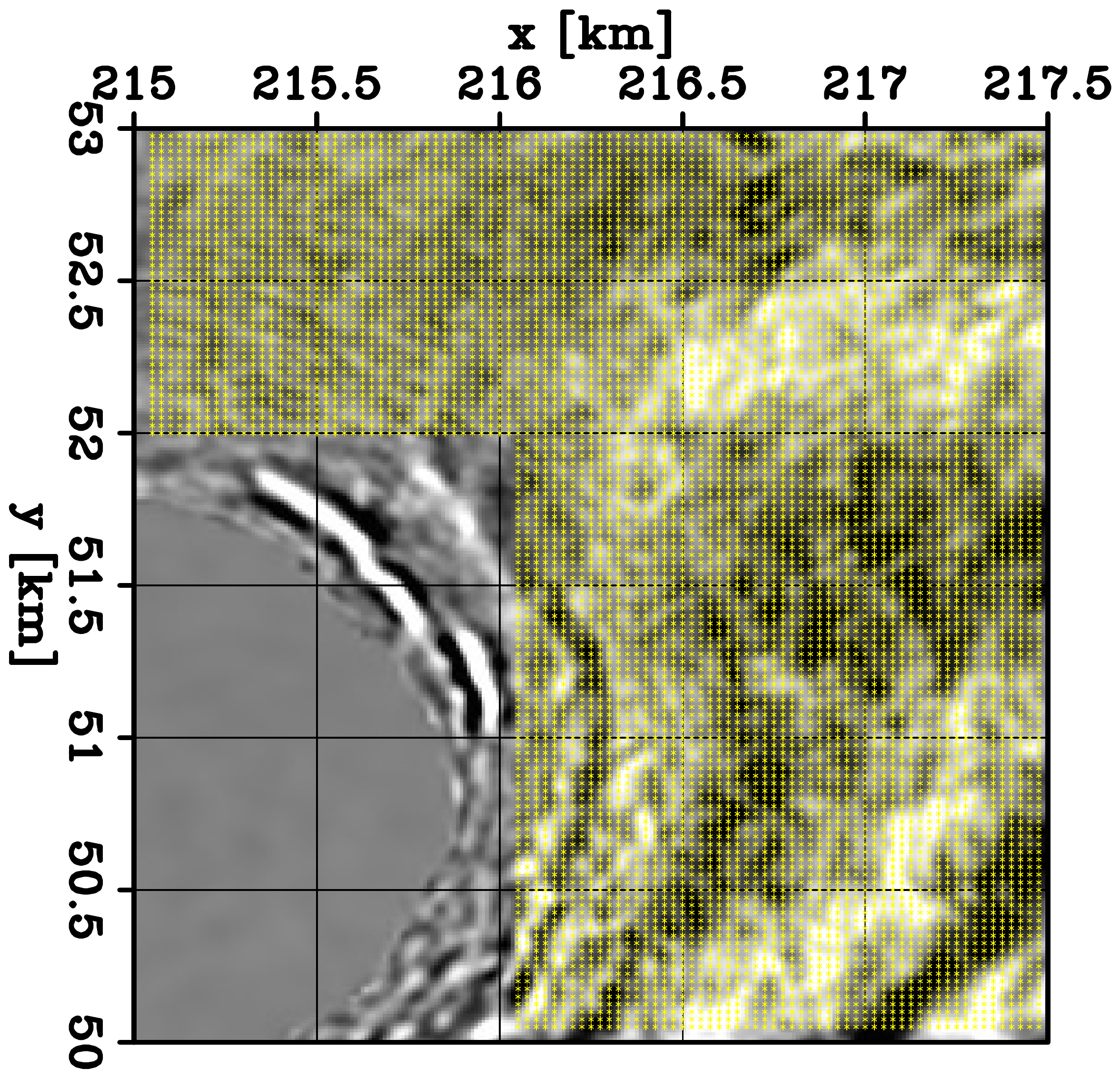}}
    
    \caption{ (a) Sources’ and (b) receivers’ x-y positions for the target-oriented inversion overlaid on the $30$ Hz RTM image depth slice extracted at $z= 2.6$ km.}
    \label{fig:CardamomTargetGeo}
\end{figure}

\begin{figure}[t]
    \centering
    \includegraphics[width=0.7\linewidth]{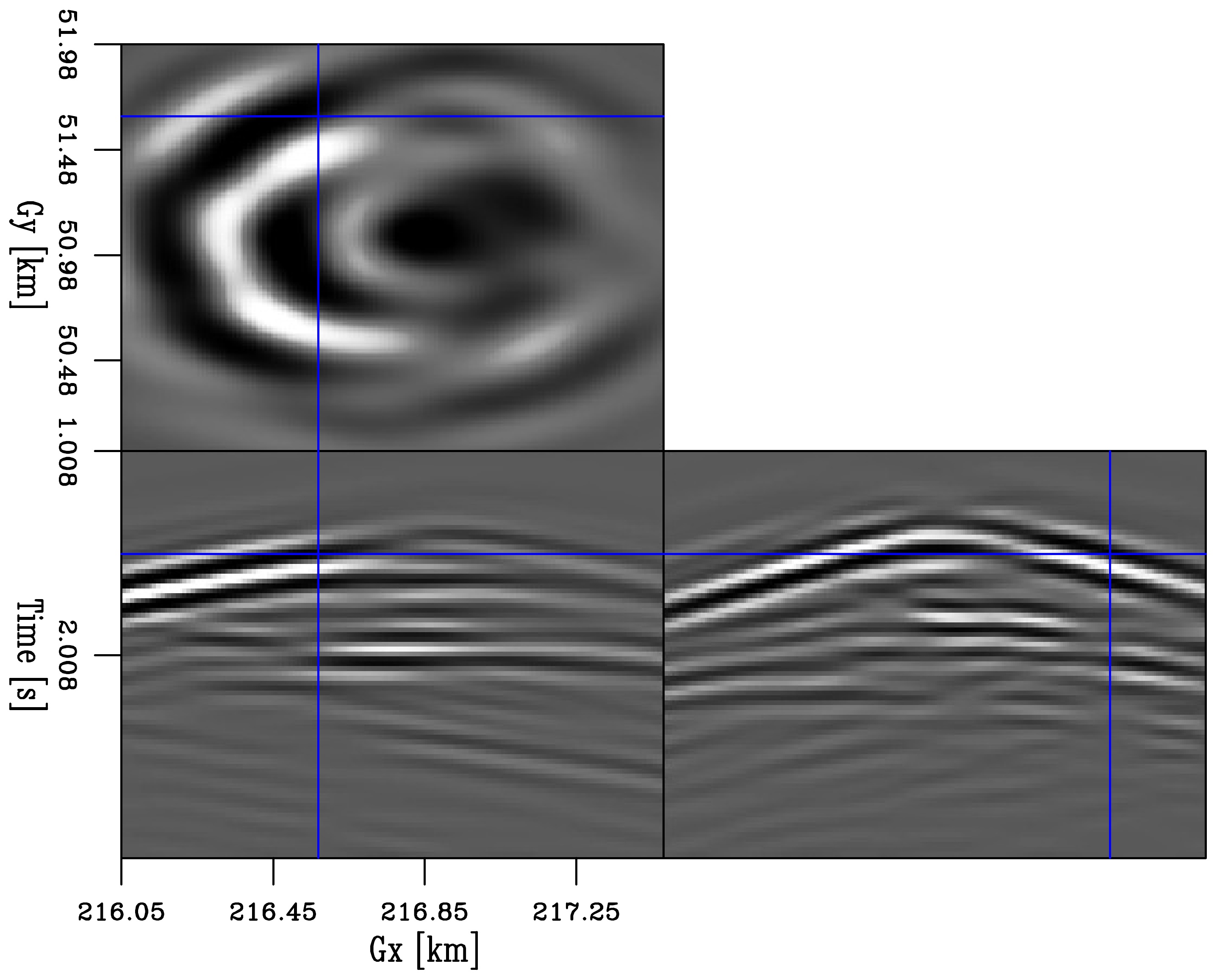}
    \caption{Representative shot gather for $Sx=216.9$ and $Sy=51.05$ km. Only part of the receivers is plotted.}
    \label{fig:CardamomDatumData}
\end{figure}

We employ this extended image volume to synthesize the elastic pressure data with a new redatumed acquisition geometry placed at $z=2.1$ km. The sources' and receivers' x-y positions are shown in the panels of Figure~\ref{fig:CardamomTargetGeo}. We employ $150$ sources and $8444$ receivers, with the latter regularly sampled and spaced by $25$ m in each direction. This new acquisition is chosen based on how the original OBN geometry has illuminated the target. We purposely avoid placing acquisition devices on the salt body, given the limited illumination of the target by the original OBN geometry present in that section of the model. Figure~\ref{fig:CardamomDatumData} shows a representative shot gather where an increase in amplitude for the first reflected event is noticeable for receivers at a further distance from the source position. This behavior is a potential indication of an AVO signature from the chosen prospect.

The entire model domain is approximately $10\times10\times4$ $\text{km}^3$, while the target domain size is approximately $1.5\times3\times1$ $\text{km}^3$, making the target computational domain approximately $67$ times smaller compared to the original one. 

\begin{figure}[t]
    \centering
    \subfigure[]{\label{fig:CardamomTargetInitElaVpZ}\includegraphics[width=0.32\columnwidth]{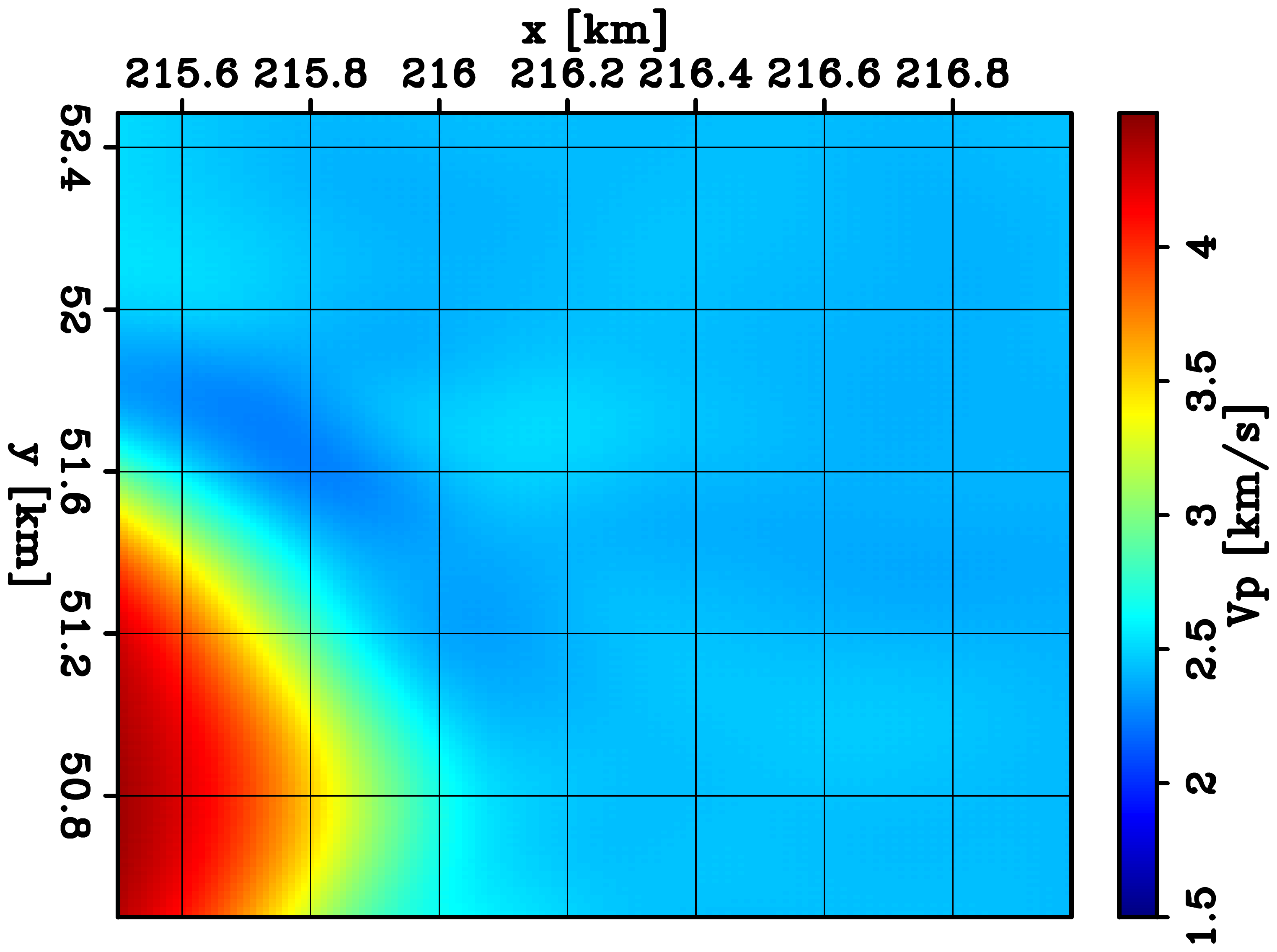}}
    \subfigure[]{\label{fig:CardamomTargetInitElaVpX}\includegraphics[width=0.32\columnwidth]{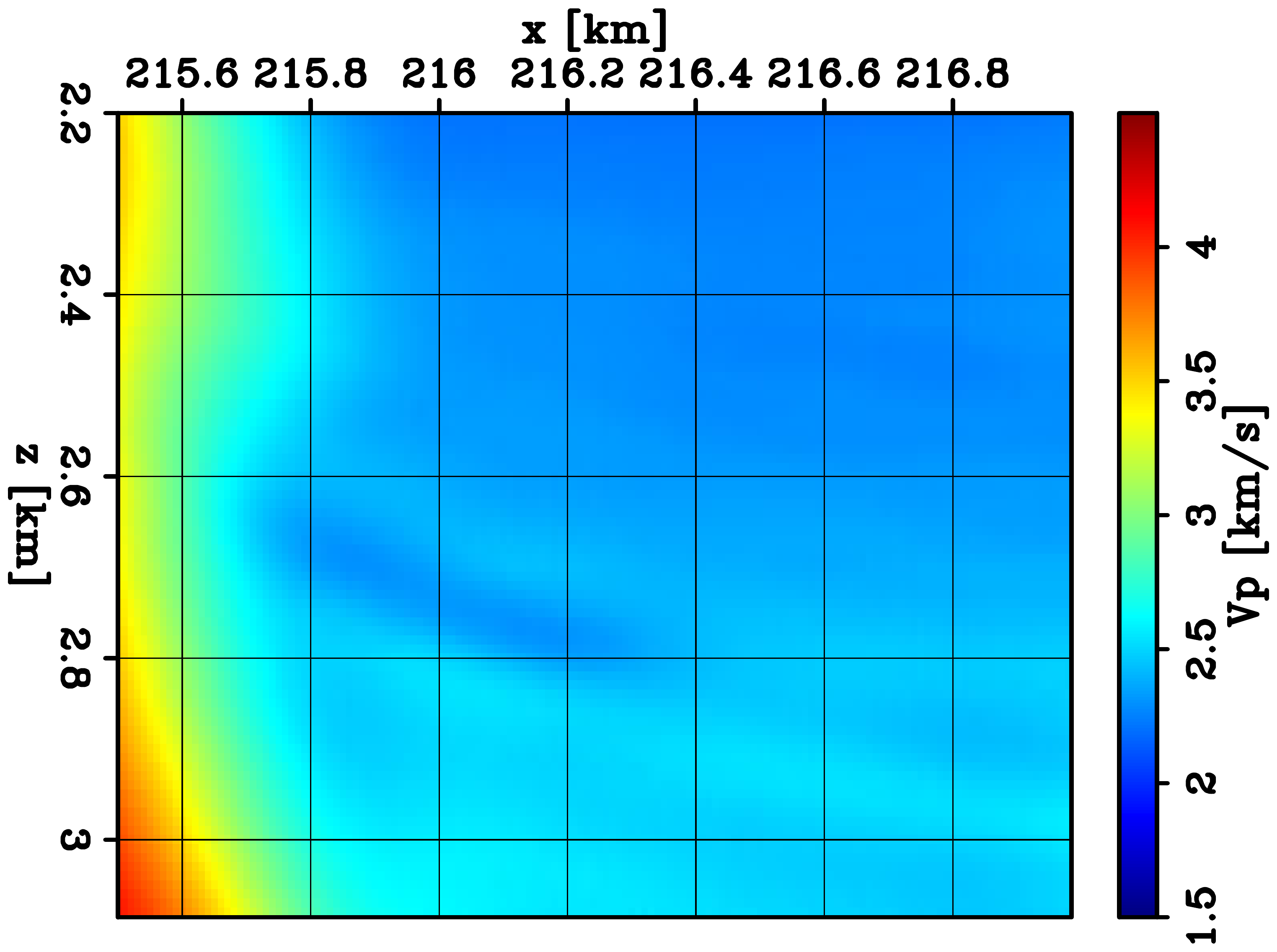}}
    \subfigure[]{\label{fig:CardamomTargetInitElaVpY}\includegraphics[width=0.32\columnwidth]{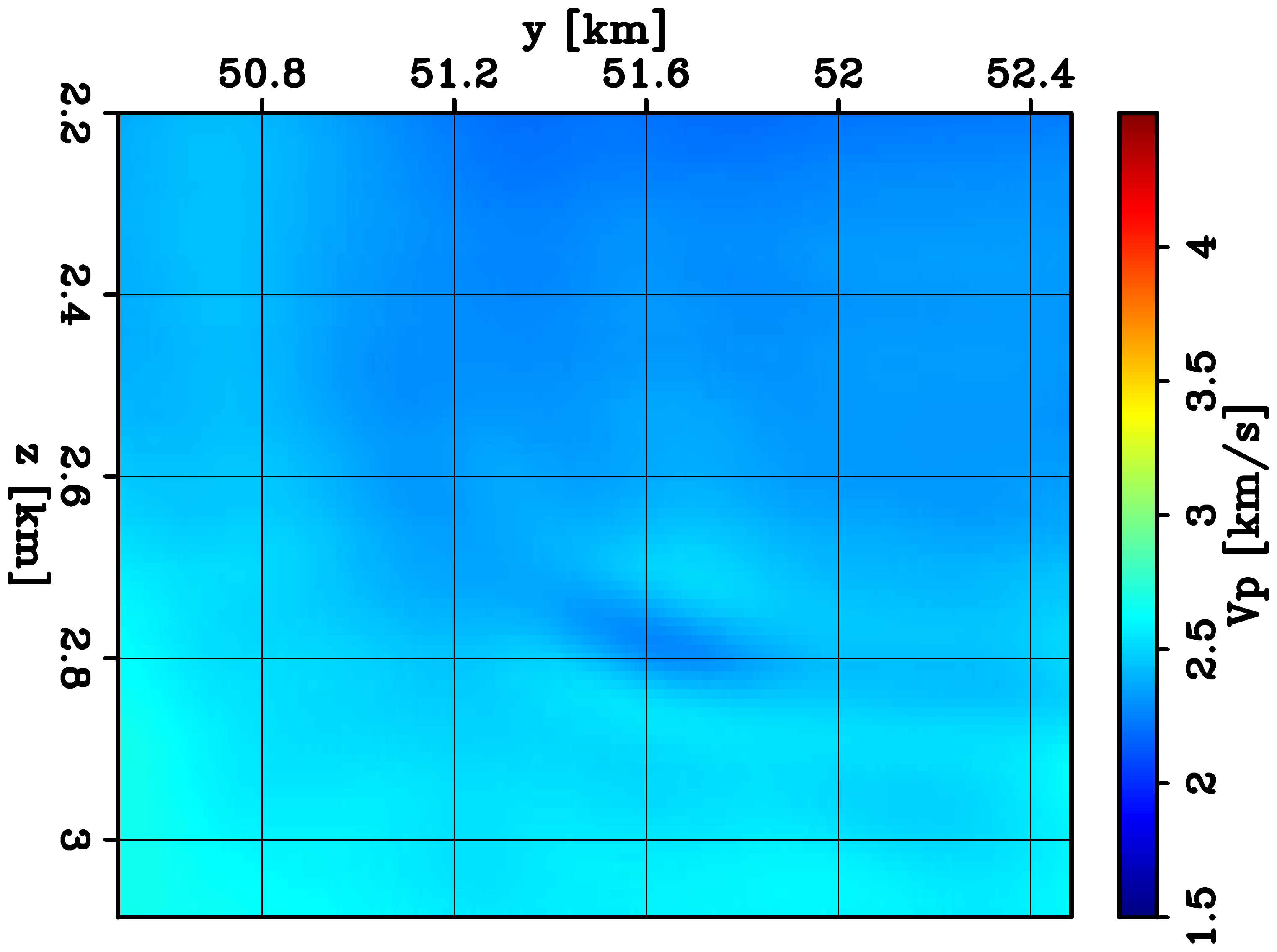}}
    
    \subfigure[]{\label{fig:CardamomTargetInitElaVsZ}\includegraphics[width=0.32\columnwidth]{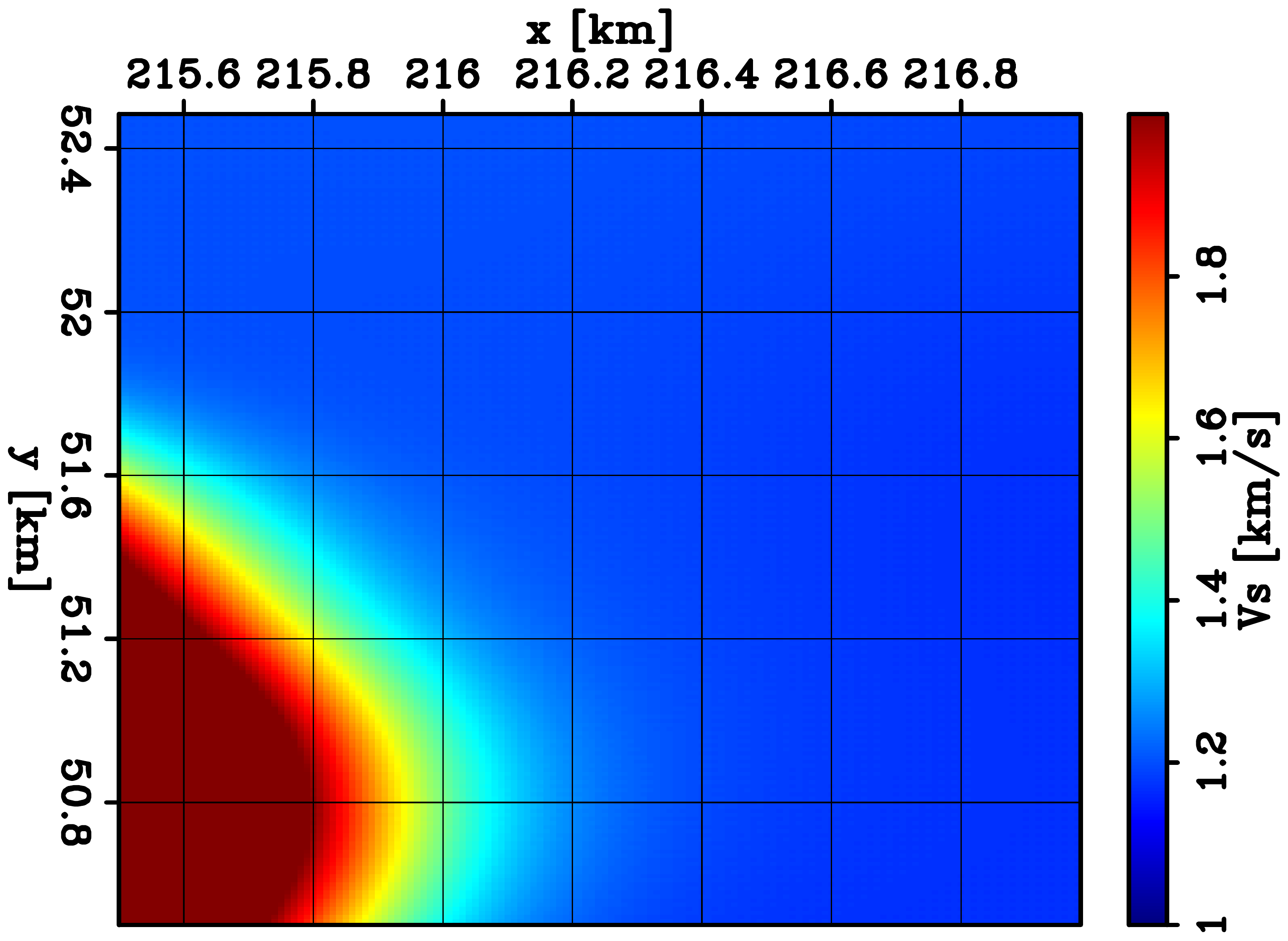}}
    \subfigure[]{\label{fig:CardamomTargetInitElaVsX}\includegraphics[width=0.32\columnwidth]{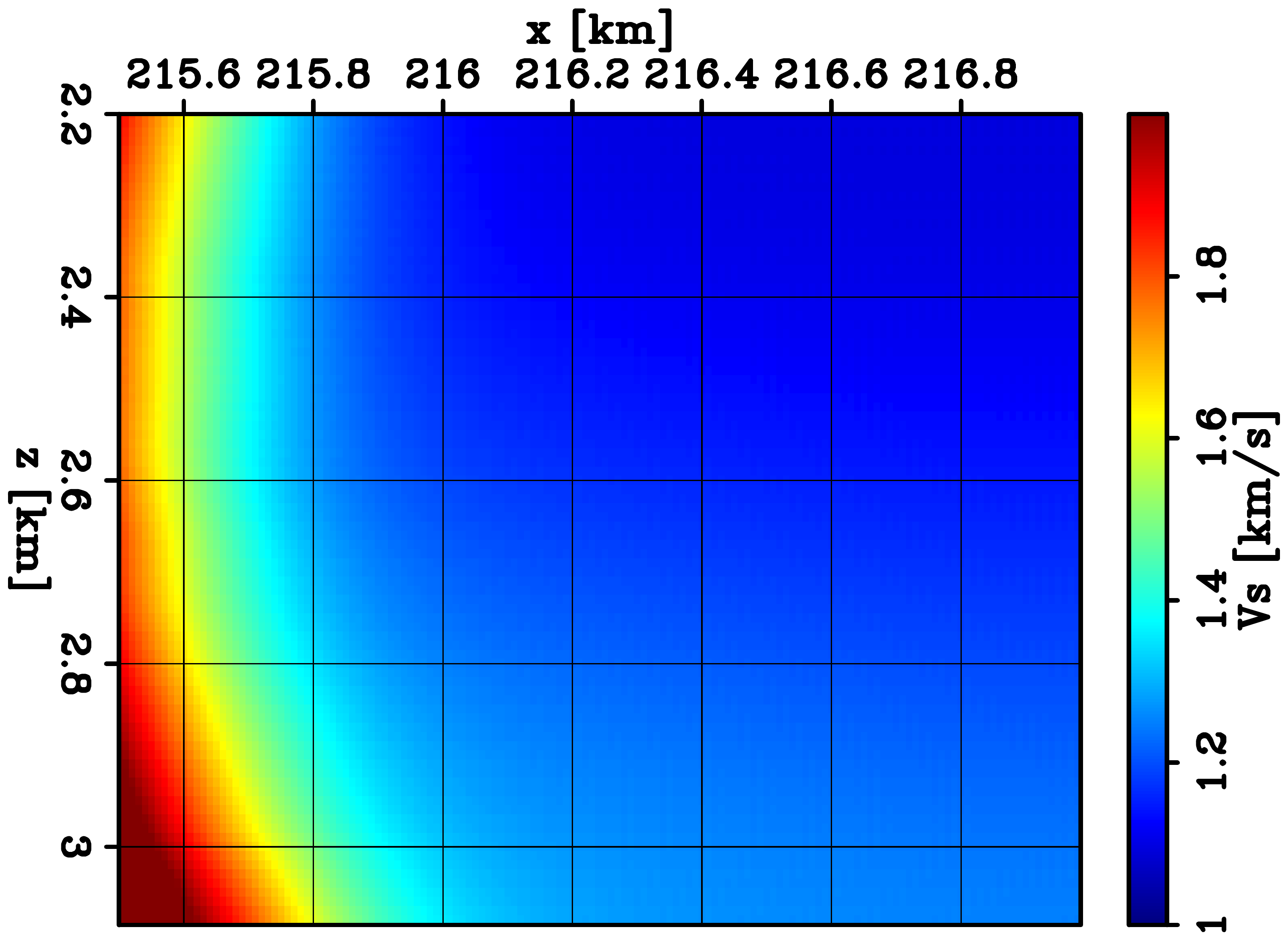}}
    \subfigure[]{\label{fig:CardamomTargetInitElaVsY}\includegraphics[width=0.32\columnwidth]{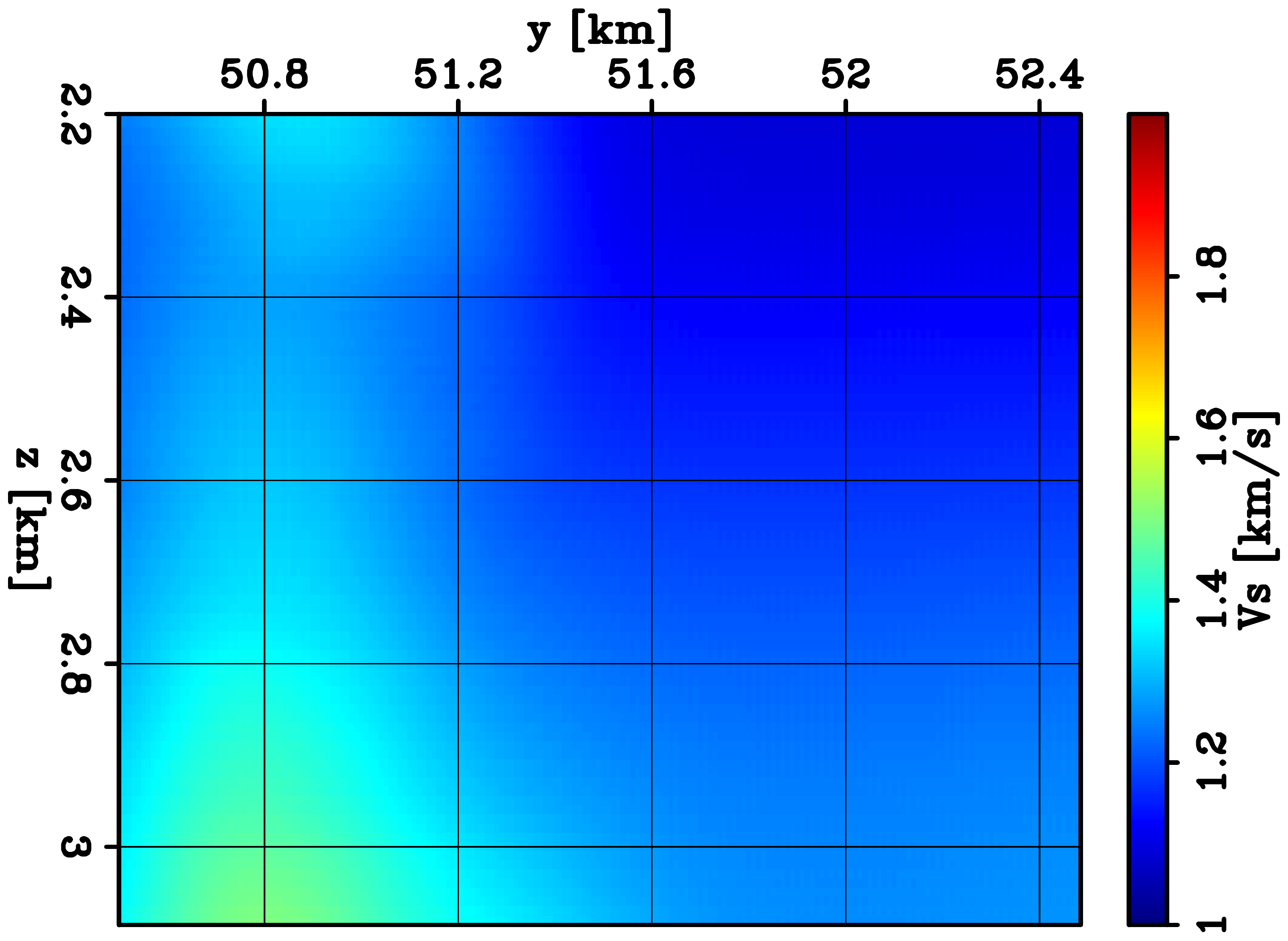}}
    
    \subfigure[]{\label{fig:CardamomTargetInitElaRhoZ}\includegraphics[width=0.32\columnwidth]{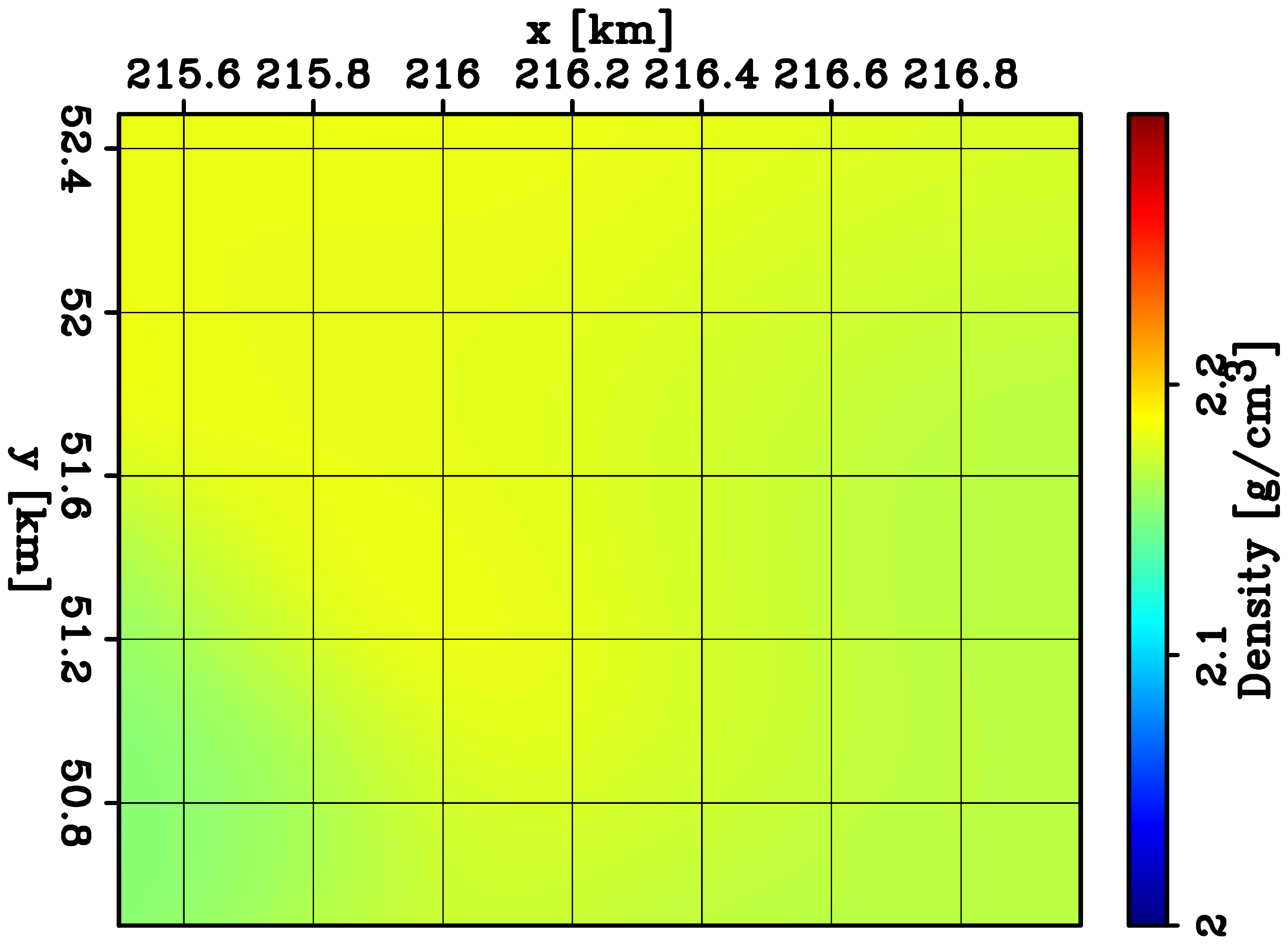}}
    \subfigure[]{\label{fig:CardamomTargetInitElaRhoX}\includegraphics[width=0.32\columnwidth]{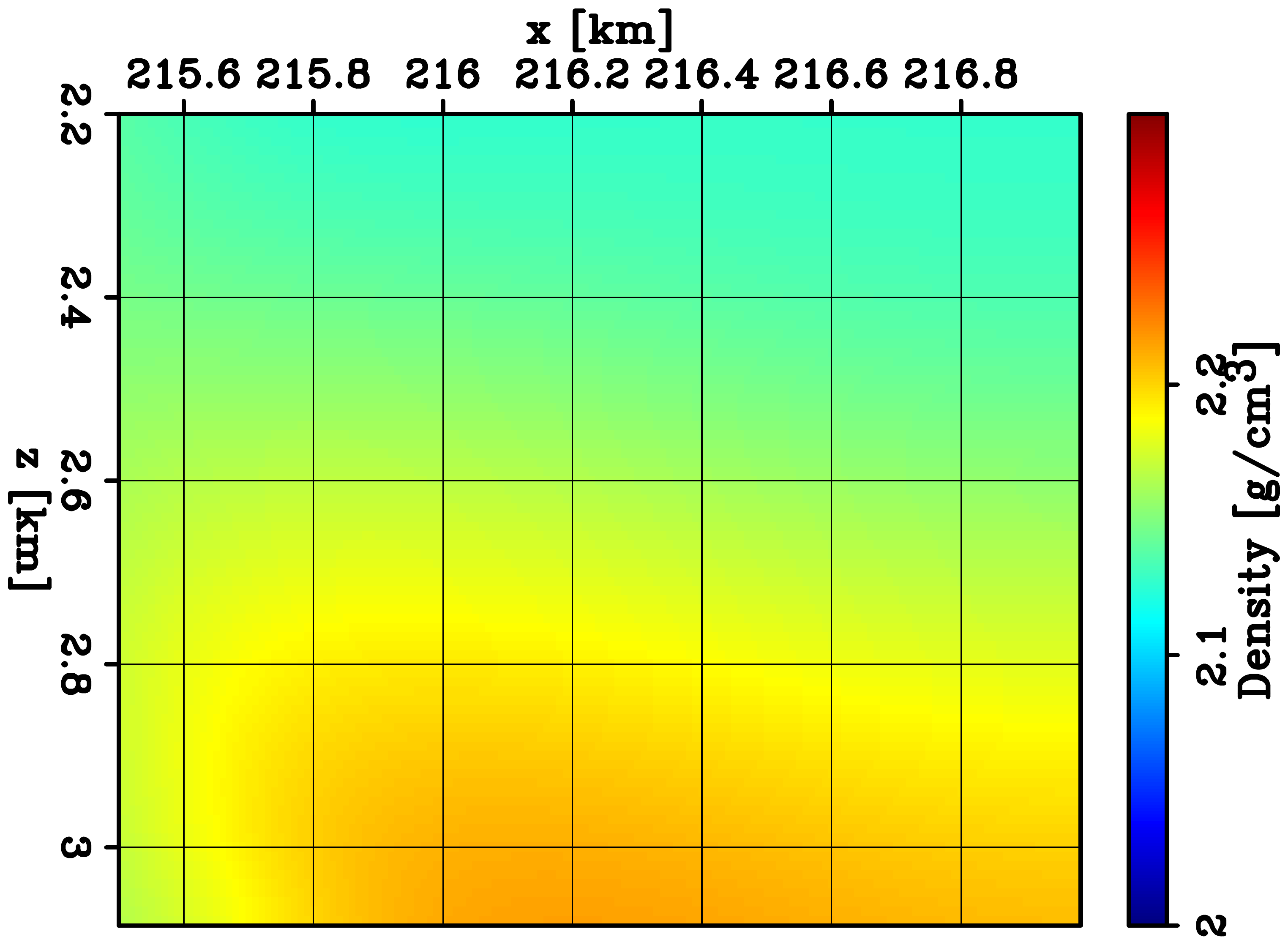}}
    \subfigure[]{\label{fig:CardamomTargetInitElaRhoY}\includegraphics[width=0.32\columnwidth]{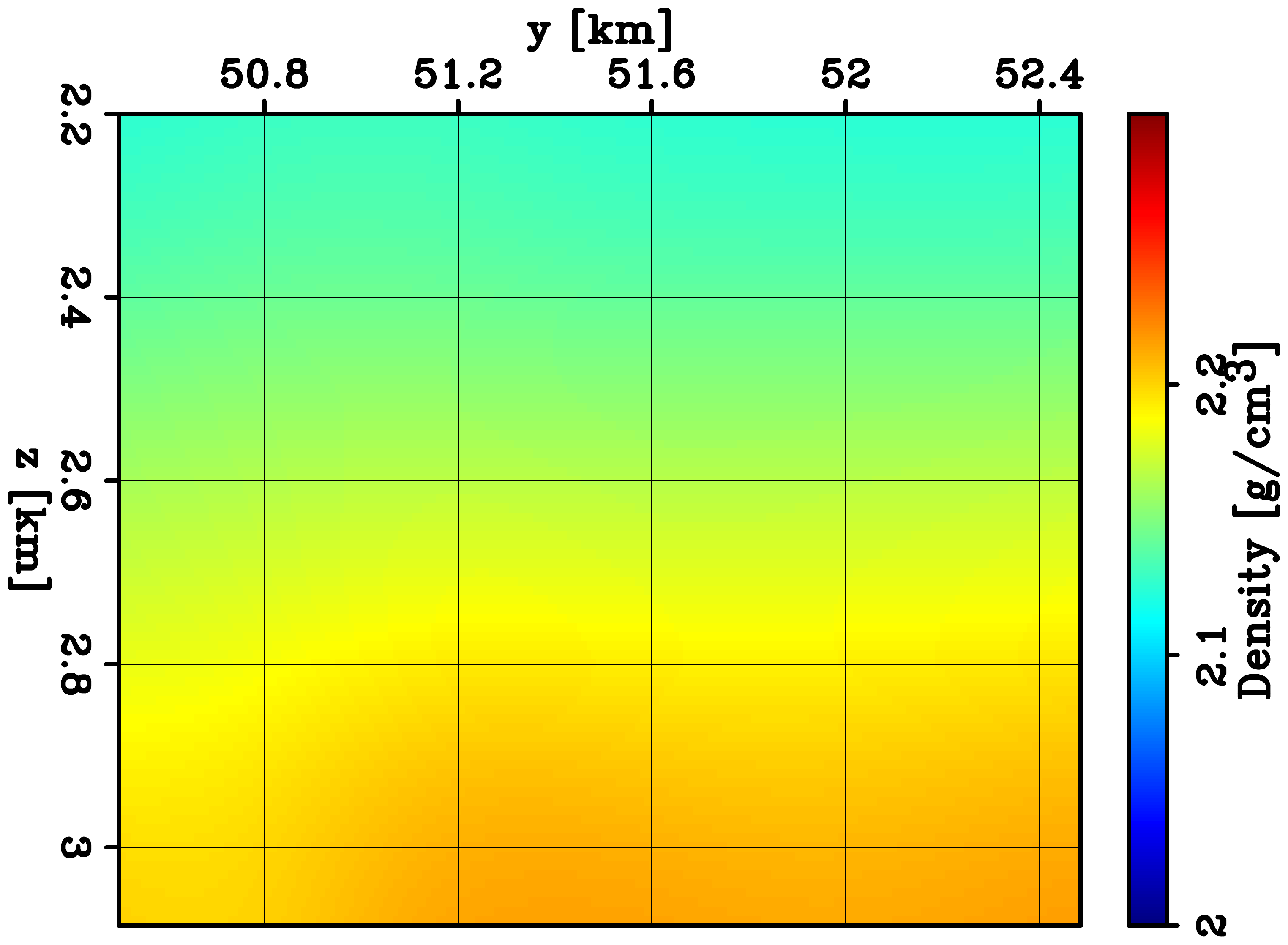}}
    
    \caption{Initial elastic parameters of the target area. The top row displays slices extracted from the P-wave velocity cube. The middle row shows panels from the S-wave velocity cube. The bottom row displays slices from the density model cube. On each row, from left to right, the panels are extracted at $z=2.6$ km, $y=51.5$ km, $x=216.1$ km, respectively.}
    \label{fig:CardamomTargetInitEla}
\end{figure}

The target area's initial P-wave velocity model is obtained by mildly smoothing the acoustically inverted FWI P-wave velocity. The initial density parameter is simply computed using Gardner's equation~\cite[]{gardner1974formation}. Finally, the starting guess for the S-wave velocity is obtained using the provided stiffness tensor components. Figure~\ref{fig:CardamomTargetInitEla} shows different panels extracted from the initial elastic parameters of the target area.

We apply an elastic FWI workflow to the redatumed dataset to estimate the target area's elastic parameters. The entire bandwidth of the reconstructed data is simultaneously injected (i.e., $3-12$ Hz), and the three elastic parameters are jointly inverted. The total recording time is $4.5$ s, which is almost half of the original $8$ s data. The elastic FD operator is based on a $20$ m grid to abide by the dispersion and stability conditions. However, the inverted model is parameterized using a spline grid of $100$ m in the x and y axes, while the z-axis has the same sampling as the FD grid. As in the acoustic FWI step, the spline parameterization effectively acts as regularization and avoids the introduction of spurious features during the inversion. By assuming the same scattering regime considered for the target-oriented inversion applied to the synthetic case, all the wave-equation operators are constructed using absorbing boundary conditions around the entire simulation domain.

\begin{figure}[t]
    \centering
    \subfigure[]{\label{fig:CardamomTargetInitResX}\includegraphics[width=0.48\linewidth]{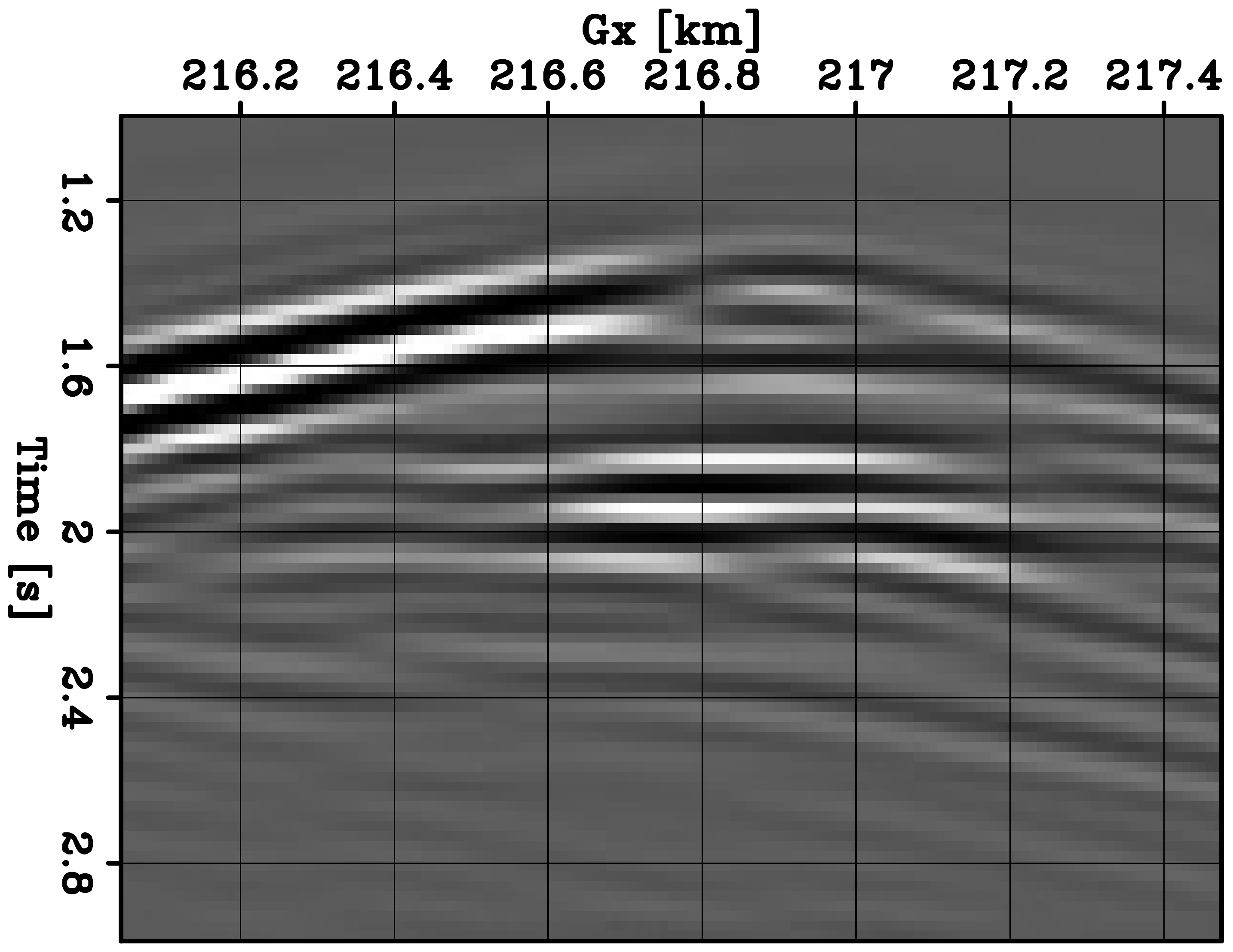}}
    \subfigure[]{\label{fig:CardamomTargetInitResY}\includegraphics[width=0.48\linewidth]{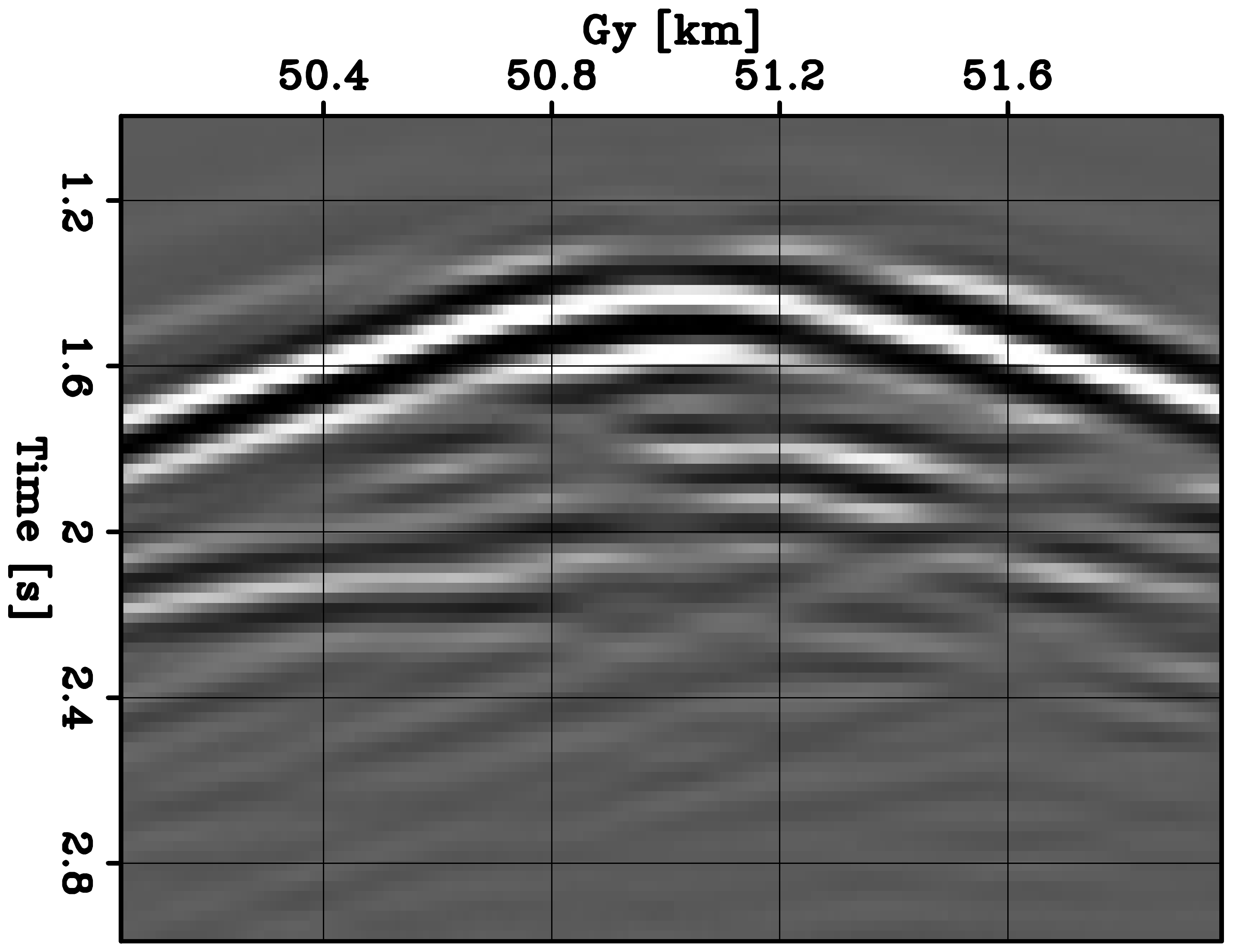}}
    
    \subfigure[]{\label{fig:CardamomTargetInvResX}\includegraphics[width=0.48\linewidth]{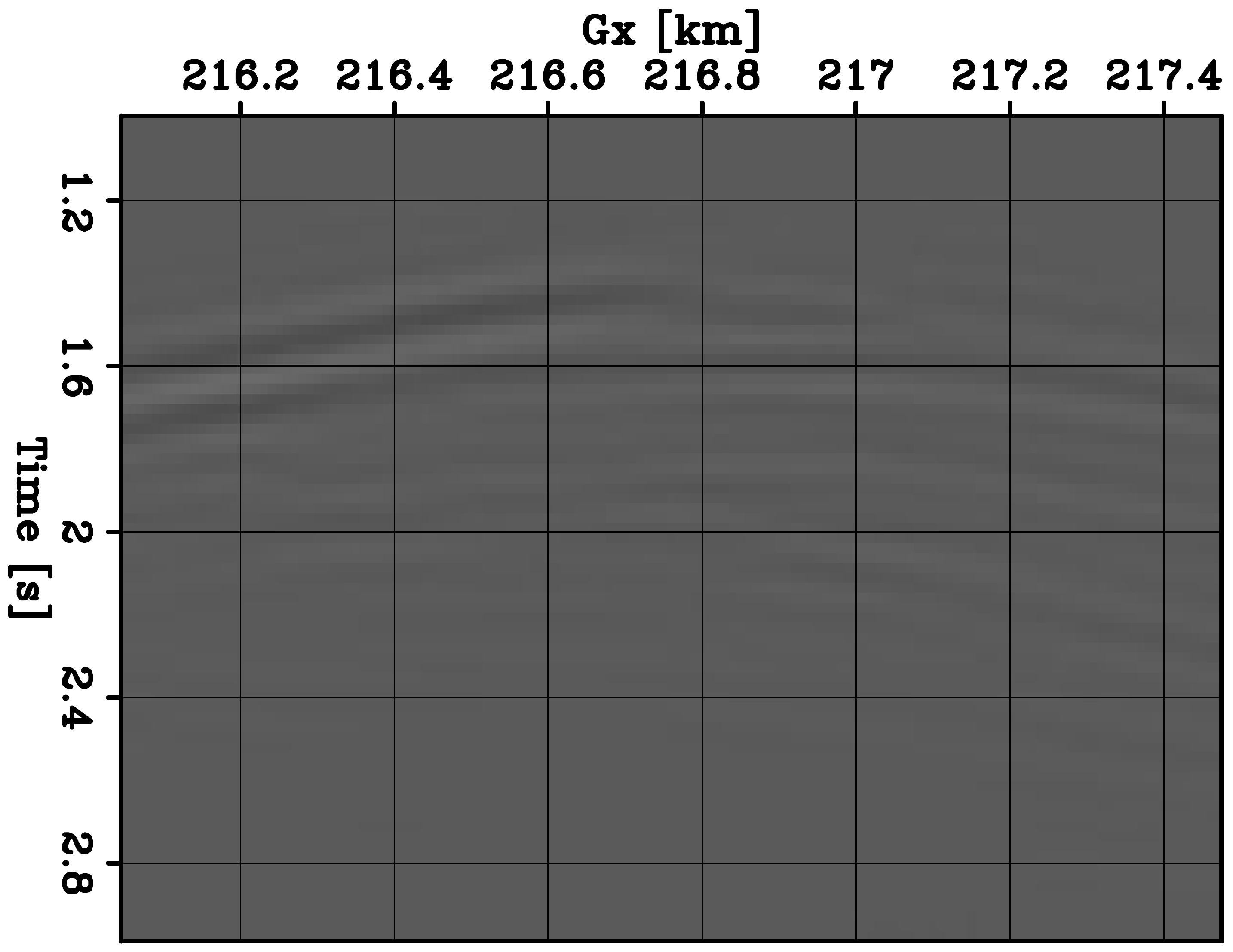}}
    \subfigure[]{\label{fig:CardamomTargetInvResY}\includegraphics[width=0.48\linewidth]{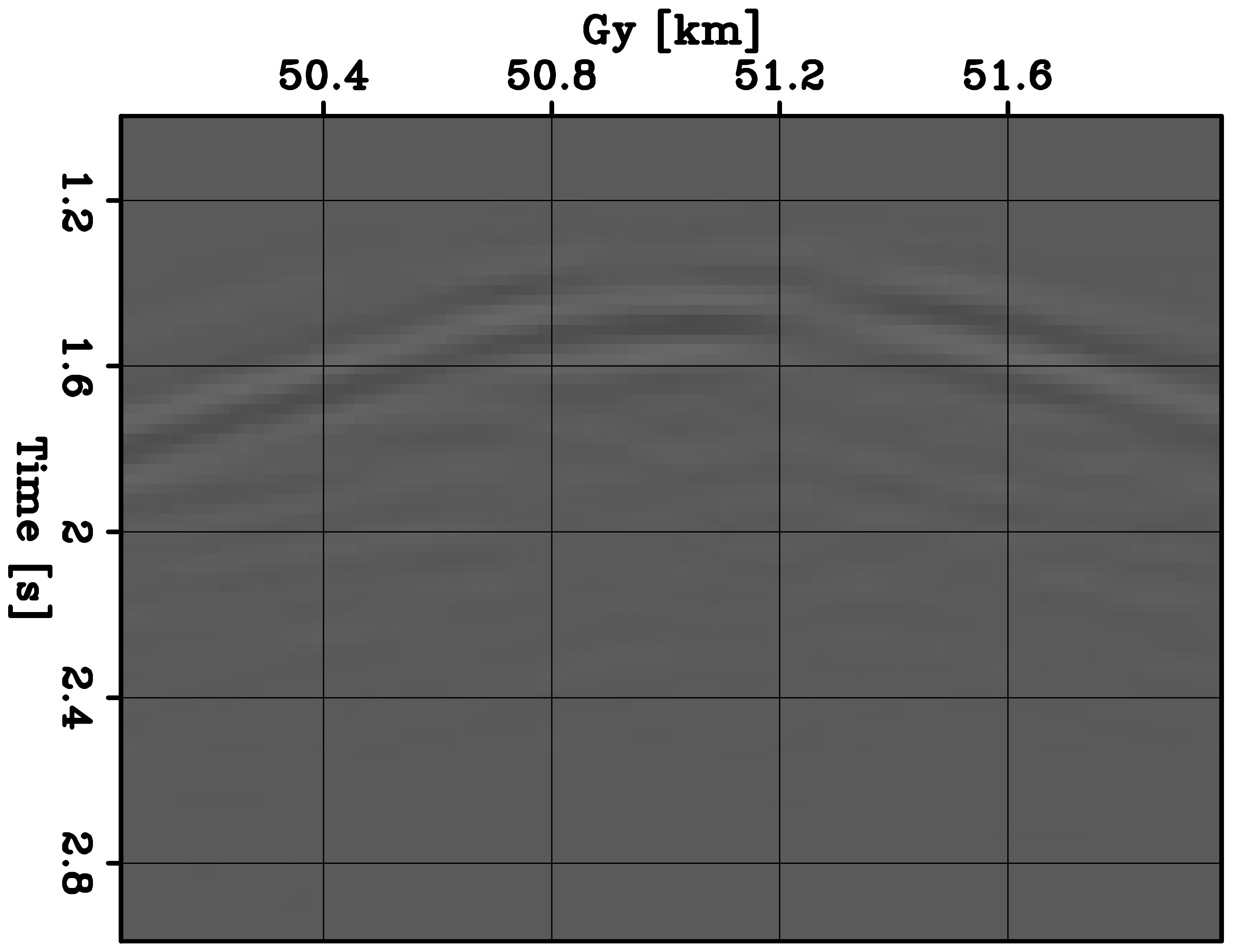}}
    
    \caption{Representative shot gather displaying pressure-data residuals of the target-oriented inversion. The panels on the left column show the the receivers at $y=51.5$ km for the initial (a) and the final (c) residuals. While the panels on the right depict the initial (b) and final (d) residuals for the receivers at $x=216.5$ km.}
    \label{fig:CardamomTargetRes}
\end{figure}

\begin{figure}[t]
    \centering
    \subfigure[]{\label{fig:CardamomTargetFinalElaVpZ}\includegraphics[width=0.32\columnwidth]{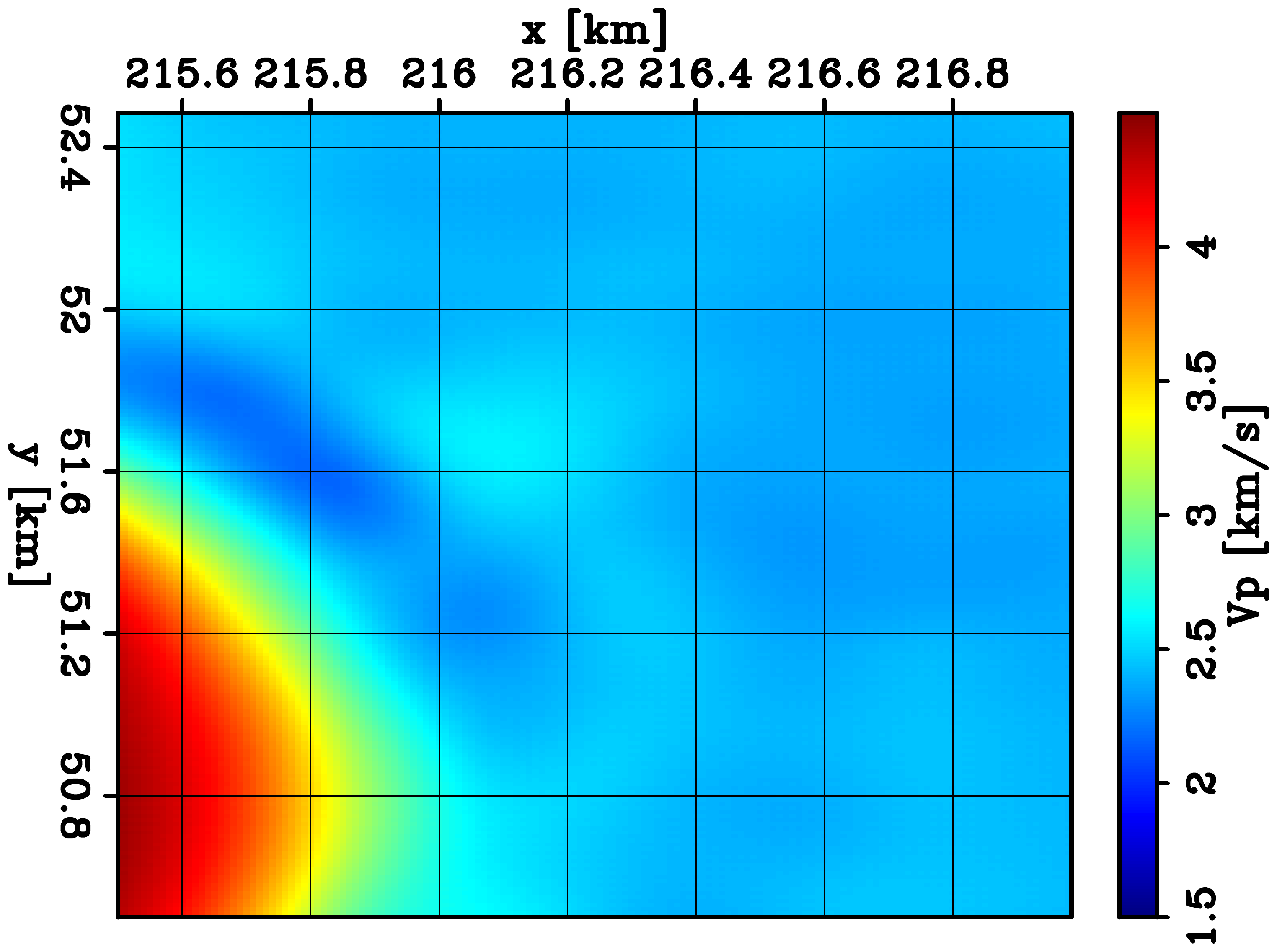}}
    \subfigure[]{\label{fig:CardamomTargetFinalElaVpX}\includegraphics[width=0.32\columnwidth]{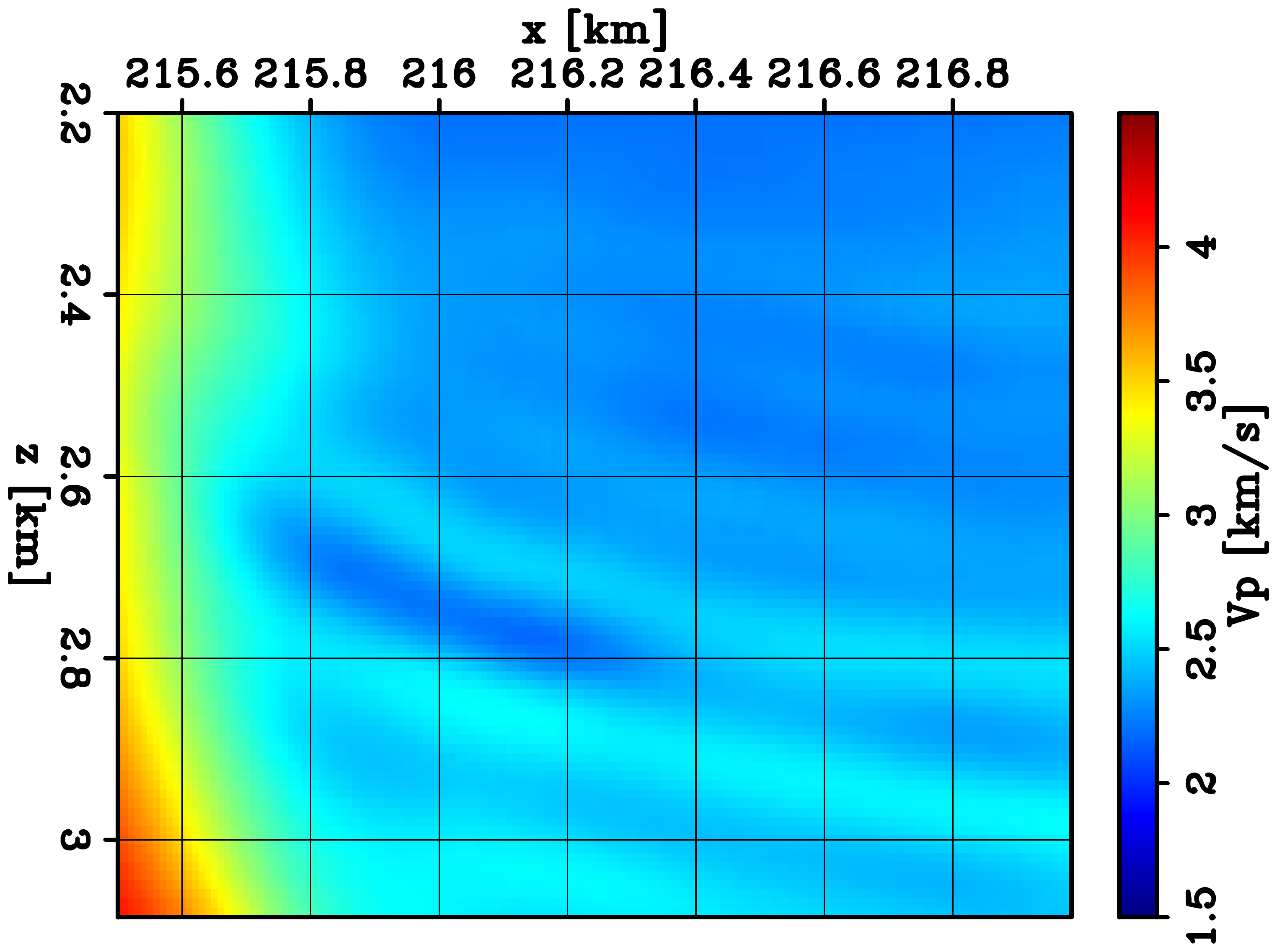}}
    \subfigure[]{\label{fig:CardamomTargetFinalElaVpY}\includegraphics[width=0.32\columnwidth]{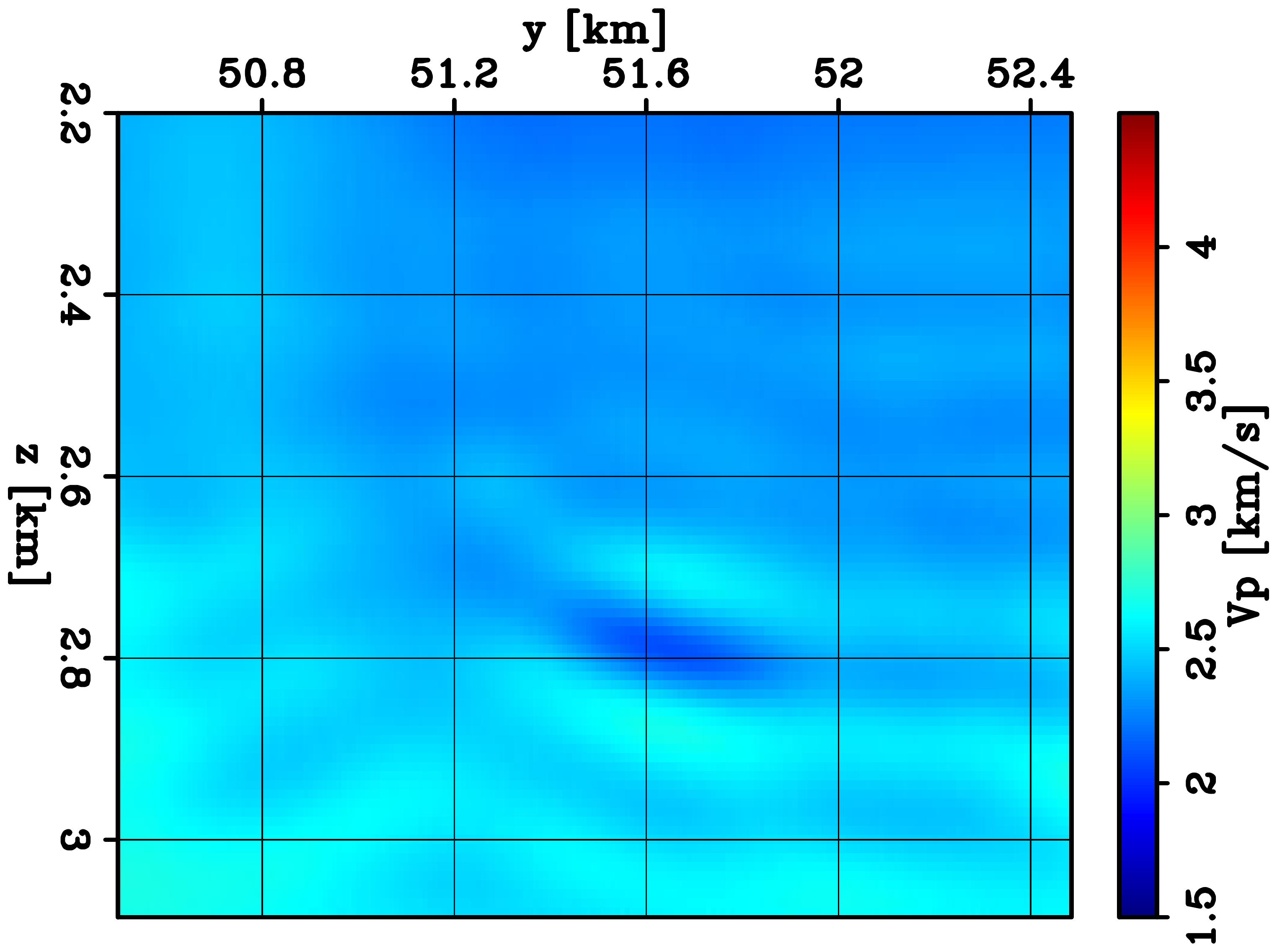}}
    
    \subfigure[]{\label{fig:CardamomTargetFinalElaVsZ}\includegraphics[width=0.32\columnwidth]{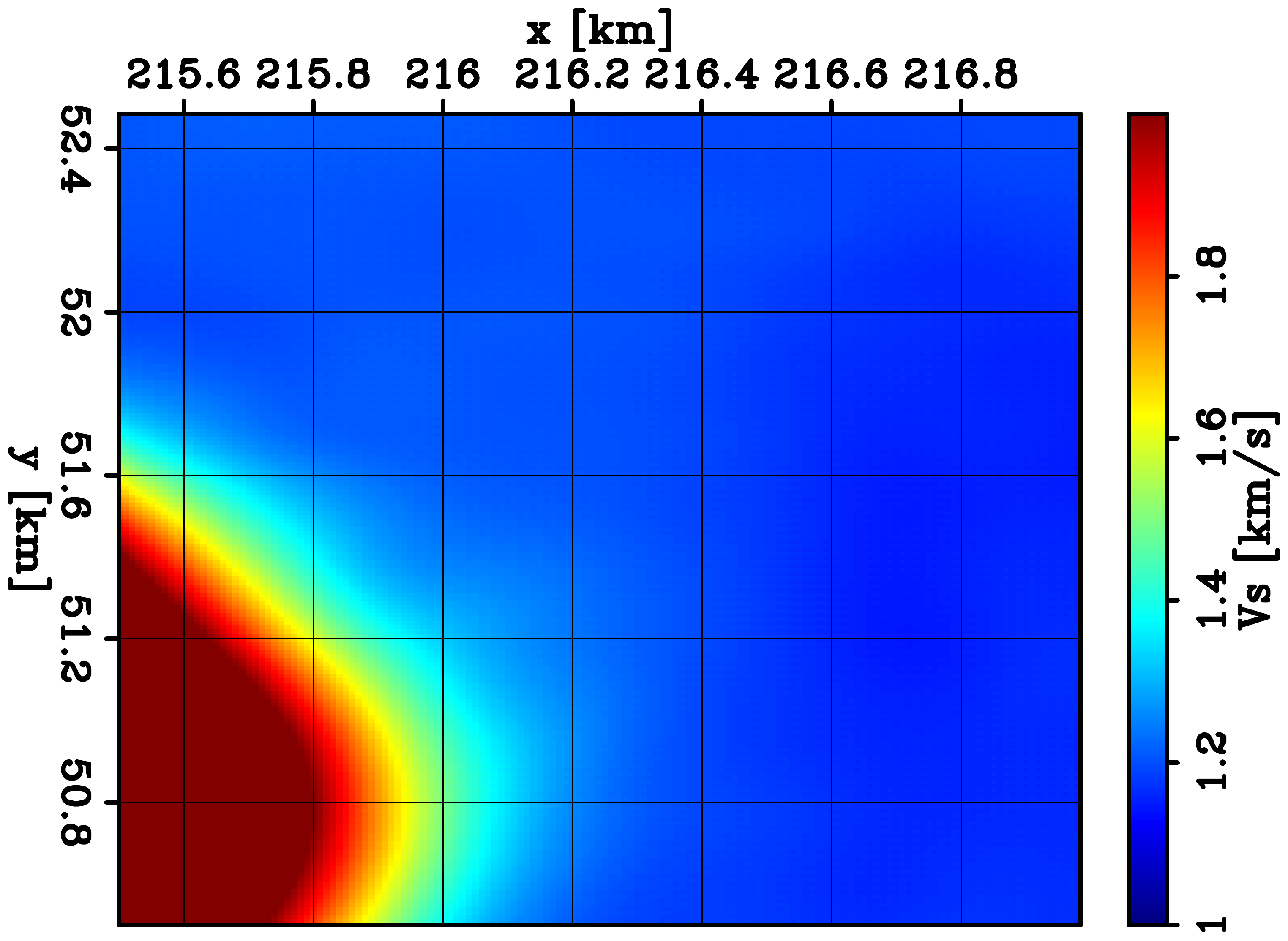}}
    \subfigure[]{\label{fig:CardamomTargetFinalElaVsX}\includegraphics[width=0.32\columnwidth]{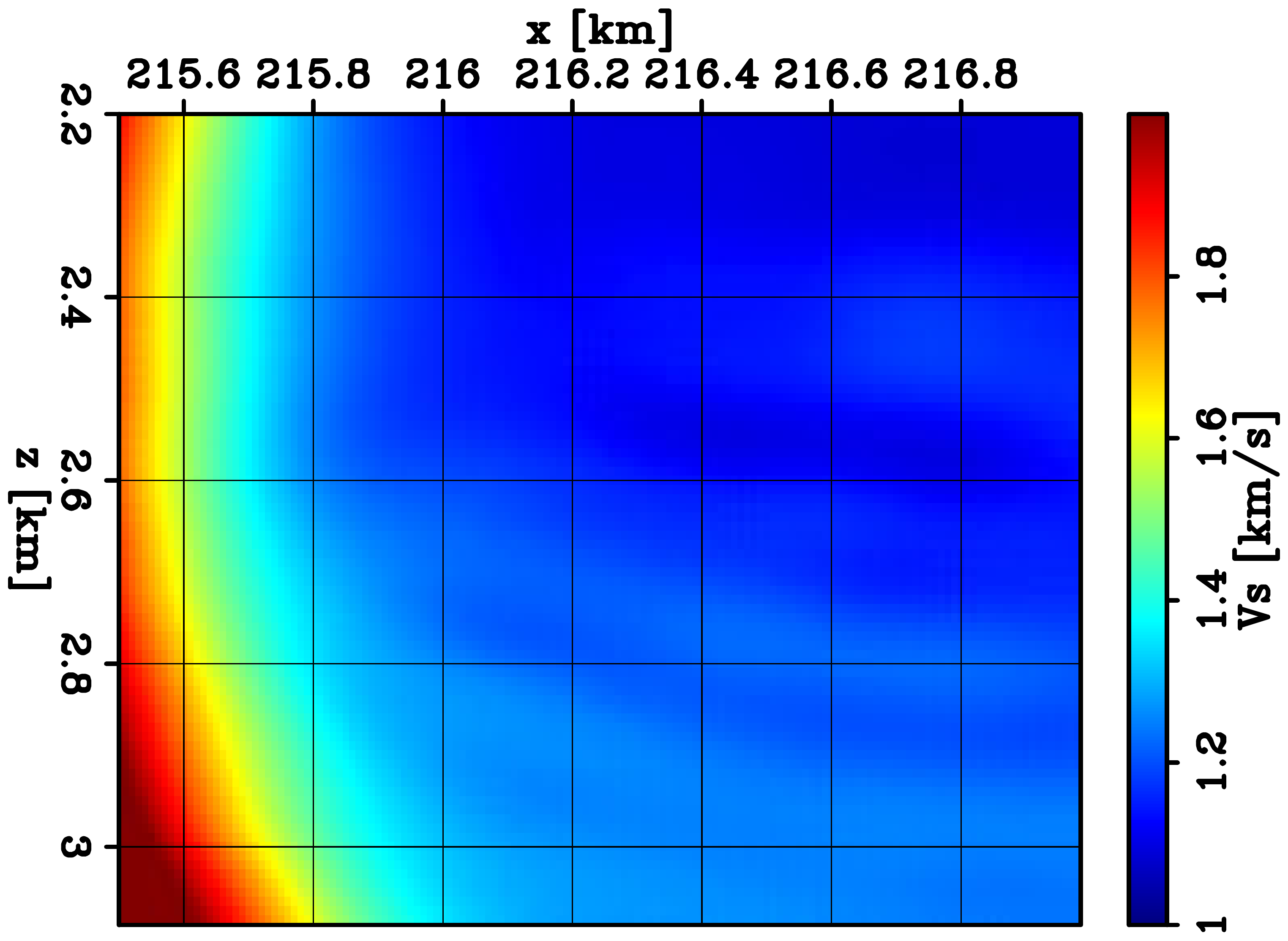}}
    \subfigure[]{\label{fig:CardamomTargetFinalElaVsY}\includegraphics[width=0.32\columnwidth]{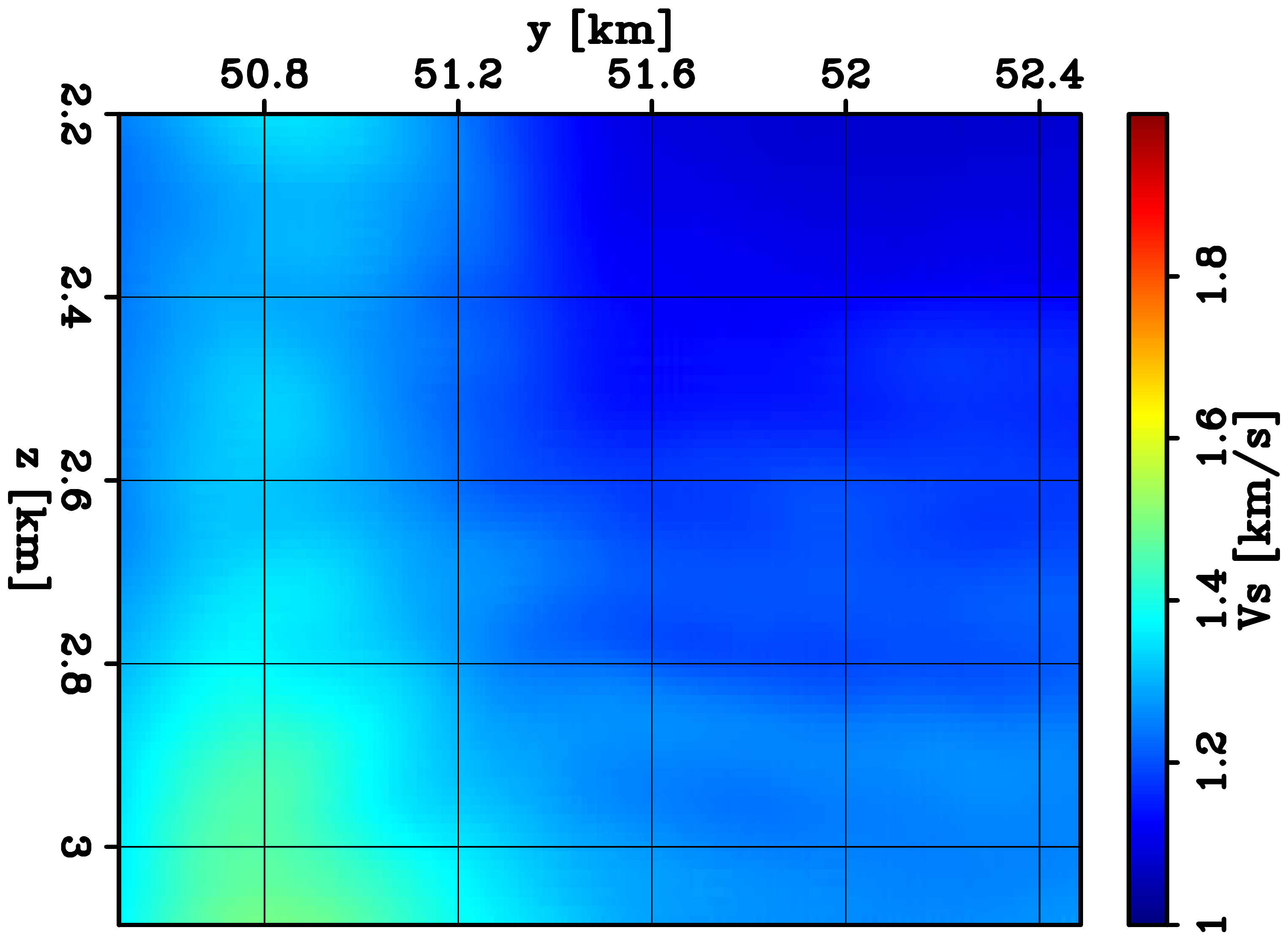}}
    
    \subfigure[]{\label{fig:CardamomTargetFinalElaRhoZ}\includegraphics[width=0.32\columnwidth]{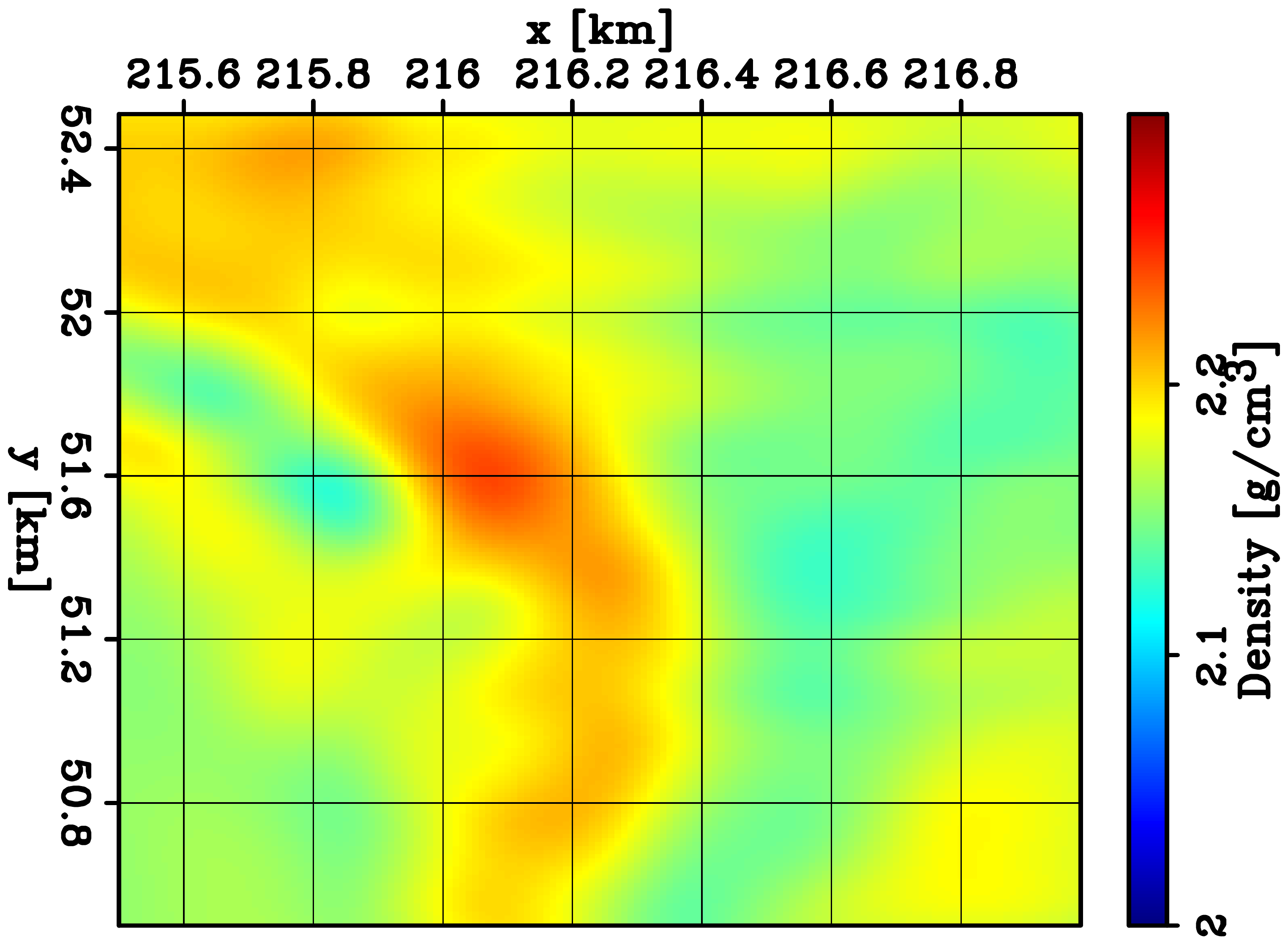}}
    \subfigure[]{\label{fig:CardamomTargetFinalElaRhoX}\includegraphics[width=0.32\columnwidth]{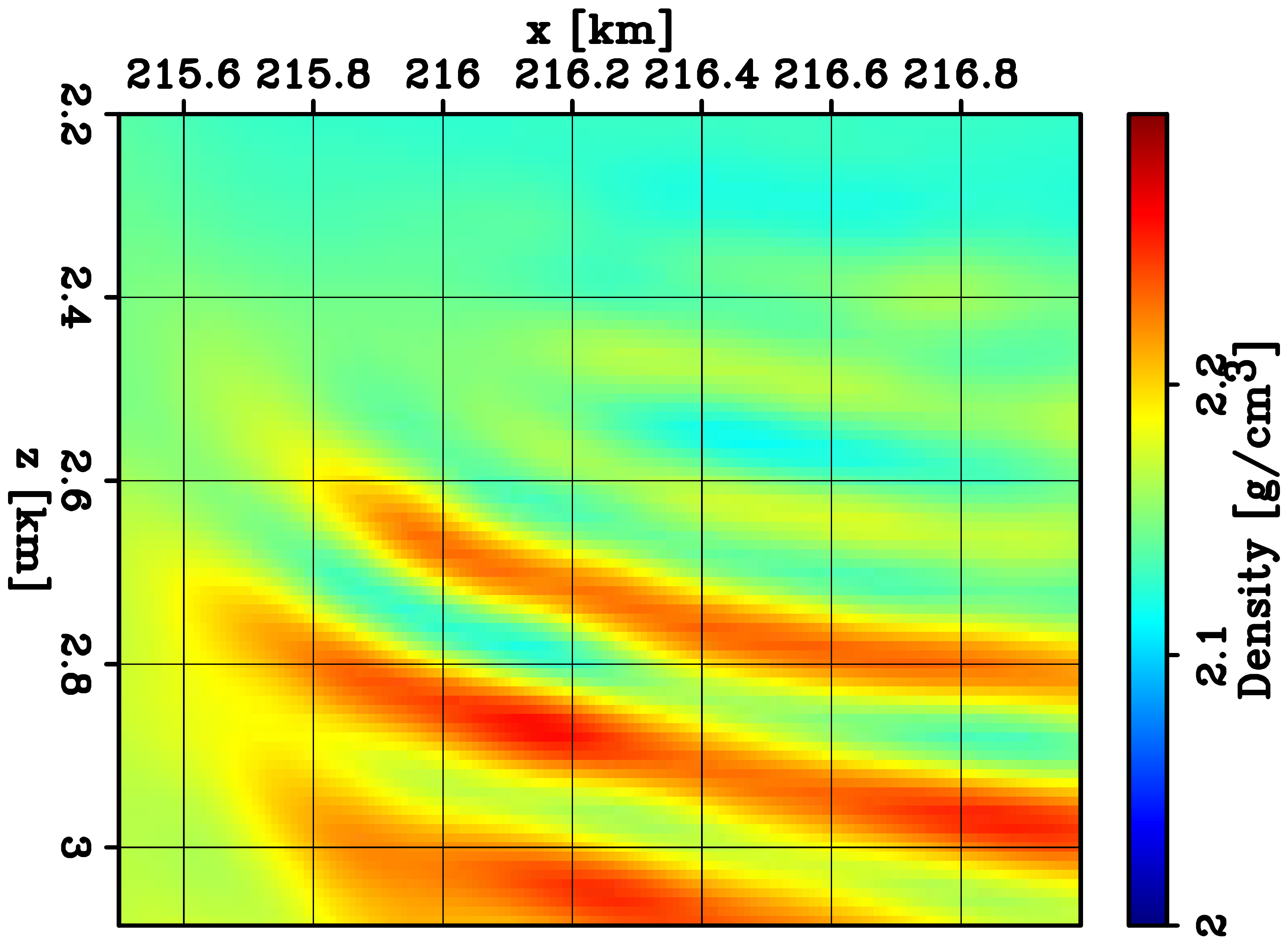}}
    \subfigure[]{\label{fig:CardamomTargetFinalElaRhoY}\includegraphics[width=0.32\columnwidth]{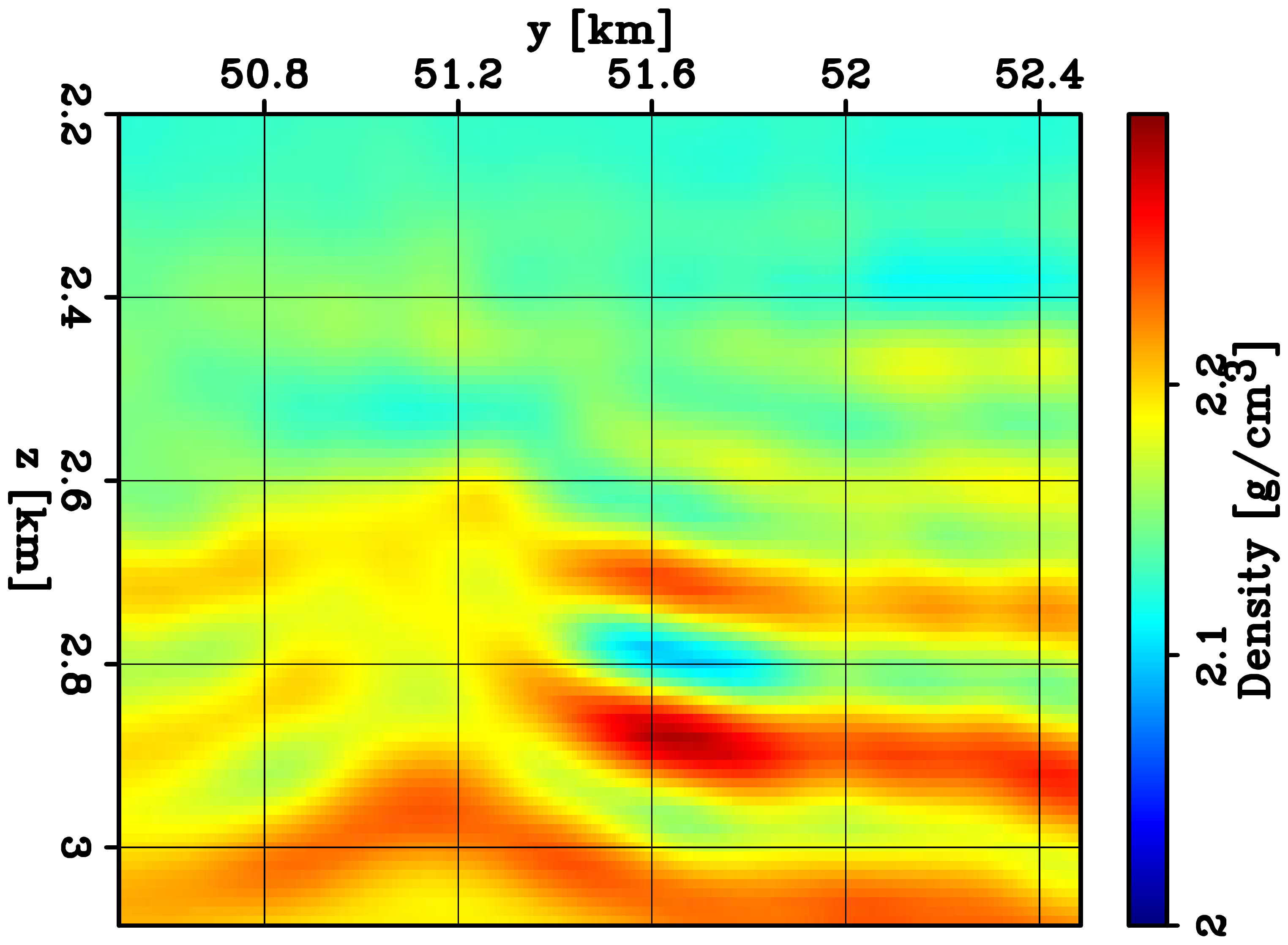}}
    
    \caption{Inverted elastic parameters of the target area using the target-oriented elastic FWI workflow. The top row displays slices extracted from the P-wave velocity cube. The middle row shows panels from the S-wave velocity cube. The bottom row displays slices from the density model cube. On each row, from left to right, the panels are extracted at $z=2.6$ km, $y=51.5$ km, $x=216.1$ km, respectively.}
    \label{fig:CardamomTargetFinalEla}
\end{figure}

After minimizing the L2-norm difference between the predicted and the synthesized elastic pressure data with a BFGS optimizer for $10$ iterations. The described inversion process achieves an accurate fit of redatum data as shown by the data-residual panels in Figure~\ref{fig:CardamomTargetRes}, and the retrieved subsurface parameter cubes are shown in Figure~\ref{fig:CardamomTargetFinalEla}. The inversion procedure introduces most of the changes within the P-wave and density parameters. A noticeable decrease in both is observed at the same position as the high-amplitude anomaly observed in the RTM image of Figure~\ref{fig:CardamomRTMFWICube}. On the contrary, no significant updates are placed within the S-wave parameter, although similar geometrical features are present within the inverted parameter.

\begin{figure}[t]
    \centering
    \subfigure[]{\label{fig:CardamomTargetDiffElaVpZ}\includegraphics[width=0.32\columnwidth]{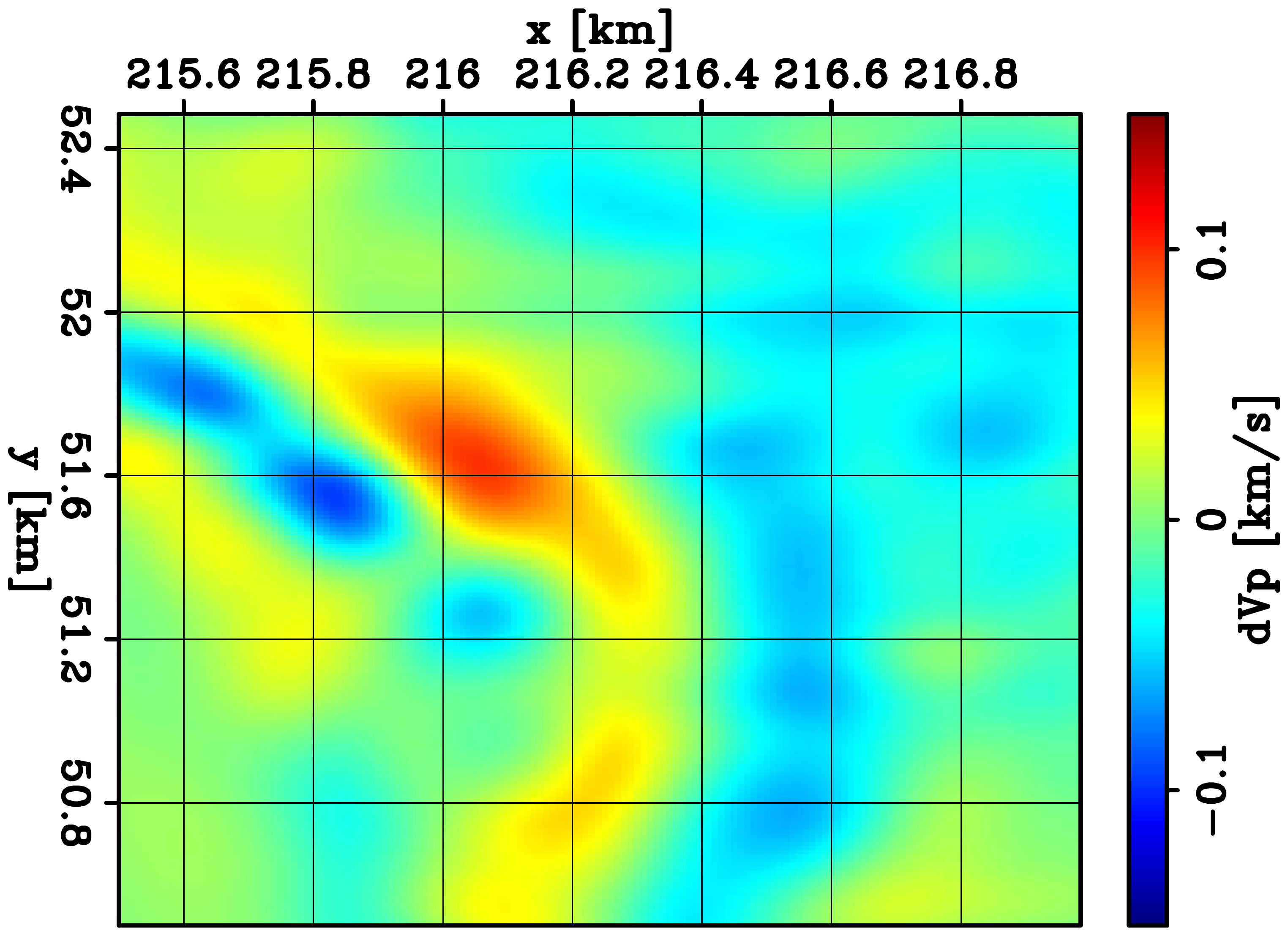}}
    \subfigure[]{\label{fig:CardamomTargetDiffElaVpX}\includegraphics[width=0.32\columnwidth]{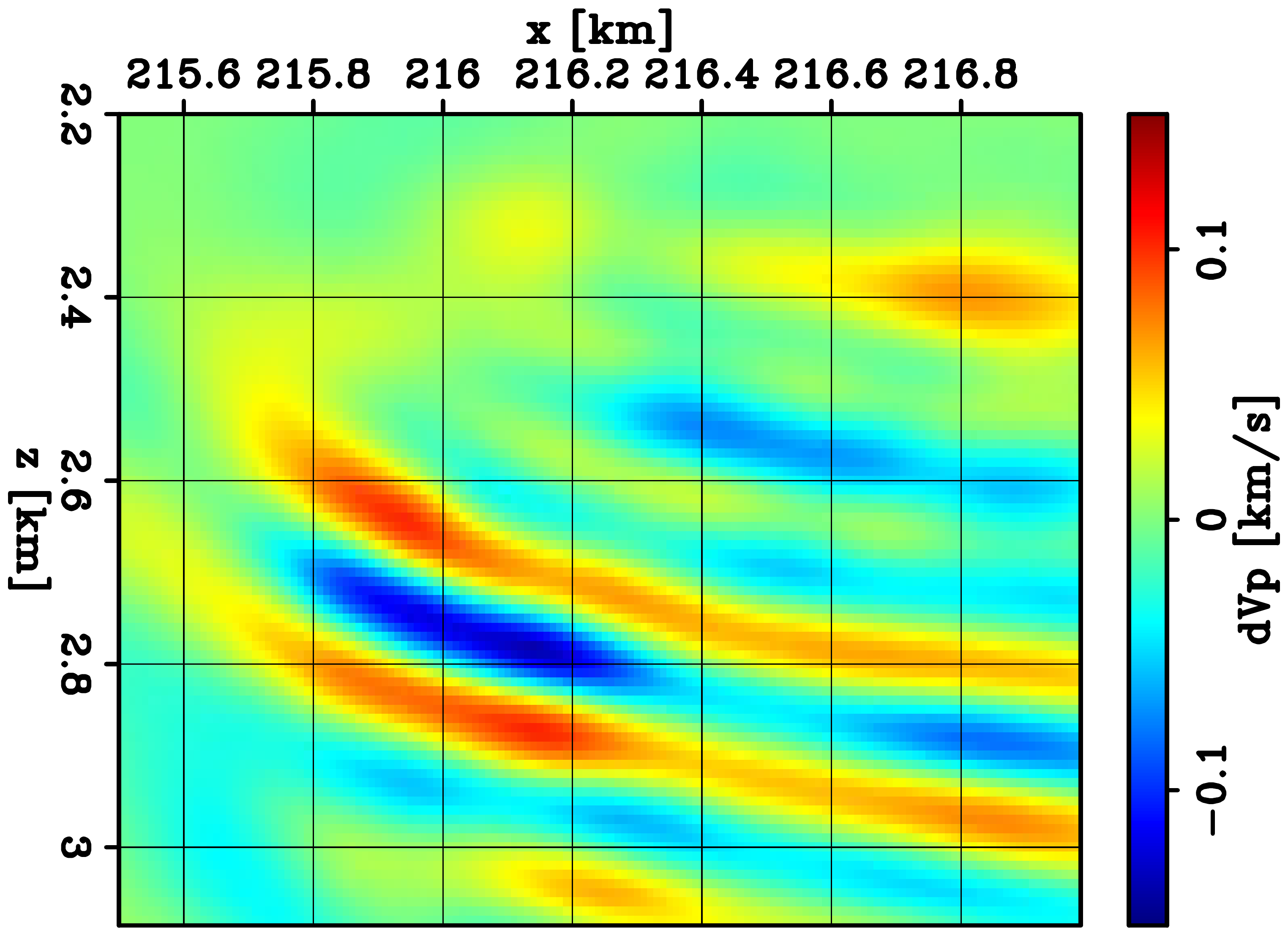}}
    \subfigure[]{\label{fig:CardamomTargetDiffElaVpY}\includegraphics[width=0.32\columnwidth]{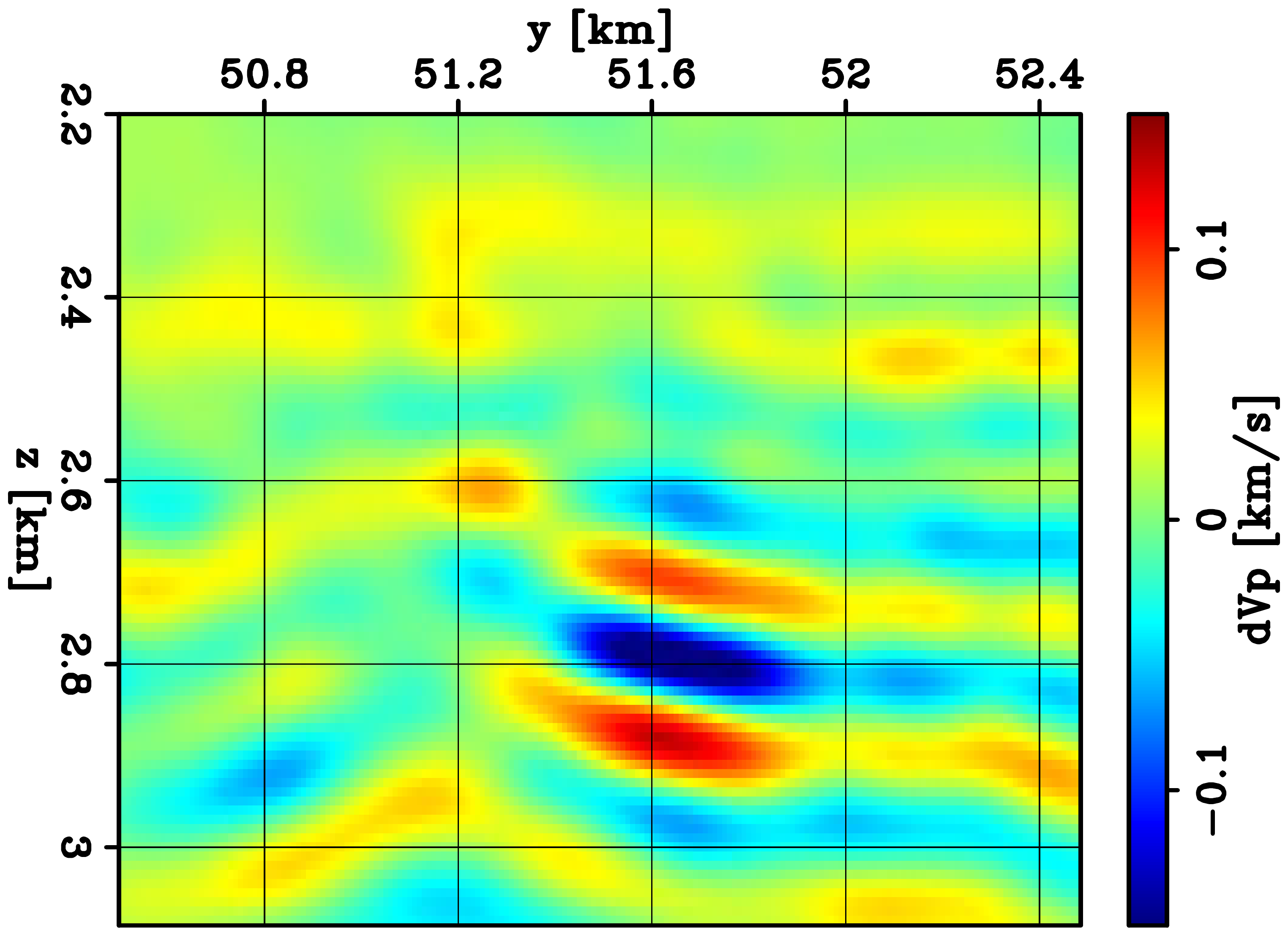}}
    
    \subfigure[]{\label{fig:CardamomTargetDiffElaVsZ}\includegraphics[width=0.32\columnwidth]{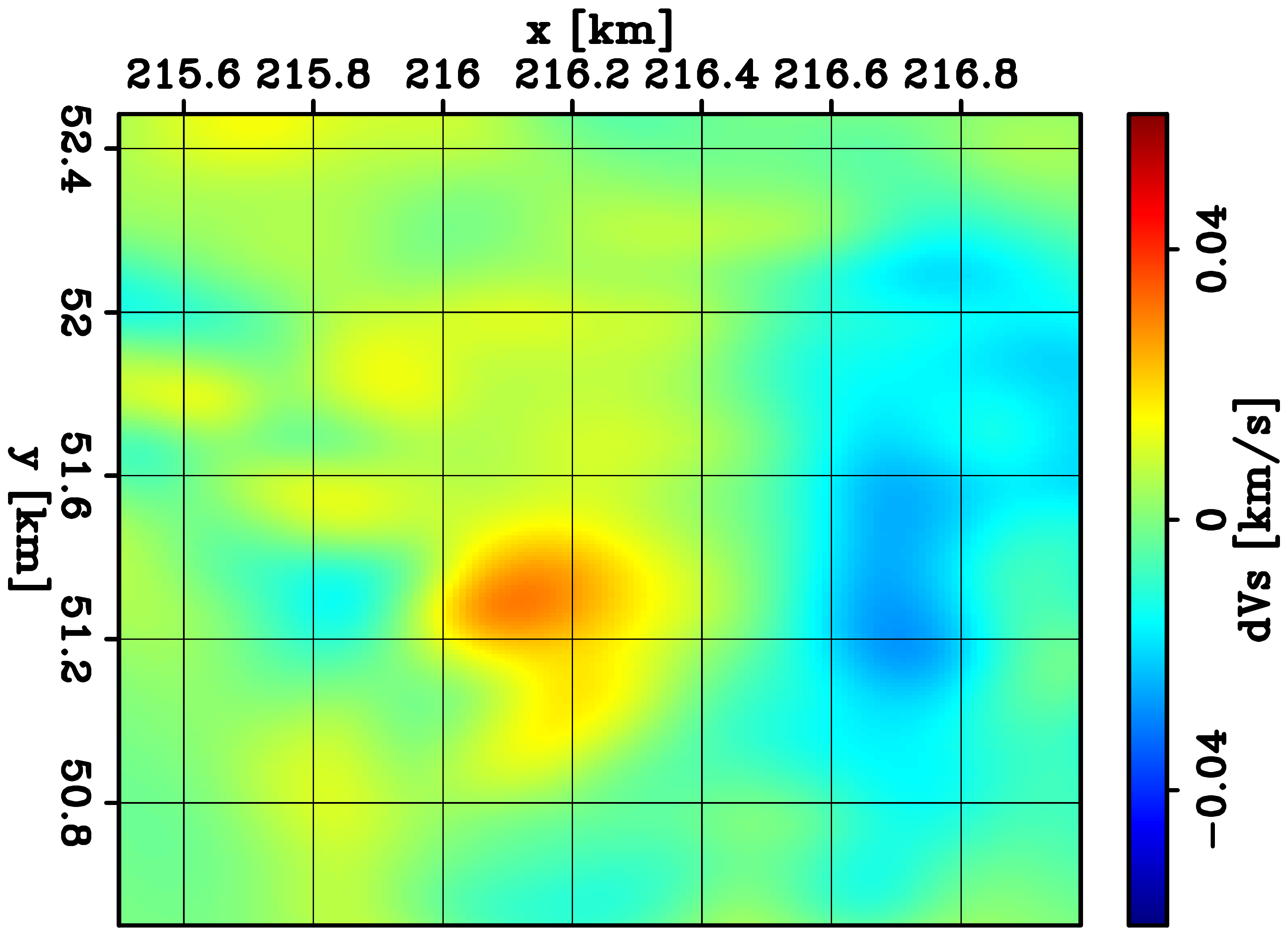}}
    \subfigure[]{\label{fig:CardamomTargetDiffElaVsX}\includegraphics[width=0.32\columnwidth]{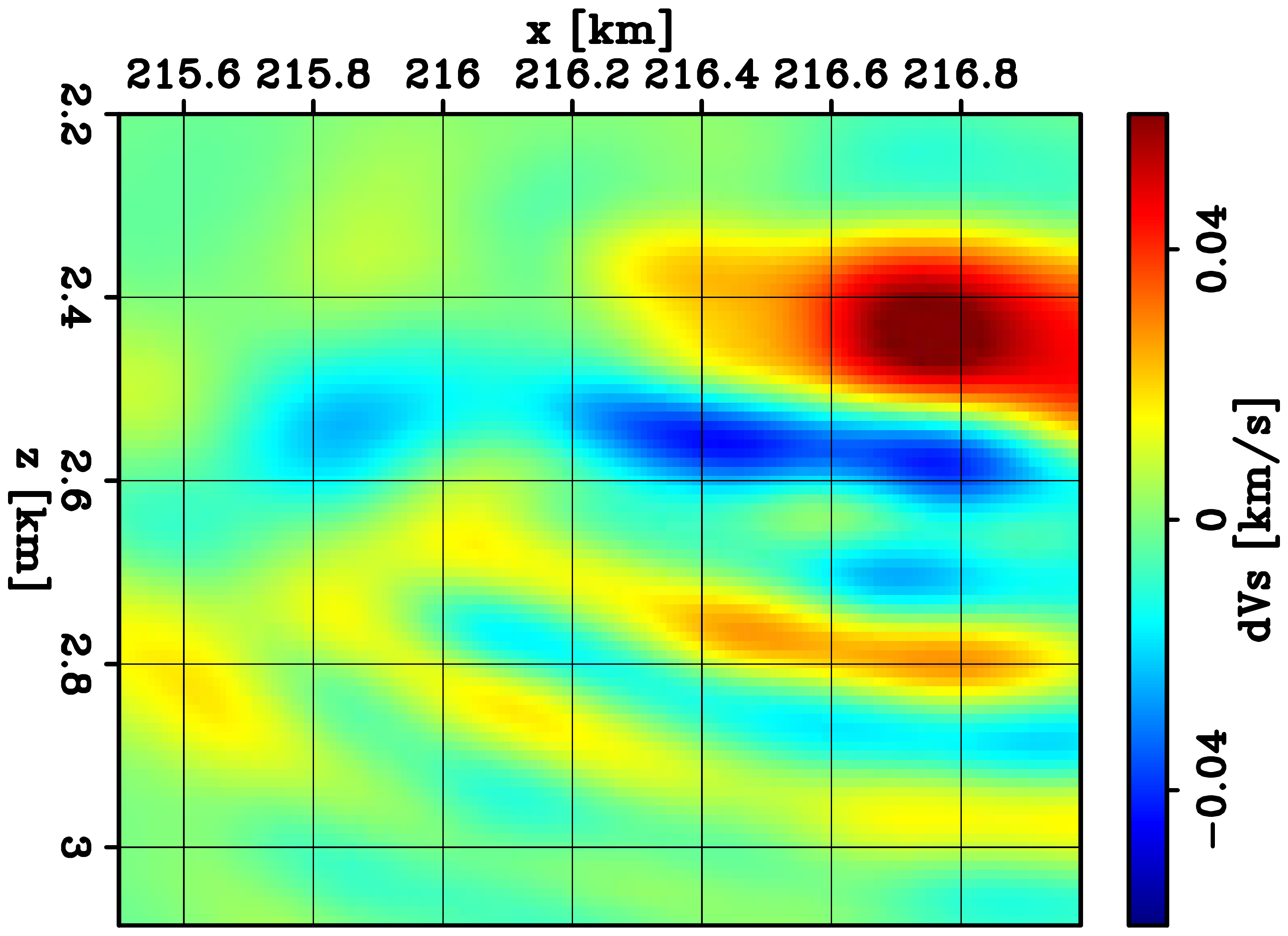}}
    \subfigure[]{\label{fig:CardamomTargetDiffElaVsY}\includegraphics[width=0.32\columnwidth]{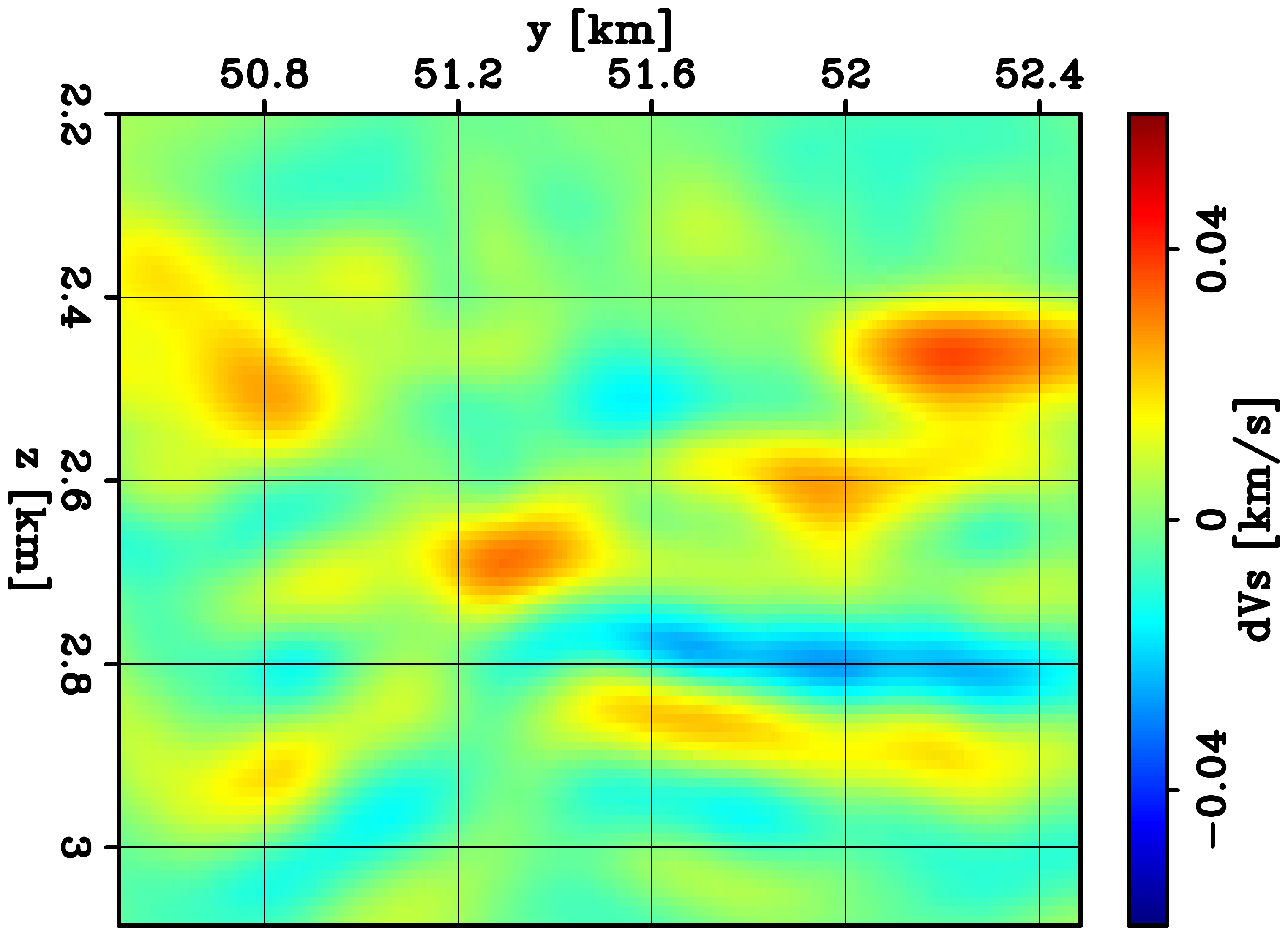}}
    
    \subfigure[]{\label{fig:CardamomTargetDiffElaRhoZ}\includegraphics[width=0.32\columnwidth]{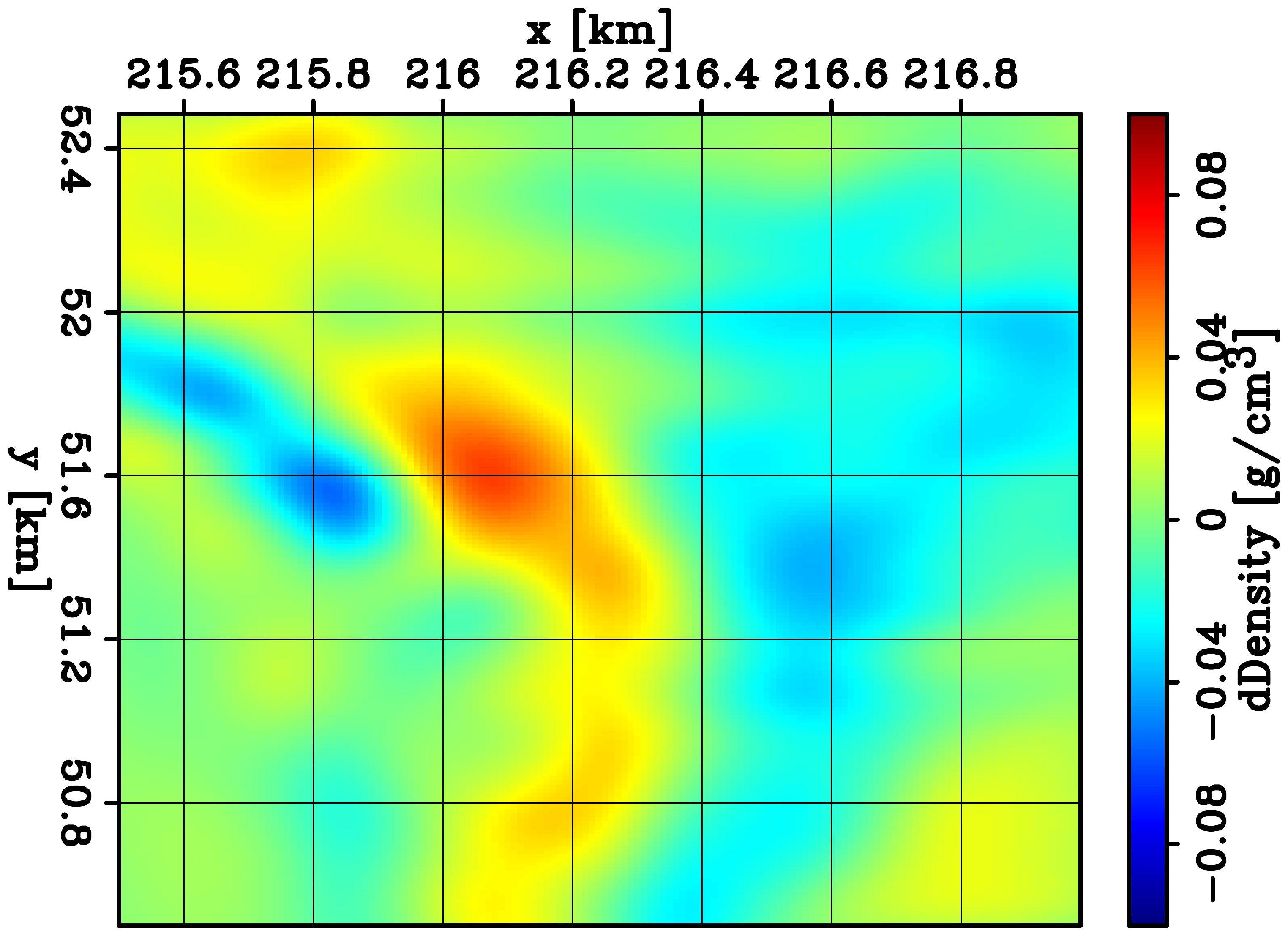}}
    \subfigure[]{\label{fig:CardamomTargetDiffElaRhoX}\includegraphics[width=0.32\columnwidth]{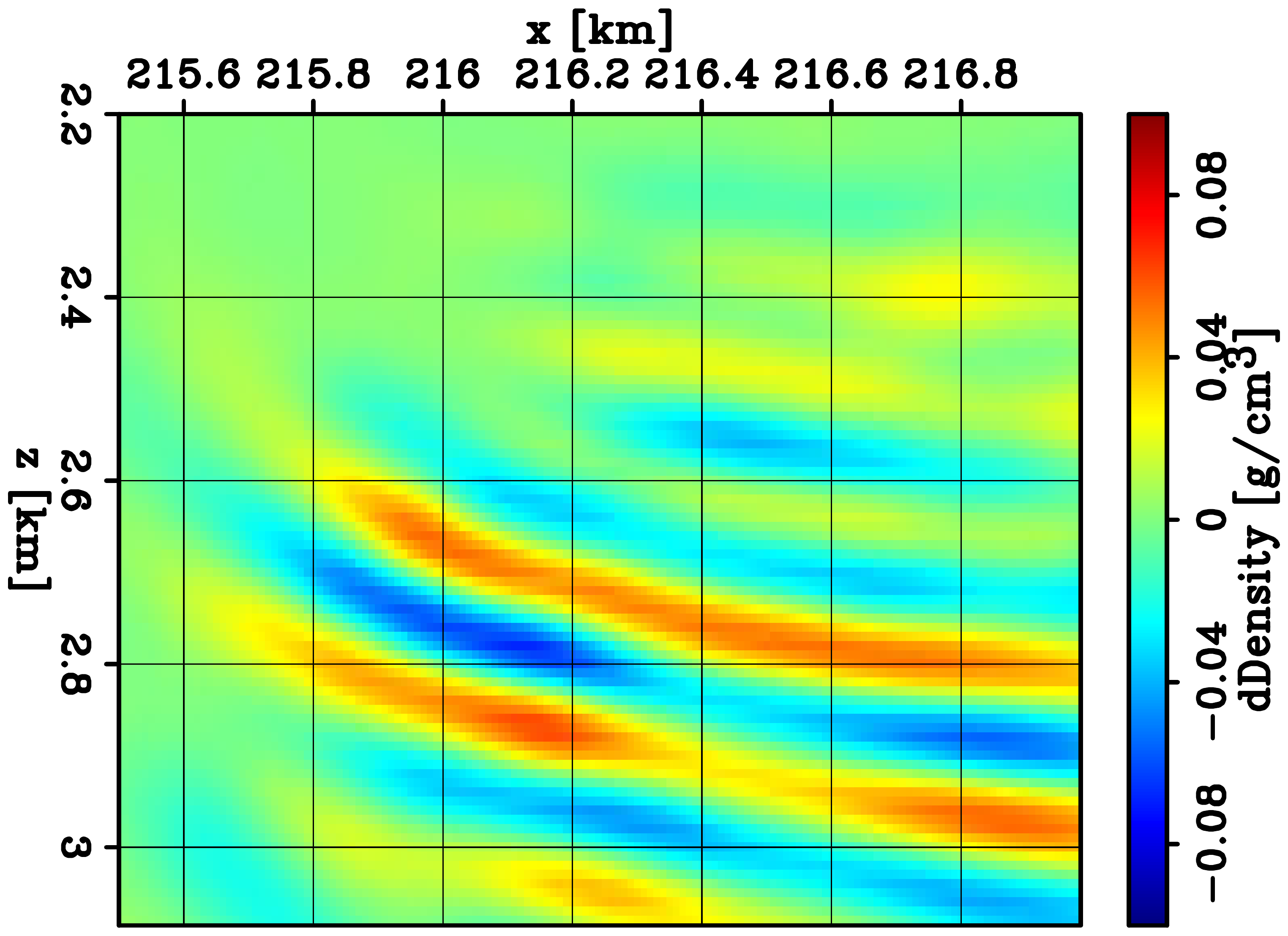}}
    \subfigure[]{\label{fig:CardamomTargetDiffElaRhoY}\includegraphics[width=0.32\columnwidth]{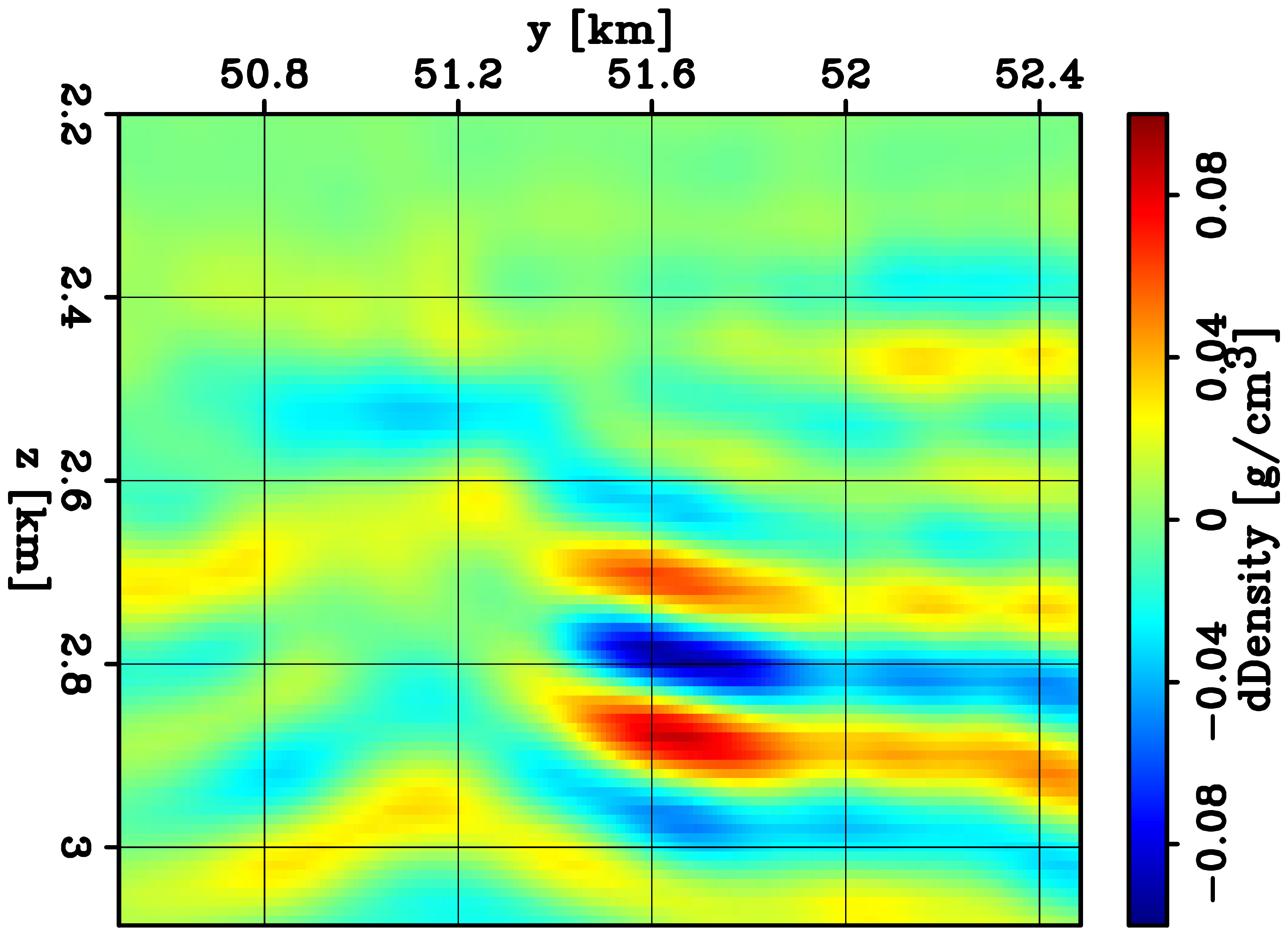}}
    
    \caption{Elastic parameter difference between the final and the initial elastic FWI model. The same slices from Figure~\ref{fig:CardamomTargetFinalEla} are shown in these panels.}
    \label{fig:CardamomTargetDiffEla}
\end{figure}

To highlight how the elastic FWI process updates the three parameters, we plot the difference between the inverted and the initial models in Figure~\ref{fig:CardamomTargetDiffEla}. As expected, an evident decrease at the target's position is observed within the P-wave and density parameters. On the other hand, the S-wave model does not present such a reduction in the same position and displays slightly different structures than the other two parameters. Moreover, the updates in the S-wave parameter are an order of magnitude smaller compared with the P-wave velocity. This behavior could be due to the limited surface offsets considered in this field-data example. However, the different structures and sensitivity provide evidence of the ability of the process not to introduce cross-talk artifacts during the elastic FWI process.

\begin{figure}[t]
    \centering
    \subfigure[]{\label{fig:CardamomTargetVpVsElaVpZ}\includegraphics[width=0.32\columnwidth]{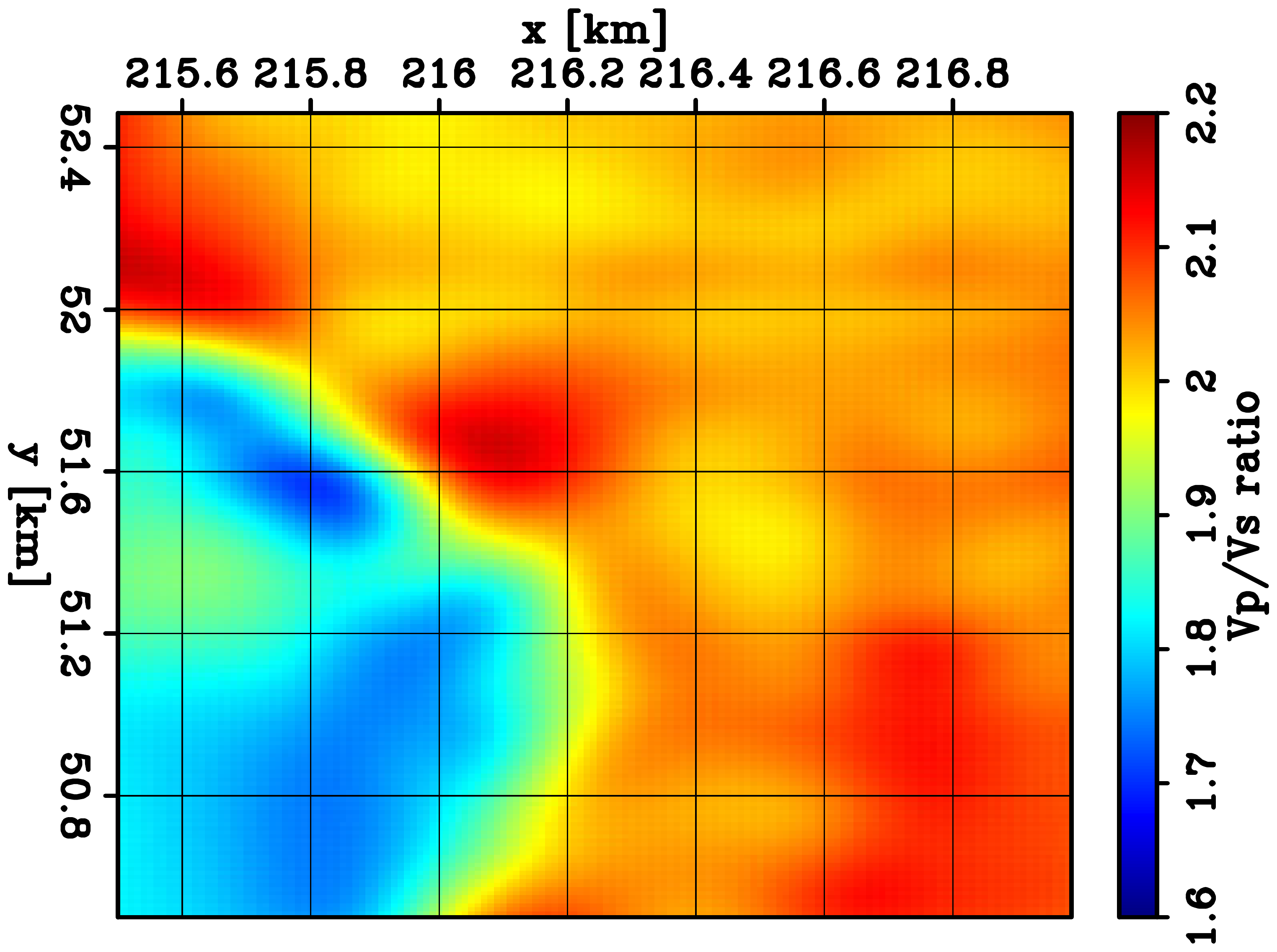}}
    \subfigure[]{\label{fig:CardamomTargetVpVsElaVpX}\includegraphics[width=0.32\columnwidth]{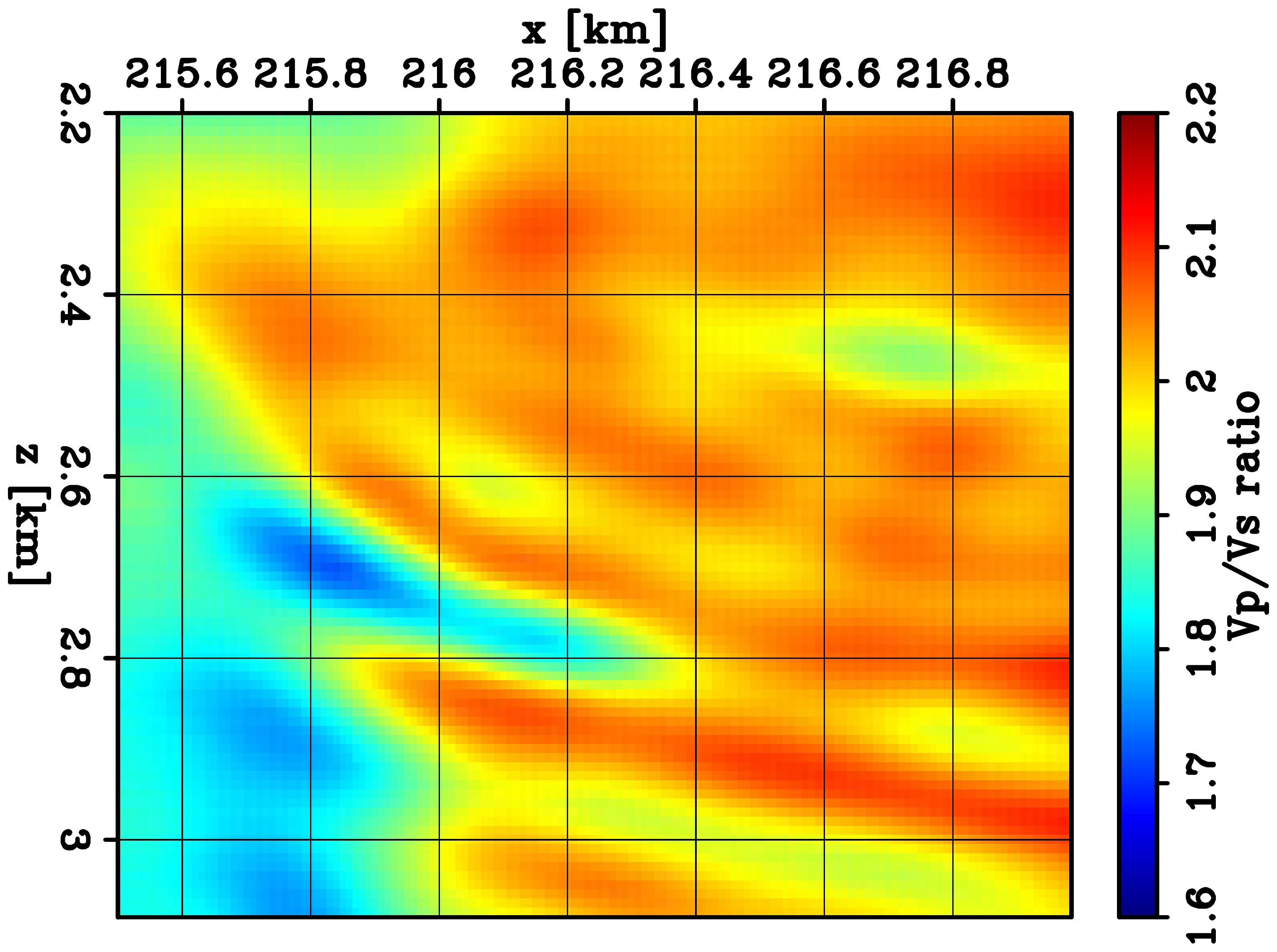}}
    \subfigure[]{\label{fig:CardamomTargetVpVsElaVpY}\includegraphics[width=0.32\columnwidth]{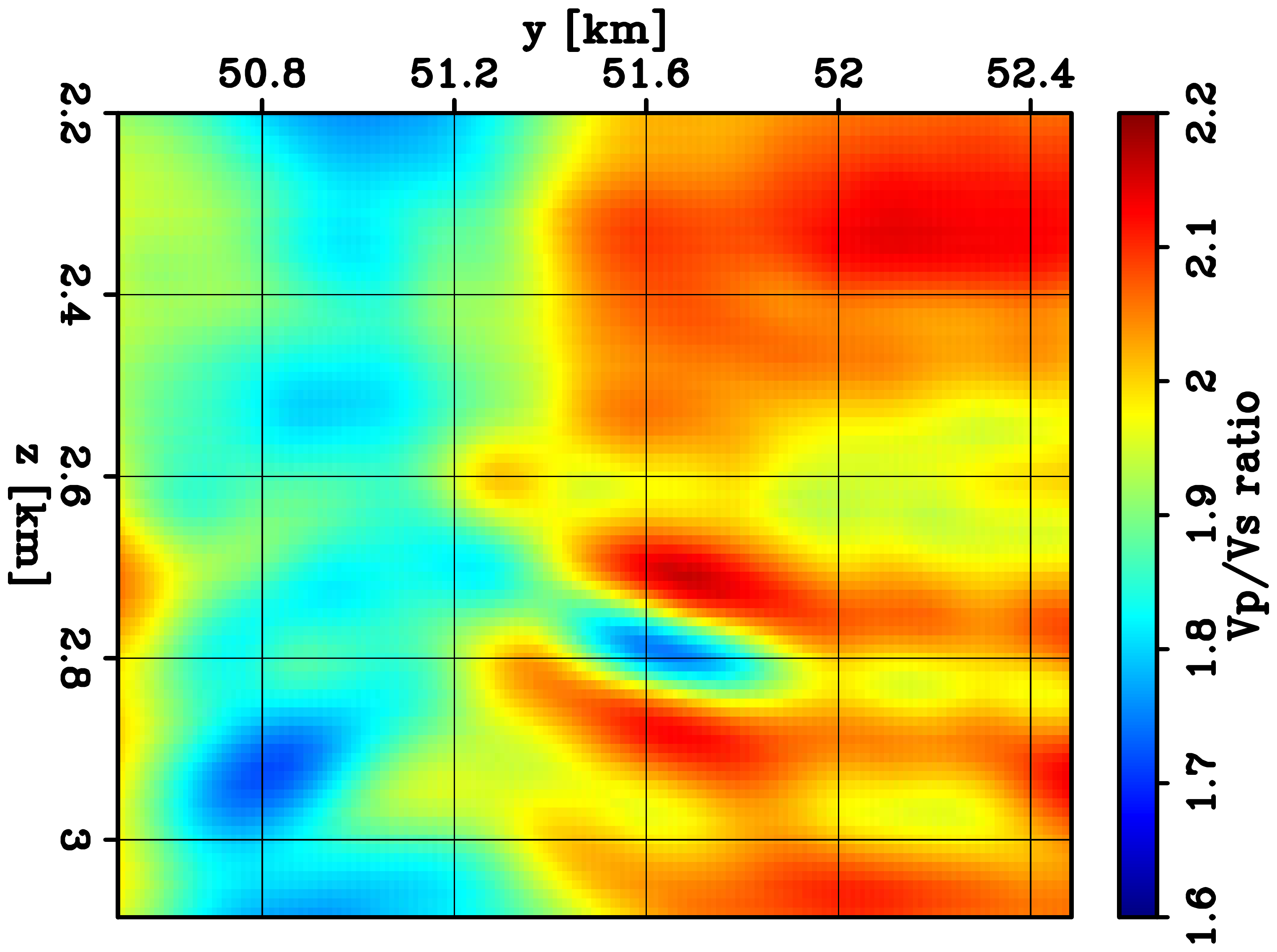}}
    
    \subfigure[]{\label{fig:CardamomTargetAIElaVpZ}\includegraphics[width=0.32\columnwidth]{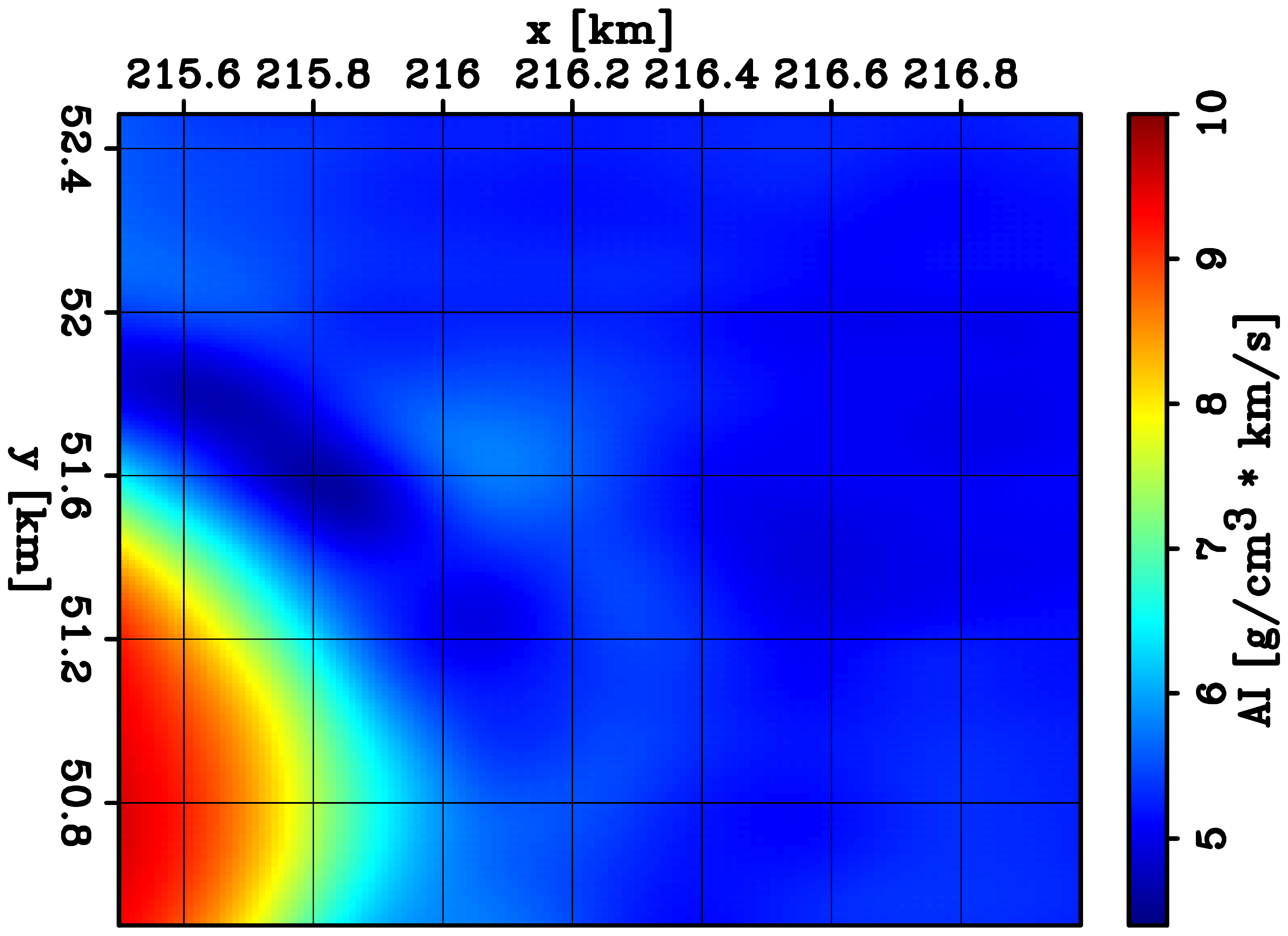}}
    \subfigure[]{\label{fig:CardamomTargetAIElaVpX}\includegraphics[width=0.32\columnwidth]{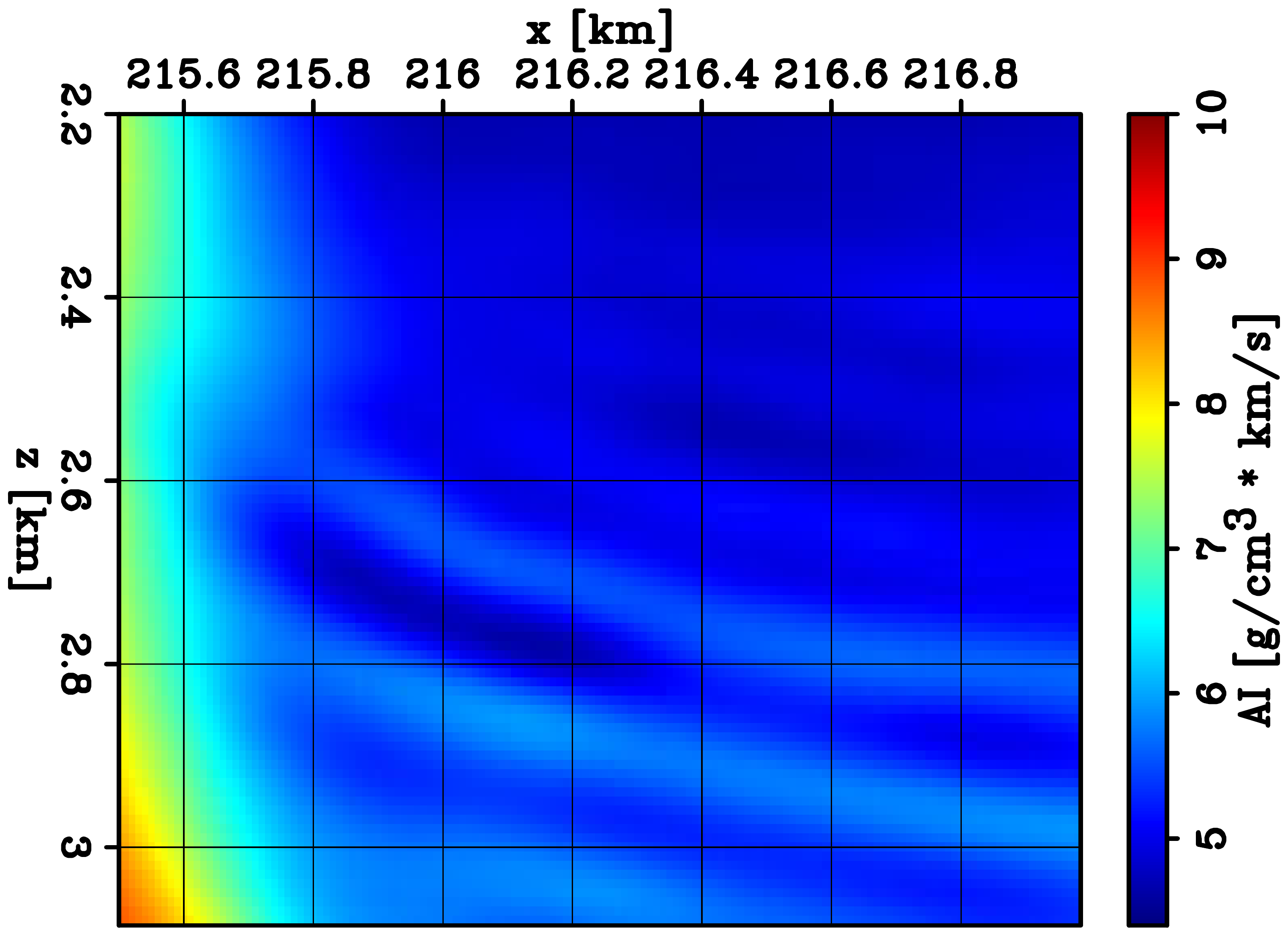}}
    \subfigure[]{\label{fig:CardamomTargetAIElaVpY}\includegraphics[width=0.32\columnwidth]{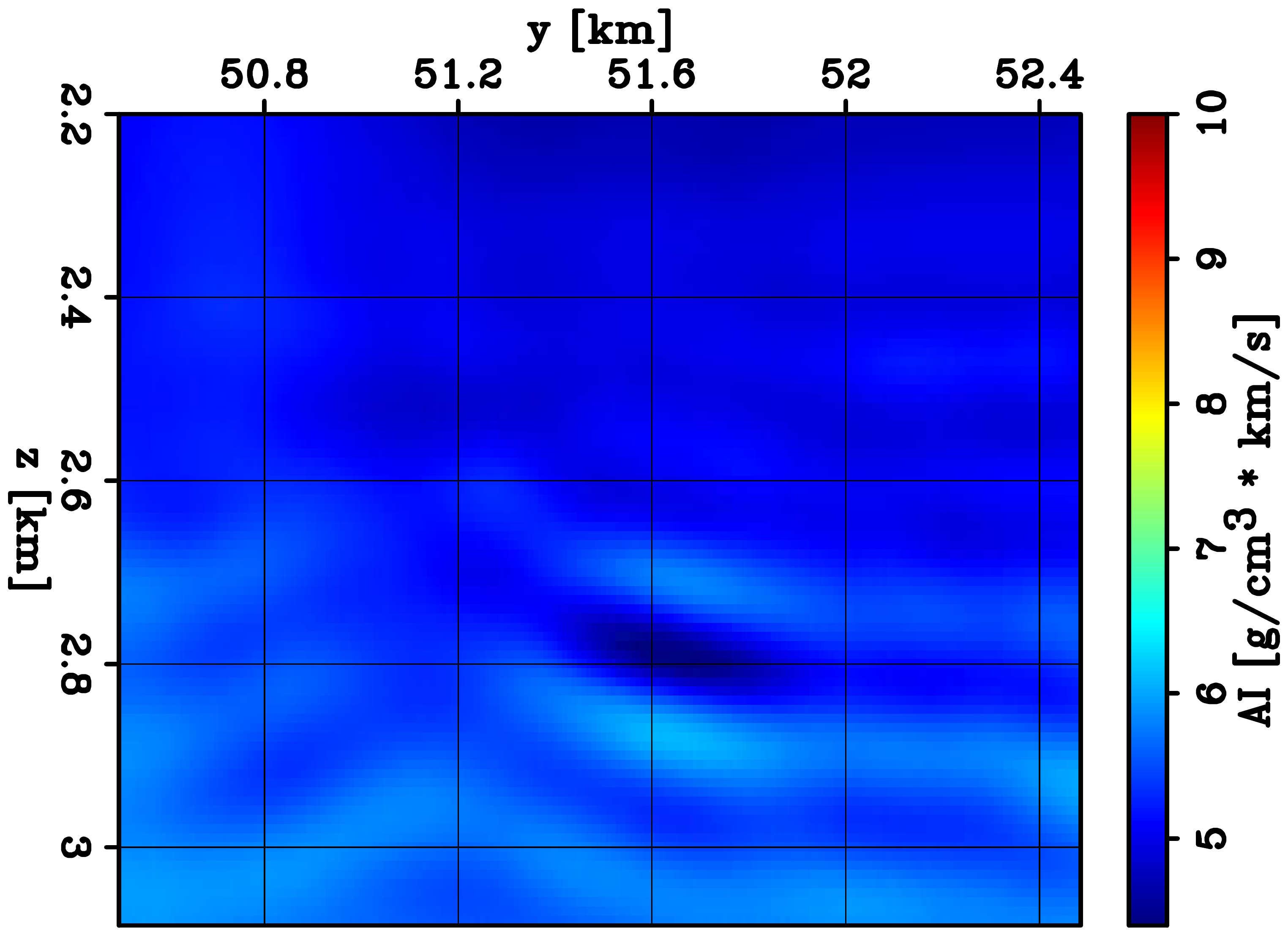}}
    
    \caption{Rock physics attributes computed using the final elastic FWI model. The top and bottom rows display the Vp/Vs ratio and the AI, respectively. The same slices from Figure~\ref{fig:CardamomTargetFinalEla} are shown in these panels.}
    \label{fig:CardamomTargetPropEla}
\end{figure}

\begin{figure}[t]
    \centering
    \subfigure[]{\label{fig:CardamomTargetVpVsProf}\includegraphics[width=0.32\columnwidth]{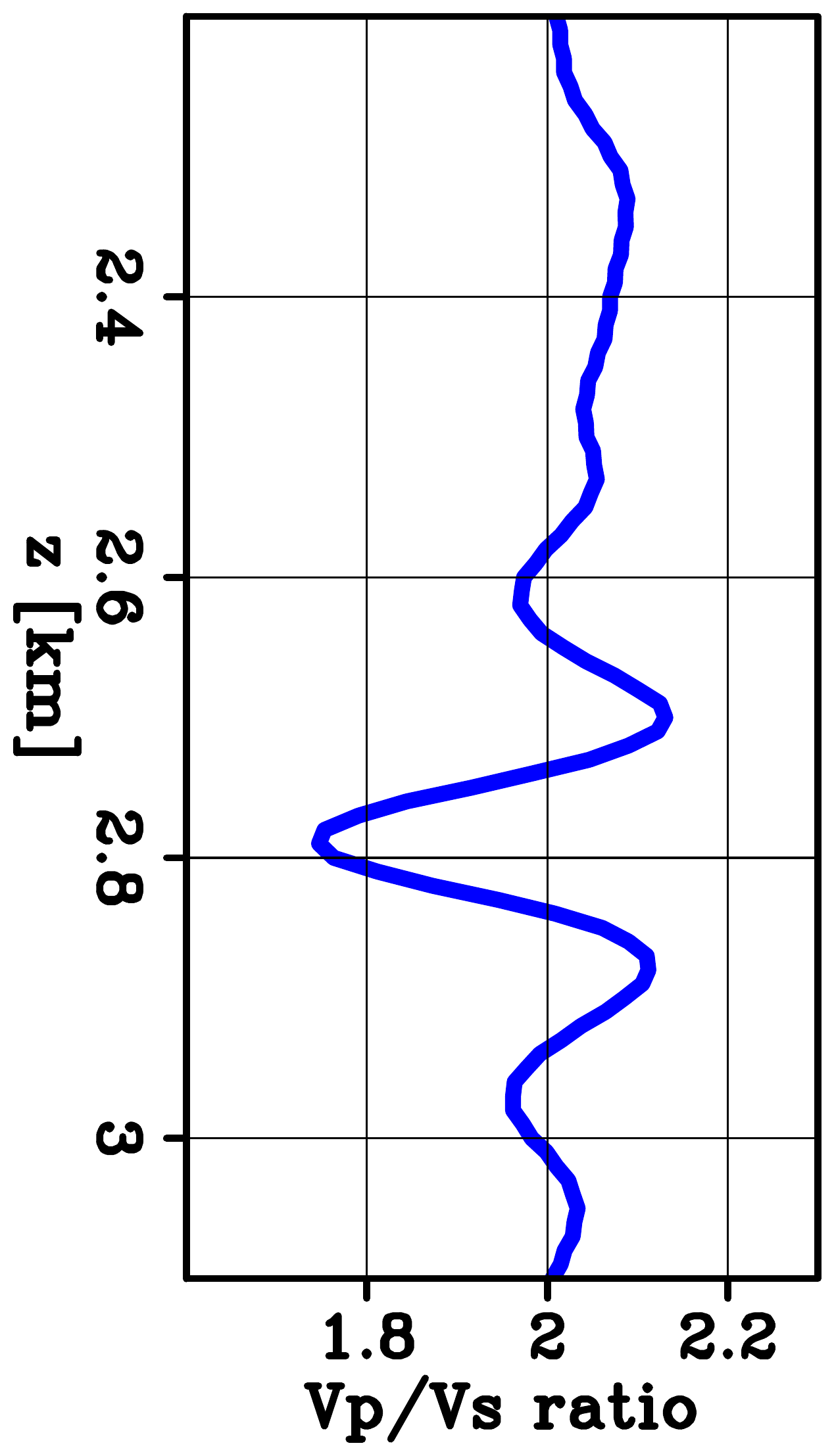}}
    \subfigure[]{\label{fig:CardamomTargetAIProf}\includegraphics[width=0.32\columnwidth]{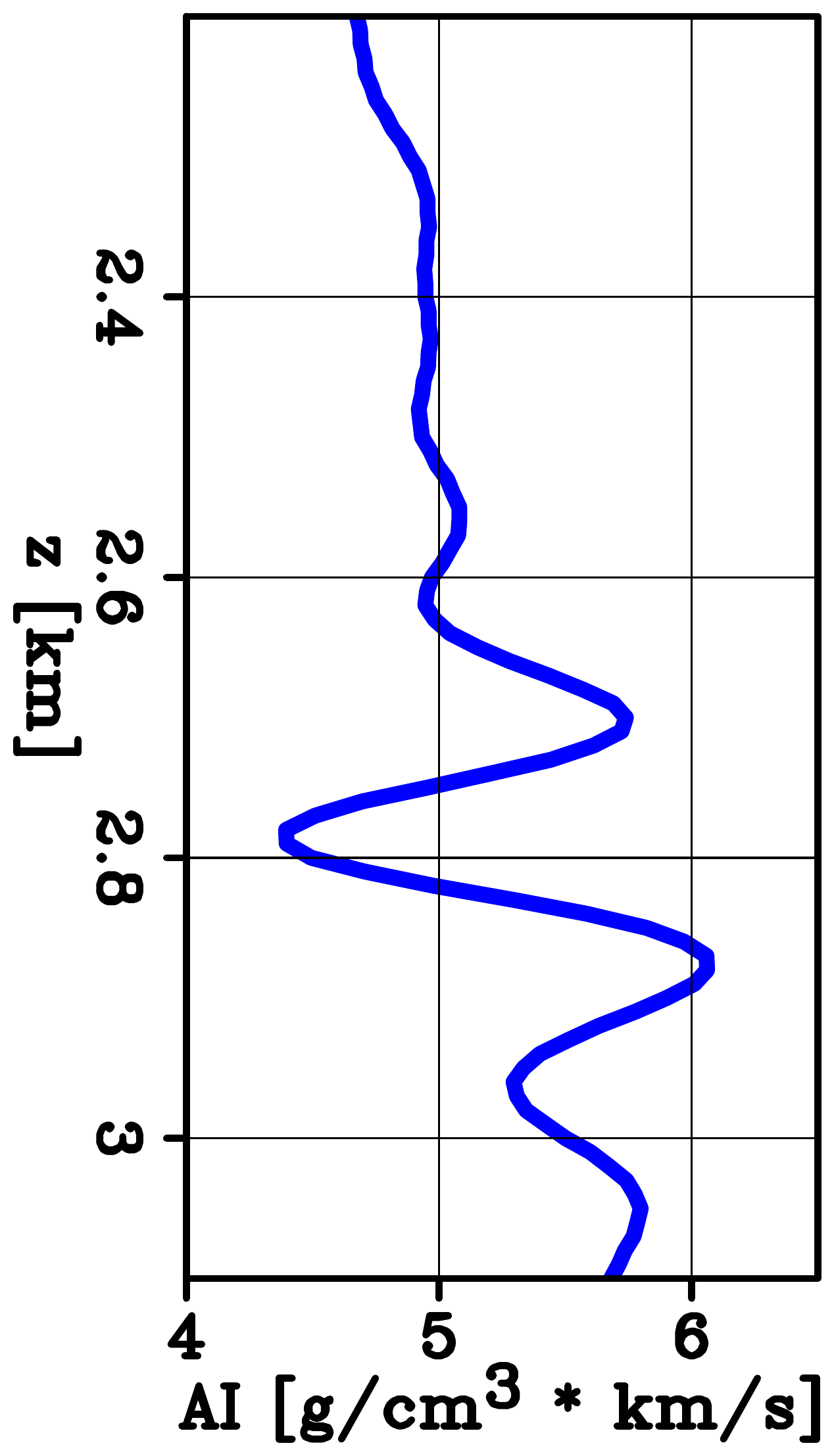}}

    \caption{Vertical profiles of the rock physics attributes shown in Figure~\ref{fig:CardamomTargetPropEla}. (a) Vp/Vs ratio profile. (b) AI profile. Both profiles are extracted at $x=216.1$ km and $y=51.6$ km.}
    \label{fig:CardamomTargetPropElaProfiles}
\end{figure}

Using the elastic parameters obtained by the target-oriented inversion, we compute two standard rock physics attributes, namely, the Vp/Vs ratio and the acoustic impedance (AI) (Figure~\ref{fig:CardamomTargetPropEla}). The average Vp/Vs ratio and AI of the low-velocity and low-density anomaly are approximately $1.7$ and $4.8$ $\text{g}/\text{cm}^3 * \text{km}/\text{s}$, respectively (Figure~\ref{fig:CardamomTargetPropElaProfiles}). These values are consistent with a potential gas-charged sand whose Vp/Vs ratio is commonly below 1.8 and its AI is ranging from 2.5 to 7 $\text{g}/\text{cm}^3 * \text{km}/\text{s}$ \cite[]{gardner1968velocity,odegaard2003interpretation}. In addition, the rock-physical parameters above the gas-bearing sand present values in agreement with a high-shale-content formation, corresponding to the necessary cap rock of this reservoir.  

This field application of the proposed target-oriented elastic FWI workflow demonstrates its ability to estimate a potential prospect's elastic properties correctly. The applied workflow retrieves the elastic parameters of a possibly gas-charged reservoir located on the flank of the salt diapir. Moreover, the method's ability to limit the computational domain to only the target area allows applying a wave-equation estimation method such as FWI. Using an elastic FWI on the entire $100$ $\text{km}^2$ domain is a challenge given the computational cost of solving the elastic wave equation. Using the same resources described within the Marmousi2 test, the elapsed time for performing a single iteration is $133$ minutes. We estimate a computational speed-up factor ranging from $500$ to $1000$ between the original and the target-oriented inversions for this specific example. This estimate does not consider the possibility of limiting the surface offsets during the elastic inversion process of the surface data. However, even considering this offset limitation, a considerable speed-up factor would be achieved by the target-oriented workflow.

\section*{Conclusions}
We describe a target-oriented elastic FWI methodology that greatly reduces the computational cost associated with the inversion of elastic pressure waves generated by a target area. Our method is based on the usage of the extended-image space to redatum the surface data to a new survey geometry placed in proximity of the identified target. To form the needed image, we solve an extended linearized waveform inversion problem and employ the resulting image to identify the potential prospects. By selecting a target area of the image, we model reflection data using an acquisition geometry placed in the subsurface. We then apply an elastic FWI procedure to estimate the target's parameters; a process that employs the full bandwidth of the original data. 

We highlight the properties of our method and apply it to synthetic data generated using the Marmousi2 model. In this test, we compare the elastic parameters estimated by applying an elastic FWI workflow on the surface and reconstructed datasets. The two inversions can correctly retrieve the elastic parameters and the extent of a gas-bearing reservoir. We also present a field application of our target-oriented elastic FWI workflow in which we employ pressure OBN data recorded in the Gulf of Mexico to retrieve the elastic properties of a potential subsurface prospect. Using an FWI model-building step to form an accurate subsurface 3D RTM volume, we identified a potential prospect positioned on the salt diapir's flank using the subsurface RTM images. On this target, we applied the proposed elastic waveform methodology to estimate its material proprieties. We show that the estimated rock-physics parameters by our method are consistent with the presence of a gas-bearing sand reservoir. This field-data application demonstrates the proposed method's potential to retrieve the elastic parameters. Our process enables the application of an elastic FWI estimation only to a limited portion of the subsurface, which results in a drastic speed-up factor compared to the inversion of the original surface dataset.

\section{ACKNOWLEDGMENTS}

We would like to thank the Stanford Exploration Project affiliate companies for their financial support. We also thank Mark Meadows for the useful suggestions made during the development stages of the method and Thomas Cullison for the help in the implementation of the 3D GPU-based finite-difference operators used in the reported field application. Finally, we would like to acknowledge Shell Exploration and Production Company for permission to show the obtained results and providing the described field dataset.

\newpage

\bibliographystyle{seg}  
\bibliography{ettore}

\newpage

\clearpage

\clearpage

\clearpage

\clearpage

\clearpage

\clearpage

\clearpage

\clearpage

\clearpage

\clearpage

\clearpage

\clearpage

\clearpage

\clearpage

\clearpage

\clearpage

\clearpage

\clearpage

\clearpage

\clearpage

\clearpage

\clearpage

\clearpage

\clearpage

\clearpage

\clearpage

\clearpage

\clearpage

\clearpage

\clearpage

\clearpage

\clearpage

\clearpage

\clearpage

\clearpage

\clearpage

\clearpage

\clearpage

\clearpage

\clearpage

\clearpage

\clearpage

\clearpage 

\clearpage

\clearpage

\clearpage

\clearpage

\clearpage

\clearpage

\clearpage

\end{document}